%% file: main.tex
\documentclass[manuscript,screen]{acmart/acmart}

\AtBeginDocument{%
  \providecommand\BibTeX{{%
    \normalfont B\kern-0.5em{\scshape i\kern-0.25em b}\kern-0.8em\TeX}}}

\setcopyright{rightsretained}

\copyrightyear{2026}
\acmYear{2026}
\acmDOI{XXXXXXX.XXXXXXX}

\usepackage{acronym}
\usepackage{listings}
\usepackage{placeins}
\usepackage{multirow,diagbox}

\usepackage{makecell}
\usepackage{siunitx}
\usepackage{vadl}

\usepackage{hyperref}
\usepackage{breakurl}
\usepackage{doi}
\usepackage{pgfplots}
\usepackage{pgfplotstable}
\usepackage{subcaption}
\usepackage{algorithm}
\usepackage{algpseudocode}
\usetikzlibrary{shapes, fit, positioning, patterns}

\tikzstyle{generator} = [rectangle, rounded corners, minimum height=1cm,text centered, draw=black, fill=yellow!30, text width=2cm]
\tikzstyle{ir} = [ellipse, rounded corners, minimum height=1cm,text centered, draw=black, text width=1.7cm]
\tikzstyle{ir_step} = [rectangle, rounded corners, minimum height=1cm,text centered, draw=black, text width=2.3cm]

\tikzstyle{internal_component} = [rectangle, rounded corners, minimum height=1cm,text centered, draw=black, text width=2.3cm]
\tikzstyle{external_component} = [rectangle, rounded corners, minimum height=1cm,text centered, draw=black, fill=red!30, text width=2.3cm]

\tikzstyle{file} = [rectangle, minimum height=2cm,text centered, draw=black, text width=1.5cm]
\tikzstyle{tool} = [rectangle, rounded corners, minimum height=1cm,text centered, draw=black, text width=1.5cm]

\tikzstyle{writenode} = [ellipse, text centered, draw=red]
\tikzstyle{availablenode} = [ellipse, text centered, draw=green]
\tikzstyle{operationnode} = [ellipse, text centered, draw=black]
\tikzstyle{readnode} = [ellipse, text centered, draw=blue]

\tikzstyle{arrow} = [thick,->,>=stealth]

\newcommand{\mytikzmark}[1]{%
  \tikz[overlay,remember picture,baseline] \coordinate (#1) at (0,0) {};}

\newcommand{\encircle}[3]{%
  \draw[#3, thick] ([yshift=0.075cm, xshift=0.75*#2]#1) ellipse (#2 and 0.15cm);
}
\newcommand{\isainstruction}[1]{\texttt{#1}}
\newcommand{\vadlresource}[1]{\texttt{#1}}
\newcommand{\vadltype}[1]{\texttt{#1}}
\newcommand{\miaspec}[1]{\texttt{#1}}

\newcommand{\highlight}[1]{\textbf{#1}}
\newcommand{\highlighttitle}[1]{\textbf{#1}}

\begin{document}

\title{The Vienna Architecture Description Language}

\author{Florian Freitag}
\email{florian.freitag@tuwien.ac.at}
\orcid{0009-0000-5621-3476}
\author{Johannes Gernedl}
\email{johannes.gernedl@student.tuwien.ac.at}
\orcid{0009-0001-3432-8456}
\author{Johannes Gisy}
\email{johannes.gisy@tuwien.ac.at}
\orcid{0009-0006-6759-7093}
\author{Linus Halder}
\email{linus.halder@student.tuwien.ac.at}
\orcid{0000-0002-1871-7347}
\author{Nathalie Herzog}
\email{nathalie.herzog@proton.me}
\orcid{0009-0000-1339-0325}
\author{Simon Himmelbauer}
\email{simon@himmelbauer.net}
\orcid{0009-0002-9727-654X}
\author{Julian Hirschler}
\email{julian.hirschler@student.tuwien.ac.at}
\orcid{0009-0002-9687-9170}
\author{Christoph Hochrainer}
\email{christoph.hochrainer@tuwien.ac.at}
\orcid{0009-0009-4253-4821}
\author{Benedikt Huber}
\email{benedikt.huber@tuwien.ac.at}
\orcid{0009-0005-7059-3555}
\author{Benjamin Kasper}
\email{benjamin.kasper@student.tuwien.ac.at}
\orcid{0009-0005-8889-1104}
\author{Benjamin Komar}
\email{vadl@tuwien.ac.at}
\orcid{0009-0003-2630-8708}
\author{Florian Malicky}
\email{florian.malicky@student.tuwien.ac.at}
\orcid{0009-0004-6025-7366}
\author{Niklas Mischkulnig}
\email{niklas@mischkulnig.de}
\orcid{0009-0004-7203-8283}
\author{Michael Nestler}
\email{michael.nestler@yahoo.com}
\orcid{0009-0001-2469-6183}
\affiliation{%
  \institution{TU Wien}
  \city{Vienna}
  \country{Austria}
}
\author{Philipp Paulweber}
\authornote{Research done at TU Wien}
\email{ppaulweber@fiskaly.com}
\orcid{0000-0001-9954-4881}
\affiliation{%
  \institution{fiskaly GmbH}
  \city{Vienna}
  \country{Austria}
}
\author{Kevin Per}
\email{kevin.per@protonmail.com}
\orcid{0009-0009-9811-3583}
\author{Matthias Raschhofer}
\email{matthias.raschhofer@student.tuwien.ac.at}
\orcid{0009-0006-0445-3738}
\author{Jakob Rathbauer}
\email{jakob.rathbauer@tuwien.ac.at}
\orcid{0009-0004-5041-3791}
\author{Alexander Ripar}
\email{alexander.ripar@student.tuwien.ac.at}
\orcid{0009-0009-4664-9837}
\author{Jonas Rudolf}
\email{jonas.rudolph@student.tuwien.ac.at}
\orcid{0009-0001-1272-8272}
\author{Kevin Schlosser}
\email{kevin.schlosser@student.tuwien.ac.at}
\orcid{0009-0009-2504-7407}
\author{Tobias Schwarzinger}
\email{tobias.schwarzinger@tuwien.ac.at}
\orcid{0009-0003-1433-2049}
\author{Daniel Seifert}
\email{daniel.seifert@student.tuwien.ac.at}
\orcid{0009-0003-1910-5910}
\author{Elias Wazek}
\email{vadl@tuwien.ac.at}
\orcid{0009-0000-3645-3034}
\author{Tobias Weiss}
\email{vadl@tuwien.ac.at}
\orcid{0009-0002-2235-4988}
\author{Johannes Zottele}
\email{johannes.zottele@tuwien.ac.at}
\orcid{0009-0001-5328-9181}
\author{Andreas Krall}
\email{andi@complang.tuwien.ac.at}
\orcid{0009-0002-7668-6259}
\affiliation{%
  \institution{TU Wien}
  \city{Vienna}
  \country{Austria}
}

\renewcommand{\shortauthors}{Freitag et al.}

\input{src-abstract} %% andi, philipp

\begin{CCSXML}
<ccs2012>
<concept>
<concept_id>10011007.10011006.10011060.10011062</concept_id>
<concept_desc>Software and its engineering~Architecture description languages</concept_desc>
<concept_significance>500</concept_significance>
</concept>
<concept>
<concept_id>10011007.10011006.10011060.10011062</concept_id>
<concept_desc>Software and its engineering~Architecture description languages</concept_desc>
<concept_significance>500</concept_significance>
</concept>
<concept>
<concept_id>10011007.10011006.10011041.10011043</concept_id>
<concept_desc>Software and its engineering~Retargetable compilers</concept_desc>
<concept_significance>500</concept_significance>
</concept>
<concept>
<concept_id>10011007.10010940.10010971.10010972.10010978</concept_id>
<concept_desc>Software and its engineering~Simulator / interpreter</concept_desc>
<concept_significance>500</concept_significance>
</concept>
<concept>
<concept_id>10010583.10010682.10010689</concept_id>
<concept_desc>Hardware~Hardware description languages and compilation</concept_desc>
<concept_significance>500</concept_significance>
</concept>
</ccs2012>
\end{CCSXML}

\ccsdesc[500]{Software and its engineering~Architecture description languages}
\ccsdesc[500]{Software and its engineering~Retargetable compilers}
\ccsdesc[500]{Software and its engineering~Simulator / interpreter}
\ccsdesc[500]{Hardware~Hardware description languages and compilation}%%

\keywords{processor description language, compiler generator, assembler generator, simulator generator, hardware generator}

\maketitle

\input{src-introduction} %% andi, philipp

\input{src-background} %% benedikt

\input{src-introlanguage} %% andi
\input{src-isasect} %% andi, philipp
\input{src-miasect} %% philipp, tobias
\input{src-abisect} %% christoph
\input{src-umesect} %% andi
\input{src-asmsect} %% tobias
\input{src-cpusect} %% johannes zottele
\input{src-lang_evaluation} %% andi, flo

\input{src-orig_implementation} %% philipp
\input{src-orig_impl_parser} %% philipp
\input{src-vir} %% philipp

\input{src-orig_compgen} %% christoph
\input{src-orig_asmgen} %% tobias
\input{src-issgen} %% andi

\input{src-orig_miasynth} %% tobias, linus
\input{src-orig_casgen} %% niklas, simon
\input{src-orig_hdlgen} %% tobias
\input{src-orig_evalcomp} %% christoph, kevin p, tobias w, johannes g
\input{src-orig_evalass} %% tobias s, benjamin
\input{src-evaliss} %% andi, niklas, simon
\input{src-evalcas} %% niklas, simon

\input{src-orig_evalhdl} %% tobias

\input{src-open_implementation} %% andi
\input{src-open_frontend} %% flo
\input{src-open_lsp} %% jakob, flo
\input{src-viam} %% johannes

\input{src-open_compgen} %% kevin, johannes g, tobias, alex
\input{src-open_asmgen} %% benjamin, tobias

\input{src-qemugen} %% johannes, florian m
\input{src-open_miasynth} %% linus, matthias, tobias
\input{src-open_casgen} %% andi, kevin s
\input{src-open_hdlgen} %% linus, matthias, tobias s
\input{src-open_cosim} %% benni
\input{src-open_testgen} %% johannes
\input{src-evalfrontend} %% florian f
\input{src-open_evalcomp} %% kevin, tobias w, johannes g
\input{src-open_evalass} %% benjamin
\input{src-evalqemu} %% johannes z, florian m
\input{src-open_evalhdl} %% linus

\input{src-comparison} %% andi

\input{src-casestudy} %% andi
\input{src-relwork-pdl} %% andi
\input{src-relwork} %% benedikt

\input{src-futwork} %% all, philipp

\input{src-conclusion} %% andi

\begin{acks}
Part of this work was supported by a grant from Huawei.
Hermann Sch{\"u}tzenh{\"o}fer developed the first version of the simulator generator \cite{schutzenhofer2020cycle}.
Alexander Graf developed the first version of the compiler generator \cite{graf2021compiler}.
Hristo Mihaylov developed the \ac{DTC} simulator generator \cite{Mihaylov2023Optimized}.
\end{acks}

\bibliographystyle{acmart/ACM-Reference-Format}
\bibliography{references}

\appendix
\section*{Appendices}

\input{src-app_float} %% andi, tobias
\input{src-app_jalr}  %% benjamin
\input{src-app_riscv} %% andi

\section{List of Abbreviations}
\begin{acronym}[VHDL] % Replace VHDL with your longest acronym
\input{src-acronyms}
\end{acronym}

\end{document}

%% file: src-abstract.tex
\begin{abstract}
The Vienna Architecture Description Language (VADL) is a powerful processor description language (PDL) that enables the concise specification of processor architectures.
From a single VADL processor specification, the VADL compiler can automatically generate a wide range of artifacts required for rapid design space exploration.
These include assemblers, compilers, linkers, functional and cycle-accurate instruction set simulators, synthesizable hardware description language (HDL) specifications, as well as test cases and documentation.

A distinctive feature of VADL is the separation of the instruction set architecture (ISA) specification from the microarchitecture (MiA) specification.
This decoupling gives users the flexibility to combine various ISAs with different MiAs, offering a versatile approach to processor design.
In contrast to existing PDLs, VADL's MiA specification operates at a higher level of abstraction, enhancing the clarity and simplicity of the design process.
Notably, using a single ISA specification, VADL streamlines compiler generation and maintenance by eliminating the need for intricate compiler-specific knowledge.

Because the original VADL implementation is subject to restrictive licensing, the open-source implementation OpenVADL was initiated.
This article introduces VADL, compares the original implementation with the ongoing OpenVADL development, describes the generator techniques in detail, and demonstrates both the expressiveness of the language and the performance of the generators in an empirical evaluation.
The evaluation highlights the conciseness of VADL and the efficiency of the generated artifacts.
Finally, the viability of OpenVADL for design space exploration is demonstrated in a case study on CRC32 checksum computation for the RISC-V ISA.
\end{abstract}

%% file: src-introduction.tex
\section{Introduction}\label{sec:introduction}

The \ac{VADL} is a %processor description language (
\ac{PDL}.
The name is inspired by the \ac{VDL} which was developed 50 years ago for the formal specification of the programming language PL/I using operational semantics~\cite{wegnerVDL72}.

Why do we need another \ac{PDL}?
The development of backends for compilers like LLVM, GCC, or \ac{JIT} compilers is cumbersome and error-prone.
The specifications are huge and very difficult to comprehend even for experienced compiler developers. Thus, we want to get all these distinct compilers automatically generated based on a single concise specification. Additionally, we also want to automatically produce an assembler, a debugger, a disassembler, an \ac{ISS}, and a linker.
To the best of our knowledge, currently there is no \ac{PDL} or compiler backend specification language that achieves this.
Furthermore, we want to do computer architecture research and teaching on a higher level of abstraction compared to what we believe other existing \acp{PDL} or \acp{HDL} currently offer.
The low-level microarchitecture specification means provided by current \acp{PDL} or \acp{HDL} lead to huge, unmaintainable specifications where the instruction's semantics and the microarchitecture are intermingled.
It should be possible that hardware, \acp{CAS} and instruction schedulers for compilers are also automatically generated from such a high-level specification.
We want to push forward the research in the area of \acp{PDL}.
For all these reasons, we designed \ac{VADL} and developed the necessary generator technologies.

\ac{VADL} permits the complete, declarative specification of a processor architecture. In contrast to prose descriptions in traditional instruction set manuals, a \ac{VADL} specification is unambiguous, explicit, and deterministic. This structure enables automated static checks and validation of critical architectural properties, such as ensuring unique instruction encodings or preventing resource write conflicts. Furthermore, \ac{VADL} codifies the functionality of a processor rather than its procedural execution: it defines instruction encodings rather than the logic of a concrete hardware decoder, and specifies instruction semantics rather than the mechanics of a specific simulator or hardware realization. Consequently, \ac{VADL} serves as a unified, high-level specification language for generating both the processor hardware and its corresponding software toolchain.

Furthermore, it is possible to control the behavior of generators which produce different artifacts from a processor specification. From a single concise \ac{VADL} processor specification, the \ac{VADL} compiler is able to automatically generate an assembler, a compiler, a linker, a functional \ac{ISS}, a \ac{CAS}, a synthesizable specification in an \ac{HDL}, test cases, and documentation. \ac{VADL} strictly separates the \ac{ISA} specification from the \ac{MiA} specification. While the \ac{ISA} specification is required by all generators, the \ac{MiA} specification is used by the \ac{HDL} and \ac{CAS} generators as well as for instruction scheduling in the compiler generator. An \ac{ISA} specification can be implemented by one or more \ac{MiA} specifications. Finally, the \ac{ABI} specification defines a programming model, providing the necessary operational boundaries for the compiler generator.

\ac{VADL} has been designed to enable concise, comprehensible specifications.
A novice should be able to understand a specification without prior knowledge of \ac{VADL}.
Redundant specifications are avoided.
\ac{VADL} is a safe language. It is strongly statically typed.
The language frontend and the artifact generators apply extensive consistency checks.
\ac{VADL} is a generator language. Executable specifications are not possible.

\subsection{Contribution}

The development of \ac{VADL} led to innovations across different domains.
Our main contributions are:

\begin{itemize}
    \item A concise and comprehensible processor description language
    \item An \ac{ISA}-independent \ac{MiA} specification language with a high level of abstraction
    \item \ac{ISA} to \ac{MiA} mapping by using inherent properties of the
          instruction's behavior for \ac{MiA} assignment % TODO: This seems unclear. What sorts of properties, and how do they influence/create the mapping?
    \item A simple pred-LL(k) parsable \emph{syntactical} \emph{pattern-based} macro system
    \item Syntactic type-safe higher-order macro templates % using \emph{models}
    \item Specification of assembly language by string expressions
    \item Assembler generation by automatic grammar inference through program inversion
    \item Compiler generation by automatic pattern inference from operational semantics specifications
    \item \ac{MiA} synthesis by reduction of the instruction's data flow graph
    \item \ac{MiA} hazard detection and optimization
\end{itemize}

See Section~\ref{sec:background} for a short explanation of the concepts listed here.
Additionally, we present a variety of smaller contributions particularly useful for our exploratory language design of \ac{VADL}:

\begin{itemize}
    \item Composable syntax types using \emph{record}s and type aliases
    \item Constraints on instruction encodings
    \item Register file aliases with different constraints
    \item Access functions for decoding and encoding of format fields
    \item Concise specification of user mode emulation
\end{itemize}

Additionally, we have initiated the development of \emph{OpenVADL}, the open-source implementation of \ac{VADL}.

\subsection{Outline}

Section~\ref{sec:background} gives some background information about the different domains touched upon in this article.
This section can be skipped by readers who already have a deep knowledge of compilers and computer architecture.
Section~\ref{sec:processor_description_language} presents the most important language elements of \ac{VADL} by examples.
Section~\ref{sec:orig_implementation} describes the implementation of the original \ac{VADL} compiler, whereas Section~\ref{sec:open_implementation} describes the OpenVADL implementation. Both sections present the intermediate representations and generators in detail, followed by a comprehensive qualitative and quantitative evaluation.
Section~\ref{sec:impl_comparison} highlights the main differences between the original \ac{VADL} and the OpenVADL implementations.
Section~\ref{sec:dse_casestudy} demonstrates the viability of OpenVADL for design space exploration.
Section~\ref{sec:related_work} compares \ac{VADL} and its implementation with related work.
Section~\ref{sec:future_work} discusses limitations and avenues for future research, focusing on potential enhancements and optimizations to both the language and its implementation.

%% file: src-background.tex
\section{Background}
\label{sec:background}

\subsection*{\ac{DSE}}

One of the most relevant applications for a \ac{PDL} is \ac{DSE} for \acp{ASIP}.
If it is possible to \emph{automatically} generate a set of
tools from an architecture description in a \ac{PDL}, the productivity of the
architecture design process can be improved by establishing short feedback
loops between design iterations.
The set of these tools should at least contain a compiler toolchain, a \ac{CAS}
and a synthesizable hardware model in an \ac{HDL}.
Additionally, it can also include a functional \ac{ISS}, test cases, documentation and other artifacts.
As described in~\cite{mishra2008processor}, this establishes two
feedback loops:
\begin{itemize}
  \item The compiler toolchain together with the \ac{CAS} can provide
        accurate performance information for a given software workload.
  \item The hardware model can provide information about the maximum
        clock frequency or chip area needed for the actual hardware
        implementation.
\end{itemize}

Due to stringent hardware constraints and the known workload, incorporating this information into the architectural design process is particularly advantageous for embedded systems and \ac{ASIP} development.

\subsection*{Compiler Toolchain}
\label{sec:background-toolchain}

\ac{VADL} enables the generation of the toolchain required to compile high-level source code (such as C) into machine-executable binaries. This toolchain primarily comprises the compiler, the assembler, and the linker.

\subsubsection*{\highlighttitle{Compiler}}

In general, a compiler translates a program written in one programming language
into a semantically equivalent program in another programming language.
In the most common case, and the case
relevant for \ac{VADL}, the compiler translates a program written in a
high-level and architecture \emph{independent} programming language into an
architecture \emph{dependent} assembly language.  An assembly language is a
direct textual representation of machine code. In other words, there is a direct
correspondence between machine instructions and assembler instructions.

Typically, high-level programming languages offer syntax constructs to describe
control flow and data structures in a manner easily understood by a
programmer. This abstraction helps to facilitate reasoning about the semantics
of an implemented algorithm and hides unnecessary details. It also helps in the
maintenance of software. An assembly language typically does not provide such
syntax features and is more difficult to read and maintain for a programmer.
The primary purpose of the compiler is to translate from the higher abstraction
level of the programming language to the lower abstraction of the assembly
language.  This translation is typically done in multiple passes. The compiler
can employ various optimization steps during translation to increase
the resulting program's performance or reduce its code size.

A \highlight{retargetable} compiler is designed to make it
easy to add support for a new target architecture.
The design of a retargetable compiler has target architecture \emph{independent} and target architecture \emph{dependent} components.
Target architecture independent components implement common transformations
and optimizations for all targets, e.g., dead code
elimination.
Target-dependent transformations, like register allocation,
are implemented such that the main algorithm is
target-independent but can be parameterized with target-specific data.

A retargetable compiler is often roughly subdivided into three parts.
\begin{itemize}
  \item The \highlight{frontend} is input language dependent. It typically
    parses and analyzes a high-level programming language as input.
    It abstracts each concrete programming language to a common
    \ac{IR}. This \ac{IR} is input language independent and it
    is used in the later stages of the compilation process.
    To support multiple high-level programming languages, the
    front end has to be able to abstract all of them to the common
    \ac{IR}.
  \item The \highlight{middleend} operates on the \ac{IR}. It performs
    transformations and optimizations that are common to all
    target architectures. It is independent of both the
    input programming language and the target architecture.
    In order to facilitate retargetability, as much functionality
    as possible should reside in the middleend.
  \item The \highlight{backend} emits code specific
    to the target architecture. It has specialized functionality
    for each supported target.
\end{itemize}

\subsubsection*{\highlighttitle{Parser}}

An important component of the frontend in a compiler is the parser. In the context
of \ac{VADL}, an important class of parsers is \highlight{predicated LL(*)} parsers. These
parsers read input from \emph{left-to-right} and derive the \emph{leftmost} non-terminal symbol,
hence \emph{LL}. The $*$ denotes arbitrary but finite lookahead. In LL(k), $k$ bounds the lookahead.
A special case of LL(k) is \highlight{LL(1)}, which is the most resource-efficient.

A \highlight{predicated} parser can also use syntactic predicates
to decide which derivation to apply. This allows the parser to resolve possible divergences,
and it can help to make the input grammar more readable and maintainable.

Also part of the frontend is the \highlight{macro system}. Macros are a well-established way to
realize language extensions. In general, a macro is a user-defined procedure that reads and
transforms program code. In its simplest form, a macro can define simple string replacements.
A complex macro, on the other hand, can be seen as a form of generative programming.
Since a macro is also defined as program code, it can also be processed by a macro.
This leads to the idea of \highlight{higher-order macros}. So, a macro is higher-order if
it takes another macro as input, or returns a macro or both.

\highlight{Lexical} macros are string operations that operate directly on the text of the program code.
As such, they are not aware of program structure.
\ac{VADL} macros are \highlight{syntactic macros} that are aware of the syntactic structure and operate
on the \ac{AST}. In addition, \ac{VADL} macros support a type system for syntax types, where each syntax element
is of a certain type. The type system helps prevent programming errors and enhances readability.
The types in \ac{VADL} macros are extensible with programmer-defined \highlight{composable syntax types},
also known as \highlight{records}.
A record is a type-checked collection of named syntax elements.
A macro can access a member of a record via its user-defined name.

For more information about the \ac{VADL} parser and macro system, see Section~\ref{sec:orig_parser} and~\ref{sec:open_frontend}.

\subsubsection*{\highlighttitle{Instruction Selection}}

An important aspect of the compiler backend is the translation from the target independent
\ac{IR} to the target specific assembly language. This is done by \emph{selecting} the
instructions offered by the target in such a way that the resulting program is functionally
equivalent to the \ac{IR} input program. Hence the name of the pass \highlight{instruction selection}.
The most common approach to achieve this is described in the standard compiler textbook~\cite{aho2006compiler}
and also implemented in LLVM which is used in \ac{VADL}'s compiler generator.
The idea is to represent the \ac{IR} input program as a dependency graph called the instruction selection graph. Each node
is an \ac{IR} instruction and each directed edge represents a dependency. Each target instruction
is represented as a graph, which represents the input, the output and the semantics: the instruction pattern.
A pattern is a potential subgraph of the instruction selection graph.
The instruction selection pass has to find a complete cover of the instruction selection graph,
by tiling the appropriate patterns.
In a \ac{VADL} \ac{ISA} specification, the semantics of an instruction is formulated as operational semantics. % TODO: Explain "operational semantics"
Thus an important task of \ac{VADL}'s \ac{LCB} is to infer
the instruction patterns to be used during instruction selection from the \ac{VADL} specification.
See Section~\ref{sec:orig_compgen} and~\ref{sec:open_compgen} on implementation details.

\subsubsection*{\highlighttitle{Register Allocation}}

The compiler has to decide which values to keep in registers for fast access, and which
values to store in memory, resulting in slower access. This is done in the register allocation pass.
The \highlight{live range} of a value in a program is the collection of all points in the program
between the definition of a value and its last use.
If the number of overlapping live ranges is greater than the number of available registers,
the compiler has to break up live ranges and insert \highlight{spill code} to store a value in memory and load it again when needed.

A common approach for register allocation is \highlight{graph coloring}, described in~\cite{chaitin1981graphcoloring}.
The idea is to construct an interference graph, representing live ranges as nodes and interference relations
as edges. Each available register is represented by a color.  The goal of the approach is to assign each node a color,
such that no two nodes with the same color are connected by an edge.

OpenVADL's QEMU Generator uses the same approach to minimize the
number of temporary variables, see Section~\ref{sec:tcg-variable-allocation}.

\subsubsection*{\highlighttitle{Instruction Scheduling}}

While respecting all dependencies and keeping the original semantics of the program,
the order in which the instructions are arranged can be subject to optimization.
This pass is called \highlight{instruction scheduling}. Skillfully rearranging the order
of the instructions can lead to better resource usage, and fewer pipeline stalls, thus
improving performance. Especially in target architectures with explicit instruction level
parallelism, like \ac{VLIW} architectures, a good instruction schedule is crucial,
as the instruction scheduler has to decide which instructions should be executed in parallel.

A \highlight{\ac{BB}} is a sequence of instructions with one entry point, and one exit point,
and no other control flow.
In most cases, the instruction scheduler operates on a single \ac{BB}. This limits the scope
of the scheduler and restricts its possibilities. One technique to improve this situation
is to merge \acp{BB} with \highlight{control flow elimination}, thus increasing the scope of the scheduler.
This is done by transforming control dependencies into data dependencies.

This can be done by using predicated instructions,
which have to be available in the target architecture.
Alternatively it can be realized by speculative execution and conditional assignment.
However, this only works if the speculated instructions do not have side effects.

\ac{VADL}'s \ac{MiA} synthesis also uses control flow elimination, see~\ref{sec:orig_miasynth} and~\ref{sec:open_miasynth}.
Again, the idea is to replace control dependencies with data dependencies.

\subsubsection*{\highlighttitle{Compiler Optimizations}}

Optimizations done by the compiler are program transformations that retain the semantics of a program,
while improving it according to some metric. Typical metrics of interest are: Execution time,
memory use, code size or power intake. Often trade-offs have to be made, because a given transformation
may improve one metric, but harm another.

In order to facilitate code reuse, it is good engineering practice to place program optimizations
in the middleend of the compiler, if possible. Due to their target independent nature, optimizations
in the middleend can benefit all supported target architectures. Often optimizations in the middleend
are not entirely target independent, but are parameterized with information about the target architecture.
These parameters can inform optimization decisions in the otherwise target-independent middleend pass.

One important aspect
regarding the middleend is the design of the \ac{IR}. The \ac{IR} should make it easy to perform
program analysis and program transformations.
An \ac{IR} in \highlight{\ac{SSA}} form can help with these goals. In \ac{SSA} form, a variable is assigned a value
\emph{exactly once} and cannot be changed thereafter. This property greatly facilitates dependency analysis,
and determining \highlight{use-def chains}, since there is a single point of definition for a variable.
Many widely used compilers employ an \ac{IR} in \ac{SSA} form, e.g. GCC~\cite{stallman2020gcc} and
LLVM~\cite{lattner2004llvm}, which \ac{VADL} uses.
The article~\cite{cytron1991ssa} describes how to efficiently generate the \ac{SSA} form of a given program,
and the \ac{SSA} book~\cite{rastello2022ssa} covers the topic comprehensively, including many optimizations.

An optimization well suited for \ac{SSA} form is \highlight{global value numbering}~\cite{briggs1997value}.
The goal of global value numbering is to remove redundant computations that compute the same value,
by replacing them with the value itself. It does so by labelling expressions that
produce the same value with the same \highlight{value number}. This facilitates analyzing inputs and outputs
of computations, thus making it possible to determine which computations yield the same value.

\highlight{\ac{CSE}} has a similar goal, as it also tries to eliminate redundant computations.
It does this by analyzing which expressions are available at which point in the program. Then it
decides where to replace identical (common) expressions with a single variable holding the computed
value of the expression~\cite{cocke1970gcse}.

Important and common compiler optimizations are those operating on constant values. A constant value
in a program does not depend on the program's input. Thus, everything needed to compute it is already
known at compile time; hence the name \highlight{compile time constant}. Programs often contain
expressions that only depend on compile time constants. Therefore, it is possible to compute their value
at \highlight{compile time}, thus moving complexity from the program's runtime to its
compile time. This optimization is called constant expression evaluation or \highlight{constant folding}.

Working closely together with constant folding is \highlight{constant propagation},
which replaces the use of a variable, which the compiler can determine to be constant, with its constant value.
This in turn may lead to new opportunities for constant folding. Both of these techniques and
many more are described in the book~\cite{muchnick1997advanced}.

Dead code is program code that is guaranteed to never be executed.
\highlight{Dead code elimination} is a compiler optimization that tries to find dead code, and remove it.
One common application arises when control flow depends on a compile time constant condition and
the compiler can guarantee that a branch target can never be reached.

Sometimes it is beneficial to move computations in a program to another point within
the program. This is called \highlight{code motion} and may serve to provide the result of a
computation to more places where this result is needed, without recomputing it.
Alternatively it can be employed to prevent recomputations inside a loop where the result does not change
between loop iterations, known as \highlight{loop invariant code motion}.
Code motion has to be done carefully, as it can lengthen live ranges, or increase code size.
Thus, sophisticated approaches like
\highlight{lazy code motion}~\cite{knoop1992lazycodemotion} have been developed to make sensible code motion decisions.

In general function calls cause an overhead, as the program has to prepare the stack frame and handle input parameters and
return values. \highlight{Function inlining} is a way to remove this overhead. The idea is to replace the
function call with an inlined copy of the callee. Thus the complete functionality of the callee
is available in the caller without the overhead of a function call. This can also increase the scope on which the
instruction scheduler may operate. However, function inlining typically leads to an increase in code size.

Some compiler optimizations are target machine \emph{dependent} and are considered to be
part of the backend.
In many target architectures it is the case that some operations need more
resources than others. \highlight{Strength reduction} tries to replace expensive
operations with cheap operations which yield the same result. A typical example
is integer multiplication by a power of two. If the value is represented
as two's complement, the compiler can choose to use a bitwise left-shift instead, which
is typically cheaper than multiplication. Strength reduction is presented
in~\cite{cocke1977strengthreduction}. This optimization is an example of a pass that can reside
directly in the target dependent backend, or, with appropriate
parameterization, in the middleend.

\subsubsection*{\highlighttitle{Assembler}}

The assembly language is independent of the concrete binary encoding of the
machine instructions.  The assembler reads a textual representation of the
machine program, i.e., a program in assembly language, and generates a binary
representation of this program that a processor can execute.  This step
consists of encoding the machine instructions as a bit pattern.  The main task
of the assembler is to apply such a binary encoding to the assembly program,
thus creating an object file.

Since the assembler has to parse a textual representation of the program,
a straightforward way is to generate a parser from a grammar describing the assembly
language.
However, sometimes it is not the most convenient way to do this. An \ac{ISA} specification
in \ac{VADL} already contains a specification of how to print each instruction
as a string in the assembly language. This functionality is also called a \highlight{pretty printer},
as it takes a binary encoded instruction and prints it in readable form.
If the assembly language is not too complex, it is possible to deduce the grammar of the
assembly language from the specification of the pretty printer. Informally
speaking, it is possible to go \emph{the other direction}, and generate a parser
from the specification of a pretty printer. This technique is called \highlight{program inversion}.
This feature helps to keep \ac{VADL} specifications succinct and reduce redundancies.
For more complex assembly languages, \ac{VADL} also supports explicit specification of a grammar
for the assembly language.
See Section~\ref{sec:orig_asmgen} and~\ref{sec:open_asmgen} for details.

\subsubsection*{\highlighttitle{Linker}}

The linker joins object files together, creating a single \highlight{executable}
native program that can be run on a processor. The linker has to resolve
symbolic addresses and assign them concrete address values from the machine's
address space. It must also place the object files containing executable code
in non-overlapping memory segments so each instruction has a unique address.
This process is called \highlight{relocation}. Often, target-specific rules for
relocations exist that have to be obeyed by the linker.

\subsection*{\highlighttitle{Processor}}

In the context of \ac{VADL}, a processor is an integrated digital electronic circuit
that reads data from memory, executes operations on these data and writes data to memory.
These operations are called \highlight{instructions}.
The representation of these instructions, together with initial values in memory, is called a \highlight{native program}.
Which instructions are available and how they are represented in memory is defined
by the \ac{ISA}. Thus, a processor \emph{implements} an \ac{ISA}.
The \ac{MiA} describes how the implementation for a processor is realized,
specifically, it defines what components make up its internal structure and how they interact.

The timing behavior of these designs is based on clock signals. While modern
processors do use multiple clocks with different frequencies, \ac{VADL}
currently focuses on designs of processors with a single clock domain.
Related to the \highlight{clock cycle},
the \highlight{machine cycle} is the time interval between the start of two instructions~\cite[p.~51]{shen2013modern}.
Depending on the \ac{MiA} this can be longer than one clock cycle, e.g.,
for multi-cycle \acp{MiA}.

Besides low-level hardware design decisions the \ac{MiA} is fundamental to reach certain
design goals, such as performance or power consumption, in the hardware implementation of a processor.
Techniques used for this include pipelining, forwarding and branch prediction,
but also advanced techniques specific to superscalar and
out-of-order processors, like register renaming, reservation stations and reorder buffers.

\highlight{Pipelining} is concerned with splitting instruction execution into smaller steps structured in stages,
which allows resulting hardware to run faster, i.e., at higher clock frequencies.
In a classical scalar pipeline this results in different steps of multiple instructions being executed at each
point in time and gives rise to data hazards, that occur when instructions require
results from previous instructions still in execution. To resolve this problem without
delaying instructions a \ac{MiA} can be equipped with \highlight{forwarding logic}, which connects stages that read values
with other stages where the needed results are already available.

Another problem that is caused by the simultaneous execution of sequential instructions
occurs in the presence of branch instructions. In order to still execute further instructions
while the result of a branch is not yet known, processors employ
\highlight{branch predictors}. These predictors speculate on the outcome of the branch by predicting its target address.
The processor continues execution of subsequent instructions at the predicted target address.
Mispredicted branches must be resolved by either not changing the state during speculative
execution or by rolling back changes.

\highlight{Superscalar} processors strive to complete more than one instruction per cycle. To achieve this
they implement \highlight{out-of-order execution}. The processor here analyzes
the data dependencies between the fetched instructions and executes the instructions
it can simultaneously. At the core of such architectures are multiple functional units
that handle the computations involved in the instructions in parallel.

For the processor to be able
to continue decoding instructions while others wait for their input operands to become
available from previous computations, it uses \highlight{reservation stations}. These are buffers
that collect the waiting instructions and their operand values and then move
instructions to functional units when their operands are available.

The \highlight{reorder buffer} keeps track of the unfinished dispatched instructions.
This buffer is used to complete instructions \highlight{in order} and is also used when
doing speculative execution.

To exploit instruction-level parallelism even more, superscalar architectures try to reduce
anti- and output dependencies by \highlight{renaming registers}. The design then contains more physical
than architectural registers. If a new instruction has an anti- or output dependence on
an instruction that has not completed yet, its output register is renamed and the instruction can be
dispatched right away. This renaming can be realized using a separate register rename file
or be integrated with the reorder buffer, recording a rename register for every instruction
in progress.

\ac{VADL} can be used to describe many of these micro-architectural aspects in
its \ac{MiA} section.
To implement a \ac{MiA}, conventionally an \ac{HDL} is used to represent the design at \ac{RTL}.

\subsubsection*{\highlighttitle{\acl{RTL}}} \label{sec:rtl} This is the level of abstraction where the design
is described as a set of registers and their connection through \highlight{combinational
logic}. The registers are the only components having memory, i.e., are solely
holding state. Their inputs are consequently outputs from other registers
and external inputs wired through logic gates. Typically the description also
involves defining a clock signal and reset logic (initial values) for
the registers. In contrast to mere combinational logic, such a design with
memory embodies \highlight{sequential logic}. Its outputs depend on both its current inputs
and the sequence of past inputs.

Since the internal state and its possible transitions in sequential logic are
not always obvious on \ac{RTL},
a useful abstraction for creating sequential logic is the definition in the form
of a \highlight{state machine} (following the formal model of finite-state machines).
State machines are defined by a set of states with corresponding output and
next state logic. Such a state machine can then be implemented using
registers to hold the state and combinational logic to feed outputs and the
next state logic.

\highlight{Verilog} and \ac{VHDL} are examples of widely used low-level \acs{HDL}s.
These languages can specify a wide range of circuits and also behavior not
implementable in hardware (non-synthesizable behavior).
Hence, they are not only capable of describing processors.

\ac{VADL} emits \highlight{Chisel}~\cite{bachrach2012chisel}, which then generates Verilog. Chisel's goal is to offer convenient
abstractions and to be easier to work with than directly writing Verilog code.
For instance, it only generates synthesizable Verilog.

\subsubsection*{\highlighttitle{Chisel}} \label{sec:chisel} Chisel is an \ac{HDL} used to describe digital circuits on the \ac{RTL}. It tries to increase
the circuit designer's productivity by providing powerful abstractions compared to the low level
languages like Verilog or \ac{VHDL}. Many of these abstractions are a result of the fact that Chisel
is a \ac{DSL} embedded in the general purpose programming language \highlight{Scala}. Thanks to this,
Chisel can directly use Scala's object oriented and functional programming features as
well as parameterized types and type inference. Chisel's goal is to use these features to
make the hardware design shorter and easier to maintain and extend. The Chisel compiler
emits Verilog code which in turn can be used to map to \acp{FPGA} or for \ac{ASIC} synthesis.

\subsection*{\highlighttitle{Simulation}}

\ac{ISA} simulation is the process of executing a program
on a software implementation of the target \ac{ISA}
instead of a hardware implementation, i.e., a processor.
This software implementation is called a \highlight{Simulator}.
For the properties and behavior under consideration,
the behavior and execution of the simulator is identical to the simulated processor.
However, certain aspects might not be simulated depending on the requirements of the simulation.
For instance, the goal of an \ac{ISS} is to match the semantics of the \ac{ISA}
but without considering the behavior of an underlying microarchitecture.
This is sufficient for executing programs written for the target \ac{ISA},
but e.g. analyzing energy consumption of the processor will not be possible.
Whether a certain property or behavior should be modeled by the simulator is an important design decision.

Hence, \ac{VADL} aims to describe various architectures and generate various
simulators depending on the user's needs. The most relevant approaches
for \ac{VADL} are described in the following paragraph.

The program that is executed by the simulator is called the \highlight{guest},
the system running the simulator is called the \highlight{host}.
A simulator that primarily takes care of the \emph{semantics} of
each simulated instruction is called an \highlight{Instruction Set Simulator (\ac{ISS})}.
In addition, a simulator can also model certain other
aspects that may be of interest, particularly the performance metrics
of the simulated processor. A simulator that can
also take the \ac{MiA} of a particular processor into account
and simulate the complete processor pipeline
is called a \highlight{\acf{CAS}}. A \ac{CAS} has to handle forwarding behavior,
pipeline stalls, cache/memory latencies and other \ac{MiA} related aspects of a processor.
This is why a \ac{CAS} is usually more complex than an \ac{ISS}
and its execution is computationally more costly.

There are several ways to implement a simulator.
The most straightforward approach is \highlight{interpretive simulation}.
Here, the simulator reads the guest
program instruction by instruction and simulates the effect of each instruction.
Since all simulation decisions happen at the time of simulation,
it is a very flexible approach, but also computationally costly.
\ac{VADL}'s \ac{ISS} generator, presented in Section~\ref{sec:issgen},
emits a simulator that falls into this category.

Another approach that tries to achieve better simulation performance
is \highlight{compiled simulation}~\cite{mills1991compiled}.
It translates the guest program
into a program that is executable directly on the host, while
keeping the same functional behavior. This way it moves
complexity to compile time and also makes it possible to apply
optimizations during compilation.
Normally compiled simulation does not allow self-modifying code
in the guest program, as it is compiled ahead of time.

\highlight{\ac{DBT}}~\cite{cmelik1994dynbintrans} tries to combine the flexibility of
interpretation with the performance of compiled simulation.
It only translates frequently executed code fragments into
executable host code. In this regard
it is conceptually similar to a \ac{JIT} compiler. Its dynamic nature
also allows it to handle self-modifying code.
QEMU~\cite{bellard2005qemu} uses \ac{DBT}, thus the simulator emitted by OpenVADL's \ac{ISS}
generator, presented in Section~\ref{sec:qemugen}, falls into this category.

Both processor specifications and simulators have to be validated.
The most common technique is \highlight{co-simulation} where the execution state of a test application on one simulator is compared against the execution state on another simulator or on real hardware~\cite{goel2017engineering}.
The comparison of the execution state can be done immediately after the execution of every instruction.
An alternative is to generate execution traces and to compare the traces.
The validation can be done at different levels of accuracy depending on the kind of execution state to be checked.
The state can just be the content of the modified registers or can contain internal processor states like pipeline registers and microarchitecture elements.

%% file: src-introlanguage.tex
\section{The Processor Description Language}
\label{sec:processor_description_language}

\subsection{Introduction}

The purpose of the \ac{PDL} \ac{VADL} is the complete specification of a processor architecture regarding the instruction set, the microarchitecture, the application binary interface, the assembler, the compiler, the linker, a functional \ac{ISS}, a \ac{CAS}, a synthesizable specification in an \ac{HDL}, test cases and documentation.
The aim of \ac{VADL} is to facilitate the development and customization of processors and their corresponding toolchains.
Thus, \ac{VADL} enables rapid \ac{DSE} of \acp{ASIP}, leading to higher quality processors and tools at reduced development costs and shorter time to market.
We want to highlight that even for existing architectures \ac{VADL} can be used solely for generating compilers for systems like LLVM or GCC, avoiding the need for LLVM- or GCC-specific know-how.

\ac{VADL} is a \ac{PDL}, that is, a \ac{DSL} in the domain of computer architecture and compiler construction.
Potential users are computer architects or compiler developers with an academic or industrial background, but they do not require extensive knowledge in both fields.
Nonetheless, all use cases should be served by a single language.
The language must provide an easily comprehensible syntax and semantics.
A user without prior knowledge of \ac{VADL} should understand a specification of a moderately complex architecture at first sight.
Therefore, the behavioral parts of \ac{VADL} are inspired by Java, C++, Rust and Chisel enriched with ideas from functional programming to provide familiarity to the users.
\ac{VADL} has a unique syntax and static semantics.

There is a long-standing debate about whether a specification language should be executable~\cite{HayesJones1989,Fuchs1992}.
Executable specifications only work well if the language has a single purpose, such as specifying simulators.
A \ac{VADL} behavior specification has to fulfill multiple purposes with the same specification, describing the semantics of a compiler's code generator, the semantics of simulators and the instruction execution in hardware.
Therefore, a \ac{VADL} processor specification cannot be executed directly, but executable artifacts are produced by generators and thus \ac{VADL} is a generator language.

\ac{VADL} is a specification language where a processor can be described at a high level of abstraction.
The goal is to have a concise specification that is easy to write by the user and, at the same time, easy to analyze by the generators.
The implementation of a concrete generator should not have any influence on the design of \ac{VADL}.

\subsubsection{Strict Separation of \ac{ISA} and \ac{MiA}}
\label{rationale_separation}

\ac{VADL} strictly separates the specification of the \ac{ISA} and the specification of a concrete \ac{MiA} implementation.
Different implementations can exist for the same \ac{ISA} specification, realized by different \ac{MiA} specifications.
This strict separation follows the best practice design process in computer architecture introduced by Fred Brooks with the architecture of the IBM System/360~\cite{AmdahlBlaauwBrooks1964} and advocated by Richard Sites, the chief architect of the Alpha AXP architecture~\cite{SitesAlpha93}.
The \ac{ISA} part specifies the register sets and the behavior, encoding and assembly language representation of the instructions, while the \ac{MiA} part describes the structure and microarchitecture of the processor.
Regarding the commonly used classification of \acp{PDL} into structural, behavioral, or mixed languages, the \ac{ISA} is related to the behavioral part, and the \ac{MiA} is related to the structural part and, therefore, \ac{VADL} is a mixed \ac{PDL}.
The \ac{MiA} description of \ac{VADL} operates on a higher level of abstraction than existing structural \acp{PDL}.

There are no references from the \ac{ISA} part to the \ac{MiA} part.
Only some references from the \ac{MiA} part to the \ac{ISA} part are allowed.
The \ac{ISA} part is sufficient to generate a purely functional \ac{ISS} or, together with an \ac{ABI}, a compiler.
The \ac{MiA} part is necessary to synthesize hardware, to generate a \ac{CAS} or to generate an instruction scheduler for a compiler.

\subsubsection{Language Safety}
\label{rationale_safety}

\ac{VADL} is designed with high productivity and type-safety in mind.
Therefore, the language is strongly statically typed, but type inference is supported to keep the specification concise.
In addition, static analysis prevents \ac{VADL} developers from writing illegal specifications.
For example, format fields are not allowed to overlap, instruction encodings must be unique, or a register write cannot occur before a read in the semantics of an instruction.
Furthermore, \ac{VADL} supports syntactic macros which are also type checked.

%% file: src-isasect.tex
\subsection{Overview}
\label{sec:pdl_overview}

\ac{VADL} provides a Chisel-like type system to represent arbitrary bit vectors.
There are two primitive data types -- {\tt Bool} and {\tt Bits<N>}.
Bool represents Boolean-typed data.
{\tt Bits<N>} represents an arbitrary bit vector data type of length {\tt N}.
Furthermore, to explicitly express signed and unsigned arithmetic operations, \ac{VADL} provides two sub-types of {\tt Bits<N>} -- {\tt SInt<N>} and {\tt UInt<N>}.
{\tt SInt<N>} represents a signed two's complement integer type of length {\tt N}.
Note that the length of this signed integer data type contains the sign bit and data bits.
{\tt UInt<N>} represents an unsigned integer type of length {\tt N}.
For all types based on bit vectors -- {\tt Bits}, {\tt SInt}, and {\tt UInt} -- \ac{VADL} will try to infer the bit size from the surrounding usage.
But for definitions, a concrete bit size has to be specified in order to determine the actual size of, e.g., a register.
In contrast to Chisel, the size of the resulting bit vector of an operation is identical to the size of the source operands.
An exception is multiplication, where two versions are available, one with a result with the same size and one with a double-sized result.
\ac{VADL} already has basic support for the specification of floating-point operations. Details are presented in Appendix~\ref{app:float_isa}.
An additional {\tt String} type is available in the assembly specification and the macro system.

\begin{lstlisting}[frame=single,float=!htb,label={lst:pdl_overview},caption={\ac{VADL} specification},language=vadl,linerange={1-21}]
constant MLen = 32

using BitsM = Bits<MLen>
using SIntM = SInt<MLen>
using UIntM = UInt<MLen>

function lessthan (a: SIntM, b: SIntM) -> Bool = a < b

import rv32i::RV32I

instruction set architecture RV32IM extending RV32I = {}

application binary interface ABI for RV32IM = {}

assembly description Assemble for ABI = {}

micro architecture FiveStage implements RV32IM = {}

processor CPU implements RV32IM with ABI  = {}

user mode emulation UME for RV32IM with ABI = {}
\end{lstlisting}

Listing~\ref{lst:pdl_overview} shows the main elements of a \ac{VADL} processor specification.
Usually, a \ac{VADL} processor specification has some global definitions in the beginning, followed by some sections describing \ac{ISA} or \ac{MiA}, which are described in more detail in the following sections.
On line 1, a constant {\tt MLen} with the value 32 is defined.
Type aliases can be defined with the keyword {\tt using} as shown on lines 3 to 5.
On line 7, a function is defined that compares two values of type {\tt SIntM} and returns the result of the comparison as a value of type {\tt Bool}.
{\tt import} allows the import of \ac{VADL} specification parts from separate files.
On line 9, a specification named {\tt RV32I} is imported from a file called {\tt rv32i.vadl}.
In this example, {\tt RV32I} refers to another \ac{ISA} specification.

An {\tt instruction set architecture} specification can extend another \ac{ISA} specification (line 11).
Section~\ref{sec:isa_section} contains a detailed description of the \ac{ISA} specification.
Lines 13 to 21 demonstrate the definition of the {\tt application binary interface} (see Section~\ref{sec:abi_section}), the {\tt assembly description} (see Section~\ref{sec:assembly_description_section}), the {\tt processor} specification (see Section~\ref{sec:cpu_section}) and the {\tt user mode emulation} (see Section~\ref{sec:ume_section}).
On line 17, a \ac{MiA} named {\tt FiveStage} implements {\tt RV32IM} (see Section~\ref{sec:mia_section}).

\tikzstyle{spec}  = [ellipse, rounded corners, minimum height=1cm,text centered, draw=black, text width=3cm]
\tikzstyle{fact}  = [rectangle, rounded corners, minimum height=1cm,text centered, draw=black, text width=2.6cm]
\tikzstyle{optia} = [thick,dotted,->,>=stealth]

\begin{figure}[h]

\begin{tikzpicture}[node distance=2cm]

\node (ASM) [fact                            ]  {assembler};
\node (CMP) [fact, right of=ASM, xshift=1.1cm]  {compiler,\\disassembler, linker};
\node (CAS) [fact, right of=CMP, xshift=1.1cm]  {cycle-accurate\\simulator};
\node (CPU) [fact, right of=CAS, xshift=1.1cm]  {hardware};
\node (ISS) [fact, right of=CPU, xshift=1.1cm]  {instruction set simulator};
\node (isa) [spec, above of=CAS              ]  {instruction set architecture};
\node (asm) [spec, above of=ASM, xshift=1.2cm]  {assembly description};
\node (ume) [spec, above of=ISS, xshift=-1.2cm] {user mode\\emulation};
\node (abi) [spec, below of=ASM, xshift=1.2cm]  {application\\binary interface};
\node (cpu) [spec, below of=ISS, xshift=-1.2cm] {processor};
\node (mia) [spec, below of=CAS] {microarchitecture};

\draw [arrow] (isa) -- (ASM);
\draw [arrow] (isa) -- (CMP);
\draw [arrow] (isa) -- (ISS);
\draw [arrow] (isa) -- (CAS);
\draw [arrow] (isa) -- (CPU);
\draw [optia] (asm) -- (ASM);
\draw [optia] (mia) -- (CMP);
\draw [arrow] (mia) -- (CAS);
\draw [arrow] (mia) -- (CPU);
\draw [arrow] (abi) -- (CMP);
\draw [arrow] (abi) -- (ASM);
\draw [arrow] (cpu) -- (CAS);
\draw [arrow] (cpu) -- (CPU);
\draw [arrow] (cpu) -- (ISS);
\draw [optia] (ume) -- (ISS);
\draw [optia] (abi) -- (ISS.south west);

\end{tikzpicture}

\caption{VADL specifications required for artifacts. Dotted arrows show optional specifications}
\label{fig:spec_for_artifact}

\end{figure}

Figure~\ref{fig:spec_for_artifact} shows which \ac{VADL} specifications are required to generate the corresponding artifact. Optional specifications are drawn as dotted arrows.
More complex examples and the \ac{VADL} grammar are available at the \href{https://openvadl.github.io/openvadl/}{OpenVADL documentation page}.

\subsection{Instruction Set Architecture Section}
\label{sec:isa_section}

The \ac{ISA} section is the major part of a processor specification.
Listing~\ref{lst:isa_basic_example} gives a small example specifying a subset of the RISC-V architecture with one branch instruction.
The section starts with the keyword {\tt instruction set architecture} followed by the name of the architecture, which we simply call {\tt RV32I}.
On line number 3, a constant {\tt Size} with the value 32 is defined.
Constant expressions can be used in a constant definition, as demonstrated on line number 4.
These constant expressions are evaluated during parsing.

Type casting is done with the keyword {\tt as}, shown on lines 28 and 30.
Type casting does zero extension, sign extension, or truncation of values if necessary.

Memory is defined on lines 12 to 14 by the mapping of a 32-bit address to an 8-bit byte.
The memory definition shows the use of annotations in square brackets.
Annotations can be applied to most of the definitions.
Annotations for memory are {\tt [little endian]} and {\tt [big endian]} which are allowed to be used in a dynamically evaluated expression, e.g., depending on the value of a configuration register.
A further memory annotation is the memory consistency model, e.g., {\tt [sequentialConsistency]}, {\tt [totalStoreOrdering]}, or {\tt [rvWeakMemoryOrdering]}.
There are only a small number of predefined consistency models.
Predefined microarchitectural elements like cache protocols know which consistency models they obey, and the correct combinations are verified by the type checker.
If more than one memory type is defined, one has to be declared as instruction memory.
It is also verified that the type of the program counter fits the address type of the instruction memory.

Declaring a \ac{PC} (line 17) is mandatory.
In most architectures, the \ac{PC}, when read, points to the start of the current instruction when used inside an instruction specification (lines 16 to 17).
This behavior can be changed by adding the annotation {\tt [next]}, which lets the \ac{PC} point to the end of the current instruction.
The ARM AArch32 architecture has the peculiar behavior that the \ac{PC} points to the end of the following instruction, which can be specified by the annotation {\tt [next next]}.
If an instruction does not explicitly modify the \ac{PC}, it is implicitly incremented by the instruction size in each execution cycle.

The RISC-V RV32I architecture has an integer register file named {\tt X} with 32 registers, which are 32 bits wide (lines 19 to 20).
The register with index 0 is hardwired to the value 0.
This can be specified with an annotation that maps the constant 0 to the specified register.

\begin{lstlisting}[frame=single,float=!htb,label={lst:isa_basic_example},caption={ISA specification basic example (RISC-V)},language=vadl,linerange={1-42,65-66}]
instruction set architecture RV32I = {

  constant Size  = 32                     // architecture size is 32 bits
  constant Size1 = Size - 1               // architecture size minus 1

  using Byte     = Bits<  8 >             // 8 bit byte
  using Inst     = Bits< 32 >             // instruction word type
  using Regs     = Bits<Size>             // register word type 
  using Addr     = Regs                   // address word type is equal to the register type
  using Index    = Bits<  5 >             // 5 bit register index type for 32 registers

  [little endian]                         // memory is accessed in little-endian order
  [rvWeakMemoryOrdering]                  // RISC-V weak memory ordering
  memory MEM : Addr -> Byte               // byte-addressed memory

  [current]
  program counter PC : Addr               // PC points to the start of the current instruction

  [zero: X(0)]                            // register with index 0 is always 0
  register         X : Index -> Regs      // integer register file with 32 registers

  format Btype : Inst =                   // Btype uses the width of Inst
    { imm    [31, 7, 30..25, 11..8]       // 12 bit immediate value
    , rs2    [24..20]                     // 2nd source register index
    , rs1    [19..15]                     // 1st source register index
    , funct3 [14..12]                     // 3 bit function code
    , opcode [6..0]                       // 7 bit operation code
    , immS   =  imm as SInt<Size> << 1    // sign-extended and shifted immediate value immS
    , imm    := immS(12..1)               // slice bits 12 to 1 from immS to encode imm
    , immS   :- immS(0) = 0 && immS as UInt + 4096 <= 8191
    }                                     // immS is a multiple of 2 and -4096 <= immS <= 4095

  [operation : BranchOp]                  // BEQ belongs to the set of branch operations
  instruction BEQ : Btype = {             // branch equal instruction
    let cond = X(rs1) = X(rs2) in
      if cond then
        PC := PC + immS
    }
  [select when : rs1 != rs2]              // source register indices have to be distinct
  encoding BEQ = {opcode = 0b110'0011, funct3 = 0b000}
  assembly BEQ = (mnemonic, " ", register(rs1), ",", register(rs2), ",", sdec(immS))

  }
\end{lstlisting}

Instruction words or system registers are commonly split into multiple fields.
The \ac{VADL} {\tt format} definition allows the specification of such instruction or register formats with their corresponding fields.
They can either be specified by connecting names with bit positions (as in Listing~\ref{lst:isa_basic_example} lines 23 to 27) or by connecting names with types (as in Listing~\ref{lst:isa_model_example} lines 2 to 6).
Sometimes fields in instruction words are not used directly, e.g., an immediate value is sign-extended or a register index can access only the higher half of a register file.
For convenient use of such fields, access functions can be defined.
In Listing~\ref{lst:isa_basic_example} line 28, the access function {\tt immS} is defined, which sign-extends the field {\tt imm} to {\tt Size} bits and shifts the sign-extended value left by one.

During instruction selection, a compiler must know what immediate values are valid and how they can be encoded.
For trivial access functions, the compiler generator can determine the validity and encoding function of values.
For nontrivial access functions, the validation and encoding functions have to be specified explicitly in the format specification.
The encoding on line 29 specifies that {\tt imm} is a bit slice of {\tt immS} from position 12 to position 1.
The validation on line 30 specifies that the value must be a multiple of 2 and must be in the range from -4096 to 4095.

With the {\tt instruction} definition, the behavior of an instruction is specified (lines 33 to 38).
Every instruction has a name ({\tt BEQ} in the example) and an instruction format type ({\tt Btype} in the example).
Between the curly braces (lines 35 to 37), statements in a style inspired by functional programming languages specify the behavior.
Assignments denoted by the assignment operator "{\tt :=}" are possible only to registers and memory locations.
To each variable, only one assignment is allowed, i.e., the behavior is \emph{single assignment}.
All reads to a register must occur before any writes to this register.
The same is true for a given memory location.
This requirement is checked by the \ac{VADL} language parser.
The {\tt let} statement defines a constant ({\tt cond} in the example) that can be used in the (block) statement after the keyword {\tt in}.
The {\tt else} part of the {\tt if} \emph{statement} is optional but required for an {\tt if} \emph{expression}.
Multiple conditional expressions can be written using a {\tt match} expression (see Listing~\ref{lst:isa_enum_funct_match_string} lines 8 to 13).
The {\tt match} statement works analogously.
The optional {\tt operation} annotation is used to assign an instruction to a set of operations.
These operation sets can be used to specify the grouping for \ac{VLIW} architectures or in the \ac{MiA} section to filter instructions for superscalar microarchitectures.

An instruction {\tt encoding} assigns fixed values, like operation codes, to certain fields of the instruction word (see Listing~\ref{lst:isa_basic_example} line 40).
To improve readability, the symbol "{\tt '}" can be inserted between the digits of a number as demonstrated with the binary number {\tt 0b110'0011}.
With annotations, it is possible to add constraints on format fields to give stronger restrictions on the encoding.
On line 39, there is the restriction that the two register indices {\tt rs1} and {\tt rs2} must be distinct (this is a reasonable constraint but not a requirement in the RISC-V architecture).

An {\tt assembly} definition specifies how an instruction is represented in a human-readable textual form as used, e.g., by a disassembler or a compiler (see Listing~\ref{lst:isa_basic_example} line 41).
The keyword {\tt assembly} is followed by one or more names with a common assembly representation.
The textual representation is specified by a string expression.
The only available string operator is the comma symbol "{\tt ,}", which does string concatenation.
\ac{VADL} offers some built-in string functions:
{\tt mnemonic} returns the identifier of the instruction as a string.
{\tt register} returns a standard representation of a register based on the name in the register (file) definition.
{\tt sdec}, {\tt udec} or {\tt hex} return their argument in a decimal (signed or unsigned) or hexadecimal string representation.

\begin{lstlisting}[frame=single,float=!htb,label={lst:isa_model_example},caption={ISA model definition and instantiation (RISC-V)},language=vadl,linerange={43-63}]
  format Itype : Inst =                   // immediate instruction format
    { imm    : Bits<12>                   // [31..20] 12 bit immediate value
    , rs1    : Index                      // [19..15] source register index
    , funct3 : Bits<3>                    // [14..12] 3 bit function code
    , rd     : Index                      // [11..7]  destination register index
    , opcode : Bits<7>                    // [6..0]   7 bit operation code
    , immS   = imm as SInt<Size>          // sign-extended immediate value
    }

  model ItypeInstr (name : Id, op : BinOp, funct3 : Bin) : IsaDefs = {
    instruction $name : Itype = {
      X(rd) := X(rs1) $op immS
      }
    encoding $name = {opcode = 0b001'0011, funct3 = $funct3}
    assembly $name = (mnemonic, " ", register(rd), ",", register(rs1), ",", sdec(imm))
    }

  $ItypeInstr (ADDI ; + ; 0b000)          // add immediate instruction
  $ItypeInstr (ANDI ; & ; 0b111)          // and immediate instruction
  $ItypeInstr (ORI  ; | ; 0b110)          // or immediate instruction
\end{lstlisting}

{\tt instruction}, {\tt encoding} and {\tt assembly} definitions are often quite similar for different instructions.
\ac{VADL}'s macro system helps to reduce redundancies caused by these similarities.
As simplicity and safety are crucial requirements for a processor description language, \ac{VADL} provides a pattern-based syntactical macro system.
The core of \ac{VADL}'s macro system consists of syntax models.
Every model has a name, a typed parameter list, a result type, and a body (see Listing~\ref{lst:isa_model_example} line 10).
Possible parameter types are syntactic elements like identifiers ({\tt Id}), binary operator symbols ({\tt BinOp}), binary constants ({\tt Bin}), or multiple \ac{ISA} definition elements ({\tt IsaDefs}) as used in Listing~\ref{lst:isa_model_example}.
Further syntactic types for general expressions ({\tt Ex}), expressions on the left-hand side of an assignment ({\tt CallEx}), statements ({\tt Stat}) or encoding elements ({\tt Encs}) are used in Listing~\ref{lst:rv32i} in the appendix.
{\tt model} parameters can be used at every position in the body which has the same syntactic type as the parameter.
The use of a parameter is indicated by a leading "{\tt \$}".
This design decision has two advantages.
Firstly, it simplifies parsing as it explicitly marks the use of a macro element.
Secondly, the "{\tt \$}" captures the model parameter names, preventing name collisions with other ISA definitions.
Similar to the parameters, the "{\tt \$}" marks the instantiation of defined syntax models.
The symbol "{\tt ;}" separates the syntax elements inside an instantiation.

Architectures like the ARM AArch64 or AMD64 \acp{ISA} are more complex and have many variants of the same instruction.
In \ac{VADL}, it is required that every variant is specified in a separate {\tt instruction} definition.
For such applications, the core macro system is not sufficient.
Therefore, \ac{VADL} supports higher-order macros (macros which take macros as arguments or which generate a macro), type aliases (e.g., the definition of a higher-order macro type), composition of syntax types (like structures in conventional programming languages), conditional macros, and built-in lexical macro functions (e.g., generating new identifiers).
Descriptions regarding these advanced features, their application and their efficient implementation have been presented in an earlier article in full detail~\cite{HochrainerKrall23}.

\begin{lstlisting}[frame=single,float=!htb,label={lst:isa_register_alias},caption={ISA register aliasing (AArch64)},language=vadl]
  register        S : Index -> Word       // general purpose register file, S31 SP
  [zero : X(31)]                          // X31 is zero register ZR
  alias register  X = S                   // general purpose register file, X31 ZR
  alias register SP : Word = S(31)        // stack pointer
  alias register ZR : Word = X(31)        // zero register
\end{lstlisting}

In the ARM AArch64 architecture, the register with index 31 of the general-purpose register file can serve two different purposes.
Depending on the instruction, it can be used as a stack pointer or zero register.
The {\tt alias} directive allows access to a register (file) with another name and different constraints, as shown in Listing~\ref{lst:isa_register_alias}.
When using the name {\tt S(31)} or {\tt SP}, the real register is accessed; when using the name {\tt X(31)} or {\tt ZR}, the zero register is accessed.
It is also possible to define an alias of the \ac{PC} to a certain register of a register file.
This is required for the ARM AArch32 architecture.

\begin{lstlisting}[frame=single,float=!htb,label={lst:isa_enum_funct_match_string},caption={ISA enumerations, functions, match and String (AArch32, AArch64)},language=vadl]
  enumeration conditions: Bits<4> =       // condition code encodings
    { EQ                                  // equal      Z == 1
    , NE                                  // not equal  Z == 0
    ...
    , AL                                  // always
    }

  function cond2string (condition: Bits<4>) -> String = match condition with
    { conditions::EQ => "eq"              // equal
    , conditions::NE => "ne"              // not equal
    ...
    , _              => "al"              // always
    }
\end{lstlisting}

Listing~\ref{lst:isa_enum_funct_match_string} shows how enumerations are defined and used in \ac{VADL}.
Enumerations are typed.
Their values can be either derived implicitly or set in the definition as shown in Listing~\ref{lst:ume_specification}.
When using an enumeration value, the full name consisting of the enumeration name and the element name separated by "{\tt ::}" has to be specified.
\ac{VADL} supports the definition of pure functions.
The function body is an expression.
Therefore, no statements or side effects are possible.
As demonstrated in Listing~\ref{lst:isa_enum_funct_match_string}, lines 8 to 13, a {\tt match} expression allows a selection between multiple alternatives and always needs the catch-all condition "{\tt \_}" in the last alternative.

\begin{lstlisting}[frame=single,float=!htb,label={lst:isa_exception},caption={ISA exception handling (simplified MIPS)},language=vadl]
instruction set architecture MIPSIV = {

  using IWord        = Bits<32>           // 32 bit instruction word
  using RWord        = Bits<64>           // 64 bit register word
  using Address      = RWord              // 64 bit register word
  using Index        = Bits<5>            // register index for 32 registers

  [next]                                  // PC points to the next instruction
  program counter PC : Address            // program pointer
  register       EPC : Address            // saved exception program counter

  [ zero : GPR(0)]                        // zero register
  register       GPR : Index -> RWord     // general-purpose registers

  format R_Type : IWord =                 // instruction word with 3 register operands
    { opcode [31..26]                     // operation code
    , rs     [25..21]                     // 1st source register
    , rt     [20..16]                     // 2nd source register
    , rd     [15..11]                     // destination register
    , shamt  [10.. 6]                     // unsigned shift amount
    , funct  [ 5.. 0]                     // function code
    }

  exception Overflow = {                  // overflow exception
    EPC := PC - 4                         // save PC of the faulting instruction
    PC  := 0xFFFF'FFFF'8000'0180          // set PC to the exception handler address
    }

  instruction add : R_Type = {            // add with overflow
    let result, status = VADL::adds(GPR(rs), GPR(rt)) in {
      if status.overflow then
        raise Overflow
      GPR(rd) := result
      }
    }
  }
\end{lstlisting}

\ac{VADL} has special notations to mark exceptional behavior.
In theory, these notations are not necessary, as every exceptional behavior can be described with the basic \ac{ISA} language constructs.
However, neither a human reader nor the compiler generator can distinguish normal behavior from exceptional behavior.
Therefore, it is required that exceptional behavior is marked by the keyword {\tt raise} as shown in Listing~\ref{lst:isa_exception} at line 32.
Exception-raising code is often quite similar.
Exceptions can be specified similarly to functions to enable code reuse.
In contrast to functions, exceptions do not return an expression but have side effects caused by assignment statements (see lines 24 to 27).
Nevertheless, it must be guaranteed that reads to a register or memory location precede all writes.

To specify exceptional behavior like overflow, the basic \ac{VADL} built-in functions exist in two flavors.
In the normal one, only the primary function result is returned.
In the exceptional one, there are two return values: the result and the status (see line 30).
The status contains information like overflow, zero, negative, or carry.
These built-in functions are used to specify instructions that handle operations with overflow or to specify architectures that have a status register, like the ARM AArch64 or AMD64 architectures.

\begin{lstlisting}[frame=single,float=!htb,label={lst:isa_tensor},caption={ISA tensor definitions},language=vadl]
instruction set architecture Tensor = {
  using Index = Bits<4>
  register X : Index -> Bits<32>
  register Y : Index -> Bits<2><2><16>
  register Z : Index -> Bits<4><16>

  format F : Bits<16> = {rs2: Index, rs1: Index, rd: Index, opcode : Bits<4>}

  instruction AddElements : F =
    Y(rd) := forall i in 0 .. 1, j in 0 .. 1 tensor Y(rs1)(i)(j) + Y(rs2)(i)(j)
//  Y(rd) := ((Y(rs1)(0)(0) + Y(rs2)(0)(0), Y(rs1)(0)(1) + Y(rs2)(0)(1)),
//            (Y(rs1)(1)(0) + Y(rs2)(1)(0), Y(rs1)(1)(1) + Y(rs2)(1)(1)))

  instruction Dot : F =
    X(rd) := forall i in 0 .. 3 fold + with Z(rs1)(i) *# Z(rs2)(i)
//  X(rd) := ( ((Z(rs1)(0) *# Z(rs2)(0)) + (Z(rs1)(1) *# Z(rs2)(1))) +
//             ((Z(rs1)(2) *# Z(rs2)(2)) + (Z(rs1)(3) *# Z(rs2)(3))) )
}
\end{lstlisting}

Listing~\ref{lst:isa_tensor} shows the two variants of the {\tt forall} definition, which enables comfortable specification of tensor operation instructions.
Despite the name {\tt forall}, this definition is not a loop.
It specifies the parallel, independent execution of operations on tensors (multidimensional representation of registers).
In the example in Listing~\ref{lst:isa_tensor}, register file {\tt Z} is a one-dimensional vector and register file {\tt Y} is a two-dimensional matrix.
{\tt Y} and {\tt Z} could also be aliases of register file {\tt X}.
The keyword {\tt forall} is followed by at least one index specifier with a given range after the keyword {\tt in}.
The {\tt tensor} expression creates a tensor with the same dimensions as the provided index ranges.
Each resulting element contains the evaluated {\tt tensor} expression.
The comments after the {\tt tensor} definition in Listing~\ref{lst:isa_tensor} line 10 show this definition's semantically equivalent unrolled version.
{\tt fold} is used to specify reductions.
The reduction operation defined by the operator after the keyword {\tt fold} is applied to reduce all results of the expression after the keyword {\tt with} to a single value.
Again, the comments after the {\tt fold} definition in Listing~\ref{lst:isa_tensor} line 15 show this definition's semantically equivalent unrolled version.
The {\tt Dot} example shows an instruction which is a building block of a dot product algorithm inspired by the same instruction of the TIC64x architecture.
\ac{VADL} additionally provides constructs for the specification of constant tensors.

\ac{VADL} also supports language features for the convenient specification of \ac{VLIW} architectures by applying regular expressions with constraints on operation sets which are described in detail in a thesis~\cite{NiklasMischkulnig2025}.

%% file: src-miasect.tex
\subsection{Micro Architecture Section}
\label{sec:mia_section}

The microarchitecture section aims to specify the processor implementation at a high level of abstraction. These abstractions enable a concise and understandable specification, as the generators handle many implementation details (e.g., hazard detection or pipeline registers).
Users would have more flexibility and control with low-level microarchitecture specifications (e.g., in an \ac{HDL}), but it would be impossible for generators to determine the purpose or the correctness of such specifications.
Therefore, only predefined elements configurable by annotations are supported.
If these predefined elements are not sufficient, further elements have to be added to the language and the affected generators have to be extended.
These extensions to the language have to be done by experts.
Their usage should be simple and can be accomplished by inexperienced users.

Firstly, this section will present the core concepts of the MiA modeling view -- pipeline stages and instructions. Then, these concepts are illustrated with the example of a 5-stage implementation of the RISC-V architecture. Finally, logic elements that model components outside of the stages (e.g., caches, control logic) are discussed.

\subsubsection{Pipeline Stage}

Pipeline stages allow users to define the hardware structure of the processor. Each stage defines cyclic behavior, which the processor executes. For example, one processor stage might fetch instructions from memory while another computes arithmetic results. Users can specify the exact behavior using syntax similar to that used to express the instruction behavior in Section~\ref{sec:isa_section}. It is easy to define a concise microarchitecture using powerful language built-ins. Section~\ref{sec:example_mia} provides some examples of pipeline stages.

In addition to the provided examples, annotations can specify a stage's restart interval and latency period. The restart interval governs the frequency at which new inputs are allowed to enter the stage. In contrast, the latency period controls the number of machine cycles required to complete a single execution. Additionally, users can assign a range to the latency, thus providing pipeline stages of varying lengths.

\subsubsection{Instruction Abstraction}
\label{sec:instruction_abstraction}

The instruction abstraction is a central concept of the \ac{MiA}. Users can leverage this concept with variables of type \vadltype{Instruction}. These variables abstract away two dimensions -- the kind of instruction and the progress of the instruction execution. The first aspect implies that the \ac{MiA} specification is not aware of the instructions present in the \ac{ISA}. Such variables may even represent VLIW bundles. The second aspect implies that the \ac{MiA} specification is not aware of the execution state. That is, it is not aware of which parts of the instruction semantics have already been computed at any point in the pipeline. The generator resolves these abstractions automatically during the microarchitecture synthesis. If the generator cannot entirely resolve the abstractions, it will raise an error. Section~\ref{sec:orig_miasynth} explains this process in more detail.

Because the \ac{MiA} is blissfully unaware of the complexity behind the \vadltype{Instruction} variable, it can solely interact with the instruction using abstract operations on the variable. For example, it can specify that the instruction should make arithmetic computations using \texttt{instr.compute}. \ac{VADL} provides a set of such operations. We will refer to them as instruction mappings, or simply mappings. Some mappings are very general (e.g., read any register), while others are more specific (e.g., read register file \vadlresource{X}). This enables users to trade off between precise control and compatibility with other \acp{ISA}.

\subsubsection{An Exemplary Pipeline}
\label{sec:example_mia}

This section describes the \texttt{IMPL} microarchitecture depicted in Listing~\ref{lst:stage_example_1}. A microarchitecture must implement an \ac{ISA}, such as the \texttt{RV32I} architecture (line 2) in our example. The pipeline consists of five stages. The specifications of each stage will be discussed in the following paragraphs. The \texttt{dataBusWidth} annotation determines the width of the memory interface. In this example, reading from and writing to memory is done in 32-bit blocks.

Listing~\ref{lst:stage_example_2} depicts the \vadlresource{FETCH} and \vadlresource{DECODE} stages of the pipeline.
All stages but the final stage have to specify the result of the stage.
The order of stages is defined by accessing the result of a previous stage.
The \vadlresource{FETCH} stage makes use of the \vadlresource{fetchNext} built-in. The result type of this operation (\vadltype{FetchResult}) abstracts the fetch size while the built-in automatically determines the next program counter. The generator determines the fetch size by analyzing the instructions in the \ac{ISA}. In the future, \ac{VADL} users may provide additional options for the fetch operation (e.g., buffers, multiple instructions). To understand the \ac{MiA} specification, it is sufficient to know that the \vadlresource{fetchNext} built-in loads enough bytes from the correct memory position to represent a single instruction.

The \vadlresource{DECODE} stage makes use of the \texttt{decode} built-in. The primary goal of this built-in is to represent a decode for the implemented \ac{ISA}. The generator will synthesize a decoder automatically. It takes a \vadltype{FetchResult} as input and produces an \vadltype{Instruction} as output. This is the origin of the instruction abstraction, which was discussed in Section~\ref{sec:instruction_abstraction}. The \vadltype{FetchResult} input is obtained from the preceding \texttt{FETCH} stage.  Note that the generator can resolve the instruction abstraction because it has access to the \ac{ISA}. The decoded instruction then reads the source operands from the \vadlresource{X} register file.

\begin{minipage}[b]{0.49\textwidth}
    \begin{lstlisting}[language=VADL, caption={Execute Stage}, label=lst:stage_example_1]
[dataBusWidth = 32]
micro architecture IMPL implements RV32I = {
    stage FETCH // ...
    stage DECODE // ...
    stage EXECUTE // ...
    stage MEMORY // ...
    stage WRITEBACK // ...
}
    \end{lstlisting}
\end{minipage}
\hfill
\begin{minipage}[b]{0.49\textwidth}
    \begin{lstlisting}[language=VADL, deletekeywords={decode, read}, caption={Fetch and Decode Stages}, label=lst:stage_example_2]
stage FETCH -> (fr : FetchResult) = {
    fr := fetchNext
}

stage DECODE -> (ir : Instruction) = {
    let instr = decode( FETCH.fr ) in {
        instr.read( @X )
        ir := instr
    }
}
    \end{lstlisting}
\end{minipage}

Listing~\ref{lst:stage_example_3} shows the specification for the \vadlresource{EXECUTE} stage. It is responsible for computing arithmetic operations and executing branches. Firstly, the stage obtains the current instruction from the \vadlresource{DECODE} stage (line 2). Then, the specification checks whether the instruction is valid (line 3). If not, the stage raises an invalid instruction exception, thus redirecting the control flow to the exception handler (line 4). If the instruction is valid, the stage computes arithmetic operations (line 6) and writes the new program counter (line 8). In addition, the stage verifies whether the instruction is on the correct program execution path (line 7). If this is not the case (branch misprediction), the control logic flushes the \vadlresource{EXECUTE} stage and all its predecessors. The \vadlresource{MEMORY} and \vadlresource{WRITE\_BACK} stages in Listing~\ref{lst:stage_example_4} complete the 5-stage pipeline.
The displayed definitions define a valid \ac{VADL} \ac{MiA} specification.

\begin{minipage}[b]{0.49\textwidth}
    \begin{lstlisting}[language=VADL, deletekeywords={compute, write}, caption={Execute Stage}, label=lst:stage_example_3]
stage EXECUTE -> ( ir : Instruction ) = {
    let instr = DECODE.ir in {
        if( instr.unknown ) then
            raise invalid
        else {
            instr.compute
            instr.verify
            instr.write( @PC )
        }
        ir := instr
    }
}
    \end{lstlisting}
\end{minipage}
\hfill
\begin{minipage}[b]{0.49\textwidth}
    \begin{lstlisting}[language=VADL, deletekeywords={write, read, compute}, caption={Memory and Write-Back Stages}, label=lst:stage_example_4]
stage MEMORY -> ( ir : Instruction ) = {
    let instr = EXECUTE.ir in {
        instr.write( @MEM )
        instr.read( @MEM )
        ir := instr
    }
}

stage WRITE_BACK = {
    let instr = MEMORY.ir in {
        instr.write( @X )
    }
}
    \end{lstlisting}
\end{minipage}

\subsubsection{Logic Elements}

\ac{VADL} uses the concept of a logic element to model microarchitectural concepts besides stages. The complexity of logic elements varies greatly depending on their semantics. An annotation determines a logic element's type and, thus, its semantics. For example, Listing~\ref{lst:forwarding} displays a logic element that allows users to define forwarding paths between stages. The generator must be aware of the logic element's semantics as it must derive the implementation in the microarchitecture synthesis.

Connecting logic elements with the instruction abstraction realizes their full potential. Listing~\ref{lst:forwarding} also shows how instructions may read and write values to the previously mentioned forwarding logic. As the generator is aware of the semantics, it can synthesize the logic of the forwarding network. Furthermore, it can also integrate this knowledge into the hazard detection logic element. After all, the control unit should not stall the pipeline if a forward can resolve the hazard.

\begin{minipage}[b]{0.49\textwidth}
    \begin{lstlisting}[language=VADL, deletekeywords={decode, forwarding}, caption={Decode Stage with Forwarding Logic}, label=lst:forwarding]
logic [forwarding] bypass

stage DECODE -> (ir : Instruction) = {
    let instr = decode( FETCH.fr ) in {
        instr.readOrForward( @X, @bypass )
        ir := instr
    }
}
    \end{lstlisting}
\end{minipage}
\hfill
\begin{minipage}[b]{0.49\textwidth}
    \begin{lstlisting}[language=VADL, deletekeywords={decode, forwarding, write}, caption={Cache Definition}, label=lst:caches]
[ write through ]
[ evict roundrobin ]
[ entries = 1024 ]
[ blocks = 4 ]
[ n_set = 2 ]
[ attached_to MEM ]
cache L1 : VirtualAddress -> Bits<8>
    \end{lstlisting}
\end{minipage}

Readers familiar with microarchitecture design may have noticed that the specification does not contain elements for the necessary control logic and hazard detection. If the generator does not find a logic element that handles these circumstances, it inserts a default hazard detection and control element into the \ac{MiA}. Later, the microarchitecture synthesis determines the necessary control logic for the processor. Letting the generator synthesize these elements changes the role of hardware designers. Instead of testing an idea in a specific processor, they can define it as a new logic element in the \ac{VADL} generator. Then, they can test this concept in many different configurations with the regular \ac{VADL} design flow.

\subsubsection{Caches}

To represent a memory subsystem, \ac{VADL} defines a \texttt{cache} definition to describe caches. The definition can be parameterized through annotations. Listing~\ref{lst:caches} defines a cache named \vadlresource{L1} with 1024 entries (cache lines). A single cache line has 4 blocks, where a single block corresponds to one addressable unit. For instance, this would be eight bits on a byte-addressable architecture. Our cache is defined to be 2-way associative (\texttt{n\_set}). Since the cache has 1024 entries and each set contains two entries, the cache has a total of 512 sets. Observe that setting \texttt{n\_set} to 1 is equivalent to a directly mapped cache, while $\texttt{n\_set} = \texttt{entries}$ makes the cache fully associative. Most importantly, the \texttt{attached\_to} annotation defines where the cache can fall back to in case of a miss. The fallback storage can be another cache (e.g., level 2), memory, or a process. The latter can be used to translate a virtual address to a physical one before accessing main memory, for instance. In addition, several behavioral aspects of the cache can be specified, such as write and eviction policies. The attribute naming and design were inspired by~\cite{patterson2017comporganddesign}.

\subsubsection{Branch Prediction}

In order to model different simple branch prediction schemes and the branch unit behavior, \ac{VADL} provides the \texttt{fetchNext} construct that automatically incorporates branch prediction and control hazard resolving. If no user-defined branch predictor can be found, a default \texttt{always\_not\_taken} branch predictor will be added to the \ac{MiA}. In general, the architecture of the \ac{MiA} synthesis is agnostic to the correctness of the branch predictor. This is done by storing the source address alongside the actual instruction variable and comparing the source address to the actual \ac{PC} at an adequate place. This place is determined by the \texttt{instruction.verify} mapping. Listings~\ref{lst:stage_example_1} and~\ref{lst:stage_example_3} show the use of a simple branch prediction scheme. The branch predictor's implementation can be changed with different annotations. Combining multiple branch prediction schemes is also possible by defining two logic elements and using appropriate instruction mappings on them.

\subsubsection{Advanced Techniques}
\label{sec:advanced_mia}

This section will introduce concepts that are required for describing superscalar and out-of-order processors. Many modern processors employ at least one of these two techniques. We will try to introduce these concepts very briefly in the next paragraph. Readers interested in this topic can find more information in~\cite{hennessy2011computer}. We would like to highlight that we have not yet implemented these constructs in the generators. Therefore, this part of the language is still a work in progress.

Superscalar processors can finish executing multiple instructions per clock cycle. As a result, the overall throughput of the processor may increase. Furthermore, out-of-order processors dynamically schedule the execution of instructions depending on the availability of their inputs. This technique allows the processor to tolerate a certain amount of latency in the instruction stream while still keeping parts of the processor busy.

While these techniques are orthogonal optimizations, i.e., they can be applied separately, we will discuss them with a single example. The following paragraphs will extend the 5-stage pipeline from earlier to a superscalar out-of-order implementation. This \ac{MiA} employs reservation stations, multiple execution units, register renaming, and a reorder buffer. Again, we will explain these concepts superficially. Interested readers can find more information in~\cite{tomasulo1967},~\cite{hennessy2011computer} and~\cite{shen2013modern}.

A superscalar out-of-order processor may be implemented as follows. Firstly, the processor fetches multiple instructions from memory, which are then decoded in parallel. This can be achieved by parameterizing the \texttt{fetchNext} and \texttt{decode} built-ins from Listing~\ref{lst:stage_example_2}. The former will include the number of bytes to fetch, while the latter includes the maximum number of instructions to decode.

After decoding the instructions, the \ac{MiA} tracks the instructions in the reorder buffer. This buffer allows the processor to reconstruct the program order after the dynamic dispatching. Listing~\ref{lst:reorder_buffer_reservation_station} shows a reorder buffer definition which is modeled with a logic element. The buffer shown is also used to rename the \vadlresource{X} register file, mapping 32 architectural registers to 64 physical registers (one per reorder buffer entry). This technique is used to eliminate anti- and output dependencies in the original program.

Usually, a superscalar processor contains multiple execution units which are specialized to execute a subset of the supported \ac{ISA}. However, the processor must ensure that all operand values are available before executing an instruction. To facilitate this, the \ac{MiA} parks instructions in a \emph{reservation station} until all their operands are ready. Listing~\ref{lst:reorder_buffer_reservation_station} depicts two example definitions of a reservation station. The \ac{MiA}'s next task is to dispatch the instructions to their correct reservation station. This step requires separating the instruction stream. For example, only integer instructions must be dispatched to the integer reservation station. Often this is done in a separate stage which we will call \vadlresource{DISPATCH}.

Listing~\ref{lst:dispatch} shows an exemplary definition of a \vadlresource{DISPATCH} stage. The \texttt{filter} built-in is used to partition the instructions into integer instructions and memory instructions. The resulting instructions are then dispatched to the corresponding reservation station. The used \texttt{operation} concept models a set of instructions.

\begin{minipage}[b]{0.49\textwidth}
    \begin{lstlisting}[language=VADL, caption={Reorder Buffer and Reservation Station Definition}, label=lst:reorder_buffer_reservation_station]
[renames X]
[size = 64]
logic [reorder buffer] ReorderBuffer

[size = 16]
logic [reservation station] IntegerQueue

[size = 8]
logic [reservation station] MemoryQueue
    \end{lstlisting}
\end{minipage}
\hfill
\begin{minipage}[b]{0.49\textwidth}
    \begin{lstlisting}[language=VADL, caption={DISPATCH Stage Definition}, label=lst:dispatch]
operation IntOps = {ADD, ADDI, SUB, ...}
operation MemOps = {LW, SW, ...}

stage DISPATCH = {
    let is = DECODE_AND_RENAME.ir in
    let ints = filter(is, @IntOps) in
    let mems = filter(is, @MemOps) in {
        IntegerQueue.dispatch(ints)
        MemoryQueue.dispatch(mems)
    }
}
    \end{lstlisting}
\end{minipage}

After dispatching, the instructions reside in the reservation station until all their operands are ready. Execution units subscribe to the reservation stations as consumers, as shown in Listing~\ref{lst:exu_units}. The example shows two simplified integer units and one memory unit that consume from corresponding reservation stations. Even though both stages use the \texttt{i.execute} mapping, the integer units implement arithmetic computations, while the memory unit implements memory access. The \ac{VADL} generator is responsible for tracking which instructions are scheduled to the respective execution units. Users may also use more specific instruction mappings (e.g., \texttt{i.read(@Mem)} in the memory unit) if they prefer.

The resulting instruction streams are then merged in the \vadlresource{COMPLETION} stage depicted in Listing~\ref{lst:retiring}. This is done by using the combine operator (\texttt{|}). In the example, the \vadlresource{all} variable joins the three instruction streams together. The processor then marks all instructions in the joined instruction stream as completed in the reorder buffer. Note that this step happens in the order of the instruction execution, not the program order.

Lastly, the \ac{MiA} must retire the instruction in the reorder buffer. This is done by using the retire mapping as shown in Listing~\ref{lst:retiring}. The example shows a possible specification that can retire up to three instructions in a cycle. Once an instruction has retired, its allocated space in the reorder buffer is freed and the architectural state of the processor is updated. Note that, contrary to the \texttt{COMPLETION} stage, this process is done in program order. As a result, the original order of the instructions is reconstructed and their architectural side effects are applied in this order.

\begin{minipage}[b]{0.49\textwidth}
    \begin{lstlisting}[language=VADL, deletekeywords={compute, read, write}, caption={Definitions of Execution Units}, label=lst:exu_units]
stage IntegerExu1 -> (ir: Instruction) = {
    let i = IntegerQueue.consume in {
        i.execute
        ir := i
    }
}

stage IntegerExu2 = // equal to IntegerExu1

stage MemoryExu -> (ir: Instruction) = {
    let i = MemoryQueue.consume in {
        i.execute
        ir := i
    }
}
    \end{lstlisting}
\end{minipage}
\hfill
\begin{minipage}[b]{0.49\textwidth}
    \begin{lstlisting}[language=VADL, deletekeywords={compute, read, write}, caption={Retiring Instruction Streams}, label=lst:retiring]
stage COMPLETION = {
    let intExu1 = IntegerExu1.ir in
    let intExu2 = IntegerExu2.ir in
    let memExu = MemoryExu.ir in
    let all = intExu1 | intExu2 | memExu in
        all.markAsCompleted(@ReorderBuffer)
}

stage RETIRE -> {
    ReorderBuffer.retire(3)
}
    \end{lstlisting}
\end{minipage}

%% file: src-abisect.tex
\subsection{Application Binary Interface Section}
\label{sec:abi_section}

The \ac{ABI} ensures consistent and well-defined interoperation between units
of object code.

The \ac{ABI} specification section in \ac{VADL} supports the definition of
\begin{itemize}
    \item special-purpose registers,
    \item stack alignments,
    \item register aliases,
    \item calling conventions and
    \item special instruction sequences.
\end{itemize}
This section provides a description and an example for each of these
definitions.

\ac{ABI} definitions are top-level elements inside a \ac{VADL} file. The section
starts with the keyword {\tt application binary interface} followed by a unique
identifier. Since most elements inside the \ac{ABI} section rely on previously
defined \ac{ISA} elements, it is required to reference an \ac{ISA} section
using the {\tt for} keyword after the identifier. Definitions from the
referenced \ac{ISA} are available inside the \ac{ABI} section.
Listing~\ref{lst:abi_section_definition} shows an empty \ac{ABI}
section for the {\tt RV32I} \ac{ISA}.

\begin{lstlisting}[language=vadl, label=lst:abi_section_definition, caption={ABI section definition}]
  application binary interface ILP32 for RV32I { }
\end{lstlisting}

Specifying the calling convention is one of the most important tasks of the
\ac{ABI}. Calling conventions describe how a function call is executed. The
specification contains information on the instructions performing the call,
which registers are used to pass arguments or return values, and which registers
are managed by the caller or callee. Additionally, it holds information on
special-purpose registers, such as a {\tt frame pointer}, {\tt stack pointer},
or {\tt return address}. Listing~\ref{lst:abi_calling_conventions} contains
\ac{ABI} code that defines a calling convention with special-purpose
registers. Each definition has the same structure, i.e., a descriptive keyword
that declares what register or register group will be specified, followed by the
"$=$" sign and one or more references pointing to the actual registers. To be more
concise, \ac{VADL} provides a special syntax to address multiple registers with
similar names. In the example, the compact expression {\tt a\{0..7\}} evaluates to
{\tt [a0, a1, a2, a3, a4, a5, a6, a7]}. Moreover, Listing
\ref{lst:abi_calling_conventions} showcases the {\tt alignment} annotation.
This is used to specify the stack alignment.

Using expressive names for registers is not only helpful for reading and
understanding the specification but can also have a positive impact on
debugging and writing correct specifications. In order to provide registers
with additional names,
the \ac{ABI} section provides the {\tt alias register} keyword. With the help
of this mechanism, it is possible to assign multiple names to registers and
use them as references throughout the \ac{VADL} specification.
The statement to declare an
alternative name for a register follows a structure similar to that of defining
special-purpose registers. First, the keywords {\tt alias register} are written,
followed by the new identifier. Next, the "$=$" operator points to the register
reference which should be extended by a new name.
Note that a single hardware
register or register cell is allowed to have multiple different names.
If multiple names are available for a specific register, you may use the
annotation {\tt [preferred alias]} to emit only the preferred name in the
generated code.
Listing~\ref{lst:abi_alias_register} showcases different alias
register statements and enforces the name {\tt fp} for register {\tt X(8)}.
In Listing~\ref{lst:abi_calling_conventions}, the alias names can be seen in
action.

\begin{minipage}{0.48\linewidth}
\begin{lstlisting}[language=vadl, label=lst:abi_alias_register, caption={ABI Register Alias}]
  alias register zero = X(0)
  alias register ra   = X(1)
  alias register sp   = X(2)
  alias register gp   = X(3)
  alias register tp   = X(4)

  // ...

  [ preferred alias ]
  alias register fp   = X(8)

  alias register s0   = fp
  alias register s1   = X(9)
  alias register a0   = X(10)
  alias register a1   = X(11)
  alias register a2   = X(12)
  alias register a3   = X(13)
\end{lstlisting}
\end{minipage}
\begin{minipage}{0.48\linewidth}
\begin{lstlisting}[language=vadl, label=lst:abi_calling_conventions, caption={ABI Calling Convention}]
  [ alignment : Bits<128> ]
  stack pointer = sp

  return address = ra

  global pointer = gp

  frame pointer = fp

  return value = a{0..1}

  function argument = a{0..7}

  caller saved = [ ra, a{0..7}, t{0..6} ]

  callee saved = [ sp, fp, s{0..11} ]
\end{lstlisting}
\end{minipage}

Finally, the \ac{ABI} section supports the definition of special instruction
sequences. An instruction sequence is a particular order in which a specific
list of instructions has to be executed. For example, a call sequence might
consist of two separate parts. One instruction loads an address to a specific
location and a second instruction jumps to this address and prepares the return
register. \ac{VADL} is able to detect a lot of sequences on its own, e.g., stack
manipulations or frame index-related loads and stores. Detecting certain
sequences can be challenging due to their explicit inclusion in the processor's
\ac{ABI} or their unreliable detection. To address this issue, the ABI section
includes and mandates the use of various sequences such as {\tt call sequence},
{\tt return sequence}, {\tt address sequence}, {\tt nop sequence} and
{\tt constant sequence}.
Listing~\ref{lst:abi_instruction_sequence} defines call and return sequences for
the RISC-V processor.
Every sequence has a predefined set of parameters with specific meanings.
In the case of the presented {\tt call sequence}, the first operand is the
target call address. The body of the definition describes how the address
is split into two parts using \ac{VADL} \emph{modifiers}. The instructions
\isainstruction{LUI} and \isainstruction{JALR} are then used to load the
address into a specific register and jump to it. In addition, the return
register {\tt X(1)} is set.
The {\tt call sequence} is appropriately named, as it outlines the steps
required to call a specific address or symbol. Similarly, the
{\tt return sequence} serves the purpose of returning from a procedure call, as
the name implies. Both sequences
are only allowed once per ABI section. The {\tt address sequence} definition is
used to specify complex address loads. At present, the sequence is designed to
load the entire address space and handle only absolute addresses.
Additional features are being planned for the {\tt load sequence} to allow for
the indication of various types of loads using different flags, such as PC-relative
or absolute addresses.
For now, the \ac{ABI} section expects a single
{\tt load sequence}. When executing the {\tt nop sequence}, no state
transformation should be performed. Finally, the {\tt constant sequence}
specifies actions to load constant integers of different sizes.
The \ac{ABI} section supports multiple {\tt nop} and {\tt constant} sequences.
All mentioned sequences are analyzed and used by the compiler generator,
introduced in Sections~\ref{sec:orig_compgen} and~\ref{sec:open_compgen}, to generate compiler backend source
code.

\begin{lstlisting}[language=vadl, label=lst:abi_instruction_sequence, caption={ABI Call and Return Sequence}]
call sequence( symbol : Address ) =
{
    LUI{ rd = 1, imm20 = hi20( symbol ) }
    JALR{ rd = 1, rs1 = 1, imm = lo12( symbol ) }
}

return sequence =
{
    JALR{ rs1 = 1, rd = 0, imm = 0 }
}
\end{lstlisting}

%% file: src-umesect.tex
\subsection{User Mode Emulation Section}
\label{sec:ume_section}

\begin{lstlisting}[frame=single,float=!htb,label={lst:ume_specification},caption={UME specification (RISC-V ECALL for Linux)},language=vadl,deletekeywords={read,write},linerange={1-19}]
import rv32i::RV32I

enumeration RVLinuxSyscall : Bits<64> =
  { sys_openat = 57
  , sys_close  = 58
  }

[os : linux] 
user mode emulation UME for RV32I with ABI = {

syscall table = RVLinuxSyscall

syscall instruction = ECALL

syscall register = a7
syscall argument = a{0..5}
return  register = a0

}
\end{lstlisting}

In the field of processor simulation, there is a distinction between two different modes of simulation: \ac{UME} (or simulation) and full system emulation (or simulation).
In \ac{UME}, the processor only simulates user mode instructions.
System call instructions of the emulated processor are mapped to system calls of the host operating system.
In contrast to this, full system emulation also virtualizes system elements of the host architecture, like storage, network interfaces, attached keyboards, or monitors.
This means that an operating system also has to be executed on the emulated processor to make these virtualized resources available to the emulated processor.
With the development of OpenVADL, the \ac{UME} section has been redesigned to be compatible with QEMU and is still a work in progress.

Listing~\ref{lst:ume_specification} shows an example \ac{UME} section for RISC-V. In line 9, the \ac{UME} section is defined. The annotation {\tt [os : linux]} specifies that the guest operating system is a Linux OS. There is an internal default system call (syscall) table which maps the syscall numbers to entry points for the selected operating system. For every supported operating system, \ac{VADL} has its own set of entry point identifiers. Using these entry point identifiers as element names in an enumeration, the mapping of syscall numbers to entry points can be changed (see lines 3 to 6). When a syscall mapping table is defined (see line 11), the mappings in the used enumeration are merged with the internal mapping table. The {\tt system call} definition in line 13 specifies that {\tt ECALL} is the system call instruction, the system call number is passed in register {\tt A7} (see line 15), arguments to the system call are passed in registers {\tt A0} to {\tt A5} (see line 16), and the result of the system call is returned in register {\tt A0} (see line 17).

%% file: src-asmsect.tex
\subsection{Assembly Description Section}
\label{sec:assembly_description_section}

To complete the compiler toolchain, a generator must be able to produce an assembler and a linker so that users can build executable programs for the specified processor. A critical aspect of this task is modeling the tools' inputs and outputs (assembly and object files) and how they relate. Establishing this relationship requires semantic knowledge. For example, a generated assembler must parse an instruction, map its textual representation to an instruction in the \ac{ISA}, structure the output object file, and emit the corresponding binary encoding in the appropriate section. Because the parser generator relies on this knowledge, \ac{VADL} must capture these aspects explicitly. This section details how \ac{VADL} provides the auxiliary information needed to generate an assembler and linker.

Fortunately, standardized object file formats (such as \ac{ELF}) have successfully unified the structural requirements of diverse processor architectures. These formats dictate the overall layout of the object file while leaving inherently architecture-specific details such as instruction encodings open. Consequently, processor description languages targeting these formats do not need to model the internal structure of object files. This design choice reduces specification effort while leveraging the mature tool ecosystems built around standard formats.

In contrast, assembly languages lack a standardized format comparable to object files. Nevertheless, most assembly languages share common structural conventions, allowing \ac{VADL} to make three simplifying assumptions that reduce specification effort. First, labels follow a fixed syntax consisting of a name followed by a colon (e.g., \texttt{loop:}). Second, each statement corresponds to an instruction or pseudo-instruction of the target architecture. Third, an assembly source file consists of a sequence of labels and statements. As a result, a \ac{VADL} specification can focus exclusively on defining the syntax of individual assembly instructions.

Listing~\ref{lst:assembly_description_element} illustrates the structure of an assembly description element, including its three subsections. Each assembly description must reference an \ac{ABI} and, by extension, the \ac{ISA} associated with that \ac{ABI}. Binding to a specific \ac{ABI} rather than directly to an \ac{ISA} is necessary to supply additional register-usage information. For example, a generated linker can use the defined global pointer to optimize variable accesses. As with any top-level element, annotations can attach supplementary metadata for the generators.

The core component of an assembly description is the grammar definition, which specifies assembly instruction syntax as a formal grammar augmented with semantic information. Specifically, users can annotate sub-elements of an instruction with type information to capture their roles (e.g., indicating that an operand refers to a register). While the syntax of the grammar section is inspired by Xtext~\cite{eysholdt2010xtext}, a full discussion of all its intricacies is beyond the scope of this work. Instead, Listing~\ref{lst:lui_grammar} provides an intuitive example of how a grammar rule captures the information required for assembler generation. The rule defines the RISC-V \texttt{LUI} (load upper immediate) instruction, where \texttt{Register} and \texttt{ImmediateOperand} are built-in non-terminals that users can override by defining rules with matching names.

\begin{minipage}[b]{0.59\linewidth}
    \begin{lstlisting}[language=vadl, deletekeywords={mnemonic}, caption={Assembly Description Element}, label=lst:assembly_description_element]
[ commentString = "#" ]
assembly description RV32I_ASM for RV32I_ABI = {
    directives = {
        ".word" -> BYTE4
    }

    modifiers = {
        "lo" -> RV32I::lo12,
        "hi" -> RV32I::hi20
    }

    grammar = { ... }
}
    \end{lstlisting}
\end{minipage}
\hfill
\begin{minipage}[b]{0.39\linewidth}
    \begin{lstlisting}[language=vadl, deletekeywords={mnemonic}, caption={Grammar for a RISC-V LUI Instruction}, label=lst:lui_grammar]
grammar = {
    ...
    // Example: lui ra, %hi(main)
    LuiInstruction:(
        mnemonic="lui"@operand
        rd=Register @operand
        ","
        imm20=ImmediateOperand
    )@instruction
    ...
}
    \end{lstlisting}
\end{minipage}

The power of the grammar system is rooted in the type system of the language because the type system also models semantic information. Typically, during assembly parsing, the parser receives tokens with primitive types from lexical analysis. At that point, the tokens carry no semantic meaning. However, an assembler must verify that these tokens satisfy context-dependent constraints, such as the requirement that the first operand of an \texttt{ADD} instruction be a valid register. \ac{VADL} captures these constraints through its type system. By annotating grammar elements with semantic types, users instruct the parser generator to insert conversion routines for the corresponding values. Depending on the input and output types, a conversion routine can both validate and transform the input. For instance, converting a primitive string to a register type checks whether a register with a matching name exists, thereby guaranteeing that the resulting value refers to a valid register. \ac{VADL}'s type system then propagates this semantic information to other parts of the grammar, while enabling the parser to emit meaningful error messages if validation fails.

Readers may wonder why \ac{VADL} requires a separate grammar for the assembly syntax, even though the \ac{ISA} section already defines assembly formatting functions. In principle, the grammar could be derived solely by inverting these formatting functions. However, we decided against this approach for three reasons. First, \ac{VADL} does not strictly require explicit grammar specifications for every instruction. By establishing naming conventions for grammar rules, generators can assist users by automatically synthesizing rules from the formatting functions. This design enables graceful degradation of the specification effort, allowing users to supply custom rules on a per-instruction basis. For example, a generator can infer the grammar rule in Listing~\ref{lst:lui_grammar} directly from its formatting function. Second, relying solely on function inversion would force generators to implement complex inversion routines capable of handling arbitrary formatting functions. Separating the grammar provides an escape hatch whenever a generator's rule-synthesis capabilities are insufficient. Finally, a single assembly instruction can map to multiple valid textual representations (e.g., varying whitespace). Inverting a formatting function would therefore require handling syntactic ambiguities to capture all valid representations. While prior work addresses this by introducing biased choice operators and specialized whitespace rules~\cite{matsuda2013flippr}, doing so significantly increases the complexity of the formatting functions. Keeping the \ac{ISA} section simple serves the broader goal of making computer architecture comprehensible. Because understanding the \ac{ISA} is more critical than capturing every assembly syntax variation, keeping this section concise helps \ac{VADL} users focus on the architecture's essential aspects.

%% file: src-cpusect.tex
\subsection{Processor Section}
\label{sec:cpu_section}

The processor definition combines an \ac{ISA} with an \ac{ABI} and specifies the execution environment of the resulting processor. It complements the architectural specification with register initialization on reset, a stop condition and memory regions with optional initial contents. This information supports the generation of software simulators and hardware artifacts for the specified processor.

Listing~\ref{lst:cpu_example} shows a RISC-V processor that implements the RV32IM \ac{ISA} with the ILP32 \ac{ABI}. The \texttt{reset} block defines the reset vector by initializing the \ac{PC} to the boot ROM address \texttt{0x1000}. It also sets the trap-vector register \vadlresource{mtvec} to \texttt{0xe000'0000}. The \texttt{stop} condition is satisfied when the \ac{PC} reaches this trap-vector address, causing the simulation to terminate.

The processor defines two memory regions within the \ac{ISA} memory address space \vadlresource{MEM}. The \texttt{base} and \texttt{size} annotations place \vadlresource{MAIN\_RAM} at \texttt{0x8000'0000} with a size of 128~MiB. If the optional \texttt{size} annotation is omitted for a \ac{RAM} region, emulation uses the maximum available \ac{RAM}, limited by the host device. The \texttt{firmware} annotation identifies this region as the destination for externally supplied firmware. The read-only region \vadlresource{BOOT\_ROM} is initialized with two boot instructions. OpenVADL derives its base address and size from the memory accesses in the initialization block. After reset, the processor starts execution in \vadlresource{BOOT\_ROM}, loads the \ac{RAM} base address into register \vadlresource{t0}, and jumps to the firmware entry point in \vadlresource{MAIN\_RAM}.

The processor-level \texttt{htif} annotation enables \ac{HTIF} support in the generated simulator for communication between the guest program and the host, including emulation termination.

\begin{lstlisting}[frame=single,float=!htb
,language=vadl,morekeywords={start,stop,reset,region},deletekeywords={description}
,label={lst:cpu_example}
,caption={Processor Definition (RISC-V)}
]
[ htif ]
processor CPU implements RV32IM with ILP32 = {
  constant ROM_BASE = 0x1000
  constant RAM_BASE = 0x8000'0000
  constant RAM_SIZE = 0x08000000
  constant TRAP_VEC = 0xe000'0000

  reset = {
    PC   := ROM_BASE
    mtvec := TRAP_VEC
  }

  stop = PC = TRAP_VEC

  [ firmware ]
  [ base : RAM_BASE ]
  [ size : RAM_SIZE ]
  memory region [RAM] MAIN_RAM in MEM

  memory region [ROM] BOOT_ROM in MEM = {
    // Jump to the firmware entry point at RAM_BASE.
    MEM<4>(0x1000) := 0x800002b7  // lui  t0, 0x80000
    MEM<4>(0x1004) := 0x00028067  // jalr zero, t0, 0
  }
}
\end{lstlisting}

%% file: src-lang_evaluation.tex
\subsection{\ac{ISA} Language Evaluation}
\label{sec:isa_lang_evaluation}

To evaluate the expressive power of \ac{VADL}, we specified several processor architectures with distinct characteristics. Table~\ref{tab:spec_stats} summarizes key metrics for the \ac{ISA} sections of these specifications. \emph{Lines of Code Orig.} denotes the number of non-empty, non-comment lines in the unprocessed \ac{VADL} specification, whereas \emph{Lines of Code Exp.} reflects the line count after macro expansion, comment removal, and pretty-printing. \emph{Model Definitions}, \emph{Record Definitions}, and \emph{Model-Type Definitions} indicate the number of \texttt{model}, \texttt{record}, and \texttt{model-type} definitions within the specification, respectively. Similarly, the table reports the number of function, format, and instruction definitions before and after macro expansion, denoted by the \emph{Orig.} and \emph{Exp.} suffixes. Finally, it lists the total counts of all definitions, statements, and expressions before and after macro expansion.

\begin{table}
  \caption{\ac{ISA} specification statistics.}
  \label{tab:spec_stats}
  \footnotesize
  \setlength{\tabcolsep}{3pt}
  \begin{tabular*}{\textwidth}{@{\extracolsep{\fill}} l rrrrrrrr}
    \toprule
    \textbf{Metric} & \textbf{RV32I} & \textbf{RV64IM} & \textbf{VRisc} & \textbf{PPC64} & \textbf{Hexagon} & \textbf{AArch64} & \textbf{AArch32} & \textbf{Huge} \\
    \midrule
    Files                    & \num{1}    & \num{2}    & \num{1}    & \num{1}     & \num{1}     & \num{1}     & \num{1}      & \num{1}       \\
    \midrule
    Lines of code (orig.)    & \num{150}  & \num{576}  & \num{320}  & \num{1245}  & \num{1061}  & \num{1444}  & \num{1120}   & \num{303}     \\
    Lines of code (exp.)     & \num{226}  & \num{759}  & \num{715}  & \num{3522}  & \num{2431}  & \num{7543}  & \num{129597} & \num{165988}  \\
    \midrule
    Model definitions        & \num{8}    & \num{17}   & \num{21}   & \num{81}    & \num{28}    & \num{111}   & \num{188}    & \num{14}      \\
    Record definitions       & \num{0}    & \num{0}    & \num{0}    & \num{16}    & \num{0}     & \num{6}     & \num{5}      & \num{3}       \\
    Model-type definitions   & \num{0}    & \num{0}    & \num{0}    & \num{0}     & \num{2}     & \num{7}     & \num{6}      & \num{2}       \\
    \midrule
    Function definitions (orig.) & \num{0} & \num{3}   & \num{0}    & \num{13}    & \num{33}    & \num{25}    & \num{22}     & \num{0}       \\
    Function definitions (exp.)  & \num{0} & \num{4}   & \num{0}    & \num{13}    & \num{33}    & \num{25}    & \num{22}     & \num{0}       \\
    \midrule
    Format definitions (orig.)   & \num{6} & \num{7}   & \num{6}    & \num{23}    & \num{42}    & \num{33}    & \num{16}     & \num{4}       \\
    Format definitions (exp.)    & \num{6} & \num{7}   & \num{6}    & \num{23}    & \num{42}    & \num{33}    & \num{16}     & \num{4}       \\
    \midrule
    Instruction definitions (orig.) & \num{8}  & \num{14}  & \num{18}  & \num{47}   & \num{29}   & \num{49}    & \num{31}     & \num{6}       \\
    Instruction definitions (exp.)  & \num{37} & \num{64}  & \num{117} & \num{214}  & \num{203}  & \num{801}   & \num{10605}  & \num{32325}   \\
    \midrule
    Total definitions (orig.) & \num{75}  & \num{424}  & \num{188}  & \num{445}   & \num{711}   & \num{553}   & \num{342}    & \num{100}     \\
    Total definitions (exp.)  & \num{162} & \num{848}  & \num{485}  & \num{973}   & \num{1407}  & \num{3177}  & \num{36819}  & \num{97059}   \\
    \midrule
    Total statements (orig.)  & \num{21}  & \num{90}   & \num{59}   & \num{275}   & \num{141}   & \num{219}   & \num{231}    & \num{19}      \\
    Total statements (exp.)   & \num{57}  & \num{176}  & \num{234}  & \num{2076}  & \num{925}   & \num{3780}  & \num{80130}  & \num{68900}   \\
    \midrule
    Total expressions (orig.) & \num{554} & \num{1987} & \num{1168} & \num{5323}  & \num{5233}  & \num{4900}  & \num{4858}   & \num{1129}    \\
    Total expressions (exp.)  & \num{1496}& \num{4207} & \num{4525} & \num{17937} & \num{12821} & \num{46058} & \num{788205} & \num{2164589} \\
    \bottomrule
  \end{tabular*}
\end{table}

The first architecture is the RV32I base instruction set of the RISC-V architecture, as specified in Listing~\ref{lst:rv32i} in the appendix. RV32I is a simple 32-bit \ac{RISC} architecture without hardware multiplication or division, resulting in the smallest specification. RV64IM specifies the 64-bit RISC-V variant. It relies on a unified specification shared between the 32-bit and 64-bit versions that is configured via macros imported from a separate file.

VRisc is a variable-length, 32-bit \ac{RISC} toy architecture comparable in complexity to RV32I, supporting 8-bit, 16-bit, and 32-bit immediate values. Its \ac{ISA} comprises all instructions required to generate an efficient compiler, with instruction encodings of 16, 32, or 48 bits. VRisc serves as a testbed for variable-length instruction sets until a \ac{VADL} specification covering a substantial subset of the AMD64 architecture is available.

PPC64 specifies the complete scalar fixed-point subset of the 64-bit Power architecture. Power is a \ac{RISC} architecture of moderate complexity featuring two execution modes, several mode-dependent flags, multiple status registers, and numerous instruction variants, making it a well-suited benchmark for this evaluation.

Hexagon is a \ac{VLIW} architecture developed by Qualcomm and serves as a testbed for \ac{VADL}'s \ac{VLIW} capabilities. It features a limited set of predicated instructions, complex addressing modes, and architectural concepts not yet supported by \ac{VADL}, such as hardware loops and intra-bundle result forwarding. Consequently, the current \ac{VADL} specification covers only a subset of Hexagon's instructions and addressing modes.

AArch64 and AArch32 specify nearly complete integer instruction subsets of ARM's 64-bit and 32-bit architectures, respectively. Both \acp{ISA} rely on a status register and exhibit numerous instruction variants due to rich sets of addressing modes, scaled operands, and predicated execution (particularly in AArch32). Both specifications make heavy use of higher-order macros, resulting in line-of-code expansion factors of approximately 5 for AArch64 and 115 for AArch32. The AArch64 specification benefits substantially from register-file aliasing with distinct constraints, which reduces the number of instruction variants by roughly 500. Meanwhile, the high count of function definitions stems from the complex computation of immediate encodings for logical operations. Whereas the corresponding encoding routines in LLVM are implemented in \texttt{C++} using loops and destructive assignments, \ac{VADL} expresses them in a pure single-assignment style via recursive divide-and-conquer functions. Similarly, the large number of format definitions reflects both the variety of instruction encodings and the numerous system registers.

Finally, Huge is a synthetic architecture designed exclusively as a stress test for OpenVADL. It employs 64-bit instruction words to accommodate an exceptionally large instruction space. Although its underlying operations resemble those of RV64IM, the specification defines over 32,000 instructions by combining 25 predication conditions, 4 operand sizes, and 9 complex operand variants for every operation. Thanks to heavy macro usage, this entire instruction set is expressed in only a few hundred lines of code.

The architectures specified in \ac{VADL} to date demonstrate the language's strong capability to produce concise yet comprehensive specifications for complex \ac{RISC} processors as well as \ac{VLIW} and \ac{SIMD} architectures. The primary language features contributing to this expressive power in the \ac{ISA} section include the syntactic macro system, type inference, two distinct mechanisms for format specification, format access functions, constrained encoding definitions, enumerations and pattern matching, pure functions, register-file aliasing with constraints, regular-expression-based \ac{VLIW} instruction grouping with constraints, and tensor definitions using \texttt{forall}.

\subsection{\ac{MiA} Language Evaluation}
\label{sec:mia_lang_evaluation}

To evaluate the expressive power of the \ac{MiA} description language, we specified multiple microarchitectures for the RV32I instruction set. Note that these implementations can be readily retargeted to other architectures, for instance, by defining the corresponding operations. Table~\ref{tab:mia_spec_stats} reports the lines of code per specification, the length of the longest pipeline, and the number of functional units.

The \miaspec{p1} microarchitecture consists of a single stage that executes one instruction per machine cycle\footnote{As noted in Section~\ref{sec:controlflow_elimination}, a single machine cycle may span multiple clock cycles in hardware.}. The \miaspec{p2} microarchitecture separates instruction fetch from decoding and execution, while \miaspec{p3} further splits decoding and execution into distinct stages. The \miaspec{p5} microarchitecture implements the classic 5-stage \ac{RISC} pipeline, and \miaspec{p5\_fw} extends it with forwarding logic. \miaspec{p5\_alt} models the same microarchitecture as \miaspec{p5} using \ac{VADL}'s alternative pipeline construct, which allows an entire pipeline to be specified within a single definition where stages are delimited by a keyword.

\miaspec{OoO} is an experimental specification of a superscalar out-of-order processor, as introduced in Section~\ref{sec:advanced_mia}. Building on this design, \miaspec{OoO\_max} widens the decoders, adds functional units, and expands the internal buffers. Finally, \miaspec{OoO\_cfe} extends \miaspec{OoO\_max} with a complex four-cycle fetch-and-decode frontend that features a two-level branch predictor. A slower, more accurate predictor can override the fast initial prediction. Overall, these results demonstrate that \ac{VADL} can concisely express a broad spectrum of microarchitectural designs, and analogous specifications can be developed for the other \acp{ISA} evaluated in this section.

\begin{table}
  \caption{Microarchitecture specification statistics. The * symbol marks experimental specifications that have not been tested for feasibility with the current \ac{VADL} generators.}
  \label{tab:mia_spec_stats}
  \footnotesize
  \setlength{\tabcolsep}{3pt}
  \begin{tabular*}{\textwidth}{@{\extracolsep{\fill}} l rrrrrrrrr}
    \toprule
    \textbf{Metric} & \miaspec{p1} & \miaspec{p2} & \miaspec{p3} & \miaspec{p5} & \miaspec{p5\_alt} & \miaspec{p5\_fw} & \miaspec{OoO}* & \miaspec{OoO\_max}* & \miaspec{OoO\_cfe}* \\
    \midrule
    Lines of code    & 19 & 23 & 31 & 52 & 44 & 57 & 82 & 126 & 147 \\
    Pipeline length  &  1 &  2 &  3 &  5 &  5 &  5 &  5 &   5 &   8 \\
    Functional units &  1 &  1 &  1 &  1 &  1 &  1 &  3 &   6 &   6 \\
    \bottomrule
  \end{tabular*}
\end{table}

%% file: src-orig_implementation.tex
\section{Original \ac{VADL} Implementation}
\label{sec:orig_implementation}

This section presents the original \ac{VADL} compiler's implementation aspects, ranging from the compiler overview, language parsing, and domain-specific \ac{IR} to the detailed descriptions of the code generators for the different \ac{PDL} artifacts and concludes with a detailed evaluation.
First, Section~\ref{sec:orig_compiler_overview} gives an overview of the compiler's architecture.
Then, each principal component of the original \ac{VADL} implementation is discussed separately.

\subsection{\ac{VADL} Compiler Overview}
\label{sec:orig_compiler_overview}

One of the first requirements for the \ac{VADL} compiler development was that it should be implemented in an efficient and safe programming language.
This demanded a strongly, statically typed language which has array bounds checking and garbage collection.
The original \ac{VADL} compiler was implemented in Xtend~\cite{Bettini2016XtextXtend}.
Xtend is an extension of Java which has type inference and excellent support for string templates.

\input{src-figures-orig_vadl_overview.tex}

Figure~\ref{fig:orig_vadl_compiler_overview} presents the complete overview of the \ac{VADL} compiler design. Each processor specification starts as a plain \ac{VADL} text file. The parser is responsible for reading the \ac{VADL} text file, applying all macros, and generating an \ac{AST}. The language compiler handles symbol resolving, type inference, type checking, annotation checking, and constant evaluation. Section~\ref{sec:orig_parser} provides an overview of the parser.

Then, the compiler transforms the \ac{AST} into an \ac{IR}. In the original \ac{VADL} compiler, this is the \ac{VIR}. The \ac{IR} is the central data structure in the compiler, as all generators operate on it. It must be able to describe behavioral aspects (e.g., instruction semantics) and structural aspects (e.g., pipeline stages).
After creating the data structure, the compiler does well-known optimizations like removal of redundant operations or dead-code elimination.
Section~\ref{sec:vir} describes the \ac{VIR} in detail.

Generators that do not require knowledge about the microarchitecture can use the \ac{IR} directly after the transformation. These generators are the assembler and linker generator (see Section~\ref{sec:orig_asmgen}) and the instruction set simulator (see Section~\ref{sec:issgen}). Furthermore, the compiler generator can do most of its work without knowing the microarchitecture. However, the generated compiler might perform better at instruction scheduling if the generator has knowledge about the microarchitecture.

The compiler executes the microarchitecture synthesis (see Section~\ref{sec:orig_miasynth}) prior to generators that require microarchitectural details. This step is responsible for integrating \ac{ISA} and \ac{MiA} while also bridging much of the semantic gap between \ac{VADL} and the generated artifacts. As already mentioned, the compiler generator might use this information to tailor the code generation to the processor implementation. Furthermore, the cycle-accurate simulator generator (see Section~\ref{sec:orig_casgen}) and hardware generator (see Section~\ref{sec:orig_hdlgen}) rely on the microarchitecture synthesis.

%% file: src-figures-orig_vadl_overview.tex
\begin{figure}[h]

\begin{tikzpicture}[node distance=3cm]

\node (vadl) [ir] {VADL \\ Specification};
\node (parser) [ir_step, right of=vadl] {Parser};
\node (vir) [ir, right of=parser] {VIR};
\node (micro_syn) [ir_step, right of=vir] {Microarchitecture Synthesis};
\node (cas) [generator, below of=micro_syn, yshift=1cm] {Cycle-Accurate Simulator};
\node (hardware) [generator, right of=micro_syn] {Hardware};
\node (compiler) [generator, above of=micro_syn, yshift=-1cm] {Compiler};
\node (asm) [generator, above of=vir, yshift=-1cm] {Assembler \& Linker};
\node (iss) [generator, below of=vir, yshift=1cm] {Instruction Set Simulator};

\draw [arrow] (vadl) -- (parser);
\draw [arrow] (parser) -- (vir);
\draw [arrow] (vir) -- (micro_syn);
\draw [arrow] (vir) -- (compiler);
\draw [arrow] (vir) -- (asm);
\draw [arrow] (vir) -- (iss);
\draw [arrow] (micro_syn) -- (cas);
\draw [arrow] (micro_syn) -- (compiler);
\draw [arrow] (micro_syn) -- (hardware);

\end{tikzpicture}

\caption{Overview of the original \ac{VADL} Compiler Architecture. Yellow boxes represent generators.}
\label{fig:orig_vadl_compiler_overview}

\end{figure}

%% file: src-orig_impl_parser.tex
\subsection{Language Parser}
\label{sec:orig_parser}

\ac{VADL}'s parser is built on top of the well-established Xtext framework
\cite{eysholdt2010xtext}. Xtext is an open-source framework for the rapid
development of programming languages and \acp{DSL}. The framework takes a
grammar file as input and generates a Java-based ANTLR~\cite{ParrFischer2011}
parser, meta-model classes for the syntax tree, and parts of an
Eclipse-Modeling-Framework project for effortless Eclipse IDE
\cite{Holzner2004Eclipse, Steinberg2008EMF} integration. This work refers
to the generated syntax tree consisting of the meta-model classes as \ac{CST}.
To gain more control over the translation and shorten the IDE feedback time, we
turned off all non-LL(k) features in the Xtext-generated parser, i.e.
\emph{backtracking}, and implemented custom semantic predicates and code
actions~\cite{ParrQuong1994}. The gained context sensitivity is mainly needed
to support \ac{VADL}'s embedded macro language
(see Section~\ref{sec:parser-macros}). After the parsing and macro
expansion, the \ac{CST} is pruned and transformed into the more abstract
\ac{AST}. Please note that the \ac{CST} contains a lot of
syntax-related information and is primarily used to handle syntactic aspects of an input specification.
All further transformations and analyses, e.g., name inference or
type inference, are performed later on the \ac{AST}.

\label{sec:parser-macros}

The \ac{VADL} macro language is embedded in \ac{VADL}. We classify the macro
system as a pattern-based and syntactically typed macro system with support for
higher-order macro patterns~\cite{HochrainerKrall23}. To benefit from the IDE
support and check syntactic correctness at parse time, we split the macro
system into the \emph{parsing} and the \emph{expansion} phases.

The first phase parses the language and collects information on macro elements.
The second phase is a recursive expansion step of the collected macro elements.
Since \ac{VADL} does not perform symbol or type inference on the \ac{CST},
semantic predicates and code actions interact with a lightweight macro API to
compare and update symbol and type information.

A more detailed description of \ac{VADL}'s macro system, its types, and
implementation can be found in previous work~\cite{HochrainerKrall23}.

%% file: src-vir.tex
\subsection{The \ac{VADL} Intermediate Representation (VIR)}
\label{sec:vir}

In order to completely decouple specification and code generation while still reusing many intermediate artifacts generated from the \ac{VADL} specification, we introduced the \ac{VIR} layer.

This compiler \ac{IR} is designed to be very close to the abstraction level of an \ac{HDL} regarding the concepts of expressing sequential and parallel computation logic.
Since the current hardware code generator emits Chisel~\cite{bachrach2012chisel} (an \ac{RTL} abstraction) code, many similar constructs can be found in the \ac{VIR}.

The goal of the \ac{VIR} is to provide both structural and behavioral elements well suited to describing the various aspects of processor design. Because of that, several ideas emerged from existing state-of-the-art \acp{IR} for hardware descriptions, like the Low-Level Hardware Description (LLHD)~\cite{schuiki2020llhd} project or the Flexible Intermediate Representation for Register Transfer Level (FIRRTL)~\cite{izraelevitz2017reusability} project.
All of those \acp{IR} allow the specification of arbitrary \ac{HDL} designs.
The \ac{VIR}, however, focuses only on the requirements of processor designs.
Thus, aspects such as register files are first-class elements in the \ac{VIR}.

Besides the structural elements in the \ac{VIR} to represent memory components, register states, signals, ports, and overall processor definitions, the main focus of the \ac{VIR} is expressing data and control flows in the behavioral elements.
Two main definitions exist to describe behavior: \emph{functions} and \emph{processes}.

A \emph{function} is a behavioral block in the \ac{VIR} that describes arithmetic (combinational) functional behavior, which would correspond to a hardware logic computation that can be performed during a single (the same) clock cycle without requiring any memorization of a state. A \emph{process}, on the other hand, describes a state-aware computational behavior that can extend over several clock cycles. LLHD~\cite{schuiki2020llhd} introduced these concepts in their \ac{IR} design.

In order to represent computations executing within the same clock cycle, one or more \ac{VIR} \emph{instructions} in \ac{SSA} form~\cite{leung1999static} describe linearized operations over virtual registers.
One \ac{BB} represents a computation that can be executed in a single clock cycle.
These operations are grouped into a \ac{BB} and subsequently into a \ac{DAG} %\cite{thulasiraman2011graphs}
of \acp{BB} to express the data and control flow.
One or more \acp{BB} can then form single or multi-cycle hardware logic.

Several classical static analysis and transformation passes are implemented on the VIR level, e.g., constant folding, constant propagation, code motion, control flow elimination, inlining, and strength reduction.

Listing~\ref{lst:vir_example} shows the \ac{VIR} representation of the RISC-V \isainstruction{ADD} instruction. Readers familiar with LLVM~\cite{lattner2004llvm} or LLHD will see the similarity with the \acp{IR} of these projects. Every instruction is implemented as a process. In this case, the process consists of a single basic block. Note how the \ac{VIR} describes some processor elements (e.g., the register file \vadlresource{@RV32.X}) as first-class citizens of the \ac{IR}.

\begin{lstlisting}[language=vir, label=lst:vir_example, caption={VIR Code for the RISC-V \isainstruction{ADD} Instruction}]
process @RV32I.ADD.execute (b5 %rs2, b5 %rs1, b5 %rd) -> () = {
  lbl %bb1:
    %1  = const u32 4
    %2  = read b32 @RV32.PC
    %3  = add u32 %2, %1
    %4  = probe b5 %rs2
    %5  = probe b5 %rs1
    %6  = probe b5 %rd
    %7  = read b32 @RV32.X, %5
    %8  = read b32 @RV32.X, %4
    %9  = cast s32 %7
    %10 = cast s32 %8
    %11 = add s32 %9, %10
    %12 = cast b32 %11
    write b32 @RV32.X, %6, %12
    write b32 @RV32.PC, %3
    halt
}
\end{lstlisting}

The example shows several features of the \ac{VIR}. The syntax is inspired by the LLVM \ac{IR}. Variables are
prefixed with the percent symbol. Names for temporary variables are numbered increasingly. Each \ac{VIR} instruction
is explicitly typed, e.g., the \vadlresource{add} instruction has the type \vadlresource{u32}, which represents an unsigned
32-bit integer. Type conversions are done by explicit cast operations, as shown in lines 11 and 12.
Examples of bit fields are the input parameters, e.g., \vadlresource{rs2}, which is a bit field of width 5.
Registers and register files are accessed with \vadlresource{read} and \vadlresource{write} \ac{VIR} instructions.

The \vadlresource{lbl} in line 2 specifies the beginning of a \ac{BB} and
is used as a jump target when specifying control flow.
Since a \ac{VIR} process represents a stateful computation with a possible duration of more than a single clock cycle,
the \vadlresource{probe} instruction is used to access stateful variables, in this example, the input parameters.
Stateful variables can change their value over time.
In this example, the \vadlresource{probe} instructions represent taking the value of the
input parameters at a certain point in time. One could think of this as \emph{probing} the input wires.

%% file: src-orig_compgen.tex
\subsection{Original \ac{VADL} Compiler Generator}
\label{sec:orig_compgen}

This section provides an overview of the design
and implementation of the compiler generation component, highlighting
its key features and functionality.

\subsubsection{Overview}

The compiler generation component closes the semantic gap between the
high-level \ac{ISA} specification of instruction semantics and the low-level compiler implementation. Similar to our structure in the \ac{VADL}
tool, modern compilers can usually be split into three main components
\cite{stallman2020gcc, lattner2004llvm}: a frontend for source-level parsing, an \ac{IR} for
target-independent optimizations and a target-specific backend for
target-specific optimizations and generating assembly or bytecode.

One of the most proven approaches for automated compiler generation is
to limit the generation process to the target-specific backend, reusing
the compiler's parsing and optimization capabilities~\cite{azevedo2005archc,
zivojnovic1996lisa}. By applying this technique, the generated implementation
is compatible with state-of-the-art compiler frameworks, enabling us to
profit from previous works in compiler research. As a proof of
concept, we implemented a \ac{VADL} compiler backend generator for
the well-established LLVM compiler toolchain~\cite{lattner2004llvm}.
In order to keep the support of additional compilers open, we have
categorized the compiler backend generation into two subtasks:
Extract generic compiler information from the specification and produce
compiler backend-specific source files. The \ac{GCB} component reduces and
transforms information provided by the \ac{VIR} into
compiler-generator-relevant information. The created \ac{IR}, mainly
consisting of \acp{DAG}, is then passed to a specific compiler toolchain
component, producing output files specific to a target compiler's backend.
In our case, we implemented the \ac{LCB} module, responsible for producing
a working LLVM backend. Figure~\ref{fig:orig_gcb_lcb_overview} gives an overview
of the main steps done by the compiler generator component.

\input{src-figures-orig_gcb_lcb_overview.tex}

\subsubsection{Generic Compiler Backend}

The \ac{GCB} module is the core component of the compiler generator.
It lifts the \ac{VIR} entities to a new abstraction, only retaining
information relevant to the compiler model. The resulting intermediate
representation is the basis for further compiler synthesis steps. While we
mainly focused on generating an LLVM backend, the \ac{GCB} \ac{IR} could
be extended to support a variety of compiler backend targets.

The \ac{GCB} \ac{IR} acts as a further abstraction layer over necessary
compiler elements. Introduced abstractions mostly behave like glue code between
\ac{VIR}, C++ sources and newly collected or synthesized information.
At the beginning, the \ac{GCB} generation starts by bundling the low-level \ac{VIR}
and generated C++ source units for relocations and immediate encoding,
decoding and predicate functions into high-level compiler elements.
During this first step, most of the core structure of the \ac{GCB} model is
created. The generated model can be seen as a processor skeleton
extended during the execution of the \ac{GCB} passes. All further passes
mainly deal with analyzing instruction semantics.

Next, the dynamic format fields are examined to recognize the instruction
operands and assign them to a specific type. The preparatory work in the
\ac{VIR} is crucial here, as it minimizes the semantics and simplifies the
recognition of register accesses or immediates used as addresses or arithmetic
operands. The categories used for instruction operands are \emph{register-}
or \emph{immediate}-operands. Constant register class access, e.g., {\tt X(0)},
single register access, or the use of register values or immediates as memory
addresses are all managed inside the instruction behavior and have no impact on
the operand type. The only additional distinction is whether the operand is used as
an input or output operand. In contrast to the LLVM specification language
\emph{TableGen}, \ac{VADL} is able to work with multiple input and output
operands. To deal with these shortcomings of target backends, the \ac{GCB} is
able to transform most instructions into a suitable form by duplicating
operands that occur as input and output operands, and generating additional
instruction operand constraints. After the operands are collected, an
additional analysis flags immediate operands that are used as relative and
absolute jump addresses, respectively.

Furthermore, the \ac{GCB} models all kinds of register-related elements as
\emph{register resource}. First, a distinction is made between single hardware
registers and register classes. While a single hardware register only contains
a \ac{VIR} type, a register class is a set of hardware registers. Second, the
register classes are separated into \emph{hardware} register classes and
\emph{virtual} register classes. A hardware register class must provide
registers for each given index. A virtual register class, on the other hand,
is a modification of a hardware register, modeling constraints and
slight modifications, e.g., replacing a single register with a zero register or
restricting specific indices. This becomes useful as some hardware instructions
that access register files have particular behaviors for specific index values.
The \ac{VADL} specification may use the {\tt alias register} mechanism
to restrict or modify the access of registers or register files. To model this behavior, the
\ac{GCB} analyzes these artificial resources, collects information on the
different indices, and creates \emph{virtual register classes} for the affected
instruction operands. Since single hardware registers are not viewed as operands,
they need special attention. A separate register analysis traverses the
instruction behavior and marks the single registers used for each instruction
individually. The information gained is helpful for instruction selection in
the backend.

The \ac{VADL} tool automatically creates a relocation symbol and function
for every specified modifier {\tt relocation}. However, this
representation usually needs to be more high-level.
The GCB creates specific low-level relocation behavior based on
the instruction's immediate operands to use relocations during linking.
This process looks at every
instruction separately, but future work to combine instructions with similar
formats into bundles is already planned. After the relocation management step,
the compiler backend has information on modifying the bit-fields of encoded
instructions to perform specific relocations.

Moreover, a significant transformation done by the \ac{GCB} is converting the
instruction semantics to a \ac{DAG} form. Alternatively, we experimented with
keeping the \ac{VIR} representation, which turned out to be unnecessarily
complex as most of the applied analyses, transformations, and especially
pattern-matching tasks are better suited for \acp{DAG}. In an iterative
process, the initially rudimentary graph nodes are merged into more complex
node patterns. This process serves a dual purpose. Firstly, it expedites the
identification of significant patterns in the various \ac{ABI} sequences, and
secondly, it guarantees a more concise representation of the instruction
semantics. The implemented \ac{DAG}
node kinds are inspired by the LLVM \emph{TableGen} nodes.
The reason for this is that \emph{TableGen} is a well-developed
language and secondly, it shortens the development to generate an
LLVM-compliant backend. This decision does not impact the generality of the
\ac{GCB}.

The LLVM backend requires C++ helper functions that produce specific
sequences of instructions to function correctly. These sequences are
responsible for copying registers, loading memory addresses, dealing
with complex immediates, or performing memory offset calculations. LLVM does not
deduce these sequences from the provided patterns.
Since statically retrieving this information from the generated \emph{TableGen}
patterns is impractical, \ac{VADL} additionally performs a simple
instruction selection for the mentioned sequences. Most of the C++ helper
functions are also relevant for particular \ac{ABI}-specific behavior.
In contrast to a simple value move, calling conventions or more complex loads
with symbols cannot be derived automatically.
C++ code which deals with more complex or ABI-related sequences is
synthesized using information from the \ac{VADL} \ac{ABI} section.
See Section~\ref{sec:abi_section} for more details.

Finally, all \ac{DAG} patterns are checked for semantically equivalent
alternative forms. A separate pass performs semantics-preserving transformations
and stores the newly generated patterns with their original instruction. This
step is beneficial to achieve better coverage of necessary comparison
patterns as the actual hardware usually only provides the minimal complete set
of compare operations.

This concludes the generation of a general processor model, which is passed
to a specific backend generator.

\subsubsection{LLVM Compiler Backend}

The \ac{LCB} starts by applying a lowering, followed by a validation pass
on the received generic processor model.

The lowering step transforms the generic model into a state where it only needs to be
output by the emitters. First, generated C++ classes and functions are
adapted to be compliant with the LLVM infrastructure, i.e., modification of
types and signatures. Second, the lowering pass tries to legalize the generated
patterns. This primarily consists of casting immediate operand types to a
suitable
size of a power of two and forcing a uniform operation bit width for generated
instruction patterns.
Finally, it removes incomplete or irrelevant information that LLVM or
\emph{TableGen} cannot process.

Since the lowering process modifies and removes information, the \ac{LCB}
needs to validate the final processor model. Currently, the validation
consists of ensuring the existence of LLVM essential sequences, specific purpose
registers or \ac{ABI}-relevant information to successfully compile simple test
programs.

The remaining part of the \ac{LCB} consists of individual emitters and
templates for
each LLVM source file. This enables us to locate files and adapt their
content quickly if needed.
After the lowering step, the processor model is no longer
transformed or modified. All needed information is contained inside
the model and is queried through the different emitting strategies.

Finally, the backend structure generated by the \ac{LCB} is designed to
be copied over an existing LLVM project. The \ac{LCB} generates a configuration script for convenience, which can be used to move the
generated backend and compile the LLVM project with suitable settings.

\subsubsection{LLVM Compiler Backend - Instruction Matching} \label{sec:originalVADL_InstructionMatching}

The \ac{LCB}'s task is to provide a mapping between the program representation and machine instructions by generating tree patterns from the \ac{VADL} specification. Programs are represented by nodes in an \emph{Instruction Selection Graph} in LLVM. Based on the tree patterns, LLVM creates a cover of the nodes in the \emph{Instruction Selection Graph} during the compilation of a program. In other words, a cover represents the instructions chosen by the compiler that are equivalent to the input program.
\ac{LCB} performs the following steps to generate the tree patterns for the \emph{Instruction Selection}:

\begin{itemize}
\item Convert the \ac{VADL} specification into \ac{VIR} representation
\item Extract the semantics from the \ac{VIR} by constructing a dataflow graph
\item \ac{GCB} matches known patterns to recognize instructions based on their semantics
\item \ac{LCB} converts the dataflow graphs into \emph{TableGen} records used by LLVM
\end{itemize}

In our example, the code from Listing~\ref{lst:vadl_add_instruction_src} is converted into \ac{VIR} which you can find in Listing~\ref{lst:vadl_add_instruction_vir}. Next, the \ac{VIR} is converted into a dataflow graph which captures the semantics, as depicted in Figure~\ref{fig:vadl_add_instruction_sem}. Lastly, \ac{LCB} emits the mappings in \emph{TableGen}, as shown in Listing~\ref{lst:vadl_add_instruction_tablegen}.
Note that it is not always possible to pattern match the semantics of an instruction. \emph{TableGen} does not support instructions with multiple results. So whenever the semantic representation is not a tree, mappings cannot be emitted. % TODO: This seems a bit unclear; Is the semantic representation the IR produced by GCB? Also, is this supposed the mean that pattern matching is not possible if and only when an instruction has multiple results? Is the "semantic representation" not a tree if and only if there are multiple results?

\begin{figure}[h]
\begin{minipage}[b]{0.51\textwidth}
    \begin{lstlisting}[language=vadl, deletekeywords={mnemonic}, caption={\ac{VADL} ADD instruction}, label=lst:vadl_add_instruction_src]
instruction ADD : Rtype =
{
    X(rd) := ((X(rs1) as SInt) + (rs2)) as Bits
}
...
\end{lstlisting}
\end{minipage}
\begin{minipage}[b]{0.45\textwidth}
    \begin{lstlisting}[language=vir, deletekeywords={mnemonic}, caption={ ADD instruction's \ac{VIR}}, label=lst:vadl_add_instruction_vir]
process @RV32I.ADD.execute.gcb
    (b5 %rs2, b5 %rs1, b5 %rd) -> () = {
    lbl %bb819:
    %6372 = probe b5 %rs2
    %6373 = probe b5 %rs1
    %6374 = probe b5 %rd
    %6375 = read b32 @RV32.X, %6373
    %6376 = read b32 @RV32.X, %6372
    %6377 = cast s32 %6375
    %6378 = cast s32 %6376
    %6379 = add s32 %6377, %6378
    %6380 = cast b32 %6379
    write b32 @RV32.X, %6374, %6380
    halt
}
\end{lstlisting}
    \end{minipage}

\begin{minipage}[b]{0.48\textwidth}
\begin{tikzpicture}
[
    node distance={35mm},
    thick,
    unused/.style =
        { draw
        , dotted
        , minimum width=3em
        , minimum height=4em
        , fill=black!10
        },
]

\tikzstyle{n} = [ellipse, rounded corners, minimum height=0.5cm,text centered, draw=black, outer sep=0mm]

\node[n] (rd) at (0, 0) {rd: Bits<5>};
\node[n] (rs1) at (2.5, 0) {rs1: Bits<5>};
\node[n] (rs2) at (5, 0) {rs2: Bits<5>};

\node[n] (rdx) at (0, -1) {X: Bits<32>};
\node[n] (rs1x) at (2.5, -1) {X: Bits<32>};
\node[n] (rs2x) at (5, -1) {X: Bits<32>};

\node[n] (add) at (3.75, -2) {add: SInt<32>};
\node[n] (set) at (1.8, -3) {set: Void};

\draw [arrow] (rd) -- (rdx);
\draw [arrow] (rs1) -- (rs1x);
\draw [arrow] (rs2) -- (rs2x);
\draw [arrow] (rdx) to[bend right] (set);
\draw [arrow] (rs1x) -- (add);
\draw [arrow] (rs2x) -- (add);
\draw [arrow] (add) -- (set);

\end{tikzpicture}

\caption[short]{\ac{DAG} representation of the ADD instruction's semantics}
\label{fig:vadl_add_instruction_sem}
\end{minipage}

\begin{minipage}[b]{0.49\textwidth}
\begin{lstlisting}[language=tablegen, deletekeywords={mnemonic}, caption={ ADD instruction's TableGen}, label=lst:vadl_add_instruction_tablegen]
def ADD : Instruction
{
    ...
    let OutOperandList = (outs X:$rd);
    let InOperandList = (ins X:$rs1, X:$rs2);
    ...
}

def : Pat<(add X:$rs1, X:$rs2),
        (ADD X:$rs1, X:$rs2)>;
\end{lstlisting}
\end{minipage}

\end{figure}

%% file: src-figures-orig_gcb_lcb_overview.tex
\begin{figure}
\begin{center}
\begin{tikzpicture}
[
    node distance={20mm},
    thick,
    unused/.style =
        { draw
        , dotted
        , minimum width=6em
        , minimum height=6em
        , fill=black!10
        },
    block/.style =
        { rectangle
        , draw=black
        , thick
        , fill=white
        , text width=5em
        , align=center
        , rounded corners
        , minimum height=4em
        },
    blockSmall/.style =
        { rectangle
        , draw=black
        , thick
        , fill=white
        , text width=5em
        , align=center
        , rounded corners
        , minimum height=2em
        },
    blockSmallInvis/.style =
        { rectangle
        , draw=white
        , thick
        , fill=white
        , text width=5em
        , align=center
        , rounded corners
        , minimum height=2em
        },
]

\path
    (-3, 0) node[unused]
        (VIR)
        {VIR}
    (0, 0) node[block]
        (VirToGcb)
        {Processor Skeleton}
    (2.5, 0) node[block]
        (OperandAnalysis)
        {Operand Analysis}
    (5, 0) node[block]
        (RegisterAnalysis)
        {Register Analysis}
    (7.5, 0) node[block]
        (RelocationManagement)
        {Relocation Management}
    (7.5, -2) node[block]
        (DAGGeneration)
        {Instruction DAG Generation}
    (5, -2) node[block]
        (DAGRefinement)
        {Instruction DAG Refinement}
    (2.5, -2) node[block]
        (SequenceDetection)
        {Sequence Detection}
    (0, -2) node[block]
        (SemanticAnalysis)
        {Semantic Analysis}
    (-2.5, -5) node[block]
        (LLVMLowering)
        {LLVM Lowering}
    (0, -5) node[block]
        (Validation)
        {Validation}
    (2.5, -4) node[blockSmall]
        (FileEmitter1)
        {LLVM File Emitter 1}
    (2.5, -5) node[blockSmallInvis]
        (FileEmitterDots)
        {. . .}
    (2.5, -6) node[blockSmall]
        (FileEmitterN)
        {LLVM File Emitter n}
    (6, -5) node[unused]
        (LLVMBackend)
        {LLVM Backend}
    ;

\node
    [ rectangle
    , draw=black!30
    , dotted
    , fit=
        (VirToGcb)
        (OperandAnalysis)
        (RegisterAnalysis)
        (RelocationManagement)
        (DAGGeneration)
        (DAGRefinement)
    , inner sep=3mm
    , label={GCB}
    ]
    (GCB) {};

\node
    [ rectangle
    , draw=black!30
    , dotted
    , fit=
        (LLVMLowering)
        (Validation)
        (FileEmitter1)
        (FileEmitterDots)
        (FileEmitterN)
    , inner sep=3mm
    , label={[shift={(-3.0,0.0)}] LCB}
    ]
    (LCB) {};

\draw[->] (VIR) -- (VirToGcb);
\draw[->] (VirToGcb) -- (OperandAnalysis);
\draw[->] (OperandAnalysis) -- (RegisterAnalysis);
\draw[->] (RegisterAnalysis) -- (RelocationManagement);
\draw[->] (RelocationManagement) -- (DAGGeneration);
\draw[->] (DAGGeneration) -- (DAGRefinement);
\draw[->] (DAGRefinement) -- (SequenceDetection);
\draw[->] (SequenceDetection) -- (SemanticAnalysis);
\draw[->] (SemanticAnalysis) -- (LLVMLowering);
\draw[->] (LLVMLowering) -- (Validation);
\draw[->] (Validation) -- (FileEmitter1);
\draw[->, dotted] (Validation) -- (FileEmitterDots);
\draw[->] (Validation) -- (FileEmitterN);
\draw[->] (FileEmitter1) -- (LLVMBackend);
\draw[->,dotted] (FileEmitterDots) -- (LLVMBackend);
\draw[->] (FileEmitterN) -- (LLVMBackend);

\end{tikzpicture}

\caption[short]{Original \ac{VADL} Compiler Generator Overview}
\label{fig:orig_gcb_lcb_overview}
\end{center}
\end{figure}

%% file: src-orig_asmgen.tex
\subsection{Original Assembler and Linker Generator}
\label{sec:orig_asmgen}

This section discusses the assembler generation within the original \ac{VADL} \ac{LCB} prototype. Its task is to emit the assembler and disassembler components of the generated backend. The architecture of the generated artifact is based on the relevant parts of LLVM's architecture, which is described in the official LLVM documentation\footnote{\href{https://llvm.org/docs/}{https://llvm.org/docs/}}. The main components in this architecture are the \emph{instruction printer}, \emph{assembly parser}, \emph{disassembler}, \emph{machine code emitter}, and \emph{linker}.

As discussed in Section~\ref{sec:background}, a native program has two important persistent representations - assembly and object code. In addition to the persistent manifestation, LLVM uses internal data structures during processing. Figure~\ref{fig:orig_asm_overview} illustrates the transformations between the representations and which component is responsible for which transformation.

The following sections discuss these components with a focus on how the \ac{VADL} tooling can extract the required information from the specification to generate these components. This section also introduces a straightforward approach to generating grammar rules from the assembly formatting function. Lastly, this section elaborates on handling relocations at the boundary between the assembler and linker.

\input{src-figures-orig_asm_components.tex}

\subsubsection{Instruction Printer}

This component is responsible for transforming the internal representation into assembly text. The compiler uses this component to emit assembly files. Furthermore, the disassembler uses it to print the decoded instructions to a command line interface. Implementing this functionality requires knowing how to express an instruction as text. \ac{VADL} captures this relation with the assembly printing functions in the \ac{ISA} section. The \ac{VADL} tool uses the regular translation path via an implemented C++ code generator to obtain an implementation for each instruction type.

\subsubsection{Assembly Parser}

The inverse to printing is parsing the assembly text into an internal representation. The assembly parser implements this transformation. The assembly description element is the primary information source for this task. Section~\ref{sec:assembly_description_section} introduced this definition. Generating a parser from a formal grammar is a well-studied problem~\cite[Chapter 3]{cooper2011engineering}. The \ac{VADL} tool generates an LL(1) recursive-descent parser from the grammar specification. These operations include recording, transforming, and validating values extracted from the text. Section~\ref{sec:orig_eval_asm_link} discusses some limitations of this parser implementation in the context of assembly languages.

After parsing, the algorithm identifies a set of named operands based on the grammar specification. Then, it compares the name and content of these operands to the instructions provided by the \ac{ISA}. For example, matching a RISC-V \texttt{ADD} instruction requires \texttt{mnemonic}, \texttt{rd}, \texttt{rs1}, and \texttt{rs2} operands. Furthermore, the \texttt{mnemonic} operand must equal \texttt{"add"}. After finding a match, the parser instantiates the corresponding internal representation. If the algorithm finds no matching instruction, the tool reports an error to the user. The grammar validation ensures that the operand names match with at least one instruction. Beyond that, the parser assures further invariants, such as asserting that a constant's value does not exceed the associated range in the matched instruction. During this process, the parser applies the necessary immediate decoding functions defined in the format. Determining the transformation functions is straightforward because the parser knows the operand name and the instruction type.

\subsubsection{Disassembler}

The central task in generating the disassembler, apart from understanding the object file format, is decoding the instructions into the internal representation. In \ac{VADL}, the instruction format and constant format fields define the decoding function. LLVM allows defining this information in a TableGen file. From this, the infrastructure can automatically synthesize the decoder.

\subsubsection{Machine Code Emitter \& Linker}

The machine code emitter encodes the internal representation in an object file format. This task involves encoding instructions and recording metadata. LLVM can synthesize the encoding function from the TableGen file. In addition, the final object file must include relocation entries. This information is necessary to convey program details to the subsequent linkage step. This information is necessary for using symbols in assembly (e.g., function names). Most importantly, a relocation entry contains a type and a symbol name. The relocation type entails information on how the linker shall resolve the symbol (e.g., relative or absolute). For example, one relocation type could describe the usage of a symbol that is resolved relative to the current instruction (e.g., RISC-V branches).

Before the assembler can record relocation definitions, the assembler and linker must agree on the supported relocation types. The \ac{VADL} generator emits declarations for the relocations from the \ac{ISA} section. In addition, the tool synthesizes generic relocations for immediate format fields. The latter type is required so that users do not have to define a relocation that applies no transformation to the value. For example, the relative RISC-V branching instructions use this feature in our processor description.

The biggest concern when generating the \emph{linker} is understanding the object file format. In the LCB, the LLVM infrastructure provides this capability. The architecture-specific code focuses on applying relocations to the encoded instructions.

\subsubsection{Grammar Inference}
\label{sec:grammar_inference}

Before generating the components mentioned above, the \ac{VADL} tooling infers grammar rules based on formatting functions from the \ac{ISA}. The problem of synthesizing a formal grammar from a pretty printer is related to program inversion, as the grammar defines the inverse operation. The function's parameters are the instruction's operands. The result of each formatting function is a plain assembly string. As a result, the inverted function computes the operands from the assembly string. Our implementation combines multiple ideas from program inversion to leverage this relationship.

The first observation is that, given an interpreter for \ac{VIR} functions, an algorithm can synthesize an inverter by trying all possible input combinations and recording their output. This result captures a unique input-output mapping if the pretty printer is injective, i.e., the computed output values are unique. The program could obtain a formal grammar by generating an alternative over the outputs from the mapping. Each choice is augmented with the initial input values, resulting in an inverse mapping from output to input values. However, this becomes impractical as the input domain size can increase quickly. In addition, the grammar structure resulting from this approach is ill-suited for many critical aspects of a parser. Essentially, the grammar boils down to expressions that check if the input text matches precisely with a particular string, such as \texttt{"add x1, x2, x3"}, and then assign specific values to the corresponding variables. Therefore, grammar rules no longer contain structural information. This information is crucial for tasks like automatic error message generation.

Another approach is synthesizing the grammar rule from the \ac{VIR} function by defining additional grammar generation semantics for each instruction. This approach results in a well-structured grammar and can handle large input domains as the algorithm does not have to interpret all possible values. However, once control structures are involved, writing a general inversion algorithm can take time and effort. The primary reason is that the inverter must be able to handle all \ac{VIR} instructions used in the formatting functions. In addition, the inverter must consider interactions between multiple instructions, for example, if the formatting function uses multiple conditional constructs with the same selection input.

The \ac{VADL} tooling uses a hybrid approach to remedy the problems of both techniques. It directly generates suitable grammar rules for widely used \ac{VIR} instructions, such as string concatenation. Once the algorithm encounters a \ac{VIR} instruction that it cannot directly process, it switches to an interpretation-based grammar inference technique. Implementing this approach allows leveraging synergies with other components that require an interpreter.

A detailed description of the assembler generator can be found in~\cite{schwarzinger2022flexible}.

%% file: src-figures-orig_asm_components.tex
\begin{figure}[h]
    \begin{tikzpicture}[node distance=3cm]

        \node (compiler) [internal_component] {Compiler IR};
        \node (assembly) [external_component, left of=compiler, xshift=-1cm] {Assembly Files};
        \node (object) [external_component, right of=compiler, xshift=1cm] {Object Files};

        \draw [arrow] (compiler) to[bend left] node[midway, below] {\emph{Instruction Printer}} (assembly);
        \draw [arrow] (compiler) to[bend right] node[midway, below] {\emph{Machine Code Emitter}} (object);
        \draw [arrow] (assembly) to[bend left] node[midway, above] {\emph{Assembly Parser}} (compiler);
        \draw [arrow] (object) to[bend right] node[midway, above] {\emph{Disassembler}} (compiler);

    \end{tikzpicture}

    \caption{Overview of Generated Components and Their Inputs and Outputs. Red Boxes Denote External Representations.}
    \label{fig:orig_asm_overview}
    %\Description{TODO}
\end{figure}

%% file: src-issgen.tex
\subsection{Instruction Set Simulator Generator}
\label{sec:issgen}

The \ac{ISS} of \ac{VADL} is a functional instruction set simulator only.
It does not emulate nonfunctional behavior such as caches as the \ac{CAS} does.
The design space for implementing an \ac{ISS} offers a vast number of options.
Because of limited resources, we decided to go for a simple and generic but efficient simulator.
Therefore, an \ac{ISS} using \ac{JIT} technology was out of scope. Instead, we opted for an implementation based on efficient interpretation.
The fastest interpretation technique available is \ac{DTC}, where the instruction memory only contains pointers to the code which emulates the instruction.
For the simulation of von Neumann architectures, \ac{DTC} requires an additional instruction memory mirroring the data memory.
Depending on the instruction size and the size of pointers, this instruction memory would be a multiple of the size of the data memory.
Furthermore, most entries of the instruction memory would be empty and the initialization overhead of these empty entries would be huge.
Therefore, the \ac{ISS} employs a hashmap where an instruction memory address is mapped to a pair comprising a pointer to the instruction's emulation code and the instruction at that address.
This design also eliminates a range check for the instruction pointer, as only valid addresses are entered into the hashmap.
When it is necessary to simulate self-modifying code, there are two possibilities:
It can be checked whether the returned instruction is equal to the instruction in the data memory.
Or it can be checked at every write to the data memory whether the write invalidates an entry in the hashmap.
\ac{VADL}'s \ac{ISS} uses the first checking technique.

\begin{minipage}[b]{0.51\textwidth}
\begin{lstlisting}[label={lst:iss_addi_vadl},caption={{\tt ADDI} instruction definition in \ac{VADL}},language=vadl,linerange={13-24}]
format Itype : Word =     // Itype
  { imm    : Bits<12>     // [31..20]
  , rs1    : Index        // [19..15]
  , funct3 : Bits<3>      // [14..12]
  , rd     : Index        // [11..7]
  , opcode : Bits<7>      // [6..0]
  , immS   = imm as SIntR 
  }

instruction ADDI : Itype = {
  X(rd) := ((X(rs1) as SInt) + immS) as Bits
  }
\end{lstlisting}
\end{minipage}
\hfill
\begin{minipage}[b]{0.48\textwidth}
\begin{lstlisting}[label={lst:iss_addi_cpp},caption={{\tt ADDI} translated to {\tt C++}},language=cpp,linerange={1-13}]
typedef unsigned int sint32;
typedef unsigned int uint32;

sint32 X[32];

inline uint32 ADDI (const uint32 PC,
                    const uint32 instr) {
  uint32 rd   = (instr >>  7) & 0x1f;
  uint32 rs1  = (instr >> 15) & 0x1f;
  sint32 immS = (sint32) instr >> 20;
  X[rd] = X[rs1] + immS;
  return PC + 4;
  }
\end{lstlisting}
\end{minipage}

In the \ac{ISS}, an instruction specification is represented as an inline {\tt C++} function which takes the program counter and the instruction word as arguments and returns the updated program counter.
Simple encoded format fields are derived via shifting and masking.
Complex encoded format fields can be predecoded and additionally stored in the hashmap and only a pointer to these elements is passed to the {\tt C++} function.
The presented translation in Listing~\ref{lst:iss_addi_cpp} is simplified.
Because of the transformations, casts, and optimizations on the \ac{VIR}, the generated code only contains assignments with a single binary expression and mangled names.
The {\tt C++} compiler optimizes and simplifies the expressions in the generated code.
Therefore, the \ac{ISS} generator only has to apply a few optimizations during {\tt C++} code generation.
The \ac{ISS} main interpreter loop consists only of a single access to the hashmap (which returns the address of the label where an invocation of the inlined translated function has been positioned) and a jump to that address.

The generation of the {\tt C++} code is straightforward.
There is just a simple analysis of the assignment to and the use of the program counter to add the correct program counter updating code.
The decoder is already available in the \ac{VIR} and can be reused.
It is combined with the function which adds new elements to the hashmap on a miss in the map.

To facilitate the validation of the generated simulators and of processor specifications, the simulator supports trace generation and co-simulation.
The amount of checked execution state can be controlled by command-line options.
Co-simulation has been done against other simulators and real hardware (RISC-V, AArch64).
The validations have been conducted using existing processor validation suites.
A detailed description of the simulator generator and an evaluation of performance and co-simulation can be found in a thesis~\cite{Mihaylov2023Optimized}.

%% file: src-orig_miasynth.tex
\subsection{Original \ac{VADL} Microarchitecture Synthesis}
\label{sec:orig_miasynth}

The Microarchitecture Synthesis is an intermediate step executed before obtaining a cycle-accurate simulator or a hardware schematic. Extracting this step is sensible because both artifacts require identical analysis and transformations. After all, the cycle-accurate simulator shall be able to emulate the hardware implementation. Before generating an artifact, the compiler must lower the high-level microarchitecture to standard VIR processes. I.e., the inputs to the microarchitecture synthesis are the \ac{ISA} specification and the \ac{MiA} specification. The output is a program in the \ac{VIR}~\ref{sec:vir} representing the unification of both the \ac{ISA} and the \ac{MiA}.  This endeavor currently consists of six major tasks:

\begin{enumerate}
    \item By splitting instructions into parts, the compiler maps the instruction semantics of the \ac{ISA} to the placeholders in the \ac{MiA}. The system then replaces the placeholders with the corresponding parts of the instruction semantics.
    \item The compiler synthesizes implementations for the \vadlresource{decode} built-ins. These are implemented as matchers for bit patterns.
    \item The system creates read ports and write ports for the resources. After that, the algorithm assigns these ports to VIR instructions that access resources.
    \item The next step lowers logic elements to VIR processes. In this stage, the compiler generates the control and hazard detection units.
    \item The compiler synthesizes the processor core itself. The primary goal is to allocate resources for pipeline registers and interconnect the control and pipeline components.
    \item The control flow is eliminated and replaced by conditional instructions and multiplexers.
\end{enumerate}

The artifact generators can take over once the compiler has lowered the abstractions. Since the microarchitecture mainly comprises standard \ac{VIR} processes after this stage, the mapping to an artifact-specific \ac{IR} is straightforward. The following section elaborates on these steps in further detail.

\subsubsection{Instruction Resolving}
In \ac{VADL}, each instruction is defined as a separate entity. However, a processor pipeline has to handle \emph{all} of the defined instructions. Thus, the system first has to establish a comprehensive view over all instructions. This analysis identifies overlaps and commonalities between instructions to make reuse of components possible. For example, the analysis determines the minimum number of read ports needed.

As a consequence of this, the most crucial step in microarchitecture synthesis is integrating the instructions' semantics into the processor pipeline.  Figure~\ref{fig:microarchitecture_synthesis_overview_mia} illustrates the idea of this step. The synthesis must map the two instruction definitions on the left-hand side to the partially displayed pipeline specification on the right-hand side. In this particular case, the algorithm maps three register read operations to the decode stage of the processor and the two additions to the execute stage. Astute readers may notice that inserting three read operations into the decode stage is unnecessary. Because an instruction cannot simultaneously be an \isainstruction{ADDI} and a \isainstruction{SW} instruction, the final decode stage should only contain two register reads. Similarly, the two instruction implementations should share the adder that computes the arithmetic operations.

\input{src-figures-microarchitecture_synthesis.tex}

The \ac{VADL} tool tackles all issues mentioned above by leveraging an augmented \ac{DFG} of all instruction semantics. Figure~\ref{fig:dfg_example_addi} depicts the \ac{DFG} for the \isainstruction{ADDI} instruction from Figure~\ref{fig:microarchitecture_synthesis_overview_mia}. Each node constitutes an operation. Incoming edges denote the input operands of a node, while outgoing edges define how the result of a node is distributed to other operations. The compiler associates each occurrence of an instruction variable in the  \ac{MiA} definition with a \ac{DFG}.
During instruction resolving, the \ac{DFG} is reduced stepwise towards the root nodes, starting from the leaves. Each reduction step results in the emission of one or more \ac{VIR} instructions and represents the current execution state of the instruction as it progresses through the processor pipeline. Thus, the graph represents the outstanding computations, since all completed computations have already been collapsed in the leaf nodes. A reduction step does not represent a control point in the \ac{MiA}.
Red and blue nodes denote read and write operations, while black nodes denote future pure computations. Green nodes represent values that the processor has already computed. We refer to these values as \emph{available} nodes. For example, the green nodes are the format fields in Figure~\ref{fig:dfg_example_addi}. Figure~\ref{fig:dfg_example_addi_v2} depicts the \ac{DFG} for the same instruction after reading the \texttt{X} register file and sign-extending the immediate value.

\input{src-figures-dfg_example_addi.tex}

The real power of this data structure comes from combining the \acp{DFG} of all instructions. The origin information is preserved by annotating the nodes with the original instructions. The compiler can then apply global optimizations like coalescing equivalent nodes and reducing the number of read and write nodes. This transformation may require the insertion of nodes that multiplex between values depending on the currently executing instruction. The resulting graph captures the execution progress for the \emph{entire} instruction set architecture. Therefore, this graph is called the Instruction Progress Graph. Figure~\ref{fig:ipg_example_addi_sw} depicts the \ac{IPG} for the \isainstruction{ADDI} and \isainstruction{SW} instructions from the example from above. The format fields \texttt{\%imm} and \texttt{\%imm12} are two different nodes because they are extracted from different parts of the instruction word. Thus, the main purpose of the \ac{IPG} is to synthesize all instructions from the \ac{ISA} definition and all processor stages from the \ac{MiA} definition into a single \ac{VIR} program. Later, this \ac{VIR} program is the input to the Hardware Generator described in~\ref{sec:orig_hdlgen}, which in turn generates the actual \ac{MiA} description in Chisel.

\input{src-figures-ipg_example_addi_sw.tex}

Now that the compiler has established a holistic view of the execution progress, it can map the \ac{IPG} to the microarchitecture. This process is called instruction resolving (see Algorithm~\ref{alg:instruction_resolving}).  The idea of this approach is to track the flow of the instruction variables across the microarchitecture. If the analysis encounters mappings on these variables (e.g., \texttt{instr.read(@X)}), the algorithm replaces the mapping with the actual VIR instructions to implement the instruction semantics. Then, the \ac{IPG} is updated to reflect the progress. When encountering the next instruction mapping, the algorithm uses the new \ac{IPG} to determine the \ac{VIR} instructions that replace the mapping. The following paragraphs delve into some of the intricacies of this procedure as it is paramount to the microarchitecture synthesis.

\begin{algorithm}
    \caption{Instruction Resolving}
    \label{alg:instruction_resolving}
    \begin{algorithmic}[1]
    \State $stages \gets \Call{ComputeStageOrder}$ \Comment{Logical order of stages; e.g., decode before execute}
    \ForAll{$stage$ in $stages$}
        \ForAll{$inst$ in $\Call{GetRelInstructions}{stage}$} \Comment{Replace instruction mapping with instruction semantics}
            \State $var \gets \Call{GetInstructionVariable}{inst}$
            \State $ipg \gets \Call{GetCurrentIPG}{var, inst}$
            \State $matching \gets \Call{MatchNodes}{ipg}$
            \State $\Call{ReplaceInMiA}{matching, inst}$
            \State $newipg \gets \Call{UpdateIPG}{ipg, matching}$
            \State $\Call{StoreIPG}{inst, newipg}$
        \EndFor
        \ForAll{$param$ in $\Call{GetInstructionOutputs}{stage}$} \Comment{Replace instruction abstraction}
            \State $ipg \gets \Call{FindIPGAtStageEnd}{stage, param}$
            \State $regs \gets \Call{ComputePipelineRegisters}{ipg}$
            \State $\Call{ReplaceInMiA}{param, regs}$
        \EndFor
    \EndFor
    \State $lastipg \gets \Call{FindLastIPG}{stages}$
    \If{$lastipg \neq \emptyset$} \Comment{Check if all semantics were realized in the microarchitecture}
        \State $\Call{IssueError}$
    \EndIf
    \end{algorithmic}
\end{algorithm}

The algorithm starts by computing a topological order of the stages and their interdependencies. This order ensures that the compiler processes a stage logically preceding another earlier (e.g., decode before execute). The algorithm then iterates over all stages. For each one, the algorithm must complete two necessary steps.

In the first step, the algorithm replaces the instruction mapping with the instruction semantics defined by the \ac{IPG}. For example, this means replacing \texttt{instr.compute} with VIR code that does addition and multiplication. The first problem is to extract the \ac{IPG} subgraph that matches the current instruction mapping. Each instruction mapping defines a predicate to distinguish between matching and non-matching nodes. This predicate is evaluated for each node, thus partitioning them into matching and non-matching sets. In addition, a node must be \emph{ready} to qualify as a matching node. A node is ready if all its input values are available or become available in this instruction mapping. As a consequence, for example, \texttt{instr.compute} cannot match an add node if one of the node's inputs is a register read that is not yet available. After replacing the instruction mapping, the algorithm updates the \ac{IPG} to reflect the progress. One of the major concerns during this update is marking the computed nodes as available. This procedure can make some nodes unnecessary as they are only inputs to other available nodes. In other words, all computations that require them as input have already been computed. Because the compiler must only keep relevant available nodes in the graph, this step also includes a clean-up process that removes unnecessary available nodes. The updated graph is then associated with the instruction mapping. This link is necessary to ensure the algorithm can access the correct \ac{IPG} for the next instruction mapping.

The second step in processing a stage is to replace the instruction stage output with the pipeline registers associated with this instruction variable. The available nodes define all computed values necessary for executing the remaining instruction semantics. Recall that the algorithm removed unnecessary available nodes in the previous step. The compiler can quickly determine the VIR instructions that must be saved in the pipeline registers via the maintained mapping. The compiler can also merge multiple available nodes into a single pipeline register to optimize the usage of resources. This transformation is possible if the nodes are not active in the context of any instruction. For example, a temporary result in an \isainstruction{ADD} instruction may not be necessary when executing a \isainstruction{SW} instruction and vice versa. Thus, the algorithm can store both temporary results in a single pipeline register and multiplex between the values. Once the algorithm computes the pipeline registers, it replaces the instruction stage output with the pipeline registers. The mapping from the \ac{IPG} to the \ac{VIR} instructions also records the pipeline register of available values. A stage that reads the instruction variable can create probes for these pipeline registers.

There are \emph{no instruction abstractions} in the pipeline definition once the compiler has executed both steps. This fact highlights the importance of this procedure in bridging the gap between the high-level \ac{VADL} \ac{MiA} model and a synthesizable microarchitecture. Before concluding, the algorithm ensures that the \ac{IPG} is trivial, i.e., empty at the logical end of the microarchitecture. If not, the synthesis did not realize some part of the instruction semantics in the microarchitecture. This circumstance is undesirable, and the compiler issues an error to the user. The user can then use the graphical representation of the \ac{IPG} to debug the issue. This concludes the first step in the microarchitecture synthesis.

While the prototype compiler does not yet support the advanced techniques presented in Section~\ref{sec:advanced_mia}, we would like to highlight the importance of the \ac{IPG} in this context. In the future, the information in the \ac{IPG} shall be used to automatically determine, for example, the layout of entries in reservation stations. The approach will be similar to computing the set of pipeline registers.

\subsubsection{Decoder synthesis} The compiler replaces the decoder built-ins with a bit pattern matching implementation. This is done by analyzing the instruction encodings in the \ac{ISA}. In the current compiler, the decoder is responsible for deducing the exact instruction (e.g., \isainstruction{ADD} or \isainstruction{SUB}) from the instruction word by matching the bit patterns of the opcode. The implementation will compare the constant parts of the instruction encodings with the instruction word. If all constant parts of an encoding match, the algorithm has found the corresponding instruction. The comparison order prioritizes tests from more specific encodings. This approach allows for a relatively simple decoder synthesis step. Optimizing away the instruction kind variable is done on the \ac{IPG} in the instruction resolving step. We rely on the synthesis tools to completely eliminate the instruction kind variable.

\subsubsection{Port Inference and Assignment} The next step in synthesizing the microarchitecture is generating and assigning register ports. The set of ports determines the functionality that a register file can provide in a single cycle. For example, a register file with two read ports allows the pipeline to initiate two read requests every clock cycle. The \ac{VADL} compiler analyzes the stages in the microarchitecture to determine the number of ports that are necessary to enable full parallelization. For example, if a pipeline requires reading two values from a register file in the decode stage and one value in the execute stage, the analysis would assign three read ports to the register. Note that this analysis incorporates information on the control flow. As a result, two mutually exclusive read instructions can share the same read port. Then, the system allocates the generated ports to individual read and write \ac{VIR} instructions. These processes are isolated as this design separates the generation and assignment procedures. Thus, extending the implementation to allow users to set a maximum number of read and write ports to limit resource usage is straightforward. In contrast to registers, for the main memory our prototype currently only supports a single read and write port.

\subsubsection{Logic Element Synthesis} The system synthesizes the logic elements after determining the read and write ports. Each type of logic element has a different synthesis procedure. We will illustrate the idea in the example of the combinational hazard detection unit. The logic of the synthesized component is based on~\cite{kroening2001}. This element is responsible for detecting hazards in the pipeline that appear due to the interleaved execution of multiple instructions. For example, an instruction in the decode stage may depend on the value of an instruction in the execute stage. However, the latter writes its result only in a later stage. One solution to this problem is letting the instruction in the decode stage wait until the result is available. However, doing this requires detecting the presence of a hazard. Observing these circumstances is the job of the hazard detection unit.

Synthesizing this unit requires two types of information. Firstly, the procedure must have access to the entire pipeline to look for operations that may cause a hazard. Secondly, it is necessary to know which parts of the system require and provide values from or to the logic element. The former information can be provided easily by sharing the VIR with the synthesis algorithm. The VIR implements the second piece of information with special logic element operation instructions that can be contained in any other VIR definition. These operations include inputs and outputs that the compiler must relay from and to the synthesized logic element. In addition, as regular VIR instructions, they also convey the location where this information comes from and is required. For example, one such operation in the hazard detection unit checks whether a stage requires stalling. After synthesizing the component, the algorithm usually translates these logic operations into inputs and outputs on a stage or logic element, including any required probing instructions. After the logic element synthesis, the VIR instructions explicitly capture the entire logic of the processor.

\subsubsection{Pipeline Synthesis} The next step is synthesizing the pipeline implementation. This step is relatively straightforward. Components are represented as \ac{VIR} processes. As shown in Section~\ref{sec:vir}, \ac{VIR} processes contain \vadlresource{probe} instructions to access input values. A \vadlresource{probe} instruction represents taking the value of input lines.  The \vadlresource{probe} instructions of components directly refer to the output parameters in other elements. These instructions indicate whether the result shall be observed in the same machine cycle (e.g., the output of the hazard detection) or the next machine cycle (e.g., computed value in the execute stage). This step first determines which values must be relayed from one component to another. For every connection, the synthesis assigns a wire (same-cycle) or a register (cross-cycle) to the value. This step thus realizes some inputs and outputs as pipeline registers.

\subsubsection{Control Flow Elimination}
\label{sec:controlflow_elimination}

To conclude the microarchitecture synthesis, the compiler tries to eliminate all control flow in the stages and synthesized logic elements.
The presence of control flow in a \ac{VIR} process is represented by it containing more than one basic block at this point. This implies that the process cannot be translated to hardware using combinational logic alone, but has to be synthesized into a state machine with one state for the execution of each basic block. This, in turn, implies that the execution of the \ac{VIR} process takes more than one clock cycle.

If the \ac{VIR} process is the description of a pipeline stage, this situation is especially undesirable, because then one machine cycle takes more than one clock cycle.
Under the described circumstances, the machine cycle increases due to the fact that the whole pipeline can only advance (i.e., each stage hands down results to the next stage) when the slowest pipeline stage is finished. It also causes the faster pipeline stages to be idle while waiting.

To remedy this situation, control flow elimination attempts to remove as much control flow as possible in order to reduce the complexity of the state machines. In the optimal case, all control flow is eliminated and no state machine is necessary. In this case, combinational logic is sufficient to implement the \ac{VIR} process.
Ideally, the algorithm can reduce each component to a single basic block. The \ac{VADL} compiler can still synthesize a functionally correct microarchitecture if this is impossible. However, each machine cycle is distributed across multiple clock cycles, drastically deteriorating performance. One example of such a problematic microarchitecture could contain two instructions that access the same port in a single stage. This construct often happens in single-stage implementations as loading the instruction and executing a load instruction require access to the same port.
This step concludes the microarchitecture synthesis.

%% file: src-figures-microarchitecture_synthesis.tex
\begin{figure}[h]

    \begin{minipage}[b]{0.49\linewidth}
        \begin{lstlisting}[language=vadl, escapechar=§]
instruction ADDI : Itype = {
    X( rd ) := §\mytikzmark{isaread1}§X(rs1)  §\mytikzmark{isacompute1}§+  immS
}

instruction SW : Stype = {
    let addr = §\mytikzmark{isaread2}§X(rs1)  §\mytikzmark{isacompute2}§+  immS in
    let result = §\mytikzmark{isaread3}§X(rs2) in
        MEM<4>( addr ) := result as Word
}
        \end{lstlisting}
    \end{minipage}
\hfill
\begin{minipage}[b]{0.49\linewidth}
    \begin{lstlisting}[language=vadl, deletekeywords={read, compute, write}, escapechar=§]
stage DECODE -> ( ir : Instruction ) = {
    let instr = decode( FETCH.fr ) in {
        §\mytikzmark{miaread}§instr.read( @X )
        ir := instr
    }
}

stage EXECUTE -> ( ir : Instruction ) = {
    let instr = DECODE.ir in
        §\mytikzmark{miacompute}§instr.compute
}
        \end{lstlisting}
    \end{minipage}
    \begin{tikzpicture}[remember picture, overlay]
        \encircle{isaread1}{0.6cm}{green}
        \encircle{isaread2}{0.6cm}{green}
        \encircle{isaread3}{0.6cm}{green}
        \encircle{isacompute1}{0.1cm}{red}
        \encircle{isacompute2}{0.1cm}{red}

        % Draw arrows
        \draw[->, thick, green] ([xshift=1.02cm, yshift=0.135cm]isaread1) to [out=40, in=180] ([yshift=0.05cm]miaread);
        \draw[->, thick, green] ([xshift=1.02cm, yshift=0.135cm]isaread2) to [out=40, in=180] ([yshift=0.05cm]miaread);
        \draw[->, thick, green] ([xshift=1.02cm, yshift=0.135cm]isaread3) to [out=40, in=180] ([yshift=0.05cm]miaread);
        \draw[->, thick, red] ([xshift=0.16cm, yshift=0.15cm]isacompute1) to [out=320, in=180] ([yshift=0.05cm]miacompute);
        \draw[->, thick, red] ([xshift=0.14cm, yshift=-0.05cm]isacompute2) to [out=0, in=180] ([yshift=0.05cm]miacompute);
    \end{tikzpicture}

    \caption{Exemplary Mapping Between \ac{ISA} and \ac{MiA}}
    \label{fig:microarchitecture_synthesis_overview_mia}
    \Description{On the left-hand side there is an ADDI and an SW instruction. On the right-hand side there are two stages, DECODE and EXECUTE. The register read operations are highlighted and point towards the instr.read( @X ) operation in the microarchitecture. The computation operations are highlighted and point towards the instr.compute operation in the microarchitecture.}
\end{figure}

%% file: src-figures-dfg_example_addi.tex
\begin{figure}[h]
\begin{minipage}[b]{0.49\linewidth}
    \center
    \scalebox{0.75}{
        \begin{tikzpicture}[node distance=1.5cm]
            \node (writex) [writenode] {write @X: b32};
            \node (rd) [availablenode, below right of=writex] {\%rd: b5};
            \node (add) [operationnode, below left of=writex] {add: i32};
            \node (cast) [operationnode, below left of=add] {cast: i32};
            \node (sext) [operationnode, below right of=add] {cast: i32};
            \node (imm12) [availablenode, below right of=sext] {\%imm12: b12};
            \node (readx) [readnode, below left of=cast] {read @X: b32};
            \node (rs1) [availablenode, below right of=readx] {\%rs1: b5};

            \draw [arrow] (rd) -- (writex);
            \draw [arrow] (add) -- (writex);
            \draw [arrow] (cast) -- (add);
            \draw [arrow] (sext) -- (add);
            \draw [arrow] (imm12) -- (sext);
            \draw [arrow] (readx) -- (cast);
            \draw [arrow] (rs1) -- (readx);
        \end{tikzpicture}
    }

\caption{Simplified \ac{DFG} for the \isainstruction{ADDI} instruction}
\label{fig:dfg_example_addi}

\end{minipage}
\hfil
\begin{minipage}[b]{0.49\linewidth}
    \center
    \scalebox{0.75}{
        \begin{tikzpicture}[node distance=1.5cm]
            \node (writex) [writenode] {write @X: b32};
            \node (rd) [availablenode, below right of=writex] {\%rd: b5};
            \node (add) [operationnode, below left of=writex] {add: i32};
            \node (cast) [availablenode, below left of=add] {cast: i32};
            \node (sext) [availablenode, below right of=add] {cast: i32};

            \draw [arrow] (rd) -- (writex);
            \draw [arrow] (add) -- (writex);
            \draw [arrow] (cast) -- (add);
            \draw [arrow] (sext) -- (add);
        \end{tikzpicture}
    }

\caption{Simplified \ac{DFG} for the \isainstruction{ADDI} instruction after reading the X register file}
\label{fig:dfg_example_addi_v2}

\end{minipage}
\end{figure}

%% file: src-figures-ipg_example_addi_sw.tex
\begin{figure}[h]

    \scalebox{0.75}{
        \begin{tikzpicture}[node distance=2.2cm, text width=1.5cm]
            \node (writex) [writenode, text width=2cm] {write @X: b32\\\{ADDI\}};
            \node (writemem) [writenode, left of=writex, text width=2.4cm, xshift=-2cm] {write @MEM: b32\\\{SW\}};

            \node (rd) [availablenode, below right of=writex] {\%rd: b5\\\{ADDI\}};
            \node (add) [operationnode, below left of=writex] {add: i32\\\{ADDI, SW\}};
            \node (cast) [operationnode, below left of=add] {cast: i32\\\{ADDI, SW\}};

            \node (mux) [operationnode, below right of=add, text width=2cm] {mux instr: i32\\\{ADDI, SW\}};

            \node (sext) [operationnode, below right of=mux, xshift=1.5cm] {cast: i32\\\{ADDI\}};
            \node (imm12) [availablenode, below of=sext, text width=2cm, yshift=0.6cm] {\%imm12: b12\\\{ADDI\}};

            \node (sext2) [operationnode, below of=mux, yshift=0.6cm] {cast: i32\\\{SW\}};
            \node (imm) [availablenode, below of=sext2, yshift=0.6cm] {\%imm: b12\\\{SW\}};

            \node (readx1) [readnode, text width=2cm, below left of=cast] {read @X: b32\\\{ADDI, SW\}};
            \node (rs1) [availablenode, below right of=readx1] {\%rs1: b5\\\{ADDI, SW\}};

            \node (readx2) [readnode, text width=2cm, xshift=-3cm, left of=add] {read @X: b32\\\{SW\}};
            \node (rs2) [availablenode, below left of=readx2] {\%rs2: b5\\\{SW\}};

            \draw [arrow] (rd) -- (writex);

            \draw [arrow] (add) -- (writemem);
            \draw [arrow] (add) -- (writex);

            \draw [arrow] (cast) -- (add);
            \draw [arrow] (mux) -- (add);

            \draw [arrow] (sext) -- (mux);
            \draw [arrow] (imm12) -- (sext);

            \draw [arrow] (sext2) -- (mux);
            \draw [arrow] (imm) -- (sext2);

            \draw [arrow] (readx1) -- (cast);
            \draw [arrow] (rs1) -- (readx1);
            \draw [arrow] (readx2) -- (writemem);
            \draw [arrow] (rs2) -- (readx2);
        \end{tikzpicture}
    }

\caption{Simplified \ac{IPG} for the \isainstruction{ADDI} and \isainstruction{SW} Instructions}
\label{fig:ipg_example_addi_sw}

\end{figure}

%% file: src-orig_casgen.tex
\subsection{Original \ac{VADL} Cycle-Accurate Simulator Generator}
\label{sec:orig_casgen}

Unlike an \ac{ISS}, a cycle-accurate simulator (\ac{CAS}) aims to model properties of the processors's microarchitecture such as pipeline stalls and latencies of memory accesses. As mentioned in~\ref{sec:orig_hdlgen}, the \ac{VADL} compiler is capable of generating Verilog via Chisel, an \ac{HDL}, which can be used for simulation by Verilator. However, \ac{HDL} generation can take considerably longer than generating a \ac{CAS}. Using a \ac{CAS} allows the simulation of microarchitectural aspects while also having the benefit of shorter test cycles for changes in the processor design. This facilitates analysis of the changes and estimation of their impact on real hardware.

Nonetheless, \ac{HDL} generation and \ac{CAS} generation are similar enough to reuse many generation steps and program transformations. This allows the behavior and properties of interest of the resulting \ac{CAS} to be as close as possible to the actual hardware design. However, instead of outputting Chisel, the \ac{VADL} compiler generates C++ code. Each resource definition, such as registers and memory, corresponds to an individual C++ class with corresponding functionality (e.g., read and write functions for memory access). In addition, each stage of the microarchitecture is implemented as a C++ class containing an \texttt{eval} function which executes the corresponding functionality. Each \ac{VIR} instruction is translated into an equivalent C++ operation. Since hardware is parallel by nature, unlike most common programming languages, the resulting C++ code might seem less idiomatic to regular software engineers. For instance, consider a feature that might only be used conditionally. Developers would usually use an \texttt{if} statement to direct control flow. However, hardware often uses `enable' signals, so a piece of hardware only activates if this signal is set to `HIGH'. The generated C++ code resembles this behavior (e.g., by using a bitmask where all bits are set to one/zero depending on whether the result of the piece of hardware it mimics is required).

\begin{figure}[h]
    \begin{tikzpicture}
    \begin{scope}[every node/.style={circle,thick,draw}]
        \node (halt) {\texttt{halt}};
        \node (imm1) [right of=halt, xshift=4cm] {\texttt{imm1}};
    \end{scope}

    \node (dummy) [left of=halt, xshift=-1cm, draw=none]{};
    \path[arrow]
        (dummy) edge (halt);

    \path[arrow, every node/.style={fill=white,rectangle}]
        (halt) edge node {\small \texttt{ADD.eval()}} (imm1)
        (imm1) edge[bend right] node {\small \texttt{ADD.eval()}} (halt);
    \end{tikzpicture}
    \caption{Example state machine for an \texttt{ADD} instruction which performs the addition and sets the carry flag in two separate machine cycles. \texttt{halt} represents the start and end state. The first execution of \texttt{eval()} performs the addition and switches the state to \texttt{imm1}, while the second execution sets the carry flag and switches back to the \texttt{halt} state, finishing the operation.}
    \label{fig:add-fsm-example}
\end{figure}

Furthermore, some operations might require multiple cycles. In order to reflect this, one can partition a single operation into multiple steps represented by states of a \ac{FSM}. As a toy example, consider an \texttt{ADD} instruction which performs the addition in one cycle and sets the carry flag in the next one. Thus, the instruction consists of two states as seen in Figure~\ref{fig:add-fsm-example}: a \texttt{halt} state and an intermediate state, which we call \texttt{imm1}. The former state denotes the start and end at the same time while the latter represents the situation after the addition but before the carry flag has been set. A transition from one state to another requires exactly one machine cycle.

Regarding the generated C++ code, the \ac{VADL} compiler generates an enum containing an entry for each state. The \texttt{eval} function applies the semantics of the instruction. In the prologue, the method checks what state the instruction is currently in. Considering our example from above, if the current state is \texttt{halt}, then the operation is in its starting state. The current state will be updated to \texttt{imm1} and \texttt{eval} will apply the addition but return before the carry flag is set. In the next invocation of \texttt{eval} (usually after one machine cycle), the method observes that the current state is \texttt{imm1} and thus sets the carry flag according to the addition which was calculated in the previous cycle. The current state will be updated to \texttt{halt}, showing that the operation has concluded.

Last but not least, other components can check whether an operation has finished by querying the \texttt{busy} method, which is also generated for every stage class. Recall that an operation is busy until it has reached the \texttt{halt} state.

%% file: src-orig_hdlgen.tex
\subsection{Original \ac{VADL} Hardware Generator}
\label{sec:orig_hdlgen}

The hardware generator's responsibility is emitting an equivalent hardware design based on a specification in \ac{VIR}. The microarchitecture synthesis covered in Section~\ref{sec:orig_miasynth} must have processed the design beforehand. The main objective of this component is to bridge the gap between the \ac{VIR} and \acp{HDL}. For example, in the \ac{VIR}, ports are global entities that read and write instructions can access. However, in an \ac{HDL}, the interface of a circuit module must make these connections explicit. The generator uses an \ac{HDL}-\ac{IR} to model the hardware design internally. Most \ac{IR} elements map directly to a concept in an \ac{HDL}. Thus, emitting files from the \ac{IR} is straightforward.

This paragraph briefly introduces \acp{HDL} for readers unfamiliar with the topic. \acp{HDL} enable engineers to model and simulate the behavior of electronic systems before physically implementing them. At the core of \acp{HDL} is the concept of a "module." A module is a self-contained unit of hardware description that encapsulates specific functionality or a component of a digital system. Modules can be interconnected to create complex systems, and \acp{HDL} provide a structured way to define the interactions and relationships between these modules. Users can use logical and arithmetic operations to describe a module's data flow. Furthermore, these languages support concise descriptions of standard hardware constructs like multiplexers and decoders.

Generating the \ac{HDL}-\ac{IR} from the \ac{VIR} is done in three steps. Firstly, the transformation creates a module for each \ac{VIR} resource definition (e.g., registers). The information on the number of read and write ports is used to generate the module's interfaces. While the process is straightforward, adhering to all constraints in the resource definition is vital. For example, the module must correctly implement hardwired register indices.

After synthesizing the resource definitions, an analysis computes the hardware design hierarchy. The analysis can obtain this information from the instantiation relations between processes. The root of this hierarchy is the behavior process of the processor. The algorithm recursively enumerates the reachable process definitions and defines a corresponding module. These components contain connectors for all used read and write ports. A parent module must provide dedicated connections for each port used by its child modules. It is mandatory to define these connections explicitly. The transformation combines multiple child connectors to the same port into a single one. This can be done because the microarchitecture synthesis guarantees that there are no instances where two child processes can access a port simultaneously.

The final step in computing the \ac{HDL}-\ac{IR} is generating the actual circuits. This task boils down to synthesizing the modules of nested processes rooted in the processor behavior. Each synthesized process instantiation requires a description of the module hierarchy, data flow, and control logic. Computing the module hierarchy and data flow is straightforward.

The derivation of the hierarchy naturally unfolds during the enumeration of nested processes. Each process module contains sub-modules that correspond to the instantiated sub-processes. None of these modules includes components for resource definitions. The task of defining the data flow hinges on the premise that most \ac{VIR} instructions coincide with an operator in the \ac{HDL}-\ac{IR}. There is a peculiarity when translating instructions that require access to resources. In such cases, the algorithm must automatically connect the inputs and outputs to the corresponding port interfaces. The \ac{HDL}-\ac{IR} does not contain modules for commonly found components such as an \ac{ALU}. All computations are done directly within the pipeline stages. Synthesizing the implementation of operations like addition and multiplication is left to the hardware synthesis tool. This approach contrasts with tools that compose the microarchitecture from a set of predefined components.

A fully functioning \ac{VADL}-generated processor design requires control logic that implements the control flow of processes. Unfortunately, this is often undesirable in a pipeline stage as executing control flow can occupy multiple clock cycles. Executing a basic block requires at least one clock cycle. Therefore, the whole pipeline halts once a pipeline stage executes control flow. The microarchitectural synthesis tries to eliminate the control flow in the pipeline stages to address this issue. However, control flow may still be necessary for the processor's initialization logic. Furthermore, given resource limitations, the hardware generator must resort to control flow for some design specifications. For example, the microarchitecture synthesis currently creates a single memory read port. If a pipeline stage requires two memory reads (e.g., instruction fetch and memory operation), control flow is necessary to distribute these accesses across multiple clock cycles.

The generator implements control flow by translating the process to a finite state machine. Each basic block corresponds to a state. The terminator of the basic block determines the state transitions. For example, an unconditional jump will result in a single transition between the two states. Each state machine has an enable and busy signal in addition to its regular inputs and outputs. Initially, the state machine remains idle until the parent module drives the enable signal. The state machine asserts the busy signal until it has finished execution. Thus, once the busy signal is low, a parent module is aware that a process has terminated, and the output is valid. The hardware computes all operations in a state simultaneously. Therefore, the implementation must handle side effects with great care once child \acp{FSM} run for multiple cycles.

After generating the hardware design hierarchy, the generator emits the corresponding Chisel files. The Chisel compiler then translates the specification into Verilog, a well-established hardware description language. After that, users can apply off-the-shelf hardware synthesis and simulation tools to the generated design. This also allows embedding \ac{VADL}-generated cores into a larger Verilog (or Chisel) design.

%% file: src-orig_evalcomp.tex
\subsection{Evaluation Infrastructure}
\label{sec:infra_evaluation}

All performance evaluations were executed on our continuous integration server.
The processor of the machine is an \emph{Intel(R) Xeon(R) W-1370P} running at 3.60 GHz, featuring 8 cores with 16 threads and 128 GiB of memory.
However, the number of cores was irrelevant, since we only measured single-threaded programs.

\subsection{Evaluation of the Original \ac{VADL}-Generated Compiler}
\label{sec:orig_evalcomp}

As the \emph{\ac{VADL} setup}, we used the original \ac{LCB} to generate an
LLVM compiler target based on the RISC-V 32-bit RV32IM \ac{VADL} instruction set specification, the same
as used in Section~\ref{sec:evaliss}.
This generated target can be selected by specifying \texttt{--target=rv32im} to the \texttt{clang} frontend.

As the \emph{upstream setup} serving as the baseline for comparison, we used LLVM's standard upstream RISC-V
target \texttt{--target=riscv32} with the \texttt{ILP32} \ac{ABI}.
Both compilers are based on LLVM version 17.0.6. Both setups use the LLVM upstream assembler and GNU linker to
generate the executable binaries.

As a workload, we used the \emph{Embench}~\cite{embench} benchmark suite.
In the evaluation of the original \ac{LCB}, for both the \ac{VADL} setup and the upstream setup,
we used the \ac{VADL}-generated \ac{CAS} RV32-P3 as the execution environment.
The metric for the comparison was the number of executed machine cycles as reported by the \ac{CAS}.

\subsubsection*{Limitations}

As \ac{LCB} is still a work in progress, there exist some limitations in the \ac{VADL} setup.
In the evaluation, the \ac{VADL} setup successfully compiled, linked, and executed 18 of the 22 benchmarks
found in Embench.

\begin{itemize}
  \item It failed to compile one benchmark because the instruction selection did not support
        some needed patterns.
  \item Of the 21 benchmarks that were successfully compiled, 3 could not
        be linked against the runtime library. At the moment, the \ac{VADL} setup does not yet support
        a complete \texttt{libc} runtime library.
  \item So 18 benchmarks could be compiled, linked, and executed without error.
  \item The original \ac{VADL} setup is not yet capable of utilizing optimization levels higher than \texttt{-O0}.
\end{itemize}

Figure~\ref{fig:lcb-speed-relative} shows the number of machine cycles for the \ac{VADL} setup relative to the
upstream setup, which is used as the baseline.
For some benchmarks, \ac{LCB} gives better results. For others, the upstream compiler gives better results.
On average, both compilers have identical performance.
We still have to analyze the cause for the differences in performance of individual benchmarks.

\pgfplotstableread[col sep=comma]{benchmark-data-rv32-lcb-cas-p3-rv32-lcb-cas-p3-relative.csv}\relative

\begin{figure}
    \begin{tikzpicture}
    \begin{axis}[
    ybar=1,
    ymax=1.5,
    xtick=data,
    xticklabel style={rotate=90},
    x=0.65cm,
    bar width=0.08cm,
    enlarge x limits=0.02,
    enlarge y limits={rel=0.05, upper},
    symbolic x coords={aha-mont64,crc32,edn,huffbench,matmult-int,md5sum,minver,nbody,nettle-aes,picojpeg,primecount,qrduino,sglib-combined,slre,st,statemate,tarfind,ud,mean},
    ymin=0
    ]
      \addplot table [col sep=comma]{benchmark-data-rv32-lcb-cas-p3-rv32-lcb-cas-p3-relative.csv};
      \addplot table [col sep=comma]{benchmark-data-rv32-clang-cas-p3-rv32-clang-cas-p3-relative.csv};
      \legend{VADL,upstream}
    \end{axis}
    \end{tikzpicture}
    \caption{Embench machine cycle counts for the \ac{VADL} setup, scaled relative to the upstream setup (lower is better)}
    \label{fig:lcb-speed-relative}
\end{figure}

%% file: src-orig_evalass.tex
\subsection{Evaluation of the Original \ac{VADL} Assembler and Linker Generator}
\label{sec:orig_eval_asm_link}

The evaluation of the generated assembler and generated linker focused on ensuring that the tools work correctly. We gathered evidence for this by including them in our testing environment. The test infrastructure uses the assembler and linker to build executables for all test programs generated for the RISC-V architecture. These tests include a RISC-V compliance suite, handwritten assembly programs, and the compiler output for small C programs. A subsequent step uses these executables to test the generated simulators. The setup exercises the assembly printer (compiler output), assembly parser, machine code emitter, and linker. If one of these components is erroneous, the subsequent simulator tests will fail if the simulator cannot execute the program correctly or the simulation returns an incorrect result.

During development, the \ac{VADL} team used the RISC-V architecture to test the assembler generator and linker generator. The resulting prototype was applied to the MIPS IV and AArch64 architectures to test the ability to capture different assembly syntaxes. The following text lays out the findings in a qualitative discussion.

Describing the canonical MIPS IV syntax posed no problems. However, abbreviated instructions that leave out some operands lead to problems due to the prototype's LL(1) parsing algorithm. A similar limitation can also be observed for abbreviated RISC-V instructions. One could extend the parsing algorithm to LL(k) or incorporate a state-of-the-art parser generator to alleviate this problem. Please note that this is a limitation of the prototype and not \ac{VADL} itself.

In addition, the test with the AArch64 instruction set showed some limitations of \ac{VADL}'s grammar definition. The problem is rooted in the separation of the parsing and matching phases, i.e., casting a grammar element to an \texttt{@instruction}. As a result, the instruction matcher is oblivious to the applied grammar rule. This is not a problem if the parser conveys the necessary information to distinguish between instructions with equivalent operands to the matcher, e.g., via the instruction's mnemonic. However, the current prototype only supports communicating this syntax information for mnemonics. As a result, distinguishing instructions based on their syntactical structure proved problematic. Providing additional data to the matcher can remedy this issue.

\begin{table}
    \caption{Number of generated grammar rules.}
    \label{tab:eval_generated_rules}
    \begin{tabular}{l rrrr}
        \toprule
        \textbf{ISA} & \textbf{\# Insts} & \textbf{\# Inverted} & \textbf{Inverted (\%)} & \textbf{\# Rules} \\
        \midrule
        RV32I   &     74 &     74 & 100.00 &     75 \\
        MIPS IV &    106 &    104 &  98.11 &    105 \\
        AArch64 & 1\,439 & 1\,437 &  99.86 & 1\,446 \\
        \bottomrule
    \end{tabular}
\end{table}

The proposed grammar rule inference was evaluated on the \ac{VADL} specifications of the RISC-V, MIPS IV, and AArch64 architectures. The focus was on identifying what types of formatting functions the approach could handle. Table~\ref{tab:eval_generated_rules} shows the number of instructions in an architecture and the proportion of successfully inferred grammar rules. The last column lists the number of generated rules. This number also includes generated helper rules from the interpretation-based inference mechanism. The system could process all instructions in the RV32I ISA and most instructions in the MIPS IV and AArch64 architectures. Two AArch64 instructions could not be inverted because of issues in the interpretation of the VIR. Therefore, the issue lies in the implementation of the interpreter and not the grammar inference.

In the MIPS IV architecture, the approach could not deal with the abbreviated syntax of the \isainstruction{syscall} and \isainstruction{ebreak} instructions. This abbreviation emits the $20$-bit code field only if it is unequal to zero. Because conditionals require switching to the interpreter, this would result in $2^{20}$ combinations requiring evaluation. Section~\ref{sec:grammar_inference} details this behavior. Fortunately, \ac{VADL} provides the escape hatch to manually define grammar rules for problematic instructions to overcome this issue quickly. Still, the experiments show room for improvement when dealing with abbreviated syntax.

%% file: src-evaliss.tex
\subsection{Evaluation of the Original \ac{VADL} Instruction Set Simulator}
\label{sec:evaliss}

We evaluated the \ac{VADL} simulators using the Embench~\cite{embench} benchmark suite. For the \ac{ISS}, we implemented the RISC-V 32-bit instruction set including the M extension (RV32IM) and the complete integer subset of the AArch64 instruction set (Armv8-A).  The RISC-V benchmarks were compiled with GCC 12.2.0 and glibc, and the AArch64 benchmarks were compiled with Clang 14.0.6 and musl libc v1.2.4.

VADL currently has incomplete support for floating-point numbers, so the floating-point operations in Embench were compiled to use software emulation for both architectures instead (which is normal for RV32IM, but the Armv8-A specification technically mandates hardware floating-point support).

The \acp{ISS} were compared against QEMU, which by default uses \ac{JIT} compilation for increased performance. Therefore, we also include QEMU with the \texttt{one-insn-per-tb} flag enabled (which JITs every instruction into its own block, negating much of the JIT speedup), and QEMU compiled to use the fallback interpreter (\texttt{{-}{-}enable-tcg-interpreter}) instead of the JIT. These are called "QEMU singlestep" and "QEMU nojit" in the figures below, respectively. All QEMU runs were performed using user mode emulation. For RISC-V, we also included the Spike reference simulator~\cite{spike}.

Figure~\ref{fig:dtc-rv32} compares the benchmark runtimes (and their geometric mean) of the described simulators (relative to QEMU). The \ac{VADL} \ac{ISS} is more than 21 % rv32-dtc:mean

times slower than QEMU in JIT mode and 75\% % rv32-dtc:mean / rv32-spike:mean
slower than Spike, but still faster than the other QEMU versions. This is expected because JIT simulation is faster than interpreting, while the interpreter Spike is written and optimized for just a single architecture.

As shown in Figure~\ref{fig:dtc-aarch64}, the performance characteristics of the AArch64 \ac{ISS} are similar to the RISC-V performance characteristics.

\pgfplotstableread[col sep=comma]{benchmark-data-results-rv32-dtc-rv32-qemu.csv}\dtcresults
\begin{figure}[h!]
  \begin{tikzpicture}
      \begin{axis}[
          ybar=1,
          xtick=data,
          xticklabels from table={\dtcresults}{benchmark},
          xticklabel style={rotate=90},
          x=0.65cm,
          bar width=0.07cm,
          enlarge x limits=0.02,
          enlarge y limits={rel=0.05, upper},
          ymin=0
      ]
    \addplot table [col sep=comma, x expr=\coordindex, y=time]{benchmark-data-results-rv32-dtc-rv32-qemu.csv};
    \addplot table [col sep=comma, x expr=\coordindex, y=time]{benchmark-data-results-rv32-dtc-rv32-qemu-nojit.csv};
    \addplot table [col sep=comma, x expr=\coordindex, y=time]{benchmark-data-results-rv32-dtc-rv32-qemu-singlestep.csv};
    \addplot table [col sep=comma, x expr=\coordindex, y=time]{benchmark-data-results-rv32-dtc-rv32-spike.csv};
    \addplot table [col sep=comma, x expr=\coordindex, y=time]{benchmark-data-results-rv32-dtc-rv32-dtc.csv};
      \legend{QEMU,QEMU nojit,QEMU singlestep,Spike,ISS,gem5}
      \end{axis}
  \end{tikzpicture}
  \caption{Embench runtime of instruction-level simulators using RV32IM (relative to QEMU, smaller is better)}
  \label{fig:dtc-rv32}
\end{figure}

\pgfplotstableread[col sep=comma]{benchmark-data-results-aarch64-dtc-aarch64-qemu.csv}\aarchdtcresults
\begin{figure}[h!]
  \begin{tikzpicture}
      \begin{axis}[
          ybar=1,
          xtick=data,
          xticklabels from table={\aarchdtcresults}{benchmark},
          xticklabel style={rotate=90},
          x=0.65cm,
          bar width=0.08cm,
          enlarge x limits=0.02,
          enlarge y limits={rel=0.05, upper},
          ymin=0
      ]
      \addplot table [col sep=comma, x expr=\coordindex, y=time]{benchmark-data-results-aarch64-dtc-aarch64-qemu.csv};
      \addplot table [col sep=comma, x expr=\coordindex, y=time]{benchmark-data-results-aarch64-dtc-aarch64-qemu-nojit.csv};
      \addplot table [col sep=comma, x expr=\coordindex, y=time]{benchmark-data-results-aarch64-dtc-aarch64-qemu-singlestep.csv};
      \addplot table [col sep=comma, x expr=\coordindex, y=time]{benchmark-data-results-aarch64-dtc-aarch64-dtc.csv};
      \legend{QEMU, QEMU nojit, QEMU singlestep, ISS}
      \end{axis}
  \end{tikzpicture}
  \caption{Embench runtime of instruction-level simulators using AArch64 soft-fp (relative to QEMU, smaller is better)}
  \label{fig:dtc-aarch64}
\end{figure}

%% file: src-evalcas.tex
\subsection{Evaluation of the Original \ac{VADL} Cycle-Accurate Simulator}
\label{sec:evalcas}

We implemented four microarchitectures for the RISC-V 32-bit RV32IM instruction set specification as used in Section~\ref{sec:evaliss}:

\begin{itemize}
    \item \textbf{RV32-P1:} This is a 1-stage pipeline with all steps (fetch, decode, execute, memory, write-back) executing within one cycle.
    \item \textbf{RV32-P2:} This 2-stage pipeline separates fetching and decoding/execution/write-back into two separate stages.
    \item \textbf{RV32-P3:} This 3-stage pipeline has fetch, decode, and execute/write-back stages.
    \item \textbf{RV32-P5:} This 5-stage pipeline has fetch, decode, execute, memory, and write-back stages.
\end{itemize}
We compare the \acp{CAS} generated by \ac{VADL} to gem5 using its \texttt{AtomicSimpleCPU}. Atomic refers to the memory subsystem, meaning that memory accesses return instantly. We chose to use this model because the \ac{CAS} does not accurately simulate the memory subsystem yet. The processor simulated by gem5, however, does not have a pipeline, which is unrealistic for real-world processors. Thus, we decided to provide a 1-stage pipeline microarchitecture in \ac{VADL} to allow for a fair comparison.

Figure~\ref{fig:cas-speed-results} shows the runtimes of the Embench benchmarks with the three \acp{CAS} and gem5 (relative to the one-stage \ac{CAS}). The \ac{CAS} becomes slower as more stages are added (in this case, the three-stage simulator is more than 5 % rv32-p3:mean
times slower than the single-stage simulator) because there is an overhead associated with simulating a longer pipeline. gem5's performance lies between the 1-stage and 2-stage \ac{CAS}.

\pgfplotstableread[col sep=comma]{benchmark-data-results-rv32-cas-rv32-p1.csv}\casresults
\begin{figure}
    \begin{tikzpicture}
    \begin{axis}[
        ybar=1,
        xtick=data,
        xticklabels from table={\casresults}{benchmark},
        xticklabel style={rotate=90},
        x=0.65cm,
        bar width=0.06cm,
        enlarge x limits=0.03,
        enlarge y limits={rel=0.11, upper},
        ymin=0,
        extra y ticks={1},
    legend columns=-1
    ]
    \addplot table [col sep=comma, x expr=\coordindex, y=time]{benchmark-data-results-rv32-cas-rv32-p1.csv};
    \addplot table [col sep=comma, x expr=\coordindex, y=time]{benchmark-data-results-rv32-cas-rv32-p2.csv};
    \addplot table [col sep=comma, x expr=\coordindex, y=time]{benchmark-data-results-rv32-cas-rv32-p3.csv};
    \addplot table [col sep=comma, x expr=\coordindex, y=time]{benchmark-data-results-rv32-cas-rv32-p5.csv};
    \addplot table [col sep=comma, x expr=\coordindex, y=time]{benchmark-data-results-rv32-cas-rv32-gem5-atomic.csv};
    \legend{CAS RV32-P1, CAS RV32-P2, CAS RV32-P3, CAS RV32-P5, gem5}
    \end{axis}\end{tikzpicture}
  \caption{Embench runtime of cycle-level simulators using RV32IM (relative to CAS RV32-P1, smaller is better)}
    \label{fig:cas-speed-results}
\end{figure}

Figure~\ref{fig:cas-cycles} compares the cycle counts of the Embench benchmarks. gem5's estimated cycle count aligns with that of the one-stage \ac{CAS}. The simulators with two or more stages each have higher cycle counts (in this case, the three-stage simulator has a 62\% % rv32-p3-cycles / rv32-p1-cycles
higher cycle count than the single-stage simulator), which is expected because the real processors would then be able to run with a (much) higher clock frequency, making the processor faster overall.

\pgfplotstableread[col sep=comma]{benchmark-data-results-rv32-cas-rv32-p1-cycles.csv}\casaccresults
\begin{figure}
    \begin{tikzpicture}
    \begin{axis}[
        ybar=1,
        xtick=data,
        xticklabels from table={\casaccresults}{benchmark},
        xticklabel style={rotate=90},
        x=0.65cm,
        bar width=0.08cm,
        enlarge x limits=0.03,
        enlarge y limits={rel=0.05, upper},
        ymin=0
    ]
    \addplot table [col sep=comma, x expr=\coordindex, y=cycles]{benchmark-data-results-rv32-cas-rv32-p1-cycles.csv};
    \addplot table [col sep=comma, x expr=\coordindex, y=cycles]{benchmark-data-results-rv32-cas-rv32-p2-cycles.csv};
    \addplot table [col sep=comma, x expr=\coordindex, y=cycles]{benchmark-data-results-rv32-cas-rv32-p3-cycles.csv};
    \addplot table [col sep=comma, x expr=\coordindex, y=cycles]{benchmark-data-results-rv32-cas-rv32-p5-cycles.csv};
    \addplot table [col sep=comma, x expr=\coordindex, y=cycles]{benchmark-data-results-rv32-cas-rv32-gem5-atomic-cycles.csv};

    \legend{CAS RV32-P1, CAS RV32-P2, CAS RV32-P3, CAS RV32-P5, gem5}
    \end{axis}
    \end{tikzpicture}
  \caption{Embench simulated cycle count using RV32IM}
    \label{fig:cas-cycles}
\end{figure}

The \ac{CAS} was created to provide a faster cycle-accurate simulation than a low-level \ac{HDL} simulation using Verilator version 5.010. This is evaluated in Figure~\ref{fig:verilator-speed}: currently, the \ac{CAS} is slower than the equivalent HDL compiled into a simulator using Verilator. This issue stems from linearizing the stage computations so that each stage must be executed only once by the \ac{CAS}. However, currently, this leads to duplicate computations in multiple stages. Verilator handles these situations better than the C++ compiler and thus outperforms the \ac{CAS}. We are working on addressing this issue. As expected, the \ac{ISS} is three orders of magnitude faster than verilated HDL or the \ac{CAS}, with gem5 being slightly faster than the HDL simulation.

\begin{figure}
    \begin{tikzpicture}
    \begin{axis}[
    ybar,
    ymin=0,
    xtick=data,
    enlarge x limits=0.1,
    ymax=3000,
    x=2.3cm,
    symbolic x coords={hdl-p3,rv32-p3,gem5-atomic,rv32-dtc},
    xticklabels={Verilator RV32-P3, CAS RV32-P3, gem5, RV32 ISS}
    ]
      \addplot table [col sep=comma]{benchmark-data-results-rv32-cas-verilator.csv};
    \end{axis}
    \end{tikzpicture}
    \caption{Embench runtime of RV32IM, compared with HDL simulated using Verilator (in milliseconds, geometric mean over all benchmarks, smaller is better)}
    \label{fig:verilator-speed}
\end{figure}

%% file: src-orig_evalhdl.tex
\subsection{Evaluation of the Original \ac{VADL} Hardware Generator}
\label{sec:evalhdl}

We evaluated the hardware generator similarly to the CAS. The \ac{VADL} tooling generated multiple Chisel implementations of an RV32IM-compliant processor. Each implementation was translated to Verilog and simulated using Verilator version 5.010. Then, the test setup validated the designs by running the Embench benchmark suite and comparing the actual output with the expected one. Furthermore, a RISC-V compliance suite was executed on every verilated design. All designs exhibited the desired behavior in all test runs.

Quality is another crucial aspect of the designs, as it impacts the realized processor's performance, power consumption, and chip area. The generated designs are compared to hand-crafted implementations of the same ISA. The Sodor standalone open-source designs (revision e5638c39e5750ea98527547fbc3f9d269c451f3a) will be a reference in this work. Once the \ac{VADL} tooling is mature enough to handle more sophisticated concepts (e.g., reorder buffers), future work may compare the generated artifacts to industry-grade processors. Before continuing with the evaluation, we want to highlight that both the \ac{VADL}-generated designs and Sodor used idealized memory. The result of the comparison may be different for real-world memory modules.

We compared the structural metrics of the corresponding 5-stage implementations using the tool \emph{Yosys} version 0.29.
To get a chip area metric for comparison, we used a simple demo cell library and mapping script found in the Yosys distribution.
In order to achieve a fair comparison, we chose a \ac{VADL} specification similar to the 5-stage Sodor.
This specification contains forwarding and does not implement the RISC-V M extension for multiplication, which is not present in Sodor.

The RISC-V \ac{ISA} specifies the possibility of including up to 4096 \acp{CSR} in a
RISC-V implementation, each 32 bits wide. The \acp{CSR} are used to manage privilege modes and to provide general information about the processor state, such as machine trap handling and debugging information.

Not all \acp{CSR} have a standardized usage, leaving the possibility for custom extensions.
Since most \acp{CSR} are optional, Sodor only implements a subset of \acp{CSR}, resulting
in a much smaller \ac{CSR} register file than the complete one used in our \ac{VADL}
specification.

Table~\ref{tab:structural-stats} shows the numbers reported by Yosys' stat utility.
Since the \ac{VADL} hardware generation does implement the \emph{full} \ac{CSR} RISC-V register file,
which is $2^{12}*32 = 131072$ bits,
we also provide the numbers without the chip area used for the \ac{CSR} modules.

\begin{table}
\caption{Comparison of chip area for \ac{VADL} RV32I and Sodor.}
\label{tab:structural-stats}
\begin{tabular}{l rrr}
\toprule
\textbf{Metric} & \textbf{VADL RV32I} & \textbf{Sodor} & \textbf{Factor} \\
\midrule
Total area ($\mu\text{m}^2$)       & 10\,620\,711.0 & 170\,811.0 & 62.18 \\
Area without CSR ($\mu\text{m}^2$) &    203\,168.0 & 114\,785.0 &  1.77 \\
\bottomrule
\end{tabular}
\end{table}

In the future, the \ac{VADL} hardware generator will be able to restrict the number of implemented registers and lift other limitations of the current prototype. For example, the memory is currently idealized, meaning read results can be served in the same cycle as they are requested. Another example is the need for a floating-point implementation and supporting advanced pipeline techniques, such as out-of-order execution. However, we see no reason why it should not be possible to add these extensions to the current prototype. Furthermore, they do not impact the primary goal of the prototype: demonstrating the feasibility of mapping the \ac{ISA} to the \ac{MiA}.

%% file: src-open_implementation.tex
\section{OpenVADL Implementation}
\label{sec:open_implementation}

Due to the restrictive licensing of the original \ac{VADL} implementation, we initiated the development of OpenVADL, an open-source alternative.
An early stage of OpenVADL was presented in~\cite{freitag2025OpenVADL}, a short article describing the QEMU-based \ac{ISS} appeared at the SAMOS 2025 conference~\cite{zottele2025QEMUOpenVADL}, and a two-page extended abstract at ASAP 2026 outlined the hardware generation~\cite{halder2026ASIP}.
Insights gained from the original framework frequently necessitated alternative implementation strategies.
Consequently, this section details the architectural aspects of the OpenVADL compiler, mirroring the structure of the original implementation while highlighting key design divergences. First, Section~\ref{sec:open_compiler_overview} provides a high-level overview of the compiler architecture. Subsequently, each principal component is analyzed in depth, followed by a comprehensive evaluation.

\subsection{OpenVADL Compiler Overview}
\label{sec:open_compiler_overview}

Consistent with the original \ac{VADL} framework, OpenVADL required a safe and efficient development environment, necessitating a strongly and statically typed programming language featuring automatic garbage collection and array bounds checking. Although Xtend initially appeared ideal, its source-to-source compiler generated Java code as an intermediate step before producing JVM bytecode, which introduced significant challenges. Other JVM languages typically compile directly to bytecode. Most notably, runtime line numbers correspond to the generated Java source code rather than the original Xtend files, severely impeding debugging. Furthermore, Xtend lacked mature IDE and tooling ecosystems. Consequently, OpenVADL was implemented directly in Java utilizing the Thymeleaf string template engine. This transition leveraged Java’s superior tooling infrastructure, including automated code style enforcement, which accelerated development and ultimately validated the language migration.

\input{src-figures-open_vadl_overview.tex}

Figure~\ref{fig:open_vadl_compiler_overview} presents an overview of the OpenVADL compiler design. Each processor specification starts as a plain \ac{VADL} text file. The frontend is responsible for reading the \ac{VADL} text file, applying all macros, and generating an \ac{AST}. The frontend handles name resolution, type inference, type checking, annotation checking, and constant evaluation. Section~\ref{sec:open_frontend} provides an overview of the frontend. Finally, the compiler transforms the \ac{AST} into the \ac{VIAM}, OpenVADL's \ac{IR}. Section~\ref{sec:viam} describes the \ac{VIAM} in detail.

Generators that do not require knowledge about the microarchitecture can use the \ac{VIAM} directly after the transformation. These generators are the assembler and linker generator (see Section~\ref{sec:open_asmgen}) and the QEMU-based instruction set simulator (see Section~\ref{sec:qemugen}). Furthermore, the compiler generator can do most of its work without knowing the microarchitecture. However, the generated compiler might perform better at instruction scheduling if the generator has knowledge about the microarchitecture.

The compiler executes the architecture synthesis (see Section~\ref{sec:open_miasynth}) prior to generators that require microarchitectural details. This step is responsible for integrating \ac{ISA} and \ac{MiA} while also bridging much of the semantic gap between \ac{VADL} and the generated artifacts. In addition to the compiler's instruction scheduler, the cycle-approximate simulator generator (see Section~\ref{sec:open_casgen}) and hardware generator (see Section~\ref{sec:open_hdlgen}) also rely on the architecture synthesis.

%% file: src-figures-open_vadl_overview.tex
\begin{figure}[h]

\begin{tikzpicture}[node distance=3cm]

\node (vadl) [ir] {VADL \\ Specification};
\node (parser) [ir_step, right of=vadl] {Frontend};
\node (viam) [ir, right of=parser] {VIAM};
\node (micro_syn) [ir_step, right of=vir] {Architecture Synthesis};
\node (cas) [generator, below of=micro_syn, yshift=1cm] {Cycle-Approximate Simulator};
\node (hardware) [generator, right of=micro_syn] {Hardware};
\node (compiler) [generator, above of=micro_syn, yshift=-1cm] {Compiler};
\node (asm) [generator, above of=viam, yshift=-1cm] {Assembler \& Linker};
\node (iss) [generator, below of=viam, yshift=1cm] {QEMU Simulator};

\draw [arrow] (vadl) -- (parser);
\draw [arrow] (parser) -- (viam);
\draw [arrow] (viam) -- (micro_syn);
\draw [arrow] (viam) -- (compiler);
\draw [arrow] (viam) -- (asm);
\draw [arrow] (viam) -- (iss);
\draw [arrow] (micro_syn) -- (cas);
\draw [arrow] (micro_syn) -- (compiler);
\draw [arrow] (micro_syn) -- (hardware);

\end{tikzpicture}

\caption{Overview of the OpenVADL Compiler Architecture. Yellow boxes represent generators.}
\label{fig:open_vadl_compiler_overview}

\end{figure}

%% file: src-open_frontend.tex
\subsection{OpenVADL Frontend}
\label{sec:open_frontend}

\begin{figure}[htb]
  \centering
  \begin{tikzpicture}[
      % Define styles to make it easy to change the look later
      node distance=10mm,
      oval/.style = {draw,         ellipse, align=center}, % For YADL spec and VIAN
      box/.style={draw, rectangle, rounded corners, align=center, minimum height=9mm}
    ]
    % --- NODE DEFINITIONS ---
    \node[oval] (vadl) {VADL\\Spec};
    \node[box, right=of vadl] (parser) {Parser};
    \node[box, above=of parser] (macro) {Macro\\expander};
    \node[box, right=of parser] (symbol) {Name\\resolver};
    %\node[box, right=of symbol] (ungrouper) {Ungrouper};
    \node[box, right=of symbol] (typecheck)  {Type checker};
    %\node[box, below=of typecheck] (remover) {Macro\\remover};
    \node[box, above=of typecheck] (const) {Constant\\evaluator};
    \node[box, right=of typecheck] (lowering) {Lowering};
    \node[oval, right=of lowering] (viam)  {VIAM};

    % --- CONNECTIONS (Edges) ---
    \draw[arrow] (vadl) -- (parser);
    \draw[arrow] (parser) -- (macro);
    \draw[arrow] (macro) -- (parser);
    \draw[arrow] (parser) -- (symbol);
    \draw[arrow] (symbol) -- (typecheck);
    %\draw[arrow] (ungrouper) -- (typecheck);
    %\draw[arrow] (remover) -- (typecheck);
    \draw[arrow] (const) -- (typecheck);
    \draw[arrow] (typecheck) -- (lowering);
    \draw[arrow] (lowering) -- (viam);
    \draw[arrow] (typecheck) -- (const);
  \end{tikzpicture}
  \caption{OpenVADL's front-end compilation pipeline.}
  \label{fig:implementation-openvadl-pipeline}
\end{figure}

The Xtext framework used in the original implementation requires generating a \ac{CST}, which adds additional passes to the parser.
Furthermore, the generated parser is based on the high-overhead Eclipse Modeling Framework, preventing ahead-of-time compilation.
Therefore, for OpenVADL, we replaced Xtext with Coco/R~\cite{Woess2003CocoR,CocoR}.
Coco/R is a pred-LL(1) parser generator based on an attribute grammar.
Coco/R does not generate a \ac{CST}. The generation of an \ac{AST} has to be specified by the user with attributes.
This makes it possible to generate an \ac{AST} with its syntax macros already expanded, all in a single pass.
Combined with ahead-of-time compilation, the OpenVADL parser is up to 80 times faster compared to the original \ac{VADL} parser.
It also introduces several language improvements, such as macros which generate macros, additional syntax types, and operator precedence for expressions (see~\cite{Nestler2024} for a detailed description).

Following parsing, the type checker verifies the types and bit widths of all expressions and infers them where needed. It also ensures the correctness of function arguments and return values, as well as valid access to tensors. Additionally, the type checker performs several semantic checks, such as verifying that format fields are valid (containing no overlapping fields or unused bits), that bit-slice indices remain within bounds, and that every instruction has a matching encoding and assembly definition.

Generator-dependent semantic checks are deferred to the back-ends, such as ensuring that all encodings are unique and no unused fields exist. Furthermore, on the \ac{VIAM}, the compiler verifies that each resource is written at most once per instruction, while the \ac{MiA} synthesis checks the correctness of the microarchitecture.

The type checker traverses the \ac{AST} in a single pass using a bidirectional approach: expected types from parent nodes are passed down to their children when available, while final types are inferred or verified bottom-up. Overall, OpenVADL's front-end pipeline consists of only 6 passes (down from 21 in the original \ac{VADL} implementation), with several passes interleaved, as indicated by the double-headed arrows in Figure~\ref{fig:implementation-openvadl-pipeline}.

%% file: src-open_lsp.tex
\subsubsection{Language Server}
\label{sec:open_lsp}

OpenVADL comes with a language server that is used by \acp{IDE} to provide semantic information and editing functionality to \ac{VADL} developers. Client-server communication is done via the \ac{LSP}~\cite{LSP}, which serves as a standardized interface between \acp{IDE} and language servers and is supported by most \acp{IDE}. Thus, language support has to be implemented only once and can then be reused across different \acp{IDE}.

Our language server implementation is built on top of the \ac{VADL} frontend, consuming the \ac{AST} provided by the parser and thereby further reducing duplicate implementation. It provides compiler diagnostics and go-to-definition functionality, as well as type information and expanded macros via hover boxes.

%% file: src-viam.tex
\subsection{The \ac{VADL} Intermediate Architecture Model (\ac{VIAM})}
\label{sec:viam}

The \ac{VIAM} serves as an internal representation of a \ac{VADL} specification used in the OpenVADL compiler. It is designed to be both generic, ensuring compatibility with various generators, and extensible, allowing each generator to customize it as required. The \ac{VIAM} is divided into two components: one defines the structural aspects of a \ac{VADL} specification, including its definitions, while the other captures the behavioral elements, such as instruction behavior.

\subsubsection{Behavior Graph}

The behavior component of the \ac{VIAM} is modeled as a multi-graph that integrates both a dependency graph and a control flow graph. The functional nature of \ac{VADL}'s behavior description enables the construction of a data dependency graph, where data flow is naturally in \ac{SSA} form due to the semantics of let assignments~\cite{Appel1998SSA,singer2003static,Beringer2022}. Control flow constructs, such as if-else, are represented within the control flow graph. This design is inspired by the \ac{IR} of the GraalVM compiler~\cite{duboscq2013anintermediate, duboscq2013graalir}.

\ac{VADL} behavior follows a sequential reading style for enhanced readability, while certain constraints enable a parallel interpretation. Semantically, all resource reads complete before instruction execution, and all writes apply only after it finishes.
By modeling side effects---such as register writes---as dependencies of end nodes, \ac{VIAM} enforces this parallel view, discarding execution order and representing only the conditions under which side effects occur. Read operations function as standard expression dependencies, ensuring each register read appears only once in the graph, regardless of multiple occurrences in the source code. Additionally, no explicit ordering is maintained between a read and a write to the same resource, providing generators with greater flexibility in managing side effects.

\begin{figure}[h]
\begin{tikzpicture}
  [
      node distance={35mm},
      thick,
      unused/.style =
          { draw
          , dotted
          , minimum width=3em
          , minimum height=4em
          , fill=black!10
          },
  ]

\tikzstyle{n} = [ellipse, rounded corners, minimum height=0.5cm,text centered, draw=black, outer sep=0mm]
\tikzstyle{c} = [rectangle, minimum height=0.5cm,text centered, draw=black, outer sep=0mm]

\node[c] (start) at (2, 0) {start};
\node[c] (end) at (0, -1) {instr end};

\node[n, dotted] (wregx) at (0, 0)      {write<X>};
\node[n] (add) at (-2, 1)  {add: Bits<64>};
\node[n] (rx1) at (-4, 2)    {read<X>: Bits<64>};
\node[n] (rx2) at (0, 2)    {read<X>: Bits<64>};
\node[n] (frd) at (2, 1)    {field<rd>: Bits<5>};
\node[n] (frs1) at (-4, 3) {field<rs1>: Bits<5>};
\node[n] (frs2) at (0, 3) {field<rs2>: Bits<5>};

\draw [arrow] (rx1) -- (frs1);
\draw [arrow] (rx2) -- (frs2);
\draw [arrow] (add) -- (rx1);
\draw [arrow] (add) -- (rx2);
\draw [arrow] (wregx) -- (frd);
\draw [arrow] (wregx) -- (add);
\draw [arrow] (end) -- (wregx);
\draw [red, arrow] (start) -- (end);

\end{tikzpicture}
\caption[short]{VIAM behavior graph of the RISC-V \isainstruction{ADD} instruction}
\label{fig:vadl_add_instruction_behavior}
\end{figure}

Figure~\ref{fig:vadl_add_instruction_behavior} illustrates the behavior graph for the RISC-V \isainstruction{ADD} instruction. Each instruction begins with a control flow start node and concludes with an end node, representing the minimal control flow structure. The end node depends on the register file write operation, which depends on the instruction's format field reference \vadlresource{rd}, indicating the destination register index. Additionally, the add operation depends on the register file reads for the operand indices.

To prevent read-write and write-write conflicts caused by the side effect relaxation described above, users are prohibited from writing to the same resource multiple times within a single instruction execution path, as this would lead to undefined behavior. While this constraint can be enforced for registers, it is not always feasible for register files and memory locations, as their exact addresses may not be statically determinable.

\subsubsection{VIAM vs. VIR}

As described in Section~\ref{sec:vir}, the \ac{VIR} is a quadruple-code \ac{IR} used in the original \ac{VADL} implementation, which fundamentally differs from the behavior component of the \ac{VIAM}. Different generators have distinct requirements for the \ac{IR}, depending on how they analyze specifications and apply transformations. During the original \ac{VADL} implementation, it became evident that the more hardware-oriented \ac{VIR} is not well suited for most generator use cases, leading generators to introduce their own intermediate representations.

The \ac{ISS} generator produces sequential code that closely resembles the quadruple-code structure of the \ac{VIR}. However, due to the uniqueness of expressions in \ac{VIAM} behavior graphs, optimizations such as common sub-expression elimination, copy propagation, and global value numbering occur naturally during graph construction.  Furthermore, in the new QEMU-based \ac{ISS}, dependency analysis and scheduling of potentially conflicting operations are more efficiently managed with \ac{VIAM} behavior graphs. By lowering the graph to custom control nodes, an operation sequence is generated that can be easily translated into C code.

For the compiler generator, the primary task is to analyze instruction semantics and generate instruction selection patterns. These patterns define the dataflow nodes that must match to emit a machine instruction. In the original \ac{VADL} implementation, the compiler generator introduced its own dataflow representation, as performing analysis on \ac{VIR} is unnecessarily complex. In the OpenVADL implementation, this additional representation is no longer required, as the behavior graph itself is directly usable to find selection patterns.

Hardware generation leverages the \ac{VIAM} to create the \ac{IPG}, which is then combined with the \ac{MiA} description in \ac{MiA} synthesis. \ac{VIAM} behavior design streamlines the process by eliminating the need to construct an intermediate \ac{DFG} from quadruple-code and enabling control flow elimination during \ac{IPG} creation. Unlike the original \ac{VIR}, where the \ac{CAS} can execute one basic block per cycle, the \ac{DFG} in \ac{VIAM} lacks this strict cycle-based operation grouping. Instead, during \ac{VIAM} behavior scheduling, dependencies must be analyzed to determine which operations can execute within a single cycle.

The \ac{VIR} can be easily dumped into human-readable files, which is particularly useful for debugging. In contrast, \ac{VIAM} behavior graphs cannot be directly represented in a text-based format that is easily readable by developers. To address this, OpenVADL exports the \ac{VIAM} as an \acs{HTML} file, embedding all graphs in DOT format. These graphs can then be visualized using an \acs{HTML}-embedded graph viewer, allowing developers to inspect and debug behavior graphs.

%% file: src-open_compgen.tex
\subsection{OpenVADL Compiler Generator}
\label{sec:open_compgen}

OpenVADL's compiler generator consumes the \ac{VIAM} as input and outputs C++ and TableGen files that are copied to a cloned LLVM repository\footnote{The compiler generator works with the git tag version llvmorg-19.1.7.}. The generated files extend or overwrite files in the repository.
Afterwards, the LLVM project can be compiled by specifying the new target backend. The compilation emits a toolchain with an option for the specified target architecture in \ac{VADL}. \\
OpenVADL's compiler generator was rewritten from scratch with lessons learned from the original \ac{VADL} implementation.
Generally, the generator has two phases and is structured as a pass-based compiler. The generic compiler backend, \ac{GCB}, contains all the
passes that are required for different compiler toolchain frameworks. In contrast, the LLVM compiler backend \ac{LCB} contains all the passes that are specific to generating an LLVM toolchain. At the moment, the LLVM compiler toolchain is the only supported compiler infrastructure project. \\
Figure~\ref{fig:open_gcb_lcb_overview} gives an overview of the different stages of the compiler generation. Generally, the compiler generator
derives target-specific information from a specification, using default values for any information that was omitted. Afterwards, the compiler generator generates instruction-selection patterns for LLVM and finally emits the files for the toolchain. The output consists of C++ classes that override methods with target-specific code and TableGen files. TableGen is a domain-specific language for specifying registers, calling conventions, instruction definitions and instruction-selection patterns in LLVM. During the compilation of the generated compiler, the TableGen files are translated to C++.

\begin{figure}[h]
\begin{center}
\begin{tikzpicture}
[
    node distance={20mm},
    thick,
    unused/.style =
        { draw
        , dotted
        , minimum width=6em
        , minimum height=6em
        , fill=black!10
        },
    block/.style =
        { rectangle
        , draw=black
        , thick
        , fill=white
        , text width=5em
        , align=center
        , rounded corners
        , minimum height=4em
        },
    blockSmall/.style =
        { rectangle
        , draw=black
        , thick
        , fill=white
        , text width=5em
        , align=center
        , rounded corners
        , minimum height=2em
        },
    blockSmallInvis/.style =
        { rectangle
        , draw=white
        , thick
        , fill=white
        , text width=5em
        , align=center
        , rounded corners
        , minimum height=2em
        },
]

\path
    (-3, 0) node[unused]
        (VIAM)
        {VIAM}
    (0, 0) node[block]
        (GenEnc)
        {Generate Encoding / Predicates}
    (2.5, 0) node[block]
        (GenOperands)
        {Operand Detection}
    (5, 0) node[block]
        (EdgeCasePruning)
        {Edge Case Pruning}

    (-3, -3) node[block]
        (InstructionMatching)
        {Instruction Matching}

    (-0.5, -3) node[block]
        (LLVMLowering)
        {LLVM Lowering}
    (2, -3) node[block]
        (GenTableGen)
        {Prepare TableGen}

    (4.5, -2) node[blockSmall]
        (FileEmitter1)
        {LLVM File Emitter 1}
    (4.5, -3) node[blockSmallInvis]
        (FileEmitterDots)
        {. . .}
    (4.5, -4) node[blockSmall]
        (FileEmitterN)
        {LLVM File Emitter n}
    (7.5, -3) node[unused]
        (LLVMBackend)
        {LLVM Backend}
    ;

\node
    [ rectangle
    , draw=black!30
    , dotted
    , fit=
        (VIAM)
        (GenEnc)
        (GenOperands)
        (EdgeCasePruning)
    , inner sep=3mm
    , label={GCB}
    ]
    (GCB) {};

\node
    [ rectangle
    , draw=black!30
    , dotted
    , fit=
        (InstructionMatching)
        (LLVMLowering)
        (GenTableGen)
        (FileEmitter1)
        (FileEmitterDots)
        (FileEmitterN)
    , inner sep=3mm
    , label={[shift={(-3.0,-3.5)}] LCB}
    ]
    (LCB) {};

\draw[->] (VIAM) -- (GenEnc);
\draw[->] (GenEnc) -- (GenOperands);
\draw[->] (GenOperands) -- (EdgeCasePruning);
\draw[->] (EdgeCasePruning) to[out=230,in=45] (InstructionMatching);
\draw[->] (InstructionMatching) -- (LLVMLowering);
\draw[->] (LLVMLowering) -- (GenTableGen);
\draw[->, dotted] (GenTableGen) -- (FileEmitterDots);
\draw[->] (GenTableGen) -- (FileEmitter1);
\draw[->] (GenTableGen) -- (FileEmitterN);
\draw[->] (FileEmitter1) -- (LLVMBackend);
\draw[->,dotted] (FileEmitterDots) -- (LLVMBackend);
\draw[->] (FileEmitterN) -- (LLVMBackend);
\end{tikzpicture}

\caption{OpenVADL Compiler Generator Overview from~\cite{PerKevin2025lcb}}
\label{fig:open_gcb_lcb_overview}
\end{center}
\end{figure}

\subsubsection{Encoding and Predicate Generation}

As shown in Figure~\ref{fig:open_gcb_lcb_overview}, the first step is the generation of encoding functions and predicates from field access functions. \ac{VADL}'s instruction format defines the binary representation of an instruction. In the instruction behavior, the decoded values can be accessed by field access functions. However, the compiler needs to encode values, which requires applying the inverse transformation of a field access function. For convenience, an encoding function can be omitted in a \ac{VADL} specification when the field access function is trivial or solely relies on arithmetic or shifting operations. The \ac{GCB} generates such encoding functions automatically from the field access function~\cite[Section 3.3]{PerKevin2025lcb}.

A predicate function is used to check during instruction selection whether a candidate value can be represented in an encoded field of an instruction format. As with encoding functions, these can also be omitted from a \ac{VADL} specification when the field access function is trivial or solely relies on arithmetic or shifting operations~\cite[Section 3.4]{PerKevin2025lcb}.

\subsubsection{Operand Detection}

After the function generation, the next step is operand detection, as shown in Figure~\ref{fig:open_gcb_lcb_overview}. An instruction may consume input and produce output values. These values may be represented by register operands that specify where values are read from or written to. However, an immediate operand is a constant value encoded directly in the instruction. Since operands are not explicitly defined in a \ac{VADL} specification, the compiler generator has to compute them from the instruction behavior. The generator derives the operands from the behavior graph by inspecting its node types. Nodes of type \textit{ReadRegTensorNode} and \textit{ReadArtificialResNode} are treated as register input operands. Similarly, nodes of type \textit{WriteRegTensorNode} and \textit{WriteArtificialResNode} are treated as register output operands. Finally, invocations of field access functions are recognized as immediate or bare symbol operands.
Bare symbol operands reference labels or methods and require special treatment at a later stage during compilation because both need to be relocated by a linker~\cite[Section 3.5]{PerKevin2025lcb}.

\subsubsection{Edge Case Pruning and Instruction Matching}

Edge case pruning is a preparation step for the instruction classification because it simplifies the recognition of patterns in the behavior graph.
The compiler generator requires knowledge about the available instructions in an ISA to generate an efficient compiler.
The generator defines a set of static properties that characterize instructions. It then iterates over all the instructions and checks these properties against their dataflow graph. A detailed overview of the predefined properties can be found in~\cite[Section 3.6]{PerKevin2025lcb}. The difficulty is that compilers rely on \textit{undefined behavior} of instructions for optimizations. However, the instruction behavior in a \ac{VADL} specification has no information about which statement implements undefined behavior. With edge case pruning, the generator reintroduces undefined behavior by eliminating branches and side effects that model cases irrelevant to the compiler selection. This simplifies instruction matching because the important dataflow properties remain in the behavior graph. Unfortunately, the relationship between pruning and instruction matching creates a circular dependency because what behavior is undefined is only clear after matching. For example, statements that raise an exception or certain conditionals are removed as they represent edge cases~\cite[Section 3.7]{PerKevin2025lcb}.

\subsubsection{LLVM Lowering}

After the instructions are labeled, the generator can use the assigned instruction label to generate instruction-selection patterns. Figure~\ref{fig:generalLoweringApproach} gives an overview of the lowering for machine instructions.

The LCB's central task is to provide the LLVM backend with the tree patterns representing the semantics of various instructions specified in the \ac{VADL} specification. During compilation, LLVM uses these patterns to completely cover a program's dataflow representation. This phase is called \emph{Instruction Selection} and LLVM's dataflow representation is a \ac{DAG} which is called the \emph{Instruction Selection Graph}. Once LLVM has found a complete cover of all the nodes from the \emph{Instruction Selection Graph}, it will emit machine instructions which have been specified in the \ac{VADL} specification. Thus, the generated code will be semantically equivalent to the LLVM IR program.

For most instructions, instruction-selection patterns closely resemble the behavior specified in a \ac{VADL} specification. Therefore, the generator constructs a pattern by recursively traversing the dataflow graph. Unconditional jumps, indirect jumps and memory accesses require specialized TableGen patterns~\cite[Section 3.7]{PerKevin2025lcb}.

A specification can also contain instructions that have no label. For example, a \ac{VADL} specification may contain an instruction that combines two instructions into one, e.g. fused multiply-add. It is infeasible to define static properties for every possible combination of operations. In cases where the generator does not recognize an instruction, the generator checks whether the dataflow graph contains nodes that cannot be lowered by the recursive generation method~\cite[Section 3.7.1]{PerKevin2025lcb}. The current algorithm cannot handle instructions that have multiple results or contain select nodes because LLVM's instruction selector requires tree-shaped patterns. If a behavior graph contains any of those, then the generation of instruction-selection patterns for this particular instruction is skipped. Otherwise, the recursive traversal method is applied.

\subsubsection{Putting It All Together}

Finally, after the generation of instruction encodings, immediate-value predicates, operand detection and lowering of instruction selectors, the compiler generator requires additional C++ and TableGen files to create a functional cross-compiler. In~\cite[Section 3.8]{PerKevin2025lcb} the operand detection is extended to support pseudo instructions. Additionally,~\cite[Section 3.9]{PerKevin2025lcb} presents the lowering of registers. \ac{VADL} specifications may define relations between register files to represent subregister relationships that need to be defined in TableGen. Further,~\cite[Section 3.12]{PerKevin2025lcb} presents constant materialization. A constant materialization sequence is a compiler-only pseudo instruction that directs the compiler to construct an immediate value by using the machine instruction sequence specified in \ac{VADL}.
Together, these steps provide the remaining information required to generate an LLVM compiler from a \ac{VADL} processor specification.

\begin{figure}[h]
\begin{tikzpicture}
  [
      node distance={35mm},
      thick,
      unused/.style =
          { draw
          , dotted
          , minimum width=3em
          , minimum height=4em
          , fill=black!10
          },
  ]

\tikzstyle{c} = [rectangle, minimum height=0.5cm,text centered, outer sep=0mm]

\node[c] (noPat) at (10, -2) {No Pattern};
\node[c] (checkRedFlags) at (10, 0) {\begin{tabular}{c}
    Check for \\
    Red Flags
\end{tabular}};
\node[c] (recursiveMethod) at (6, -3) {\begin{tabular}{c}
    Recursive \\
    Lowering \\
    Method
\end{tabular}};
\node[c] (knownPat) at (1, 0) {\begin{tabular}{c}
    Pattern is known \\
    to have a certain \\
    form
\end{tabular}};
\node[c] (check) at (6, 3) {\begin{tabular}{c}
    Check properties \\
    of dataflow
\end{tabular}};

\draw [arrow,bend right=20] (check) edge node[left, font=\itshape]{\begin{tabular}{c}
    Unconditional jumps \\
    Indirect jumps \\
    Memory write / load \\
    ...
\end{tabular}} (knownPat);
\draw [arrow,bend left=20] (check) edge node[right, font=\itshape]{Found no match} (checkRedFlags);
\draw [arrow,bend left=10] (checkRedFlags) edge node[above, font=\itshape]{Ok} (recursiveMethod);
\draw [arrow] (checkRedFlags) edge node[right, font=\itshape]{Not ok} (noPat);
\draw [arrow] (check) edge[loop above] node[above, font=\itshape]{Repeat for all instructions} (check);
\draw [arrow] (check) edge node[left, font=\itshape]{\begin{tabular}{c}
    Arithmetic \\
    Logical \\
    Conditionals \\
    ...
\end{tabular}} (recursiveMethod);
\end{tikzpicture}
\caption{General Lowering Approach for Machine Instructions from~\cite[Figure 3.2]{PerKevin2025lcb}}
\label{fig:generalLoweringApproach}
\end{figure}

%% file: src-open_asmgen.tex
\subsection{OpenVADL Assembler and Linker Generator}
\label{sec:open_asmgen}

As in the original \ac{VADL} implementation (Section~\ref{sec:orig_asmgen}), OpenVADL's assembler and linker generator is situated in the \ac{LCB}.
Its generated artifacts therefore also conform to the structure mandated by LLVM's architecture.
We refrain from detailing each component here, as their design largely mirrors that of the corresponding components in the original \ac{VADL} implementation.
Instead, we will focus on the key improvement of OpenVADL, the predicated-LL(1) parsing algorithm of its generated assembly parser.

As detailed in the evaluation of the original \ac{VADL} assembler generator (Section~\ref{sec:orig_eval_asm_link}), the combination of its LL(1) parsing algorithm with the separated parsing and matching phases leads to limitations with abbreviated syntaxes and distinguishing instructions based on syntactical structure.
One example of a problematic abbreviated syntax is the RISC-V \texttt{JALR} instruction, which has multiple syntactical forms such as \texttt{JALR rd, rs} and \texttt{JALR rs} (\texttt{rd} defaulting to register \texttt{x1}).
Given an instruction like \texttt{JALR x1, x2}, upon encountering the first register operand \texttt{x1}, it is not clear to the parser which form is being used.
Since an LL(1) parser makes decisions based solely on the next token, it cannot disambiguate between these forms at this point.
Therefore, the original \ac{VADL} assembly parser is not able to parse such abbreviated syntaxes.

OpenVADL's solution for this limitation is the introduction of semantic predicates as elements of the assembly grammar.
Semantic predicates are a well-known concept used in popular parser generators such as ANTLR~\cite{ParrFischer2011} and Coco/R (called resolvers)~\cite{Woess2003CocoR}.
In OpenVADL, semantic predicates are boolean expressions written in \ac{VADL}, guiding the parser's decision-making process at points where multiple derivations are possible.
At such decision points, the parser evaluates semantic predicate expressions in their order of specification and selects the first derivation whose semantic predicate evaluates to true.

Semantic predicates must be placed precisely at decision points where LL(1) conflicts arise.
However, the burden of identifying conflict points is not offloaded to the assembly grammar designer.
OpenVADL's frontend identifies LL(1) conflicts and provides detailed conflict locations and descriptions.
The same mechanism also prohibits grammar designers from over-specifying semantic predicates, ensuring that they are only introduced where necessary to resolve actual LL(1) conflicts.

To enable semantic predicates to evaluate the syntactical structure beyond the current token, OpenVADL provides the built-in function \texttt{LaIdEq} (short for \textit{Lookahead Identifier Equals}).
It takes two arguments, where the first is a natural number \texttt{n}
specifying that the token to consider is \texttt{n} tokens ahead of the current token, and the second is an identifier to compare against the token at that position.
\texttt{LaIdEq} evaluates to \texttt{true} if the token \texttt{n} positions ahead of the current token matches the specified identifier, and \texttt{false} otherwise.
An example showing an assembly grammar rule for the RISC-V \texttt{JALR} instruction using semantic predicates and the \texttt{LaIdEq} function can be found in Listing~\ref{lst:asm_jalr} in Appendix~\ref{app:jalr}.

%% file: src-qemugen.tex
\subsection{QEMU Generator}
\label{sec:qemugen}

OpenVADL introduces a new \ac{ISS} generator to overcome the limited performance of \ac{VADL}'s \ac{DTC}-based \ac{ISS}.
The new \ac{ISS} is based on QEMU, an open-source emulator and virtualizer that enables hardware virtualization and full-system emulation for various architectures.

QEMU enables programs compiled for a guest architecture to run on a different host architecture using dynamic binary translation for high performance. Its modular design simplifies the addition of new guest and host architectures. To decouple guest and host implementations, QEMU employs the \ac{TCG}.
As shown in Figure~\ref{fig:qemu_translation}, the guest frontend reads and decodes instructions from a \ac{TB} (a basic block of target code) and translates them into \ac{TCG} ops, QEMU's architecture-independent \ac{IR}. The \ac{TCG} then optimizes this \ac{IR} before passing it to the host backend, which translates it into machine instructions.

\begin{figure}[t]
    \centering
    \begin{tikzpicture}[node distance=1.8cm, thick]
        \tikzstyle{box} = [rectangle, draw, fill=gray!20, minimum width=2cm, align=center, minimum height=1cm]
        \tikzstyle{arrow} = [thick,->,>=stealth]

        % Nodes
        \node (guest) [box] {RISC-V\\ \texttt{lb a11, 8(a10)}};
        \node (tcg) [box, right of=guest, xshift=4cm] {
            TCG IR \\
            \texttt{add\_i64 loc3,x10,8} \\  \texttt{q\_ld\_i64 x11,loc3}
        };
        \node (host) [box, right of=tcg, xshift=4cm] {
            x86\_64 \\
            \texttt{lea rdi, [r10 + 8]} \\
            \texttt{mov r11, qword ptr [rdi]}
        };

        % Arrows
        \draw [arrow] (guest.east) -- (tcg.west) node[midway, above] {Guest Frontend};
        \draw [arrow] (tcg.east) -- (host.west) node[midway, above] {Host Backend};

    \end{tikzpicture}
    \caption{\ac{TCG} Translation Process}
    \label{fig:qemu_translation}
\end{figure}

To generate a minimal QEMU target, the generator must define four key components: a \textit{\acs{CPU} state}, which stores all register values and \acs{CPU}-related states; a \textit{machine}, responsible for memory initialization and firmware loading; an \textit{instruction decoder}; and the \textit{TCG translation}. The \acs{CPU} definition can be generated straightforwardly, as it is directly derivable from the \ac{VIAM}.

The machine definition is relatively generic, incorporating only memory definitions from the \ac{VADL} ISA specification. When defining the processor, which serves as the entry point for \ac{ISS} generation, users can annotate it with \texttt{[enable htif]}. This enables support for the \ac{HTIF}, a simple protocol used in the RISC-V Spike simulator to facilitate communication with the simulation host. \ac{HTIF} operates by mapping user-defined memory addresses to callbacks that interpret commands sent by the simulated program. For example, this mechanism allows the simulated program to end full-system emulation with a specific exit code, which is particularly useful for running self-verifying tests on the \ac{ISS}.

QEMU provides its own decode tree format, allowing frontend developers to define a readable instruction format specification, similar to what the \ac{VADL} language offers. The build system then generates C functions and structs, which the frontend uses to decode and process instructions. While the \ac{VDT} generator can emit this format, the format itself lacks support for variable-length instructions. To overcome this limitation, the \ac{ISS} generates its own C-based decoder instead.

The core functionality of the generated \ac{TCG} frontend lies in \ac{TCG} operation generation. The \ac{ISS} generator produces a translate function for each instruction, which generates a sequence of \ac{TCG} operations. These operations work on strongly typed variables, where each operation follows a fixed format: a number of leading output variable operands, followed by input or constant variables. Examples of such translate functions can be seen in Listing~\ref{lst:qemu_trans_addi} and Listing~\ref{lst:qemu_trans_beq}, which show implementations for the \texttt{ADDI} and \texttt{BEQ} RISC-V instructions, respectively.

There are different types of variables in \ac{TCG}, defined by their lifespan and modifiability. Global variables persist across all \ac{TB}s and correspond to memory in the \acs{CPU} state. For example, a register in the \acs{CPU} state is represented as a global variable, and any write to the \ac{TCG} variable during instruction execution is propagated to the corresponding register. Constant variables exist throughout a \ac{TB} but are immutable singletons. They are allocated on demand during translation, only if a constant for the given value does not already exist. Temporary \ac{TB} variables live for the duration of a \ac{TB} but are discarded upon any exit.

\begin{figure}
\begin{minipage}[b]{0.49\textwidth}
\begin{lstlisting}[label={lst:qemu_trans_addi},caption={QEMU translate function for the {\tt ADDI} instruction},language=vadl,morekeywords={bool,DisasContext,int,TCGv_i64,arg_addi},linerange={1-24}]
bool trans_addi(
    DisasContext *ctx, 
    arg_addi *a) 
{
    TCGv_i64 x_rs1 = get_x(ctx, a->rs1);
    TCGv_i64 x_rd = dest_x(ctx, a->rd);
    TCGv_i64 const_immS = 
        tcg_constant_i64(a->immS);
        
    tcg_add_i64(x_rd, x_rs1, const_immS);
    return true;
}
\end{lstlisting}
\end{minipage}
\hfill
\begin{minipage}[b]{0.49\textwidth}
\begin{lstlisting}[label={lst:qemu_trans_beq},caption={QEMU translate function for the {\tt BEQ} instruction},language=vadl,morekeywords={bool,DisasContext,int,TCGv_i64,TCGLabel,arg_beq},linerange={1-20}]
bool trans_beq(
    DisasContext *ctx, 
    arg_beq *a) {
    TCGv_i64 x_rs1 = get_x(ctx, a->rs1);
    TCGv_i64 x_rs2 = get_x(ctx, a->rs2);
    TCGLabel *l_else_0 = gen_new_label();
    tcg_brcond_i64(TCG_COND_EQ, x_rs1, 
        x_rs2, l_else_0);
    gen_goto_tb(ctx, 1, 
        ctx->base.pc_next + a->immS);
    gen_set_label(l_else_0);
    ctx->base.is_jmp = DISAS_CHAIN;
    return true;
}
\end{lstlisting}
\end{minipage}
\end{figure}

The following sections outline the key passes of the \ac{ISS} generator, presented in the order of their execution, which are essential for generating \ac{TCG} translation functions.

\subsubsection{Operation Decomposition}
While the \ac{VADL} specification allows arbitrary bit widths, QEMU imposes a 64-bit limit for most operations. This becomes problematic when an instruction specification requires types larger than 64 bits. For example, the \isainstruction{MULH} instruction in the \texttt{RV64IM} specification performs a long multiplication of two 64-bit values and extracts the upper half of the 128-bit result. To handle such cases, the Operation Decomposition pass splits these operations into multiple logically equivalent operations that only accept and return values with a maximum size of 64 bits.

\subsubsection{Side Effect Scheduling}
As discussed in Section~\ref{sec:viam}, the \ac{VIAM} behavior graph represents expressions and side effects using a dependency graph. However, since \ac{TCG} ops execute sequentially, this dependency graph must be scheduled. The first step in this process is scheduling side effects, such as register writes. Additionally, the pass analyzes whether a side effect causes an instruction exit by modifying the program counter. Non-exit side effects are scheduled at the start of the control flow branch, while program counter manipulations are placed immediately before the branch end. This ensures that a jump out of the instruction does not occur before all other side effects have been applied.

\subsubsection{Safe Resource Read}
To ensure that writes do not occur before reads to the same resource, potentially conflicting reads must be scheduled before any writes to that resource. Since register file indices and memory addresses are not statically known, all reads to these resources must be conservatively treated as potential conflicts with all writes to the same resource.

\subsubsection{TCG Expression Scheduling}
Before lowering to \ac{TCG} operations, it must be determined which expressions are evaluated during \ac{TCG} translation and which are executed at runtime when the translated \ac{TCG} code is executed. Expressions evaluated at runtime must be converted into \ac{TCG} operations and scheduled accordingly. This includes all expressions that depend on the \acs{CPU} state or memory, such as register reads.

Conversely, expressions that depend only on immediate values, such as format field values, can be computed at translation time and represented as constant \ac{TCG} variables. These expressions do not require scheduling, as their dependencies can be directly translated into C expressions.

After this pass, all dependency nodes corresponding to \ac{TCG} operations are correctly scheduled.

\subsubsection{TCG Branch Lowering}
At this stage, branches within the instruction are represented in the control flow using if-else nodes. However, \ac{TCG} implements jumps within a \ac{TB} using goto-like operations, such as \texttt{set\_label}, \texttt{br} and \texttt{brcond}. This pass analyzes which if-else control flow must be converted into \ac{TCG} operations—specifically, those where the condition expression was previously scheduled as a \ac{TCG} operation. These control flow structures are then transformed into a linear sequence of \ac{TCG} operations using labels and conditional branching.

If-else control flow that does not require \ac{TCG} translation remains unchanged and is later directly translated into if statements in C.

\subsubsection{TCG Op Lowering}
The scheduled dependency nodes are lowered into control nodes, each corresponding to one or more \ac{TCG} operations. During this process, a node retrieves its destination and input \ac{TCG} variables from a context that generates variables on demand and attaches them to the dependency node.
Once lowering is complete, all dependency nodes are removed from the graph. The resulting structure is a \ac{CFG} consisting of \ac{TCG} op nodes in \ac{SSA} form.

\subsubsection{TCG Variable Allocation} \label{sec:tcg-variable-allocation}
To minimize the number of temporary \ac{TCG} variables, a variable allocation pass is applied to the \ac{TCG} \ac{CFG}. First, the live ranges of previously created temporary variables are determined. Then, graph coloring is used to compute an optimized \ac{TCG} variable assignment. The primary objective is to maximize the reuse of written registers, reducing unnecessary temporary allocations.

\subsubsection{Putting It All Together}
After the final pass before code generation, the instruction behavior graphs directly reflect the structure of the C code to be generated. \ac{TCG} operations correspond to control nodes with a single successor, if-nodes translate to C if statements, and expression nodes map to C expressions. A code generator processes this graph, producing a C function named \texttt{trans\_<mnemonic>}, which takes a \ac{TCG} context and a struct containing all format fields of the instruction. This translation function is then invoked by the decoder during QEMU execution.

%% file: src-open_miasynth.tex
\subsection{OpenVADL Architecture Synthesis}   % linus
\label{sec:open_miasynth}

The architecture synthesis uses the \ac{VIAM} produced by the frontend to generate a processor implementation from the description of the \ac{MiA} and the \ac{ISA}. Primarily, it converts the MiA definition so that it references the \ac{ISA} the \ac{MiA} should implement and includes all instruction behavior. The result is the processor implementation represented in the \ac{VIAM}, which can be used by the hardware generator to output \ac{HDL}, by the compiler to supplement instruction selection and by the cycle-approximate simulator.

\subsubsection{\ac{IPG} Creation}

The central data structure for the architecture synthesis is the \acf{IPG}. It captures the behavior of all instructions in a single graph and is used to represent the intermediate values during an instruction's execution in the stages of the \ac{MiA}. This approach is inspired by the original \ac{VADL} implementation (Section~\ref{sec:orig_miasynth}), where a similar data structure is used for MiA synthesis. In the \ac{IPG}, instruction behavior is represented by the instruction's dataflow. Because of this, the illustrations in this subsection show VIAM dependency graphs with reversed edges, i.e., dataflow graphs. Each node is labeled with the set of instructions it is used for and the stage it will be placed in. Upon creation, the stage label remains empty.

In order to build this graph, first, control flow must be eliminated from all instruction behavior. As a part of the frontend, all control flows in \ac{VIAM} behavior graphs are traversed, branch conditions are collected and enable conditions for side effects are computed accordingly. This already yields a representation free from control flow by simply disregarding control flow nodes.
In the MiA implementation, read nodes also have side effects, because they use ports on register files and affect hazard detection and forwarding. Read nodes are also assigned enable conditions by analyzing the write nodes a read value flows into. This concludes the preprocessing.

\begin{figure}[h]
\begin{tikzpicture}
  [
      node distance={35mm},
      thick,
      unused/.style =
          { draw
          , dotted
          , minimum width=3em
          , minimum height=4em
          , fill=black!10
          },
  ]

\tikzstyle{n} = [rectangle, rounded corners, minimum height=0.5cm,text centered, draw=black, outer sep=0mm, align=center]

\node[n, dotted] (wregx1) at (-5, -1.5)      {write<X>\\\footnotesize \{ADDI\}};
\node[n] (rmem) at (2, 0)      {read<MEM>: Bits<64>\\\footnotesize \{LW\}};
\node[n, dotted] (wregx2) at (2, -1.5)     {write<X>\\\footnotesize \{LW\}};

\node[n] (add) at (-2, 1)  {add: Bits<64>};
\node[n] (rx) at (-4, 2)    {read<X>: Bits<64>};
\node[n] (sext) at (0, 2)    {signExt: Bits<64>};
\node[n] (frd) at (-2, 0)    {field<rd>: Bits<5>};
\node[n] (frs) at (-4, 3) {field<rs>: Bits<5>};
\node[n] (fimm) at (0, 3) {field<imm>: Bits<12>};

\draw [arrow] (frs) -- (rx);
\draw [arrow] (fimm) -- (sext);
\draw [arrow] (rx) -- (add);
\draw [arrow] (sext) -- (add);
\draw [arrow] (frd) -- (wregx1);
\draw [arrow] (add) to[bend right] (wregx1);

\draw [arrow] (add) -- (rmem);
\draw [arrow] (rmem) -- (wregx2);
\draw [arrow] (frd) -- (wregx2);

\end{tikzpicture}
\caption[short]{Simplified IPG of the RISC-V \isainstruction{ADDI} and \isainstruction{LW} instructions after creation}
\label{fig:ipg_addi_lw_created}
\end{figure}

Next, the \ac{IPG} is constructed by copying all instructions' dataflows into one graph. The behavior graph implementation deduplicates identical nodes in this step. Figure~\ref{fig:ipg_addi_lw_created} shows an example of an IPG after the creation step for the RISC-V \isainstruction{ADDI} and \isainstruction{LW} instructions. The nodes that only belong to one of the instructions are annotated with this instruction. The rest of the nodes are annotated with both instructions (this is omitted in the illustration to simplify it).

To allow the \ac{IPG} to be the single representation of behavior inlined into the stages, it must also be extended with an implementation for the \texttt{fetchNext} built-in. Generally, this entails reading an instruction word from memory at the address stored in the program counter and incrementing it, or using a branch predictor to predict the next program counter value.

\subsubsection{Mapping \ac{IPG} Nodes to Stages}

For associating the IPG nodes with the stages, a linear order of the stages is determined by examining the dataflow between the stages. Then the stages and their built-in calls are examined in order. Each of them matches a selection of nodes from the IPG. This selection of nodes and all its dependencies not already mapped to a previous stage are associated with the current stage using the IPG data structure. The most important built-in calls in the MiA match the following nodes:

\begin{itemize}
    \item \texttt{fetchNext}: generated implementation with the instruction word read and next program counter write
    \item \texttt{decode}: maximal set of nodes such that all nodes (i) do not depend on
    read nodes and (ii) are not writes
    \item \texttt{compute}: all built-in calls that operate on inputs more than one bit wide
    \item \texttt{read(@R)}: all reads from the resource R
    \item \texttt{write(@R)}: all writes to the resource R
\end{itemize}

Note that the rule for the \texttt{decode} built-in implies that all nodes computable from the instruction word alone are associated with the stage that contains this built-in.

In general, some of the IPG nodes cannot be selected with these rules, e.g., select nodes. All such nodes are associated with a stage solely because they are a dependency of a node that was selected by a rule. These nodes are treated in a particular manner. They are not considered to be fixed to the specific stage and may be reassigned to adjacent stages as long as the data dependencies permit it. Optimization passes utilize this to reduce the number of stage output registers, or in terms of the IPG, to reduce the number of inter-stage dependencies.

After the instruction built-in calls are processed in all stages, it is ensured that every IPG node is associated with a stage, making sure that all instruction behavior will be implemented.

\subsubsection{Optimization of the \ac{IPG} and its Mapping}

In order to simplify the behavior graphs and aid the later generation of the hardware description, the read and write nodes referencing the same resources are merged in each stage. This is done by checking that they are not annotated with overlapping instructions and therefore cannot be active simultaneously. The inputs of the merged nodes are replaced with select nodes that are controlled by the output of the decoder detecting which instruction we actually execute. Similar to the select nodes, the condition inputs of the read and write nodes are masked with decoder outputs that are only true if an instruction that the read/write node is annotated with was decoded.

What the resulting IPG for our example with the \isainstruction{ADDI} and \isainstruction{LW} instructions looks like after mapping and optimization is depicted in Figure~\ref{fig:ipg_addi_lw_mapped}. This example uses the five-stage \ac{MiA} described in Section~\ref{sec:example_mia}. The \isainstruction{FETCH} stage is not present in the graph to keep it simple. The \texttt{decode} built-in maps the field reference and sign-extension nodes to the decode stage, while the \texttt{read(@X)} built-in binds the read from the register file. The \texttt{compute} built-in in the execute stage maps the add node, and the \texttt{read(@MEM)} built-in in the memory stage maps the read from memory. The write to the register file is placed in the writeback stage because of the \texttt{write(@X)} call there.

This example also shows two optimizations performed during architecture synthesis. Compared to Figure~\ref{fig:ipg_addi_lw_created}, the graph only contains one write node. The two nodes were merged by placing the new select node at the input of a single write node. The select node selects based on the decoded instruction. The second optimization concerns the placement of this select node. Putting the select node in the writeback stage would require pipeline registers for two 64-bit values. Because the select node is not fixed to a stage, it is moved to the memory stage to reduce pipeline registers.

\begin{figure}[h]
\begin{tikzpicture}
  [
      node distance={35mm},
      thick,
      unused/.style =
          { draw
          , dotted
          , minimum width=3em
          , minimum height=4em
          , fill=black!10
          },
  ]

\tikzstyle{n} = [rectangle, rounded corners, minimum height=0.5cm,text centered, draw=black, outer sep=0mm, align=center]

\node[n, dotted] (wregx1) at (-2, -2)      {write<X>};
\node[n] (sel) at (-2, -0.5)      {select: Bits<64>};
\node[n] (rmem) at (2, 0)      {read<MEM>: Bits<64>\\\footnotesize \{LW\}};

\node[n] (add) at (-2, 1)  {add: Bits<64>};
\node[n] (rx) at (-4, 2)    {read<X>: Bits<64>};
\node[n] (sext) at (0, 2)    {signExt: Bits<64>};
\node[n] (frd) at (-7.5, 3)    {field<rd>: Bits<5>};
\node[n] (frs) at (-4, 3) {field<rs>: Bits<5>};
\node[n] (fimm) at (0, 3) {field<imm>: Bits<12>};

\draw [arrow] (frs) -- (rx);
\draw [arrow] (fimm) -- (sext);
\draw [arrow] (rx) -- (add);
\draw [arrow] (sext) -- (add);
\draw [arrow] (frd) to[bend right] (wregx1);

\draw [arrow] (add) -- (rmem);
\draw [arrow] (add) -- (sel);
\draw [arrow] (rmem) -- (sel);
\draw [arrow] (sel) -- (wregx1);

\draw[dashed,transform canvas={yshift=-2mm}] (rmem.east |- rx.south) -- (frd.west |- rx.south)
    node [anchor=south west] (decodelbl) {DECODE};
\draw[dashed,transform canvas={yshift=-2mm}] (rmem.east |- add.south) -- (frd.west |- add.south)
    node [anchor=south west] (executelbl) {EXECUTE};
\draw[dashed,transform canvas={yshift=-2mm}] (rmem.east |- sel.south) -- (frd.west |- sel.south)
    node [anchor=south west] (memorylbl) {MEMORY};
\node [anchor=west] at (frd.west |- wregx1.north) (writebacklbl) {WRITEBACK};

\end{tikzpicture}
\caption[short]{Simplified IPG of the RISC-V \isainstruction{ADDI} and \isainstruction{LW} instructions after mapping and optimization (stages separated with dashed lines)}
\label{fig:ipg_addi_lw_mapped}
\end{figure}

\subsubsection{Inlining Instruction Behavior into the Stages}

The IPG can now be inlined directly into the stages in the VIAM. Subgraphs following from the stage annotations in the IPG are copied to the respective stages. For every output to a node outside the stage's subgraph, stage registers are created that hold the intermediate values during an instruction's execution. For every missing input, the appropriate stage register of the previous stage is read. The result is a preliminary MiA implementation without forwarding or control logic.

\subsubsection{Control and Forwarding Logic}

The key component for generating control logic is an analysis of the data hazards in the pipeline. It results from considering all pairs of reads and writes on the same resource where the read occurs in an earlier stage than the write. For these pairs, all stages between the read and write need to be analyzed for the intermediate value of the write condition and address. If any of the values is not available, this situation is treated as a hazard. To inform forwarding, the analysis also assesses if the value is available in an intermediate stage. Writes to the program counter are an exception: a hazard is not detected until the write condition is true and a value is available to allow speculative execution.

The IPG is used to evaluate if an input to a write is available in a given stage. This is done by traversing select nodes up in the dependency graph while collecting the select nodes' conditions under which the input is available in a previous stage.

Consider the example in Figure~\ref{fig:ipg_addi_lw_mapped}: there is only one pair of read/write nodes to evaluate (the two nodes accessing register file \texttt{X}). The address input to the write node is available from the decode stage on, via the field reference \texttt{rd}. The write value is available in the memory stage and afterwards. By following the select node that delivers the value, we can also determine that the value is available in the execute stage in case we are executing an \isainstruction{ADDI} instruction. This becomes part of the forwarding conditions for the read node in the decode stage.

Forwarding logic is generated based on the MiA definition. If forwarding is not specified in the MiA, the generator assumes bypass on all feasible paths. To represent forwarding in the VIAM, the forwarding logic element is extended with behavior that connects necessary signals from the stages to detect forwarding conditions and values. They are connected to the read nodes receiving forwarded values. The read node outputs are replaced by a select, choosing between the read value and the forwarded value.

The control logic consists of registers tracking if there is an instruction in a stage, enable signals for each stage and internal stall signals. It is based on~\cite{kroening2001}. Data hazards are detected based on the results of the hazard analysis and the signals from the forwarding logic. External stalls from memory are connected to the stall logic. Each stage's reads and writes are masked with the enable signal. When a write to the program counter occurs apart from the speculative write during instruction fetch, all previous stages are rolled back. The resulting behavior is incorporated into the MiA definition in the form of a control logic element, and this step yields the MiA implementation, the result of the architecture synthesis.

%% file: src-open_casgen.tex
\subsection{OpenVADL Cycle Approximate Simulator Generator}
\label{sec:open_casgen}

The original \ac{VADL} \ac{CAS} implementation prioritizes hardware conformance and reduced development effort over execution speed, mirroring the underlying hardware generator closely. However, this architectural design severely limits simulation performance. To mitigate this limitation, the emerging OpenVADL \ac{CAS} leverages QEMU, optimizing execution speed by statically computing specific processor states during dynamic translation. The development of the OpenVADL \ac{CAS} is currently ongoing, and comprehensive architectural details are not yet finalized. Users requiring cycle-accurate data in the interim can utilize the hardware-generated \ac{CAS}.

%% file: src-open_hdlgen.tex
\subsection{OpenVADL Hardware Generator}
\label{sec:open_hdlgen}

The MiA implementation produced by the architecture synthesis is a model of the processor at \acf{RTL}. The hardware generator takes this model and emits a description in a \acf{HDL}. HDLs define modules, their inputs and outputs (ports), and their behavior. To prepare the MiA implementation for HDL generation, each of the following elements is first wrapped in an HDL module data structure:
stage implementations and
logic elements, including forwarding and control logic elements.

A top-level module combining these HDL modules is also created. Additionally, the top module contains register files and registers from the implemented \ac{ISA}. Next, the behavior graphs in the stages and logic elements are analyzed for nodes that require a connection to another module. These are operations accessing registers or memory, as well as signals connecting the stages and logic elements, e.g., enable or forwarding signals. For such operations, ports are created on the HDL modules where necessary and connected to the correct destination.

During HDL emission, code is generated for the determined ports and their connections. The behavior graphs in the HDL modules are broken down into HDL statements that either implement node semantics directly or connect to ports of the module. Internal signals are introduced to reuse expressions in multiple statements.

The OpenVADL hardware generator outputs the hardware description in Chisel, an HDL embedded in Scala that generates Verilog code for simulation and synthesis. The generator mostly uses basic Chisel constructs, which should also map easily to other HDLs.
The generated Chisel project also contains files to build a Verilator-based, cycle-accurate simulator capable of executing ELF binaries directly for testing.

%% file: src-open_cosim.tex
\subsection{OpenVADL Co-simulator}
\label{sec:open_cosim}

To validate the functional correctness of the \ac{ISS} generated by OpenVADL, a lockstep co-simulation framework uses an independent upstream QEMU \ac{ISS} as a reference. A co-simulation test case passes only when no architectural state divergence is detected between the two simulators throughout the entire execution of the test program.

Architecturally, this framework is realized via a custom QEMU \ac{TCG} plugin loaded at startup, operating in tandem with a standalone process designated as the Co-simulation Broker. The plugin registers instrumentation callbacks at the instruction or \ac{TB} granularity and traces all subsequent memory read and write operations. Upon invocation, these callbacks record the complete architectural state into a concurrent shared-memory ring buffer. The recorded state comprises all general-purpose and floating-point registers, the program counter, all status and control registers, and memory accesses. The Co-simulation Broker continuously polls the buffer and compares the recorded states. By default, the entire architectural state is validated, while a configurable register mask allows developers to restrict the comparison scope during incremental bring-up. In the event of a divergence, the broker automatically generates a diagnostic report detailing the offending instruction alongside the pre- and post-execution architectural states of both \acp{ISS}.

%% file: src-open_testgen.tex
\subsection{OpenVADL Test Data Generator}
\label{sec:open_testgen}

OpenVADL generates instruction-level tests to validate its synthesized \acp{ISS} against an independent reference simulator (upstream QEMU) using the co-simulator described in Section~\ref{sec:open_cosim}.

The test framework supports both semi-automatic and fully automatic test generation. The semi-automatic approach, used for our RISC-V validation alongside the official RISC-V architectural conformance suite (covering integer and floating-point extensions), instantiates handwritten assembly templates with randomly sampled registers, immediate operands, and initial register values, while the templates define the program structure and operand constraints.

For AArch64 scalar instructions, the fully automatic approach derives tests directly from the \ac{VADL} specification without handwritten assembly templates or instruction-specific generation routines. To achieve broad input coverage across valid instruction encodings, the generator samples non-constant encoding fields within their declared bit widths and selects register indices from configurable register sets, deriving source-register initialization from register-access metadata in the processed \ac{VIAM} behavior graph. Candidate assignments that satisfy an instruction's \texttt{undefined when} predicate are automatically discarded. For example, AArch64 encodes the shift amount of a shifted-register addition in six bits, even though its 32-bit variant only permits shift amounts below 32; the predicate excludes invalid shift amounts even when they fit the encoding field. Accepted assignments are then encoded and disassembled using the instruction's assembly specification.

Nevertheless, the automatic generator does not yet infer all runtime constraints required for arbitrary control-flow and memory operations: branch and jump instructions still rely on semi-automatic templates to ensure valid targets and terminating control flow, while load/store tests require templates to constrain memory addresses, alignment, and initialization. More fundamentally, a limitation of specification-driven test generation is that unspecified or omitted architectural behavior cannot be synthesized and therefore cannot be detected by generated tests alone. Detecting such omissions requires independent external test suites such as the official RISC-V conformance tests.

%% file: src-evalfrontend.tex
\subsection{Evaluation of the OpenVADL Frontend}
\label{sec:open_eval_frontend}

\begin{figure}[H]
  \centering
  \resizebox{\linewidth}{!}{%
    \begin{tikzpicture}
      \begin{axis}[
          ybar,
          bar shift=0pt,
          bar width=0.415cm,
          x=0.5185185185185185cm, height=8cm,
          xtick={1.000,4.800,8.100,10.900,14.200,18.000,21.300,24.100},
          xticklabels={{RV32I},{RV64IM},{VRisc},{PPC64},{Hexagon},{AArch64},{AArch32},{Huge}},
          xticklabel style={anchor=north,font=\small},
          enlarge x limits={abs=1.2},
          xlabel={Specification},
          ylabel={Parsing time},
          legend style={at={(1.02,1)},anchor=north west,nodes={anchor=west},font=\small},
          ymajorgrids=true,
          grid style={dashed,gray!30},
          ymin=0,
          tick align=inside,
          xtick align=inside,
          ymode=log, log basis y=10,
          ymin=0.1,
          ytick={0.1,0.5,1,5,10,50,100,500,1000,5000},
          yticklabels={0.1ms,0.5ms,1ms,5ms,10ms,50ms,100ms,500ms,1s,5s},
          yticklabel style={font=\small},
          log origin=infty,
        ]
        \addplot[fill=blue!40!white, postaction={pattern=north east lines}]
        plot[error bars/.cd, y dir=both, y explicit]
        coordinates {(0.000,432.40) +- (0,19.73) (3.800,500.10) +- (0,35.86) (13.200,877.80) +- (0,26.88) (17.000,1251.70) +- (0,33.21)};
        \addlegendentry{original}
        \addplot[fill=red!40!white, postaction={pattern=north west lines}]
        plot[error bars/.cd, y dir=both, y explicit]
        coordinates {(1.000,35.87) +- (0,1.07) (4.800,62.28) +- (0,1.34) (7.600,54.48) +- (0,1.48) (10.400,88.28) +- (0,2.11) (14.200,95.88) +- (0,4.45) (18.000,115.78) +- (0,3.41) (20.800,713.39) +- (0,8.41) (23.600,1585.58) +- (0,24.77)};
        \addlegendentry{graalvm}
        \addplot[fill=green!50!black!40, postaction={pattern=crosshatch}]
        plot[error bars/.cd, y dir=both, y explicit]
        coordinates {(2.000,0.26) +- (0,0.00) (5.800,0.27) +- (0,0.00) (8.600,0.27) +- (0,0.00) (11.400,7.27) +- (0,0.00) (15.200,9.28) +- (0,0.00) (19.000,14.18) +- (0,0.31) (21.800,241.48) +- (0,1.93) (24.600,618.78) +- (0,14.25)};
        \addlegendentry{native}
      \end{axis}
    \end{tikzpicture}%
  }
  \caption{Parsing time per specification (logarithmic scale, lower is better).}
  \label{fig:frontend_total_time}
\end{figure}

High performance of the OpenVADL frontend is required for its use in a language server. Therefore, we evaluated the time spent in the frontend with several available \ac{VADL} processor specifications of different sizes and complexity.
The evaluation was done on a MacBook Pro with an Apple M3 Max processor with 36 GB of memory. OpenVADL was checked out at git commit b0cfd55441ce.
Each specification was measured over 10 timed runs after 3 warmup runs. From the
measured runs, we calculated the arithmetic mean and the standard error.

Figure~\ref{fig:frontend_total_time} shows three different configurations.
"original" is the original \ac{VADL} implementation, which runs on OpenJDK 17.
"graalvm" is OpenVADL on GraalVM 25, and "native" is the same OpenVADL commit built as a GraalVM Native Image.

Some specifications were developed with OpenVADL in mind and are no longer compatible with original \ac{VADL}, since the language evolved over time.

The benchmark includes all compilation stages of the frontend except the lowering to VIAM.
The lowering is also not needed for the language server, which is the biggest motivator for a fast frontend.

In all benchmarks, "graalvm" is faster than "original", and "native" tops "graalvm".
We attribute the former speedup to a more performant parser framework, a simplified compilation pipeline, and OpenVADL's ability to generate the AST directly.
The "native" builds have especially low startup times, which disproportionately benefit short-lived programs, as can also be observed in the figure.
For all \emph{organic} specifications, the frontend finished in $<250ms$, and even the stress test "Huge" required only $<620ms$.
We regard these numbers as fast enough for the language server.

%% file: src-open_evalcomp.tex
\subsection{Evaluation of the OpenVADL-Generated Compiler}
\label{sec:open_evalcomp}

All experiments in this evaluation are based on OpenVADL~0.4.
Unlike the original \ac{LCB}, which could not handle long jumps due to missing LLVM target hooks, OpenVADL's \ac{LCB} supports long jumps, enabling compiler optimizations that were previously impossible.
Additionally, we refactored the materialization of constants that exceed a machine instruction's immediate operand range, yielding substantial improvements in frame lowering.
Through the labeling of machine instructions introduced in OpenVADL, large constants can now be constructed using significantly fewer instructions.
Whereas the original \ac{VADL} emitted a sequence of \texttt{ADDI} instructions, each restricted to the immediate range $[-2048, 2047]$, OpenVADL's \ac{LCB} emits an \texttt{LUI} and \texttt{ADDI} pair, drastically reducing instruction counts.
Consequently, OpenVADL outperforms the original \ac{VADL} \ac{LCB} when large arrays are allocated on the stack. However, because the evaluated benchmarks do not allocate large stack frames, this optimization is not particularly visible in Figure~\ref{fig:lcb-speed-relative}.
Furthermore, OpenVADL now generates additional LLVM target hooks to facilitate improved code generation at higher optimization levels.
This has a pronounced effect on array loops, where \emph{loop-invariant code motion} can now hoist invariants, such as array base-address computations, out of the loop body.

As a performance metric, we measured the number of dynamically executed instructions reported by the functional simulator QEMU.
This was necessary because OpenVADL does not yet provide a \ac{CAS}, and the original \ac{CAS} is incompatible with the OpenVADL \ac{LCB}.
Therefore, a direct comparison between Figure~\ref{fig:lcb-speed-relative} and Figure~\ref{fig:openvadl-lcb} is not possible.

In contrast to the original \ac{LCB}, OpenVADL's \ac{LCB} successfully compiles, links, and executes all benchmarks for both \texttt{RV32IM} and \texttt{RV64IM} without error at optimization level \texttt{-O3}.
Figure~\ref{fig:openvadl-lcb} shows that the mean number of executed instructions across all benchmarks is approximately 6.5\% higher than that of upstream LLVM for \texttt{RV32IM} and 8.3\% higher for \texttt{RV64IM}.
Whereas upstream LLVM emits machine instructions via multiple handwritten C++ passes that cannot be expressed in TableGen, OpenVADL's \ac{LCB} generates only TableGen patterns alongside a small set of target-specific hooks without manual C++ optimizations.
Nevertheless, the automatically generated target description suffices to achieve competitive performance even at the highest optimization level.

Finally, Figure~\ref{fig:openvadl-lcb-binary-size} illustrates the code-size overhead for the same \texttt{-O3}-compiled benchmarks, showing that the mean size of the executable's \texttt{.text} section is approximately 10.3\% higher than that of upstream LLVM for \texttt{RV32IM} and 13.1\% higher for \texttt{RV64IM}.

\pgfplotstableread[col sep=tab]{benchmark-data-results-rv32im-openvadl-lcb-exec-rel-openvadl_O3_rv32im_rv64im.dat}\openvadlLcbExecResults
\begin{figure}[h!]
    \begin{tikzpicture}
        \begin{axis}[
            ybar=1,
            ymax=1.5,
            xtick=data,
            xticklabels from table={\openvadlLcbExecResults}{benchmark},
            xticklabel style={rotate=90},
            x=0.65cm,
            bar width=0.08cm,
            enlarge x limits=0.02,
            enlarge y limits={rel=0.09, upper},
            ymin=0
        ]
        \addplot table [col sep=tab, x expr=\coordindex, y=rv32im_rel]{benchmark-data-results-rv32im-openvadl-lcb-exec-rel-openvadl_O3_rv32im_rv64im.dat};
        \addplot table [col sep=tab, x expr=\coordindex, y=rv64im_rel]{benchmark-data-results-rv32im-openvadl-lcb-exec-rel-openvadl_O3_rv32im_rv64im.dat};
        \addplot table [col sep=tab, x expr=\coordindex, y=upstream]{benchmark-data-results-rv32im-openvadl-lcb-exec-rel-openvadl_O3_rv32im_rv64im.dat};

        \legend{OpenVADL RV32IM, OpenVADL RV64IM, Upstream}
        \end{axis}
    \end{tikzpicture}
    \caption{OpenVADL LCB RV32IM and RV64IM: Embench instruction counts from QEMU relative to upstream (lower is better)}
    \label{fig:openvadl-lcb}
\end{figure}

\pgfplotstableread[col sep=tab]{benchmark-data-results-rv32im-openvadl-lcb-exec-rel-openvadl_O3_rv32im_rv64im_binary_size.dat}\openvadlLcbExecResults
\begin{figure}[h!]
    \begin{tikzpicture}
        \begin{axis}[
            ybar=1,
            ymax=1.5,
            xtick=data,
            xticklabels from table={\openvadlLcbExecResults}{benchmark},
            xticklabel style={rotate=90},
            x=0.65cm,
            bar width=0.08cm,
            enlarge x limits=0.02,
            enlarge y limits={rel=0.09, upper},
            ymin=0
        ]
        \addplot table [col sep=tab, x expr=\coordindex, y=rv32im_rel]{benchmark-data-results-rv32im-openvadl-lcb-exec-rel-openvadl_O3_rv32im_rv64im_binary_size.dat};
        \addplot table [col sep=tab, x expr=\coordindex, y=rv64im_rel]{benchmark-data-results-rv32im-openvadl-lcb-exec-rel-openvadl_O3_rv32im_rv64im_binary_size.dat};
        \addplot table [col sep=tab, x expr=\coordindex, y=upstream]{benchmark-data-results-rv32im-openvadl-lcb-exec-rel-openvadl_O3_rv32im_rv64im_binary_size.dat};

        \legend{OpenVADL RV32IM, OpenVADL RV64IM, Upstream}
        \end{axis}
    \end{tikzpicture}
    \caption{OpenVADL LCB RV32IM and RV64IM: Embench executable text section size relative to upstream (lower is better)}
    \label{fig:openvadl-lcb-binary-size}
\end{figure}

%% file: src-evalqemu.tex
\subsection{Evaluation of the OpenVADL Instruction Set Simulator}
\label{sec:evalqemu}

For the \ac{ISS} evaluation, we used specifications covering the 64-bit RISC-V instruction set with the M, F, and D extensions (RV64IMD) and with only the M extension (RV64IM), the 32-bit RISC-V instruction set with the M extension (RV32IM), the complete integer subset of AArch64 (Armv8-A), and the AArch32 instruction set (Armv6).

The generated RISC-V \ac{ISS} passes the official RISC-V Architectural Certification Tests~\cite{riscvArchTests} for the base integer set (I) as well as the M, F, and D extensions. These tests thoroughly exercise instruction semantics across register combinations and operand boundary values, including integer division by zero. For floating-point instructions, they additionally validate rounding modes, exception flags, and NaN handling, providing strong evidence of semantic correctness beyond standard arithmetic operations.

Furthermore, we evaluated the \ac{VADL}-generated simulators using the Embench~\cite{embench} benchmark suite. To ensure an independent baseline, the RISC-V benchmarks were compiled with upstream GCC~16.1.0 and \texttt{newlib}, while the AArch64 and AArch32 benchmarks were compiled with upstream GCC~14.2.1 and \texttt{newlib}. All benchmarks were executed on an otherwise idle x86-64 host equipped with an Intel Xeon W-1370P processor (8 cores, 16 threads, 3.60\,GHz base frequency) and 128\,GiB of RAM, running Debian GNU/Linux~11 (Bullseye) with Linux kernel 6.0.0-0.deb11.6-amd64.

Currently, RV64IMD is the only specification in our evaluation with hardware floating-point support. Accordingly, only Embench workloads containing substantial single- or double-precision floating-point computations were compiled and executed on RV64IMD (see Figure~\ref{fig:iss-float-results}). Because the standard Armv8-A ABI mandates hardware floating-point support, benchmarks requiring floating-point operations were omitted for AArch64. For RV64IM, RV32IM, and AArch32, floating-point operations were compiled using software emulation.

Since OpenVADL's \ac{ISS} targets QEMU, our evaluation compares the generated simulators directly against handwritten, upstream QEMU guest frontends. We restrict this comparison to QEMU because prior evaluations already demonstrated that QEMU significantly outperforms both the original \ac{DTC} \ac{ISS} and Spike (see Section~\ref{sec:evaliss}).

For the four configurations in Figure~\ref{fig:iss-results}, we used a nominal benchmark frequency of 1\,000\,MHz, corresponding to Embench's workload-scaling option.
Each workload and simulator pair was executed with one warm-up iteration followed by ten timed trials, and we computed the runtime ratio by dividing the minimum OpenVADL runtime by the minimum upstream QEMU runtime for the same architecture and workload.
Across all ten timed trials, the sample standard deviation remained below 2\% of the mean runtime for every combination of workload, architecture, and simulator.
The rightmost group in Figure~\ref{fig:iss-results} reports the arithmetic mean of these workload ratios for each architecture.

The generated OpenVADL simulators ran faster than upstream QEMU on 19 of 22 benchmarks for RV32IM and 17 of 22 for RV64IM.
Runtime reductions reached 43\%.
The arithmetic mean runtime was 17.3\% lower for RV32IM and 16.1\% lower for RV64IM.
The largest RV64IM overhead was 9.9\% on \texttt{nettle-aes}.

For AArch32 and AArch64, the generated simulators ran within 10\% of upstream QEMU's runtime on most workloads.
The arithmetic mean runtime overhead was 1.3\% for AArch32 and 4.8\% for AArch64.
Both generated Arm simulators exhibited their highest overhead on \texttt{primecount}, at 18\% for AArch32 and 19\% for AArch64.

These overheads stem from missing optimizations in OpenVADL's \ac{TCG} lowering, which for certain instructions emits longer \ac{TCG} operation sequences than QEMU's handwritten frontends.
Such suboptimal sequences occur more frequently for AArch32 and AArch64 because their instructions have substantially higher semantic complexity than those of RISC-V.

Conversely, the generated simulators executed \texttt{crc32} in 21\% to 43\% less time across all four architectures.
We attribute these speedups to tight instruction sequences that frequently recompute \acs{CPU} state flags or modes.
Such state updates are more costly in upstream QEMU, whose handwritten frontends must accommodate a broad range of architectural extensions and system modes, whereas the generated simulators restrict state handling strictly to the features specified in \ac{VADL}.

Overall, the generated simulators achieve execution performance competitive with, and in many cases superior to, handwritten upstream QEMU frontends across all evaluated architectures.

\begin{figure}
  \centering
  \pgfplotstableread[col sep=comma]{benchmark-data-results-rv32im-iss-rv32im-qemu.csv}\issresults
  \def\issbenchmarks{}
  \pgfplotstableforeachcolumnelement{benchmark}\of\issresults\as\issbenchmark{%
    \edef\issbenchmarks{\issbenchmarks\ifnum\pgfplotstablerow>0 ,\fi\issbenchmark}%
  }
  \begin{tikzpicture}
    \begin{axis}[
        ybar=1,
        symbolic x coords/.expanded={\issbenchmarks},
        xtick=data,
        xticklabel style={rotate=90},
        width=\linewidth,
        height=7.2cm,
        bar width=0.08cm,
        enlarge x limits=0.05,
        enlarge y limits={rel=0.15, upper},
        ymin=0,
        ytick distance=0.2,
        ymajorgrids=true,
        major grid style={draw=black!15, thin},
        legend columns=-1
    ]
    \addplot table [col sep=comma, x=benchmark, y=time]{benchmark-data-results-rv32im-iss-rv32im-open-vadl.csv};
    \addplot table [col sep=comma, x=benchmark, y=time]{benchmark-data-results-rv64im-iss-rv64im-open-vadl.csv};
    \addplot table [col sep=comma, x=benchmark, y=time]{benchmark-data-results-aarch32-iss-aarch32-open-vadl.csv};
    \addplot table [col sep=comma, x=benchmark, y=time]{benchmark-data-results-aarch64-iss-aarch64-open-vadl.csv};
    \coordinate (qemuBaseline) at (axis cs:{[normalized]0},1);
    \draw[black, dashed, thick]
      ({rel axis cs:0,0} |- qemuBaseline) -- ({rel axis cs:1,0} |- qemuBaseline);
    \addlegendimage{black, dashed, thick, sharp plot, line legend}
    \legend{RV32IM, RV64IM, AArch32, AArch64, QEMU baseline}
    \end{axis}
  \end{tikzpicture}
  \caption{Minimum OpenVADL runtimes for Embench workloads, normalized to the minimum upstream QEMU runtime for the same workload and architecture. The dashed line at 1 marks the QEMU baseline. Lower is better. The \texttt{mean} group shows the arithmetic mean of the available ratios for each architecture. AArch64 bars appear only for the workloads we ran on AArch64.}
  \label{fig:iss-results}
\end{figure}

\pgfplotstableread[col sep=comma]{benchmark-data-results-rv64imd-iss-rv64imd-qemu-float.csv}\floatissresults
\begin{figure}
  \begin{tikzpicture}
    \begin{axis}[
        ybar=1,
        xtick=data,
        xticklabels from table={\floatissresults}{benchmark},
        x=2.3cm,
        enlarge x limits=0.3,
        enlarge y limits={rel=0.1, upper},
        ymin=0,
        ytick distance=0.2,
        ymajorgrids=true,
        major grid style={draw=black!15, thin},
        legend columns=-1,
        legend style={at={(0.5,1.03)}, anchor=south}
    ]
    \addplot table [col sep=comma, x expr=\coordindex, y=time]{benchmark-data-results-rv64imd-iss-rv64imd-open-vadl-float.csv};
    \draw[black, dashed, thick]
      ({rel axis cs:0,0} |- {axis cs:0,1}) -- ({rel axis cs:1,0} |- {axis cs:0,1});
    \addlegendimage{black, dashed, thick, sharp plot, line legend}
    \legend{OpenVADL, QEMU baseline}
    \end{axis}
  \end{tikzpicture}
  \caption{Runtimes of Embench benchmarks with significant floating-point workloads on the RV64IMD instruction-level simulator generated by OpenVADL. Each runtime is normalized to QEMU for the same benchmark (dashed line at 1). Lower is better.}
  \label{fig:iss-float-results}
\end{figure}

%% file: src-open_evalhdl.tex
\subsection{Evaluation of the OpenVADL Hardware Generator}
\label{sec:open_evalhdl}

For the evaluation of the hardware generated from a \ac{VADL} specification, we compare the chip area and maximum clock frequency of synthesized RV32I cores featuring three- and five-stage pipelines. Due to the current microarchitectural scope of the hardware generator, we selected two well-documented educational cores whose pipeline structures can be modeled directly in \ac{VADL}: Sodor\footnote{\href{https://github.com/ucb-bar/riscv-sodor}{https://github.com/ucb-bar/riscv-sodor}} and Wildcat~\cite{schoeberl2025wildcat}. Each reference core is compared against a core synthesized from a \ac{VADL} specification that replicates the respective \ac{MiA} as closely as possible. Specifically, we modeled the pipeline stages, register read/write ports, and forwarding paths according to each core's documentation and \ac{HDL} implementation. To ensure a fair comparison across identical architectural baselines, \acp{CSR} were omitted from all evaluated cores, as Sodor and Wildcat implement substantially different CSR subsets.

All evaluated cores were synthesized with the SkyWater SKY130 high-density standard-cell library using Yosys~\cite{wolf2013yosys} and OpenROAD~\cite{kahng2021openroad}. We determined the maximum achievable clock frequency of each design via a binary search over the target clock period to identify the shortest period meeting timing constraints, using timing-driven placement and an input/output delay of 1\,ns. Figure~\ref{fig:open_evalhdl_area} shows the post-placement core areas: the generated cores are between 3.0\% and 13.9\% larger than their handwritten counterparts.

\pgfplotstableread[col sep=comma]{benchmark-data-results-rtl-sky130-openvadl.csv}\rtlOpenVadlData
\pgfplotstableread[col sep=comma]{benchmark-data-results-rtl-sky130-other.csv}\rtlOtherData
\pgfplotstablecreatecol[copy column from table={\rtlOtherData}{area}]{otherarea}{\rtlOpenVadlData}
\pgfplotstablecreatecol[copy column from table={\rtlOtherData}{mhz}]{othermhz}{\rtlOpenVadlData}
\begin{figure}[tb]
    \pgfplotsset{
        common/.style={
                ybar,
                bar width=0.35cm,
                xtick=data,
                xticklabels from table={\rtlOpenVadlData}{config},
                xticklabel style={rotate=30, anchor=east},
                enlarge x limits=0.25,
                tick pos=left,
                ymin=0,
                ymajorgrids=true,
                height=7.5cm,
                width=\linewidth,
        },
        handwritten/.style={fill=blue!40!white, postaction={pattern=north east lines}, bar shift=-0.175cm-2pt},
        generated/.style={fill=red!40!white, postaction={pattern=north west lines}, bar shift=0.175cm+2pt},
        % invisible bar for the relative difference
        reldiff/.style={draw=none, fill=none, bar shift=0pt, forget plot, point meta=explicit,
            nodes near coords={\pgfmathprintnumber[fixed, fixed zerofill, precision=1, showpos]{\pgfplotspointmeta}\,\%},
            nodes near coords align=above,
            every node near coord/.append style={font=\footnotesize, yshift=1pt}},
    }
    \centering
    \pgfplotslegendfromname{rtlLegend}\par\medskip\medskip
    \begin{subfigure}[t]{0.49\linewidth}
        \centering
        \begin{tikzpicture}
            \begin{axis}[
                common,
                ymax=110,
                ylabel={Area in $10^3\,\mu\mathrm{m}^2$},
                legend to name=rtlLegend,
                legend columns=-1,
                legend style={/tikz/every even column/.append style={column sep=0.5cm}},
            ]
            \addplot[handwritten, area legend] table [x expr=\coordindex, y expr=\thisrow{otherarea}/1000] {\rtlOpenVadlData};
            \addplot[generated, area legend] table [x expr=\coordindex, y expr=\thisrow{area}/1000] {\rtlOpenVadlData};
            \addplot[reldiff] table [x expr=\coordindex, y expr={max(\thisrow{area},\thisrow{otherarea})/1000},
                meta expr={(\thisrow{area}/\thisrow{otherarea}-1)*100}] {\rtlOpenVadlData};
            \legend{Hand-written (Sodor/Wildcat), OpenVADL-generated}
            \end{axis}
        \end{tikzpicture}
        \caption{Placement area (lower is better)}
        \label{fig:open_evalhdl_area}
    \end{subfigure}\hfill
    \begin{subfigure}[t]{0.49\linewidth}
        \centering
        \begin{tikzpicture}
            \begin{axis}[
                common,
                ymax=160,
                ylabel={Frequency in MHz},
            ]
            \addplot[handwritten] table [x expr=\coordindex, y=othermhz] {\rtlOpenVadlData};
            \addplot[generated] table [x expr=\coordindex, y=mhz] {\rtlOpenVadlData};
            \addplot[reldiff] table [x expr=\coordindex, y expr={max(\thisrow{mhz},\thisrow{othermhz})},
                meta expr={(\thisrow{mhz}/\thisrow{othermhz}-1)*100}] {\rtlOpenVadlData};
            \end{axis}
        \end{tikzpicture}
        \caption{Maximum frequency (higher is better)}
        \label{fig:open_evalhdl_freq}
    \end{subfigure}
    \caption{OpenVADL-generated hardware compared to the Sodor and Wildcat RV32I cores. Percentages give the difference for the generated core relative to the handwritten one.}
    \label{fig:open_evalhdl_data}
\end{figure}

Figure~\ref{fig:open_evalhdl_freq} presents the corresponding maximum clock frequencies. For three of the four configurations, the generated core achieves a slightly higher maximum frequency than the handwritten baseline (by 4.5\% to 7.3\%). However, in the five-stage Sodor configuration, the generated core exhibits a 15.8\% lower maximum clock frequency. Timing analysis reveals that the critical path in this design traverses the branch decision logic in the execute stage into the control unit. This timing limitation could be mitigated by optimizing the generated comparison logic.

Currently, the hardware generator does not perform automated timing-driven optimizations. Consequently, manually tuned designs such as the five-stage Sodor core can achieve higher clock frequencies. Conversely, the three-stage Sodor pipeline illustrates an instance where the generator trades area for performance: whereas the handwritten Sodor implementation routes memory address calculations through the shared execution \ac{ALU} (forming its critical path), the generated core instantiates a dedicated adder for address computation. At present, such resource-sharing trade-offs cannot be explicitly controlled from within the \ac{VADL} specification.

%% file: src-comparison.tex
\section{Implementation Comparison}
\label{sec:impl_comparison}

Table~\ref{tab:compare-implementation} highlights the key differences and shared architectural concepts between the original \ac{VADL} and OpenVADL implementations.
The two versions pursued distinct development goals: while the original implementation prioritized flexibility to explore new language features with minimal development effort at the expense of performance, OpenVADL focuses on an efficient implementation built on a largely stable feature set.

\begin{table}[!ht]
\centering
\caption{Implementation comparison between original \ac{VADL} and OpenVADL}
\label{tab:compare-implementation}
\begin{tabular}{lcc}
\toprule
\textbf{Aspect} & \textbf{Original \ac{VADL}} & \textbf{OpenVADL} \\
\midrule
Open source                     & No                   & Yes                          \\
Implementation language         & Xtend                & Java                         \\
Parser generator                & Xtext                & Coco/R                       \\
\ac{IDE} support                & Eclipse              & \acp{IDE} supporting \ac{LSP}\\
Number of frontend passes       & 21                   & 6                            \\
Operator precedence in expressions & No                & Yes                          \\
\midrule
Name of \ac{IR}                 & \ac{VIR}             & \ac{VIAM}                    \\
Type of \ac{IR}                 & Three-address code   & Sea-of-nodes graph           \\
Decoding representation         & List / table         & \ac{VDT}                     \\
\midrule

Compiler generator \ac{IR}      & Instruction \ac{DAG} & \ac{VIAM}                    \\
Optimization levels supported   & O0                   & O0 -- O3                     \\
Assembly grammar class          & LL(1)                & pred-LL(1)                   \\
\midrule
\ac{ISS} execution              & Interpreted          & \ac{DBT}-based               \\
Virtualization framework        & None                 & QEMU                         \\
Co-simulation approach          & Interpreted          & \ac{DBT}-based               \\
\midrule
\ac{MiA} representation         & \ac{IPG}             & \ac{IPG}                     \\
Hardware generator \ac{IR}      & Data flow \ac{DAG}   & \ac{VIAM}                    \\
\bottomrule
\end{tabular}
\end{table}

A primary motivation for initiating OpenVADL was to provide an open-source implementation that can be freely shared with the community.
Migrating from Xtend to Java simplified debugging by leveraging Java's mature tooling ecosystem, including automated code-style enforcement, which accelerated development and validated the language choice.
The same rationale applies to the transition from Xtext to Coco/R. Avoiding the explicit creation of a \ac{CST} and the overhead of the Eclipse Modeling Framework, combined with a redesigned pass architecture, reduced the number of frontend passes from 21 to 6. This makes the frontend fast enough to power a language server. Furthermore, the language was extended to support expressions with standard operator precedence rules.

The most significant architectural change was replacing the quadruple three-address code \ac{VIR} with the sea-of-nodes graph of the \ac{VIAM}, which combines data and control dependencies and simplifies all downstream generators. For instruction decoding, the \ac{VDT} representation is more expressive than the list- and table-based approach in the original \ac{VADL} implementation, enabling the generation of differently optimized decoders.

The compiler generator retains its high-level architecture, which comprises the \ac{GCB} and \ac{LCB}. The underlying implementations differ substantially. Whereas the original compiler generator required an additional \ac{IR} to convert quadruple code into a data flow graph, OpenVADL's \ac{VIAM} already embeds this graph directly. In addition, the OpenVADL \ac{GCB} features a more sophisticated instruction classification algorithm that employs edge-case pruning. For the assembler generator, the supported grammar class was extended from LL(1) to pred-LL(1).

The original \ac{VADL} implementation provides an interpreted \ac{ISS} that is substantially slower than OpenVADL's \ac{DBT}-based simulator. While the linear quadruple code of the \ac{VIR} is well suited for generating a sequential, interpreted \ac{ISS}, the dependency-graph-based \ac{VIAM} is essential for efficiently scheduling parallel operations within \ac{VADL} statements into sequential \ac{TCG} operations for QEMU. Because OpenVADL's co-simulation framework also builds on QEMU, it achieves much higher performance than the original co-simulator.

Although both the original and OpenVADL hardware generators rely on an \ac{IPG}, they construct it differently. In OpenVADL, the data flow graph corresponds to the inverse dependency graph and is implicitly contained within the \ac{VIAM}. By contrast, the original \ac{VADL} requires an additional compilation step to compute the data flow \ac{DAG} from the \ac{VIR}.

In summary, these design choices unify the compilation pipeline around the \ac{VIAM} and substantially improve both compile-time and runtime efficiency.

%% file: src-casestudy.tex
\section{Design Space Exploration Case Study}
\label{sec:dse_casestudy}

To demonstrate the viability of OpenVADL in \ac{DSE}, we selected the \texttt{crc32} benchmark from the Embench suite as our workload and the \texttt{RV32I} \ac{ISA} as the baseline target architecture. The complete case study, including all scripts and specifications to reproduce the results, is publicly available at \url{https://github.com/OpenVADL/dse}. To avoid benchmarking pseudorandom number generation itself, the \texttt{crc32} workload was slightly modified so that the kernel loop computes the CRC32 checksum over a pre-filled array of 1,024 words. Although application hotspots are typically identified via LLVM profiling tools or \acp{IDE} such as VS~Code and IntelliJ IDEA, the hotspot in \texttt{crc32} is immediately apparent without profiling. This kernel loop (with expanded macros) is shown in Listing~\ref{lst:crc_kernel_c}.

\begin{lstlisting}[frame=single,float=!htb,label={lst:crc_kernel_c},caption={crc32 kernel in C},language=cpp]
for (i = 0; i < 1024; i++) {
  oldcrc32 = (crc_32_tab[((oldcrc32) ^ ((BYTE)random_numbers[i])) & 0xff] ^ ((oldcrc32) >> 8));
}
\end{lstlisting}

Compiling the baseline benchmark at optimization level \texttt{-O3} with a compiler generated by OpenVADL~0.4 triggers standard loop optimizations, including induction variable elimination, strength reduction, and loop-invariant code motion. Listing~\ref{lst:crc_kernel_asm} shows the resulting optimized assembly for the inner loop body. Here, the loop index \texttt{i} is replaced by a pointer advancing linearly through the \texttt{random\_numbers} array in register \texttt{a1}, while register \texttt{a3} holds the end pointer governing the loop exit condition. Register \texttt{a0} maintains \texttt{oldcrc32}, register \texttt{a2} holds the base address of \texttt{crc\_32\_tab}, and register \texttt{a4} serves as temporary storage for the loaded word and intermediate values.

\begin{lstlisting}[frame=single,float=!htb,label={lst:crc_kernel_asm},caption={crc32 loop body in RV32I assembler},language=asm]
.LBB0_1:                # inner loop header
  lw      a4, 0(a1)
  xor     a4, a4, a0  # crc0
  andi    a4, a4, 255 # crc0
  slli    a4, a4, 2   # crc1
  add     a4, a4, a2  # crc1
  lw      a4, 0(a4)   # crc1
  srli    a0, a0, 8   # crc2
  xor     a0, a4, a0  # crc2
  addi    a1, a1, 4
  bne     a1, a3, .LBB0_1
\end{lstlisting}

RISC-V provides reserved opcode spaces specifically intended for custom instruction extensions. For R-type instructions, the \texttt{CUSTOM-0} opcode space accommodates up to 1,024 custom instructions via the 3-bit \texttt{funct3} field (8 combinations) and the 7-bit \texttt{funct7} field (128 combinations). Adding R-type instructions that read two source registers and write one destination register requires no modifications to the register file ports. Consequently, we constrain our custom instructions to this three-register format. This architectural constraint yields three candidate instructions (\texttt{crc0}, \texttt{crc1}, and \texttt{crc2}). Each fuses an instruction sequence annotated in the comments of Listing~\ref{lst:crc_kernel_asm}. Although fusing the \texttt{lw} in line~2 with the pointer increment \texttt{addi} in line~10 is theoretically possible, doing so would require two destination registers and thus an additional register-file write port. Moreover, loop unrolling offers a purely software-based alternative to amortize the \texttt{addi} and \texttt{bne} loop overhead without hardware changes.

These three custom instructions are integrated into the baseline RV32I specification (Listing~\ref{lst:rv32i} in the Appendix) using the macro system shown in Listing~\ref{lst:crc_vadl_spec}. First, we define a parametric \texttt{RtypeCustomInstr} model accepting the instruction mnemonic, the \texttt{funct7} encoding, and the right-hand side expression assigned to \texttt{rd} (Lines~1--6). Using this model, each custom instruction is instantiated in a single line: \texttt{crc0} computes the bitwise XOR of both source registers and zero-extends the lowest byte; \texttt{crc1} performs a word load using a scaled index register; and \texttt{crc2} shifts the second operand right by 8~bits before XORing it with the first operand.

\begin{lstlisting}[frame=single,float=!htb,label={lst:crc_vadl_spec},caption={Specification of custom crc32 instructions for RV32I},language=vadl]
model RtypeCustomInstr (name : Id, funct7 : Lit, rhs : Ex) : IsaDefs = {
  instruction $name : Rtype =                        // 3 register operand instructions
    X(rd) := $rhs
  encoding $name = { opcode = 0b000'1011, funct3 = 0, funct7 = $funct7}
  assembly $name = (mnemonic, " ", register(rd), ",", register(rs1), ",", register(rs2))
}

$RtypeCustomInstr (crc0 ; 0 ; (X(rs1) ^ X(rs2)) & 255 )
$RtypeCustomInstr (crc1 ; 1 ; MEM<4>(X(rs1) + (X(rs2) << 2)) )
$RtypeCustomInstr (crc2 ; 2 ; X(rs1) ^ X(rs2) >> 8 )
\end{lstlisting}

Once the extended \ac{ISA} specification is complete, OpenVADL generates the full toolchain and hardware artifacts.
While OpenVADL's own code generation completes in milliseconds, downstream C++ compilation, linking, and \ac{RTL} synthesis account for the majority of the turnaround time.
We benchmarked these generation and build times on an Apple MacBook~Pro (M5~Pro, 64\,GB RAM) using a native build of OpenVADL~0.4 (compiled with GraalVM~25.0.2+10.1) and 16 parallel build jobs.
The \ac{LCB} targets LLVM~19.1.7 and the \ac{ISS} targets QEMU~9.2.2, both compiled with AppleClang~21.0.0; hardware synthesis uses Chisel~7.16.0 and Yosys~0.70+post with the Sky130 PDK.
Table~\ref{tab:dse_generation_time} reports the measured times in seconds, showing the minimum and maximum across five runs for OpenVADL generation and cached rebuilds.
Initial builds include downloading upstream LLVM and QEMU sources and therefore exhibit high variance, whereas cached rebuilds show negligible difference between minimal and maximal execution times.

\begin{table}
\caption{Generation and build times of OpenVADL artifacts measured in seconds.}
\label{tab:dse_generation_time}
\begin{tabular}{l rr}
\toprule
 & \multicolumn{2}{c}{\textbf{Time (s)}} \\
\cmidrule(lr){2-3}
\textbf{Generation / Build Task}       & \textbf{Minimal} & \textbf{Maximal} \\
\midrule
OpenVADL ISS generation                                   &   0.052 &   0.057 \\
ISS build (first time)                                    &  47.698 &  61.125 \\
ISS build (second time, cached)                           &  22.605 &  23.527 \\
\midrule
OpenVADL LLVM generation                                  &   0.095 &   0.098 \\
LLVM toolchain build (first time)                         & 399.549 & 486.298 \\
LLVM toolchain build (second time, cached)                &  32.631 &  34.236 \\
\midrule
OpenVADL hardware generation                              &   0.076 &   0.078 \\
Verilog emission, synthesis, placement \& timing analysis &   4.958 &   5.064 \\
\bottomrule
\end{tabular}
\end{table}

Generating the complete LLVM toolchain (compiler, assembler, disassembler, and linker) takes less than 100\,ms. While the initial build of LLVM takes from under 7 to more than 8~minutes, incremental rebuilds leveraging cached intermediate artifacts complete in roughly 32.6\,s. Generation and rebuild times for the QEMU \ac{ISS} (22.6\,s) and \ac{RTL} hardware synthesis (5.0\,s) are even faster, enabling rapid iterative exploration across compiler, simulator, and hardware targets.

Once the custom compiler is generated, the \texttt{crc32} benchmark is recompiled at optimization level \texttt{-O3}. Listing~\ref{lst:crc_kernel_custom_asm} presents the optimized loop body utilizing the custom \texttt{crc*} instructions. Consequently, the loop body shrinks from 10 instructions to 6, and the total dynamic instruction count of the application decreases from 10,488,012 to 6,392,012 (a 39.1\% reduction, or a factor of $1.64\times$). While a branch incurs a 2-cycle penalty on the 3-stage \ac{MiA} pipeline, the 5-stage \ac{MiA} pipeline exhibits a 2-cycle load latency and a 3-cycle branch penalty. Whereas the memory load latency in the baseline loop is hidden by the subsequent \texttt{srli} and \texttt{addi} instructions, the custom-instruction loop requires single-factor unrolling on the 5-stage \ac{MiA} to enable effective instruction scheduling. Ultimately, the custom kernel achieves a $1.57\times$ speedup (11 vs. 7 cycles) on the 3-stage \ac{MiA} and a $1.67\times$ speedup (20 vs. 12 cycles across two iterations) for the unrolled loop on the 5-stage \ac{MiA}.

\begin{lstlisting}[frame=single,float=!htb,label={lst:crc_kernel_custom_asm},caption={crc32 loop body in RV32I custom assembler},language=asm]
.LBB0_1:                # inner loop header
  lw      a4, 0(a1)
  crc0    a4, a4, a0  # a4 = (a4 ^ a0) & 255
  crc1    a4, a2, a4  # a4 = lw (a2 + (a4 << 2))
  crc2    a0, a4, a0  # a0 = a4 ^ a0 >> 8
  addi    a1, a1, 4
  bne     a1, a3, .LBB0_1
\end{lstlisting}

Because \ac{VADL} strictly decouples the \ac{ISA} from the \ac{MiA} specification, evaluating multiple \ac{ISA} variants across different \acp{MiA} is straightforward. To demonstrate this capability, the RV32I \ac{ISA} specification detailed in the evaluation section served as the baseline and was extended in four configurations: adding solely \texttt{crc0}, \texttt{crc1}, \texttt{crc2}, or all three instructions simultaneously (denoted as \texttt{crc0/1/2} in Table~\ref{tab:dse_hardware_results}). These five \ac{ISA} variants were synthesized across both the 3-stage and 5-stage \acp{MiA} using Yosys and OpenROAD with the SkyWater SKY130 high-density standard-cell library.

Table~\ref{tab:dse_hardware_results} reports the post-placement decode-stage area, total core area, and maximum achievable clock frequency for each design. Overall, introducing the custom instructions has a negligible impact on chip area (at most $+2.6\%$) and clock frequency. In a few configurations, the reported frequency increases slightly, which is an artifact of the timing-characterization flow: maximum clock frequency is determined via a binary search over target clock periods (terminating at a step size below 0.1\,ns) using Yosys and OpenROAD, both of which apply timing-driven optimization heuristics. Both the discrete 0.1\,ns search resolution and heuristic optimization choices across slightly altered netlists account for identical frequencies across variants as well as minor frequency gains.

\begin{table}
\caption{Area and frequency of two MiAs implementing four different custom ISAs, absolute and relative to RV32I.}
\label{tab:dse_hardware_results}
\small
\setlength{\tabcolsep}{3.5pt}
\begin{tabular*}{\textwidth}{@{\extracolsep{\fill}} ll rr rr rr rr rr}
\toprule
 & & \multicolumn{2}{c}{\textbf{RV32I}} & \multicolumn{2}{c}{\textbf{crc0}} & \multicolumn{2}{c}{\textbf{crc1}} & \multicolumn{2}{c}{\textbf{crc2}} & \multicolumn{2}{c}{\textbf{crc0/1/2}} \\
\cmidrule(lr){3-4} \cmidrule(lr){5-6} \cmidrule(lr){7-8} \cmidrule(lr){9-10} \cmidrule(lr){11-12}
\textbf{MiA} & \textbf{Metric} & \multicolumn{1}{c}{abs} & \multicolumn{1}{c}{rel} & \multicolumn{1}{c}{abs} & \multicolumn{1}{c}{rel} & \multicolumn{1}{c}{abs} & \multicolumn{1}{c}{rel} & \multicolumn{1}{c}{abs} & \multicolumn{1}{c}{rel} & \multicolumn{1}{c}{abs} & \multicolumn{1}{c}{rel} \\
\midrule
\multirow{3}{*}{3-stage}
 & Decode area ($\mu\text{m}^2$) &  8\,977 & 1.000 &  9\,020 & 1.005 &   8\,946 & 0.997 &  8\,786 & 0.979 &  8\,995 & 1.002 \\
 & Chip area ($\mu\text{m}^2$)   & 80\,532 & 1.000 & 80\,657 & 1.002 &  82\,465 & 1.024 & 80\,793 & 1.003 & 82\,377 & 1.023 \\
 & Frequency (MHz)               &   86.05 & 1.000 &   84.63 & 0.983 &    91.43 & 1.062 &   91.43 & 1.062 &   89.82 & 1.043 \\
\midrule
\multirow{3}{*}{5-stage}
 & Decode area ($\mu\text{m}^2$) &  7\,966 & 1.000 &  8\,133 & 1.021 &   8\,254 & 1.036 &  8\,175 & 1.026 &  8\,225 & 1.032 \\
 & Chip area ($\mu\text{m}^2$)   & 91\,112 & 1.000 & 91\,238 & 1.001 &  93\,398 & 1.025 & 91\,690 & 1.006 & 93\,513 & 1.026 \\
 & Frequency (MHz)               &   90.62 & 1.000 &   90.62 & 1.000 &   101.37 & 1.119 &   91.43 & 1.009 &   90.62 & 1.000 \\
\bottomrule
\end{tabular*}
\end{table}

In this case study, the \ac{DSE} candidates were selected manually to illustrate the end-to-end workflow. Rather than embedding fixed search heuristics, OpenVADL is intentionally designed as an open, unified generation infrastructure that external \ac{DSE} frameworks and researchers can build upon. For example, automated custom-instruction discovery frameworks such as CIDRE~\cite{Rezunov2025CustumInstructionDesign} currently rely on proprietary \acp{PDL} like nML. These frameworks could directly target OpenVADL to automatically synthesize and evaluate compilers, simulators, and \ac{RTL} hardware for candidate instruction extensions.

%% file: src-relwork-pdl.tex
\section{Related Work}
\label{sec:related_work}
\subsection*{Processor Description Languages}

The comprehensive volume edited by Mishra and Dutt~\cite{mishra2008processor} provides an exhaustive overview of virtually all \acp{PDL} developed up to 2008, covering MIMOLA, nML, LISA, Expression, ASIP Meister, TIE, MADL, ADL++, ArchC, MAML, and GNR in full detail while briefly outlining eight additional languages. Furthermore, Dreesen~\cite{Dreesen2011Generating} compares several \acp{PDL} (ISP, nML, ASIP Meister, LISA, ISDL, Expression, and TIE) with ViDL in his PhD thesis across criteria such as simplicity, complexity, expressiveness, and abstraction. We refer the interested reader to Dreesen's thesis for that exhaustive qualitative evaluation. Finally, a concise summary of MIMOLA, Expression, nML, LISA, PEAS (ASIP Meister), TIE, ARC APEX, CodAL, and Andes ACE, as well as the \ac{HDL} Chisel, is provided in the book chapter on \acp{ADL} by Chattopadhyay et al.~\cite{Chattopadhyay2022ADL}.

A \ac{PDL} is a domain-specific \ac{ADL} designed for the specification and development of processor architectures, typically facilitating a succinct description of a processor's attributes and properties. According to Mishra and Dutt~\cite{Mishra2005architecture}, \acp{PDL} can be classified based on their \emph{content} (what the language describes) and their \emph{objective} (the intended application of the specification).

Content-based classification distinguishes between \emph{structural}, \emph{behavioral}, and \emph{mixed} \acp{PDL}. A structural \ac{PDL} primarily describes constituent hardware components, such as registers and pipeline stages, alongside their interconnections. Conversely, a behavioral \ac{PDL} focuses on specifying the semantics of the processor's instruction set. Although a structural \ac{PDL} captures hardware details absent in a purely behavioral description, inferring instruction semantics from a strictly structural model is generally challenging. Nonetheless, Leupers~\cite{leupers1998retargetable} and Brandner~\cite{brandner2007compiler} demonstrated that this is feasible with dedicated analysis. Consequently, most \acp{PDL} adopt a mixed approach, combining structural and behavioral elements with varying degrees of emphasis.

Common objectives of a \ac{PDL} include \emph{compilation}, \emph{simulation}, \emph{synthesis}, and \emph{validation}. When targeting compilation, the \ac{PDL} serves as input to a compiler generator: it must provide precise instruction semantics (for which behavioral abstractions are well suited) alongside structural or microarchitectural information, such as pipeline stages and functional units, to derive an accurate cost model for instruction scheduling. For simulation, the required level of detail depends on the simulator type: a purely behavioral \ac{PDL} suffices to generate a functional \ac{ISS}, whereas generating a \ac{CAS} additionally requires structural or cycle-level microarchitectural information. For hardware synthesis, a structural or synthesizable microarchitectural description is required. Finally, formal verification and automated test-case generation benefit from both behavioral and structural models, depending on the target validation scenario.

Tables~\ref{tab:compare-pdls-artifacts} and~\ref{tab:compare-pdls-characteristics} present a comparative analysis of 19~\acp{PDL}. Table~\ref{tab:compare-pdls-artifacts} reports their initial publication years and provides repository references for open-source implementations. Since several of these languages are primarily of historical significance, the table distinguishes between languages that remain actively developed or utilized and those that are discontinued. Furthermore, it summarizes the specific artifacts produced by each toolchain's code generators. Note that in the \emph{cycle accurate} column, listing an \ac{HDL} (such as Chisel or a generic \ac{HDL}) indicates that cycle-accurate execution is only supported indirectly via slow \ac{RTL} hardware simulation rather than a dedicated software simulator. Detailed discussions of each language follow in the subsequent subsections.

\begin{table}
\caption{Comparison of processor description languages regarding generated artifacts.}
\label{tab:compare-pdls-artifacts}
\footnotesize
\setlength{\tabcolsep}{3pt}
\begin{tabular*}{\textwidth}{@{\extracolsep{\fill}} l l cccccc}
\toprule
\textbf{PDL} & \textbf{Published} & \textbf{Open}   & \textbf{Active} & \textbf{Generated} & \textbf{Cycle-}   & \textbf{Generated} & \textbf{Generated} \\
             &                    & \textbf{source} &                 & \textbf{simulator} & \textbf{accurate} & \textbf{compiler}  & \textbf{hardware}  \\
\midrule
VADL         & 2020~\cite{schutzenhofer2020cycle} & Yes~\cite{OpenVADL2026Source} & Yes & \ac{DBT}    & Yes      & LLVM       & Full       \\
xADL         & 2007~\cite{brandner2007compiler}   & No                            & No  & \ac{DBT}    & Yes      & LLVM       & Full       \\
xDSPcore ADL & 2005~\cite{Farfeleder2005Ultra}    & No                            & No  & Compiled    & Yes      & OCE        & Decoder    \\
\midrule
TAIDL        & 2025~\cite{taidl-test2025taidliss} & Yes~\cite{TAIDL2026Source}    & Yes & Interpreted & No       & XLA        & --         \\
ISADL        & 2023~\cite{Xiao2023ISADL}          & No                            & No  & Interpreted & No       & Assembler  & --         \\
CoreDSL2     & 2021~\cite{CoreDSL2}               & Yes~\cite{CoreDSL2026Source}  & Yes & Source JIT  & No       & Extensions & Extensions \\
\midrule
StoneCutter  & 2019~\cite{leidel2019StoneCutter}  & No                            & No  & None        & Chisel   & --         & Partial    \\
Sail         & 2015~\cite{gray2015Sail}           & Yes~\cite{Sail2026Source}     & Yes & Interpreted & No       & --         & --         \\
CodAL        & 2012~\cite{Charvat2012CodAL}       & No                            & Yes & Multiple    & Yes      & LLVM       & Full       \\
\midrule
ViDL         & 2011~\cite{Dreesen2011Generating}  & No                            & No  & Interpreted & \ac{HDL} & --         & Full       \\
ISAC         & 2006~\cite{Masarik2006Language}    & No                            & No  & Multiple    & Yes      & LLVM       & --         \\
ArchC        & 2003~\cite{Rigo2003ArchC}          & Yes~\cite{ArchC2016Source}    & No  & Multiple    & Yes      & LLVM       & --         \\
\midrule
MAML         & 2001~\cite{FischerTeichWeper2001}  & No                            & No  & Mixed       & Yes      & LCC        & Full       \\
TIE          & 2000~\cite{Gonzalez2000Xtensa}     & No                            & No  & \ac{RTL}    & Yes      & GCC        & Extensions \\
Expression   & 1998~\cite{grun1998expression}     & Yes~\cite{Expression2008Source} & No  & Interpreted & Yes      & EXPRESS    & Full       \\
\midrule
ISDL         & 1997~\cite{hadjiyiannis1997isdl}   & No                            & No  & Interpreted & Yes      & AVIV       & Full       \\
LISA         & 1996~\cite{zivojnovic1996lisa}     & No                            & No  & Compiled    & Yes      & CoSy       & Full       \\
nML          & 1991~\cite{freericks1991nml}       & No                            & Yes & Interpreted & Yes      & CHESS      & Full       \\
MIMOLA       & 1976~\cite{zimmermann1976mimola}   & No                            & No  & Events      & Yes      & MSSQ       & Full       \\
\bottomrule
\end{tabular*}
\end{table}

Table~\ref{tab:compare-pdls-characteristics} outlines key language characteristics, building upon the previously discussed \emph{language classification} into behavioral, structural, and mixed categories. While generating a compiler backend, modeling an \ac{ISS}, and synthesizing a processor in an \ac{HDL} all rely on the same underlying instruction semantics, their target abstractions differ significantly. Certain \acp{PDL} can automatically derive these distinct targets from a \emph{single} behavioral source. Conversely, other \acp{PDL} necessitate separate behavioral descriptions per target, which introduces redundancy and increases the risk of inconsistencies or specification errors, whereas pure \ac{ISA} languages avoid redundancy simply by targeting a single abstraction level (e.g., \emph{ISA only}). This trait is captured in the \emph{redundancy of specifications} column.

Furthermore, instructions typically read input operands at the start of execution and update multiple destination resources upon completion. While each individual resource may receive only a \emph{single} assignment or undergo \emph{multiple} intermediate updates within an instruction, distinct resources are updated simultaneously. Although architecture manuals and \acp{ISS} often express these updates sequentially, language paradigms differ: most \acp{PDL} enforce either strictly \emph{sequential} semantics or simultaneous (\emph{parallel}/\emph{concurrent}) execution. Some languages support multiple paradigms, including MIMOLA, standard \acp{HDL} like Verilog, and formal specification languages like CASM~\cite{Lezuo2013CASM,Lezuo2014CASM}. These properties are detailed in the \emph{single/multi assignment} and \emph{assignment order} columns.

Finally, \ac{MiA} specifications range from \emph{low}-level structural descriptions to \emph{high}-level abstractions. Depending on the language, they can be synthesized \emph{automatic}ally using architectural hints or omitted entirely for pure \ac{ISA} modeling (see the \emph{\ac{MiA} abstraction} column). Regarding instruction encoding, some languages decouple instruction formats from individual instructions so that formats can be reused (\emph{separate format}), whereas others require repeating encoding bit patterns per instruction or rely on dedicated bidirectional \emph{mapping} clauses, as in Sail.

\begin{table}
\caption{Comparison of processor description languages regarding language characteristics.}
\label{tab:compare-pdls-characteristics}
\footnotesize
\setlength{\tabcolsep}{3.5pt}
\begin{tabular*}{\textwidth}{@{\extracolsep{\fill}} l cccccc}
\toprule
\textbf{PDL} & \textbf{Language}       & \textbf{Specification} & \textbf{Single/multi-} & \textbf{Assignment} & \textbf{\ac{MiA}}    & \textbf{Separate} \\
             & \textbf{classification} & \textbf{redundancy}    & \textbf{assignment}    & \textbf{order}      & \textbf{abstraction} & \textbf{format}   \\
\midrule
VADL         & Mixed      & Single           & Single     & Parallel     & High      & Yes     \\
xADL         & Structural & Single           & Single     & Parallel     & Low       & Yes     \\
xDSPcore ADL & Mixed      & Partial          & Multiple   & Sequential   & Low       & No      \\
\midrule
TAIDL        & Behavioral & Tensor ISA only  & Multiple   & Sequential   & --        & No      \\
ISADL        & Encoding   & Encoding only    & --         & --           & --        & Yes     \\
CoreDSL2     & Behavioral & ISA only         & Multiple   & Sequential   & --        & No      \\
\midrule
StoneCutter  & Behavioral & ISA only         & Multiple   & Sequential   & Low       & Yes     \\
Sail         & Behavioral & ISA only         & Multiple   & Sequential   & --        & Mapping \\
CodAL        & Mixed      & Single           & Multiple   & Sequential   & Medium    & Partial \\
\midrule
ViDL         & Mixed      & Single           & Single     & Parallel     & Automatic & No      \\
ISAC         & Mixed      & Single           & Multiple   & Sequential   & Medium    & Partial \\
ArchC        & Mixed      & Multiple         & Multiple   & Sequential   & Low       & Yes     \\
\midrule
MAML         & Mixed      & Multiple         & Multiple   & Sequential   & Medium    & No      \\
TIE          & Structural & Single           & Single     & Concurrent   & Low       & No      \\
Expression   & Mixed      & Single + mappings& Single     & Parallel     & Low       & No      \\
\midrule
ISDL         & Behavioral & Single           & Multiple\textsuperscript{*} & Sequential\textsuperscript{*} & Low & Yes \\
LISA         & Mixed      & Multiple         & Multiple   & Sequential   & Medium    & Partial \\
nML          & Mixed      & Single           & Single     & Concurrent   & Low       & No      \\
MIMOLA       & Structural & Single           & Multiple   & Par./con./seq. & Low     & No      \\
\bottomrule
\addlinespace[2pt]
\multicolumn{7}{@{}l}{\scriptsize\textsuperscript{*} Inferred from examples, as the publicly available literature does not explicitly specify these semantics.}
\end{tabular*}
\end{table}

\emph{VADL}~\cite{schutzenhofer2020cycle,freitag2025OpenVADL,zottele2025QEMUOpenVADL,halder2026ASIP} is a mixed \ac{PDL} comprising a single specification partitioned into distinct \ac{ISA} and \ac{MiA} descriptions. This unified model contains all information required by the downstream generators, completely eliminating specification redundancy. Key characteristics of the language include a syntactic macro system and a high-level \ac{MiA} modeling layer that enforces strict separation from the \ac{ISA}. Furthermore, \emph{VADL} facilitates clean, modular specifications by supporting comprehensive type, format, constant, and function definitions, alongside enumerations, register aliasing, and specialized format-field access functions. Leveraging its macro system, \emph{VADL} achieves substantially more concise processor descriptions than traditional \acp{PDL}. As the third \ac{PDL} developed within our research group, \emph{VADL} inherits fundamental computational models from its predecessors while introducing these high-level abstractions.

The first was an \ac{ADL} designed for the development of the xDSPcore processor, a compiler-driven configurable \ac{DSP}~\cite{Krall2004xDSPcore,Farfeleder2005Ultra}. Utilizing an \ac{XML}-based syntax, an \emph{xDSPcore ADL} specification defines the architectural state, including individual pipeline stages. Instruction behavior is modeled via micro-operations represented as C++ code snippets, while an execution-cycle specification maps these micro-operations to the start or end of specific pipeline stages. However, a subset of the behavior must be duplicated as tree patterns to drive the compiler generator, introducing partial redundancy into the specification. The generation of the optimizing compiler and the compiled instruction set simulator are detailed in~\cite{Farfeleder2006Effective} and~\cite{farfeleder2007ultra}, respectively. The generated compiler supports autovectorization~\cite{Pryanishnikov2007Compiler}, if-conversion of predicated instructions, \ac{DSP} addressing modes, and hardware loops. Furthermore, the instruction specification enables the automated generation of an instruction decoder in an \ac{HDL}. To overcome the redundancy limitations experienced with xDSPcore ADL, \ac{VADL} was designed to derive all targets from a single, non-redundant specification.

The structural \ac{PDL} \emph{xADL} represents the second-generation language developed by our research group~\cite{brandner2007compiler,Brandner2009phdthesis}. It describes processor structures at the \ac{RTL} using an \ac{XML}-based syntax. From this low-level model, an exceptionally fast, tiered \ac{CAS} featuring three optimization layers is generated: simulation begins with interpretation and dynamically translates basic blocks and ultimately entire execution regions to the host architecture using basic-block duplication~\cite{brandner2009fast}. Notably, despite the low abstraction level, the toolchain synthesizes both full hardware descriptions and an LLVM compiler backend~\cite{brandner2007compiler,Brandner2012spe}. However, a key limitation of this purely structural approach is that modifying the underlying microarchitecture requires rewriting large parts of the processor specification. These constraints directly motivated the design of \ac{VADL} and its core principle of strictly separating the \ac{ISA} from the \ac{MiA}.

The Tensor Accelerator ISA Definition Language (\emph{TAIDL}) is specifically designed to model the semantics of tensor accelerator instructions. These specifications serve as the foundation for automatically generating a functional instruction set simulator acting as a test oracle~\cite{taidl-test2025taidliss} and a compiler backend integrated with the XLA~\cite{OpenXLA2026} machine learning compiler~\cite{taidl-act2026taidlcomp}.
A TAIDL specification comprises two primary elements: the data model definition and the behavioral semantics. The latter is specified using XLA High-Level Optimizer (HLO) operations, which offer a higher level of abstraction than the lower-level \texttt{forall} and \texttt{fold} constructs in \ac{VADL}. To achieve high execution throughput, the generated simulator interprets the guest instruction stream while invoking pre-compiled XLA host kernels that offload expensive tensor operations natively to host \acsp{CPU} or \acsp{GPU}. Furthermore, defining instructions at this high level of abstraction streamlines compiler generation, as TAIDL does not need to bridge a large semantic gap to XLA IR. In contrast, OpenVADL's compiler generator must lower high-level behavioral descriptions down to LLVM IR, which operates close to scalar \acs{CPU} assembly. Finally, because not every XLA operation in a workload has a direct one-to-one hardware counterpart, the generated compiler backend uses equality saturation (e-graphs) to perform IR-to-IR rewrites into instruction sequences supported by the accelerator.

\emph{ISADL} is an instruction-encoding description language targeting \ac{VLIW} architectures~\cite{Xiao2023ISADL}. Its primary purpose is either to specify an instruction encoding explicitly or to automatically synthesize one from high-level encoding requirements, such as the available functional units, encoding fields, instruction count, and default instructions. From this encoding model, an optimizing assembler can be generated by defining both assembly syntax and instruction-matching rules via regular expressions. Although the authors mention the generation of an instruction set simulator, the published literature provides no details on how instruction semantics are specified. Moreover, the reliance on explicit regular-expression rules and detailed encoding constraints results in verbose specifications that can be challenging to read and maintain. Unlike \ac{VADL}, ISADL focuses strictly on encoding and assembly, offering no support for compiler or hardware generation.

\emph{CoreDSL~2.0}~\cite{CoreDSL2} is a lightweight \ac{ISA} description language characterized by imperative, sequential semantics. Its simulation environment generates C~source code for instruction blocks, which is then compiled at runtime via a source-level just-in-time (JIT) infrastructure~\cite{Kappes2023Effective}. Although CoreDSL~2.0 lacks native language primitives for specifying compiler configurations or detailed microarchitectural structures, it has been successfully utilized to define custom \ac{ISA} extensions for the semi-automated synthesis of LLVM compiler backends~\cite{VanKempen2024Seal5} and hardware accelerators~\cite{Oppermann2024Longnail}. In terms of simulation, \ac{VADL}'s integration with QEMU's dynamic binary translation (\ac{DBT}) engine avoids the runtime overhead of invoking an external C~compiler for JIT compilation, enabling consistently high instruction throughput.

\emph{StoneCutter}~\cite{leidel2019StoneCutter,Leidel2020StoneCutter,Leidel2021Toward,Kabrick2022Rapid} is a behavioral language designed to specify instruction semantics alongside explicit pipeline stages. Developed specifically to generate synthesizable Chisel hardware descriptions for processor execution pipelines, it lacks support for generating compilers or standalone software simulators. Instead, cycle-accurate simulation relies indirectly on the generated Chisel model. Furthermore, StoneCutter requires manually partitioning instruction behavior across individual pipeline stages, which increases specification effort and complicates maintenance when modifying the microarchitecture. The language follows an imperative style with sequential semantics, permitting multiple assignments to local and global variables as well as C-style loops. In contrast, \ac{VADL} enforces single-assignment semantics, provides higher-level microarchitectural abstractions, and natively generates both optimizing compilers and fast simulators alongside full hardware descriptions.

\emph{Sail}~\cite{armstrong2019isa} is a mature, purely behavioral \ac{PDL} designed primarily for formal \ac{ISA} specification and architectural verification. Unlike traditional \acp{PDL} that provide built-in abstractions for instructions, encodings, and microarchitectural components, Sail models instruction semantics in an imperative, operational style closer to general-purpose programming. While the Sail toolchain generates executable emulators in C and OCaml that serve as authoritative golden models, its primary strength lies in exporting definitions to theorem provers (such as Isabelle, HOL4, and Coq) to establish formal correctness. Consequently, automated downstream generation of synthesizable hardware or complete compiler backends is outside its primary scope. A defining feature of Sail is its expressive type system based on lightweight dependent types, which guarantees bit-vector safety at the cost of computationally intensive type checking. In contrast, \ac{VADL} employs a streamlined type system specifically engineered for low-latency, real-time feedback within modern \acp{IDE}. While \ac{VADL} does not yet offer a formal verification ecosystem comparable to Sail's, it provides higher-level domain abstractions tailored to the automated, end-to-end generation of high-performance simulators, optimizing compilers, and hardware descriptions. These capabilities reflect the complementary design goals of the two languages.

\emph{ViDL}, the versatile instruction set architecture description language, formally specifies processor \acp{ISA} in a functional style inspired by Standard ML (SML)~\cite{Dreesen2011Generating,Dreesen2012ViDL}. A defining characteristic of ViDL is its ability to automatically derive all microarchitectural structures solely from a user-defined target clock frequency, thereby eliminating the need for manual microarchitectural specification. From this high-level description, the ViDL framework synthesizes both structural \ac{VHDL} hardware descriptions and functional C++ interpreting simulators, while cycle-accurate execution relies on simulating the generated \ac{HDL}. Within instruction specifications, state updates to storage elements, such as memories and registers, occur via parallel assignments, whereas multiple assignments to intermediate variables are strictly prohibited. To bind expressions to local identifiers, the language employs \texttt{let} expressions analogous to those in SML and \ac{VADL}. Furthermore, the primitive operator set can be extended by explicitly defining backend code generation rules for both C++ and \ac{VHDL}. Notably, the toolchain lacks support for automated compiler generation. In terms of behavioral semantics, ViDL closely resembles \ac{VADL}, distinguishing itself primarily by supporting higher-order functions and omitting a macro system. However, whereas \ac{VADL} allows designers to configure the microarchitecture at a high level of abstraction, ViDL automates this synthesis completely, abstracting the microarchitectural implementation away from the designer.

The \emph{ISAC} (Instruction Set Architecture C) language is a mixed \ac{PDL} largely inspired by LISA and developed at the Brno University of Technology~\cite{Masarik2006Language,prikryl2011design}. Similar to LISA, instruction behavior is specified in a subset of the C~programming language. A fast, cycle-accurate interpreted simulator leveraging finite state machines is presented in~\cite{prikryl2009fast}, and subsequent research expanded this ecosystem to include both compiled and JIT-compiled simulation frameworks. To avoid duplicating behavioral semantics during compiler generation, semantic information is automatically extracted from the assembly grammar of the instructions~\cite{husar2011Automatic}. In contrast, \ac{VADL} avoids such behavioral extraction by enforcing an explicit, distinct definition for each instruction.

Building upon the architectural principles of \emph{ISAC}, \emph{CodAL} was subsequently developed~\cite{Charvat2012CodAL,Prikryl2014CodAL} and commercialized by Codasip~\cite{codasip}. A representative example of using CodAL to define custom \ac{ISA} extensions for RISC-V is documented in~\cite{Amor2022CodAL}. Beyond these publications, further architectural implementation details and internal mechanics of \emph{CodAL} remain proprietary and are not publicly available.

\emph{ArchC}~\cite{Rigo2003ArchC,azevedo2005archc} is an extension to \emph{SystemC}~\cite{panda2001systemc}. Both languages are embedded as libraries within C++ and thus act as \acp{eDSL}. While SystemC alone can model processors, its low abstraction level prevents the automated derivation of processor structure, instruction semantics, or assembly syntax. ArchC raises the abstraction level to enable these capabilities. However, instruction behavior, including all \ac{MiA} details such as program counter updates and pipeline stage transfers, is written directly in SystemC (C++). This intermingling of \ac{ISA} and \ac{MiA} concerns results in verbose specifications that are difficult to adapt when the \ac{MiA} changes. From an ArchC model, an interpreting simulator (as well as compiled variants in later extensions) can be generated. When full \ac{MiA} details are included, this simulation is cycle-accurate. Additionally, ArchC generates an assembler and linker for the GNU toolchain, and later work added compiler backend generation~\cite{Auler2012ArchC}. Because the C++ behavioral model is unsuitable for direct compiler extraction, an additional semantic description in a separate compiler-specific language is required, introducing redundancy and increasing the risk of inconsistencies. \ac{VADL} avoids these shortcomings through a unified, single-source specification.

The MAchine Markup Language (\emph{MAML}) was designed for the rapid development of application-specific instruction set processors, with a particular focus on multiprocessor systems~\cite{FischerTeichWeper2001,mishra2008processor}. Its syntax is \ac{XML}-based, while structural specifications can also be constructed via a graphical integrated development environment. Instruction behaviors are written in C++, which necessitates separate pattern definitions for the compiler's instruction selector. From this model, a mixed compiled/interpreted simulator and synthesizable hardware are generated, and the machine model for the compiler's optimizer and scheduler is extracted automatically. In \ac{VADL}, the \ac{MiA} is specified at a higher level of abstraction and neither requires nor supports graphical editing. Moreover, unlike MAML, \ac{VADL} automatically synthesizes the compiler's instruction selector directly from the instruction semantics, avoiding redundant specifications.

\emph{TIE}, the Tensilica Instruction Extension language, was designed to streamline custom instruction set extensions for the Xtensa processor architecture~\cite{Gonzalez2000Xtensa,Wang2001Hardware}. The base instruction set and architectural state remain fixed and cannot be altered. TIE is based on a subset of the Verilog \ac{HDL}, placing its behavioral descriptions at a lower level of abstraction than \ac{VADL}. The toolchain generates extensions for a GNU-based compiler, assembler, and linker, where custom instructions are primarily exposed via compiler intrinsics or targeted mapping directives, while simulation is performed at the \ac{RTL} level. In contrast, \ac{VADL} is broader in scope, as it enables the complete specification of arbitrary processor architectures from scratch rather than being limited to extending a single fixed base architecture.

\emph{Expression}~\cite{grun1998expression,halambi2008expression,mishra2008processor} is a mixed \ac{PDL} featuring a Lisp-like syntax, primarily targeting the development of \ac{SoC} architectures. It enables the generation of a \ac{CAS}, an \ac{ILP}-optimizing compiler, and synthesizable hardware, alongside extensive verification support via model checking. The language emphasizes low-level structural specifications that describe ports, connections, pipelines, and caches, while behavioral aspects are captured in operation specifications defining operands, semantics, and encoding. Additionally, operation-to-operation mappings can be specified to facilitate compiler generation and optimization. Although the toolchain and its graphical user interface were released as open source, they target legacy platforms from the early 2000s (such as Windows~XP), limiting usability on modern systems. Compared to Expression, \ac{VADL} offers a more readable, domain-specific syntax and models the \ac{MiA} at a higher level of abstraction, yielding more concise and maintainable descriptions. Currently, however, OpenVADL does not match Expression's extensive formal verification capabilities.

The syntax of \emph{ISDL}~\cite{hadjiyiannis1997isdl,hadjiyiannis2000isdl,hadjiyiannis2003isdl} builds upon an extended early version of nML and specializes in modeling \ac{VLIW} processor architectures. As a behavioral \ac{PDL}, it focuses on formalizing instruction semantics to generate assemblers, compilers, simulators, and synthesizable hardware. Similar to nML, however, ISDL tightly couples instruction behavior with microarchitectural timing and resource constraints, making architectural exploration more cumbersome than in \ac{VADL}.

When \emph{LISA}~\cite{zivojnovic1996lisa,pees1999lisa,schliebusch2002architecture,hohenauer2009ccompilers} was initially developed, it focused on efficient simulation of \ac{DSP} architectures. As a mixed \ac{PDL}, it supports the generation of a broad range of tools, including compilers, linkers, profilers, \acp{CAS}, and low-level \ac{HDL} descriptions for hardware synthesis. Its behavioral section embeds arbitrary C code fragments within individual pipeline stages, which leads to large specifications and necessitates separate semantic descriptions for compiler generation. This interleaving significantly increases specification size and maintenance effort, whereas \ac{VADL} mitigates these issues by strictly separating \ac{ISA} semantics from \ac{MiA} details.

\emph{nML}~\cite{freericks1991nml,fauth1995nml,mishra2008processor} is a mixed \ac{PDL} built around an attributed grammar that composes complete instructions from partial instruction rules. Instruction semantics are defined via \emph{action} attributes, while additional attributes capture cycle counts and stalling behavior. A structural \emph{skeleton} specifies state-holding elements and their interconnections. While early versions lacked pipeline support, pipelining was later integrated into the action attributes. From an nML description, users can generate compilers, \acp{CAS}, low-level hardware descriptions, and test-program generators. Now integrated into Synopsys's ASIP Designer \ac{EDA} suite, nML remains actively maintained. Like LISA, however, nML intertwines \ac{ISA} and \ac{MiA} aspects, resulting in complex specifications that are difficult to modify. \ac{VADL} addresses this limitation by strictly separating the \ac{ISA} and \ac{MiA}.

\emph{MIMOLA} represents the pioneering \ac{PDL}, establishing the foundation for structural hardware-software co-design~\cite{zimmermann1976mimola,marwedel1984mimola,mishra2008processor}. Characterized by a Pascal-inspired syntax, this structural \ac{PDL} prioritizes hardware synthesis to generate low-level architectural descriptions. Notably, despite its strictly structural paradigm, instruction-level data-flow graphs can still be extracted from the specification to support retargetable compilation within the MSSQ compiler framework~\cite{leupers1998retargetable}. A distinguishing capability of MIMOLA is its support for workload-driven optimization, allowing developers to supply representative benchmarks alongside the processor description to guide architectural profiling and synthesis. Consequently, MIMOLA is widely recognized as a precursor to modern \ac{HLS} tools and heavily influenced standard \acp{HDL} such as Verilog and \ac{VHDL}. In contrast, \ac{VADL} eschews purely structural specifications, avoiding the drawback that microarchitectural modifications force developers to rewrite large parts of the processor description.

%% file: src-relwork.tex
\subsection*{Retargetable Compilers}

One of the primary artifacts generated by \ac{VADL} is a complete compiler backend, produced by the \ac{LCB}.
General principles of compiler construction and retargetability are covered in standard textbooks such as Aho et al.~\cite{aho2006compiler}, while Leupers~\cite{leupers2001retargetable} specifically addresses retargetable compilers in the context of embedded systems (see Section~\ref{sec:background-toolchain} for a brief overview).

Early retargetable compiler frameworks include \emph{lcc}~\cite{fraser1991lcc}, followed by well-established open-source toolchains such as \emph{GCC}~\cite{stallman2020gcc}, which supports a broad range of target architectures.
Another widely adopted infrastructure is \emph{LLVM}~\cite{lattner2004llvm}, which was designed around retargetability from the ground up. This is reflected in its common, well-defined \ac{IR}, which enables a rich suite of target-independent transformations and optimizations. To specify target-specific properties and available instructions succinctly, LLVM relies on the \emph{TableGen} domain-specific language.
Because the \ac{VADL} \ac{LCB} emits a compiler backend as an LLVM target, one of its central tasks is automatically generating the corresponding TableGen definitions and instruction-selection patterns.

\subsection*{Simulation}

Instruction set simulators play a vital role during processor development when physical hardware is not yet available and the architecture is still evolving~\cite{brandner2013dsp}, as well as during design space exploration~\cite{mishra2008processor} (see Section~\ref{sec:background} for a brief overview of simulation in the context of \ac{VADL}).
Constructing a simulator involves balancing two primary goals: accuracy and performance. Accuracy measures how closely the simulator's behavior and reported metrics match those of the physical hardware, whereas performance reflects the computational resources required for execution. Because higher accuracy generally demands greater computational effort, there is a fundamental trade-off between these two objectives~\cite{yi2005characterizing,yi2006simulation}.
Beyond hardware development, simulators enhanced with graphical user interfaces also serve as valuable educational tools for teaching processor internals~\cite{giorgi2019webrisc}. This is also a potential application for future versions of \ac{VADL}.

A key technique for improving simulator performance is \emph{caching}. For instance, decoded instructions that execute repeatedly can be cached within a threaded-code interpreter~\cite{bedichek1990some}, an approach also adopted in the \emph{SimICS} simulator~\cite{magnusson1997efficient,magnussion2002simics} and in the \ac{ISS} generated by the original \ac{VADL} implementation. More advanced schemes cache not only decoded instructions but also fetched data values~\cite{ratsiambahotra2009versatile} or adapt caching mechanisms from \ac{JIT} compilers, as demonstrated in \emph{Pydgin}~\cite{lockhart2015pydgin}.
However, self-modifying code often complicates instruction caching. To handle this safely, the original \ac{VADL} \ac{ISS} employs the \emph{value-checking} strategy investigated by Keppel~\cite{keppel2009detectselfmodifying}.

Another important design aspect of a simulator is the handling of \emph{system calls} to the operating system kernel, for which two common approaches exist~\cite{brandner2013dsp}.
The first is \acf{UME}, also known as syscall delegation. This approach forwards system calls from the simulated program directly to the host operating system, as implemented in \emph{SimpleScalar}~\cite{austin2002simplescalar}.
The second is \emph{full-system simulation}, which models the entire target platform, including operating system functionality, input/output, peripheral devices, and the memory management unit~\cite{wagstaff2015high}.
While full-system simulation is considerably more complex to implement than \ac{UME}, it achieves higher modeling precision for the examined processor~\cite{cain2002precise}.
\emph{QEMU}~\cite{bellard2005qemu} supports both modes. Currently, both the original \ac{VADL} \ac{ISS} and the OpenVADL \ac{ISS} support \ac{UME}, though targeting QEMU makes extending OpenVADL to full-system simulation straightforward.

With respect to execution techniques, interpretive simulation represents the most basic model.
Prior studies compare classical switch-based interpreters with direct-threaded~\cite{bell1973threaded} and indirect-threaded~\cite{dewar1975indirect} interpreters~\cite{klint1981interpretation}, as well as techniques for reducing branch mispredictions in threaded code~\cite{ertl2001threaded,ertl2001behavior}.
The original \ac{VADL} \ac{ISS} is implemented as a \ac{DTC} interpreter.

To overcome the overhead of interpretation, \emph{compiled simulation} translates target instructions into host code ahead of time~\cite{mills1991compiled}, with subsequent optimizations focusing on reducing generated code size~\cite{bartholomeu2004optimizations} and detecting code modifications to support self-modifying code~\cite{reshadi2003instruction}.
Furthermore, hybrid approaches such as the \ac{CAS} by Farfeleder et al.~\cite{farfeleder2007ultra} compile basic blocks for higher performance while dynamically switching between interpreted and compiled execution modes.

To combine the flexibility of interpretation with the speed of compiled simulation, \ac{DBT} translates blocks of target instructions into host code dynamically at runtime.
This technique was pioneered in \emph{Shade}~\cite{cmelik1994dynbintrans}, an early \ac{ISS} and profiler, and forms the execution backbone of \emph{QEMU}~\cite{bellard2005qemu}, which the OpenVADL \ac{ISS} targets.

Beyond functional and \ac{DBT}-based simulation, cycle-level architectural simulators such as \emph{gem5}~\cite{binkert2011gem5,lowe2020gem5} are widely used for microarchitectural evaluation. The \emph{gem5} simulator has also been extended to co-simulate processors at the \ac{RTL}~\cite{lopez2021gem5}. We employ \emph{gem5} in the evaluation of \ac{VADL}.
Finally, to reduce the high computational overhead of full \acp{CAS}, \emph{cycle-approximate} simulation techniques trade a controlled amount of timing accuracy for substantially higher execution speed~\cite{hwang2008cycle,franke2008fast}.
Developing a cycle-approximate simulator based on QEMU is planned as future work for OpenVADL.

\subsection*{HDLs}

In the context of \ac{VADL}, an \ac{HDL} is a domain-specific language used to specify digital electronic circuits.
While these circuits can be instruction set processors as in \ac{VADL}, \acp{HDL} equally target arbitrary \acp{ASIC} and digital designs.
Compared to a \ac{PDL}, an \ac{HDL} is far more general but consequently lacks processor-specific abstractions.

The primary abstraction level at which \acp{HDL} operate is the \ac{RTL} (see Section~\ref{sec:rtl} for a brief introduction).
Logic synthesis tools transform an \ac{RTL} description into a gate-level netlist before physical placement and routing~\cite{hachtel2005logic}.
Although most industrial synthesis suites are commercial closed-source products, \emph{Yosys}~\cite{wolf2013yosys} provides an open-source synthesis framework targeting both \acp{FPGA} and \acp{ASIC}.
\ac{VADL}'s hardware generator emits synthesizable \ac{RTL} code that can be directly processed by such downstream synthesis flows.

The most widely established industry-standard \acp{HDL} are \emph{\ac{VHDL}}~\cite{ieee2004vhdlstandard} and \emph{Verilog}~\cite{ieee2002verlogrtlstandard}, the latter serving as the primary input format for Yosys.
\emph{SystemVerilog}~\cite{ieee2018systemverilog} extends Verilog with enhanced verification capabilities, more expressive syntax, and object-oriented abstractions.

Alternatively, modern \acp{HDL} are often embedded as domain-specific libraries within general-purpose programming languages.
A prominent example is \emph{SystemC}~\cite{panda2001systemc}, which is embedded in C++ and provides object-oriented modeling alongside strong system-level simulation support.
In contrast, the \ac{VADL} hardware generator targets \emph{Chisel}~\cite{bachrach2012chisel}, an \ac{HDL} embedded in Scala that leverages functional programming constructs and compiles down to Verilog (see Section~\ref{sec:chisel} for a short overview).

%% file: src-futwork.tex
\section{Limitations and Future Work}
\label{sec:future_work}

Both the original \ac{VADL} implementation and OpenVADL remain research prototypes, leaving both important methodological limitations to acknowledge and numerous avenues for evolving OpenVADL into a robust, widely usable research platform.

A primary limitation of the current work is the absence of a mechanized operational semantics for the complete \ac{VADL} language and formal soundness proofs for its lowering transformations across backends. Consequently, correctness across the generated compiler, simulator, and hardware artifacts is currently established empirically through differential co-simulation against independent upstream baselines (such as upstream QEMU and LLVM) and architectural compliance suites rather than formal equivalence proofs. Long-term research directions include developing a mechanized formal semantics for \ac{VADL} and enabling formal translation validation. This work can build on prior experience with Abstract State Machine specifications~\cite{Lezuo2013CASM,Lezuo2014CASM} and theorem proving.

While the core \ac{VADL} language design is largely complete, we plan to extend it to support heterogeneous multiprocessor systems, which requires modeling bus protocols. Currently, \ac{VADL} relies on co-simulation backed by automatically generated, instruction-based self-tests to validate both specifications and generated artifacts. Going forward, we aim to expand the language with richer constructs for automated test generation and specification validation.

Beyond language extensions, several opportunities exist to advance the tool generators. First, all generators must support the entire \ac{VADL} specification language, including full coverage of floating-point arithmetic for IEEE and non-IEEE floating-point formats.

For the compiler generator, ongoing efforts focus on efficient code generation for multi-result instructions and the automatic vectorization of nested loops into tensor instructions. Furthermore, to fully demonstrate the flexibility of the \ac{GCB}, we plan to implement backends targeting \ac{JIT} compilers as well as the GNU Compiler Collection (GCC) and its toolchain.

Regarding simulator generation, we are actively adding support for \ac{VLIW} architectures and reducing simulation overhead in \acp{CAS}. To achieve this, we plan to build OpenVADL's \ac{CAS} on top of QEMU and offload statically determinable computations to the \ac{JIT} compilation phase. In parallel, we are developing accurate cache and memory-subsystem simulation that covers cache coherence protocols, memory consistency models, atomic instructions, and address translation with \ac{TLB} support. Partial implementations of these features already exist in the original \ac{VADL} prototype~\cite{Himmelbauer2024} and will be integrated into OpenVADL.

Finally, hardware synthesis must be extended to cover all \ac{VADL} microarchitectural components, including reservation stations, reorder buffers, load/store queues, and fetch buffers. Further optimizations are also needed to generate competitive hardware designs.

%% file: src-conclusion.tex
\section{Conclusion}
\label{sec:conclusion}

This article presented the \acl{VADL} (\ac{VADL}) and provided a comparative evaluation of its two primary implementations: the original closed-source framework and the emerging open-source iteration, OpenVADL. Processor specifications across diverse target architectures, including RISC-V, AArch64, AArch32, PowerPC, and Hexagon, demonstrate that the language's expressive constructs and syntactic macro system enable concise and comprehensible architectural descriptions. Furthermore, insights gained from the initial implementation led to a substantial architectural redesign in OpenVADL, most notably the transition of the intermediate representation from quadruple-based three-address code to a sea-of-nodes style combined data- and control-flow graph.

The automatically generated artifacts exhibit performance metrics competitive with manually engineered tools. Specifically, binaries produced by OpenVADL-generated compilers execute about 6.5\% (RV32IM) and 8.3\% (RV64IM) more dynamic instructions than those generated by upstream LLVM compilers, while the synthesized \acp{ISS} outperform baseline QEMU by up to 17.3\% in the arithmetic mean across the Embench benchmark suite. In terms of hardware synthesis, the generated designs achieve comparable maximum clock frequencies (ranging from $-15.8\%$ to $+7.3\%$) at the expense of a 3.0\% to 13.9\% increase in silicon area. The viability of OpenVADL for rapid design space exploration was validated through an end-to-end case study evaluating CRC32 checksum computation on the RISC-V \ac{ISA}. Although OpenVADL remains a research prototype under active development, future efforts will focus on further optimizing the generated toolchains and expanding hardware synthesis to out-of-order superscalar architectures. The framework is open to community contributions, with source code and documentation hosted at \url{https://openvadl.org}.

%% file: src-app_float.tex
\section{Floating-Point Instruction Specification}
\label{app:float_isa}

The instruction specification for floating-point operations is designed to be highly generic to accommodate arbitrary floating-point formats. However, the current implementation in OpenVADL restricts them to 32-bit and 64-bit IEEE formats. Central to this framework is the definition of a floating-point type via the {\tt float-type} keyword, which is supplemented by annotations that specify its behavioral properties, such as bit width, format, and the representation of signaling or non-signaling NaNs (see Listing~\ref{lst:float}, lines 1–4). Furthermore, floating-point operations can trigger both sticky and non-sticky exception flags. Because these flags are conventionally stored within registers in most \acp{ISA}, targeted register fields can be marked via annotations to explicitly designate them as destinations for floating-point status flags.

\begin{lstlisting}[float=htb,label={lst:float},caption={Instruction definitions with floating-point operations},language=vadl]
[ IEEE : 32 ]
float-type Binary32
[ IEEE : 64 ]
float-type Binary64

constant RoundNE : Bits<3> = 0

using SInt32 = SInt<32>
 
instruction set architecture floatISA = {

register X    : Bits<32>   // 32 bit integer register
register F32A : Binary32   // 32 bit float register
register F32B : Binary32   // 32 bit float register
register F64  : Binary64   // 64 bit float register
 
format F : Bits<8> = {opcode : Bits<8>}

instruction singleAdd : F =                      // 32 bit float addition
    F32A := VADL::fadd (F32A, F32B, RoundNE)
instruction double2single : F =                  // 64 bit float to 32 bit float conversion
    F32A := VADL::fcvt<Binary32>(F64, RoundNE)
instruction single2sint   : F =                  // 32 bit float to 32 bit signed int conversion
    X   := VADL::fcvt<SInt32>  (F32A, RoundNE)
}
\end{lstlisting}

The floating-point builtins accept {\tt float-type} arguments. Some also require a rounding mode. They provide common operations like addition or subtraction, accumulated multiplication, comparison, testing, or type conversions. Depending on the {\tt float-type}, the corresponding operation is selected (see the definition of an add instruction in Listing~\ref{lst:float}, lines 19 and 20). For type conversions, the types of the arguments are not sufficient to determine the corresponding operation. Therefore, type conversions take a type parameter in angle brackets before the operation's arguments (see the definition of two type conversion instructions in Listing~\ref{lst:float}, lines 21-24).

More details on floating-point instruction specification and its implementation are available in the online reference manual (\url{https://openvadl.github.io/openvadl/}) and a thesis~\cite{Malicky2026}.

%% file: src-app_jalr.tex
\section{JALR assembly grammar example}
\label{app:jalr}

Listing~\ref{lst:asm_jalr} shows the definition of an assembly grammar rule for the RISC-V JALR instruction. This example shows how semantic predicates and the special lookahead function \texttt{LaIdEq} are used to distinguish between different syntactic forms of one instruction.

\begin{lstlisting}[float=htbp,label={lst:asm_jalr},caption={Assembly grammar rule for the RISC-V JALR instruction},language=vadl]
assembly description RV32I_AD for RV32I_ABI = {

  function x1_string -> String = "x1"
  function zero_const -> SInt<64> = 0

  grammar = {
    
    JalrInstruction @instruction:
      var rdTmp  = x1_string @register        // default for jalr rs
      var rs1Tmp = null @register
      var immTmp = zero_const @operand        // default for jalr rs / jalr rd, rs

      mnemonic = "JALR" @operand
      rs1Tmp = Register
      [ "," rdTmp = rs1Tmp                    // first register was actually rd
        ( ?(LaIdEq(1, "(") )
            immTmp = ImmediateOperand         // jalr rd, imm(rs)
            "(" rs1Tmp = Register ")"
          |
            rs1Tmp = Register                 // jalr rd, rs
            [ "," immTmp = ImmediateOperand]  // jalr rd, rs, imm

        )
      ]

      rd    = rdTmp @operand
      rs1   = rs1Tmp @operand
      immS  = immTmp
    ;
  }
}
    
\end{lstlisting}

%% file: src-app_riscv.tex
\FloatBarrier % prevents a float from being placed inside the long listing

\section{RISC-V RV32I example}
\label{app:appendix_a}

This appendix shows a complete specification of the RISC-V RV32I instruction set architecture.
This specification contains all the necessary instructions and definitions to generate an \ac{ISS} as used in the evaluation part of this article.

\begin{lstlisting}[label={lst:rv32i},caption={Complete RISC-V RV32I \ac{ISA} specification},language=vadl,linerange={1-161}]
instruction set architecture RV32I = {

  using Byte     = Bits< 8>               // 8 bit Byte
  using Half     = Bits<16>               // 16 bit half word type
  using Word     = Bits<32>               // 32 bit word type
  using Index    = Bits< 5>               // 5 bit register index type for 32 registers
  using SIntR    = SInt<32>               // register word signed type 
  using UIntR    = UInt<32>               // register word unsigned type 
  using UInt5    = UInt< 5>               // 5 bit unsigned shift amount

  [zero: X(0)]                            // register with index 0 is always 0
  register         X : Index -> Word      // integer register file with 32 registers of 32 bits
  program counter PC : Word               // PC points to the start of the current instruction
  memory         MEM : Word -> Byte       // byte-addressed memory

  format Rtype : Word =                   // Rtype register 3 operand instruction format
    { funct7 : Bits<7>                    // [31..25] 7 bit function code
    , rs2    : Index                      // [24..20] 2nd source register index / shamt
    , rs1    : Index                      // [19..15] 1st source register index
    , funct3 : Bits<3>                    // [14..12] 3 bit function code
    , rd     : Index                      // [11..7]  destination register index
    , opcode : Bits<7>                    // [6..0]   7 bit operation code
    , shamt  = rs2 as UInt                // 5 bit unsigned shift amount
    }
  format Itype : Word =                   // Itype immediate instruction format
    { imm    : Bits<12>                   // [31..20] 12 bit immediate value
    , rs1    : Index                      // [19..15] source register index
    , funct3 : Bits<3>                    // [14..12] 3 bit function code
    , rd     : Index                      // [11..7]  destination register index
    , opcode : Bits<7>                    // [6..0]   7 bit operation code
    , immS   = imm as SIntR               // sign-extended immediate value
    }
  format Utype : Word =                   // Utype upper immediate instruction format
    { imm    : Bits<20>                   // [31..12] 20 bit immediate value
    , rd     : Index                      // [11..7]  destination register index
    , opcode : Bits<7>                    // [6..0]   7 bit operation code
    , immU   = (imm as UIntR) << 12       // shifted unsigned immediate value
    }
  format Stype : Word =                   // Stype store instruction format
    { imm    [31..25, 11..7]              // 12 bit immediate value
    , rs2    [24..20]                     // 2nd source register index
    , rs1    [19..15]                     // 1st source register index
    , funct3 [14..12]                     // 3 bit function code
    , opcode [6..0]                       // 7 bit operation code
    , immS   = imm as SIntR               // sign-extended immediate value
    }
  format Btype : Word =                   // Btype branch instruction format
    { imm    [31, 7, 30..25, 11..8]       // 12 bit immediate value
    , rs2    [24..20]                     // 2nd source register index
    , rs1    [19..15]                     // 1st source register index
    , funct3 [14..12]                     // 3 bit function code
    , opcode [6..0]                       // 7 bit operation code
    , immS   = (imm as SIntR) << 1        // sign-extended and shifted immediate value immS
    }
  format Jtype : Word =                   // Jtype jump and link instruction format
    { imm    [31, 19..12, 20, 30..21]     // 20 bit immediate value
    , rd     [11..7]                      // destination register index
    , opcode [6..0]                       // 7 bit operation code
    , immS   = (imm as SIntR) << 1        // sign-extended and shifted immediate value immS
    }

  model RtypeInstr (name: Id, op: BinOp, fu3: Bin, fu7: Bin, lhsTy: Id, rhsTy: Id) : IsaDefs = {
    instruction $name : Rtype =                        // 3 register operand instructions
      X(rd) := (X(rs1) as $lhsTy $op X(rs2) as $rhsTy) as Word
    encoding $name = { opcode = 0b011'0011, funct3 = $fu3, funct7 = $fu7}
    assembly $name = (mnemonic, " ", register(rd), ",", register(rs1), ",", register(rs2))
    }
  model IShftInstr (name : Id, op : BinOp, funct3 : Bin, funct7 : Bin, lhsTy : Id) : IsaDefs = {
    instruction $name : Rtype =                        // shift immediate instructions
      X(rd) := X(rs1) as $lhsTy $op shamt
    encoding $name = {opcode = 0b001'0011, funct3 = $funct3, funct7 = $funct7}
    assembly $name = (mnemonic, " ", register(rd), ",", register(rs1), ",", decimal(rs2))
    }
  model ItypeInstr (name : Id, op : BinOp, funct3 : Bin, exTy : Id) : IsaDefs = {
    instruction $name : Itype =                        // immediate instructions
      X(rd) := (X(rs1) as $exTy $op immS as $exTy) as Word
    encoding $name = {opcode = 0b001'0011, funct3 = $funct3}
    assembly $name = (mnemonic, " ", register(rd), ",", register(rs1), ",", decimal(imm))
    }
  model UtypeInstr (name : Id, opcode : Bin, rhsEx : Ex) : IsaDefs = {
    instruction $name : Utype =                        // upper immediate instructions
      X(rd) := $rhsEx
    encoding $name = {opcode = $opcode}
    assembly $name = (mnemonic, " ", register(rd), ",", hex(imm))
    }
  model LtypeInstr (name : Id, funct3 : Bin, memEx : CallEx, exTy: Id) : IsaDefs = {
    instruction $name : Itype =                        // load instructions
      let addr = X(rs1) + immS in
        X(rd) := $memEx as $exTy
    encoding $name = {opcode = 0b000'0011, funct3 = $funct3}
    assembly $name = (mnemonic, " ", register(rd), ",", decimal(imm), "(", register(rs1), ")")
    }
  model StypeInstr (name : Id, funct3 : Bin, memEx : CallEx, exTy: Id) : IsaDefs = {
    instruction $name : Stype =                        // store instructions
      let addr = X(rs1) + immS in
        $memEx := X(rs2) as $exTy
    encoding $name = {opcode = 0b010'0011, funct3 = $funct3}
    assembly $name = (mnemonic, " ", register(rs2), ",", decimal(imm), "(", register(rs1), ")")
    }
  model BtypeInstr (name : Id, relOp : BinOp, funct3 : Bin, lhsTy : Id) : IsaDefs = {
    instruction $name : Btype =                        // conditional branch instructions
      if X(rs1) as $lhsTy $relOp X(rs2) then
        PC := PC + immS
    encoding $name = {opcode = 0b110'0011, funct3 = $funct3}
    assembly $name = (mnemonic, " ", register(rs1), ",", register(rs2), ",", decimal(imm))
    }
  model JLinkInstr (name : Id, iFormat : Id, reg : Ex, opcode : Encs, asm : Ex) : IsaDefs = {
    instruction $name : $iFormat =                     // jump and link (register)
      let retaddr = PC.next in {
        PC    := ($reg + immS) as UInt & 0xffff'fffe   // $reg could be equal to X(rd)
        X(rd) := retaddr                               // when rs1 is equal to rd
        }
    encoding $name = {$opcode}
    assembly $name = (mnemonic, " ", register(rd), ",", decimal(imm), $asm)
    }

  $RtypeInstr (ADD  ; +  ; 0b000 ; 0b000'0000 ; Bits ; Bits ) // add
  $RtypeInstr (SUB  ; -  ; 0b000 ; 0b010'0000 ; Bits ; Bits ) // subtract
  $RtypeInstr (AND  ; &  ; 0b111 ; 0b000'0000 ; Bits ; Bits ) // and
  $RtypeInstr (OR   ; |  ; 0b110 ; 0b000'0000 ; Bits ; Bits ) // or
  $RtypeInstr (XOR  ; ^  ; 0b100 ; 0b000'0000 ; Bits ; Bits ) // exclusive or
  $RtypeInstr (SLT  ; <  ; 0b010 ; 0b000'0000 ; SInt ; SInt ) // set less than
  $RtypeInstr (SLTU ; <  ; 0b011 ; 0b000'0000 ; UInt ; UInt ) // set less than unsigned
  $RtypeInstr (SLL  ; << ; 0b001 ; 0b000'0000 ; UInt ; UInt5) // shift left  logical
  $RtypeInstr (SRL  ; >> ; 0b101 ; 0b000'0000 ; UInt ; UInt5) // shift right logical
  $RtypeInstr (SRA  ; >> ; 0b101 ; 0b010'0000 ; SInt ; UInt5) // shift right arithmetic

  $IShftInstr (SLLI ; << ; 0b001 ; 0b000'0000 ; UInt)  // shift left  logical immediate
  $IShftInstr (SRLI ; >> ; 0b101 ; 0b000'0000 ; UInt)  // shift right logical immediate
  $IShftInstr (SRAI ; >> ; 0b101 ; 0b010'0000 ; SInt)  // shift right arithmetic immediate

  $ItypeInstr (ADDI ; +  ; 0b000 ; SInt)               // add immediate
  $ItypeInstr (ANDI ; &  ; 0b111 ; SInt)               // and immediate
  $ItypeInstr (ORI  ; |  ; 0b110 ; SInt)               // or immediate
  $ItypeInstr (XORI ; ^  ; 0b100 ; SInt)               // exclusive or immediate
  $ItypeInstr (SLTI ; <  ; 0b010 ; SInt)               // set less than immediate
  $ItypeInstr (SLTIU; <  ; 0b011 ; UInt)               // set less than immediate unsigned

  $UtypeInstr (AUIPC; 0b001'0111 ; PC + immU)          // add upper immediate to PC
  $UtypeInstr (LUI  ; 0b011'0111 ;      immU)          // load upper immediate

  $LtypeInstr (LB   ; 0b000 ; MEM   (addr) ; SIntR)    // load byte signed
  $LtypeInstr (LBU  ; 0b100 ; MEM   (addr) ; UIntR)    // load byte unsigned
  $LtypeInstr (LH   ; 0b001 ; MEM<2>(addr) ; SIntR)    // load half word signed
  $LtypeInstr (LHU  ; 0b101 ; MEM<2>(addr) ; UIntR)    // load half word unsigned
  $LtypeInstr (LW   ; 0b010 ; MEM<4>(addr) ; SIntR)    // load word
  $StypeInstr (SB   ; 0b000 ; MEM   (addr) ; Byte )    // store byte
  $StypeInstr (SH   ; 0b001 ; MEM<2>(addr) ; Half )    // store half word
  $StypeInstr (SW   ; 0b010 ; MEM<4>(addr) ; Word )    // store word

  $BtypeInstr (BEQ  ; =  ; 0b000 ; Bits)               // branch equal
  $BtypeInstr (BNE  ; != ; 0b001 ; Bits)               // branch not equal
  $BtypeInstr (BGE  ; >= ; 0b101 ; SInt)               // branch greater than or equal
  $BtypeInstr (BGEU ; >= ; 0b111 ; UInt)               // branch greater than or equal unsigned
  $BtypeInstr (BLT  ; <  ; 0b100 ; SInt)               // branch less than
  $BtypeInstr (BLTU ; <  ; 0b110 ; UInt)               // branch less than unsigned

  $JLinkInstr (JAL  ; Jtype ; PC     ; opcode = 0b110'1111 ; "") // jump and link
  $JLinkInstr (JALR ; Itype ; X(rs1) ; opcode = 0b110'0111,      // jump and link register
                                       funct3 = 0b000 ; ("(", register(rs1), ")"))
  }
\end{lstlisting}

%% file: src-acronyms.tex
\acro{ABI}{Application Binary Interface}
\acro{ADL}{Architecture Description Language}
\acro{ALU}{Arithmetic Logic Unit}
\acro{ASIC}{Application Specific Integrated Circuit}
\acro{ASIP}{Application Specific Instruction Set Processor}
\acro{AST}{Abstract Syntax Tree}

\acro{BB}{Basic Block}

\acro{CAS}{Cycle Accurate Simulator}
\acro{CFG}{Control Flow Graph}
\acro{CPU}{Central Processing Unit}
\acro{CSE}{Common Subexpression Elimination}
\acro{CSR}{Control and Status Register}
\acro{CST}{Concrete Syntax Tree}

\acro{DAG}{Directed Acyclic Graph}
\acro{DBT}{Dynamic Binary Translation}
\acro{DFG}{Data Flow Graph}
\acro{DSE}{Design Space Exploration}
\acro{DSL}{Domain Specific Language}
\acro{DSP}{Digital Signal Processor}
\acro{DTC}{Direct Threaded Code}

\acro{EDA}{Electronic Design Automation}
\acro{eDSL}{embedded Domain Specific Language}
\acro{ELF}{Executable and Linkable Format}

\acro{FPGA}{Field Programmable Gate Array}
\acro{FSM}{Finite State Machine}

\acro{GCB}{Generic Compiler Backend}
\acro{GPU}{Graphics Processing Unit}

\acro{HDL}{Hardware Description Language}
\acro{HLS}{High-Level Synthesis}
\acro{HTIF}{Berkeley Host-Target Interface}
\acro{HTML}{HyperText Markup Language}

\acro{IDE}{Interactive Development Environment}
\acro{ILP}{Instruction Level Parallelism}
\acro{IPG}{Instruction Progress Graph}
\acro{IR}{Intermediate Representation}
\acro{ISA}{Instruction Set Architecture}
\acro{ISS}{Instruction Set Simulator}

\acro{JIT}{just-in-time}

\acro{LCB}{LLVM Compiler Backend}
\acro{LSP}{Language Server Protocol}

\acro{MiA}{Microarchitecture}

\acro{PC}{Program Counter}
\acro{PDL}{Processor Description Language}

\acro{RAM}{Random Access Memory}
\acro{RISC}{Reduced Instruction Set Computer}
\acro{RTL}{Register-Transfer Level}

\acro{SIMD}{Single Instruction Multiple Data}
\acro{SoC}{System-on-Chip}
\acro{SSA}{Static Single Assignment}

\acro{TB}{Translation Block}
\acro{TCG}{Tiny Code Generator}
\acro{TLB}{Translation Lookaside Buffer}

\acro{UME}{User Mode Emulation}

\acro{VADL}{Vienna Architecture Description Language}
\acro{VDL}{Vienna Definition Language}
\acro{VDT}{VADL Decode Tree}
\acro{VHDL}{Very High Speed Integrated Circuit Hardware Description Language}
\acro{VIAM}{VADL Intermediate Architecture Model}
\acro{VIR}{VADL Intermediate Representation}
\acro{VLIW}{Very Long Instruction Word}

\acro{XML}{eXtensible Markup Language}

%% file: main.bbl

\begin{thebibliography}{161}

%%% the Web address.

\ifx \showCODEN    \undefined \def \showCODEN     #1{\unskip}     \fi
\ifx \showISBNx    \undefined \def \showISBNx     #1{\unskip}     \fi
\ifx \showISBNxiii \undefined \def \showISBNxiii  #1{\unskip}     \fi
\ifx \showISSN     \undefined \def \showISSN      #1{\unskip}     \fi
\ifx \showLCCN     \undefined \def \showLCCN      #1{\unskip}     \fi
\ifx \shownote     \undefined \def \shownote      #1{#1}          \fi
\ifx \showarticletitle \undefined \def \showarticletitle #1{#1}   \fi
\ifx \showURL      \undefined \def \showURL       {\relax}        \fi
% invisible to TeX
\providecommand\bibfield[2]{#2}
\providecommand\bibinfo[2]{#2}
\providecommand\natexlab[1]{#1}
\providecommand\showeprint[2][]{arXiv:#2}

\bibitem[iee(2002)]%
        {ieee2002verlogrtlstandard}
 \bibinfo{year}{2002}\natexlab{}.
\newblock \showarticletitle{IEEE Standard for Verilog Register Transfer Level
  Synthesis}.
\newblock \bibinfo{journal}{\emph{IEEE Std 1364.1-2002}}
  (\bibinfo{year}{2002}), \bibinfo{pages}{1--108}.
\newblock
\href{https://doi.org/10.1109/IEEESTD.2002.94220}{doi:\nolinkurl{10.1109/IEEESTD.2002.94220}}


\bibitem[iee(2004)]%
        {ieee2004vhdlstandard}
 \bibinfo{year}{2004}\natexlab{}.
\newblock \showarticletitle{IEEE Standard for VHDL Register Transfer Level
  (RTL) Synthesis}.
\newblock \bibinfo{journal}{\emph{IEEE Std 1076.6-2004 (Revision of IEEE Std
  1076.6-1999)}} (\bibinfo{year}{2004}), \bibinfo{pages}{1--118}.
\newblock
\href{https://doi.org/10.1109/IEEESTD.2004.94802}{doi:\nolinkurl{10.1109/IEEESTD.2004.94802}}


\bibitem[iee(2018)]%
        {ieee2018systemverilog}
 \bibinfo{year}{2018}\natexlab{}.
\newblock \showarticletitle{IEEE Standard for SystemVerilog--Unified Hardware
  Design, Specification, and Verification Language}.
\newblock \bibinfo{journal}{\emph{IEEE Std 1800-2017 (Revision of IEEE Std
  1800-2012)}} (\bibinfo{year}{2018}), \bibinfo{pages}{1--1315}.
\newblock
\href{https://doi.org/10.1109/IEEESTD.2018.8299595}{doi:\nolinkurl{10.1109/IEEESTD.2018.8299595}}


\bibitem[LSP(2022)]%
        {LSP}
 \bibinfo{year}{2022}\natexlab{}.
\newblock \bibinfo{booktitle}{\emph{Language Server Protocol Specification -
  3.17}}.
\newblock
\urldef\tempurl%
\url{https://microsoft.github.io/language-server-protocol/specifications/lsp/3.17/specification/}
\showURL{%
\tempurl}
\newblock
\shownote{[Online; accessed 14-October-2026]}.


\bibitem[emb(2024)]%
        {embench}
 \bibinfo{year}{2024}\natexlab{}.
\newblock \bibinfo{booktitle}{\emph{Embench: A Modern Embedded Benchmark
  Suite}}.
\newblock
\urldef\tempurl%
\url{https://www.embench.org/}
\showURL{%
\tempurl}
\newblock
\shownote{[Online; accessed 15-January-2024]}.


\bibitem[spi(2024)]%
        {spike}
 \bibinfo{year}{2024}\natexlab{}.
\newblock \bibinfo{booktitle}{\emph{{S}pike {RISC-V} {ISA} {S}imulator}}.
\newblock
\urldef\tempurl%
\url{https://github.com/riscv-software-src/riscv-isa-sim}
\showURL{%
\tempurl}
\newblock
\shownote{[Online; accessed 15-January-2024]}.


\bibitem[cod(2025)]%
        {codasip}
 \bibinfo{year}{2025}\natexlab{}.
\newblock \bibinfo{booktitle}{\emph{Codasip}}.
\newblock
\urldef\tempurl%
\url{https://codasip.com}
\showURL{%
\tempurl}
\newblock
\shownote{[Online; accessed 7-February-2025]}.


\bibitem[Aho et~al\mbox{.}(2006)]%
        {aho2006compiler}
\bibfield{author}{\bibinfo{person}{Alfred~V. Aho}, \bibinfo{person}{Monica~S.
  Lam}, \bibinfo{person}{Ravi Sethi}, {and} \bibinfo{person}{Jeffrey~D.
  Ullman}.} \bibinfo{year}{2006}\natexlab{}.
\newblock \bibinfo{booktitle}{\emph{{Compilers: Principles, Techniques, and
  Tools (2nd Edition)}}}.
\newblock \bibinfo{publisher}{Addison-Wesley Longman Publishing Co., Inc.},
  \bibinfo{address}{USA}.
\newblock
\showISBNx{0321486811}


\bibitem[Amdahl et~al\mbox{.}(1964)]%
        {AmdahlBlaauwBrooks1964}
\bibfield{author}{\bibinfo{person}{Gene~Myron Amdahl},
  \bibinfo{person}{Gerrit~Anne Blaauw}, {and}
  \bibinfo{person}{Frederick~Phillips Brooks}.}
  \bibinfo{year}{1964}\natexlab{}.
\newblock \showarticletitle{Architecture of the {IBM System/360}}.
\newblock \bibinfo{journal}{\emph{IBM Journal of Research and Development}}
  \bibinfo{volume}{8}, \bibinfo{number}{2} (\bibinfo{year}{1964}),
  \bibinfo{pages}{87--101}.
\newblock
\href{https://doi.org/10.1147/rd.82.0087}{doi:\nolinkurl{10.1147/rd.82.0087}}


\bibitem[Amor et~al\mbox{.}(2022)]%
        {Amor2022CodAL}
\bibfield{author}{\bibinfo{person}{Hela~Belhadj Amor},
  \bibinfo{person}{Carolynn Bernier}, {and} \bibinfo{person}{Zden\v{e}k
  P\v{r}ikryl}.} \bibinfo{year}{2022}\natexlab{}.
\newblock \showarticletitle{A {RISC-V ISA} Extension for Ultra-Low Power {IoT}
  wireless Signal Processing}.
\newblock \bibinfo{journal}{\emph{IEEE Trans. Comput.}} \bibinfo{volume}{71},
  \bibinfo{number}{4} (\bibinfo{year}{2022}), \bibinfo{pages}{766--778}.
\newblock
\href{https://doi.org/10.1109/TC.2021.3063027}{doi:\nolinkurl{10.1109/TC.2021.3063027}}


\bibitem[Appel(1998)]%
        {Appel1998SSA}
\bibfield{author}{\bibinfo{person}{Andrew~W. Appel}.}
  \bibinfo{year}{1998}\natexlab{}.
\newblock \showarticletitle{{SSA} is Functional Programming}.
\newblock \bibinfo{journal}{\emph{SIGPLAN Notices}} \bibinfo{volume}{33},
  \bibinfo{number}{4} (\bibinfo{date}{April} \bibinfo{year}{1998}),
  \bibinfo{pages}{17–20}.
\newblock
\showISSN{0362-1340}
\href{https://doi.org/10.1145/278283.278285}{doi:\nolinkurl{10.1145/278283.278285}}


\bibitem[Armstrong et~al\mbox{.}(2019)]%
        {armstrong2019isa}
\bibfield{author}{\bibinfo{person}{Alasdair Armstrong}, \bibinfo{person}{Thomas
  Bauereiss}, \bibinfo{person}{Brian Campbell}, \bibinfo{person}{Alastair
  Reid}, \bibinfo{person}{Kathryn~E. Gray}, \bibinfo{person}{Robert~M. Norton},
  \bibinfo{person}{Prashanth Mundkur}, \bibinfo{person}{Mark Wassell},
  \bibinfo{person}{Jon French}, \bibinfo{person}{Christopher Pulte},
  \bibinfo{person}{Shaked Flur}, \bibinfo{person}{Ian Stark},
  \bibinfo{person}{Neel Krishnaswami}, {and} \bibinfo{person}{Peter Sewell}.}
  \bibinfo{year}{2019}\natexlab{}.
\newblock \showarticletitle{{ISA Semantics for ARMv8-a, RISC-v, and
  CHERI-MIPS}}.
\newblock \bibinfo{journal}{\emph{Proc. ACM Program. Lang.}}
  \bibinfo{volume}{3}, \bibinfo{number}{POPL}, Article \bibinfo{articleno}{71}
  (\bibinfo{date}{Jan.} \bibinfo{year}{2019}), \bibinfo{numpages}{31}~pages.
\newblock
\href{https://doi.org/10.1145/3290384}{doi:\nolinkurl{10.1145/3290384}}


\bibitem[Armstrong and Team(2026)]%
        {Sail2026Source}
\bibfield{author}{\bibinfo{person}{Alasdair Armstrong} {and}
  \bibinfo{person}{Team}.} \bibinfo{year}{2026}\natexlab{}.
\newblock \bibinfo{booktitle}{\emph{The Sail ISA specification language}}.
\newblock
\urldef\tempurl%
\url{https://github.com/rems-project/sail}
\showURL{%
\tempurl}


\bibitem[Auler et~al\mbox{.}(2012)]%
        {Auler2012ArchC}
\bibfield{author}{\bibinfo{person}{Rafael Auler}, \bibinfo{person}{Paulo~Cesar
  Centoducatte}, {and} \bibinfo{person}{Edson Borin}.}
  \bibinfo{year}{2012}\natexlab{}.
\newblock \showarticletitle{{ACCGen}: An Automatic {ArchC} Compiler Generator}.
  In \bibinfo{booktitle}{\emph{2012 IEEE 24th International Symposium on
  Computer Architecture and High Performance Computing}}.
  \bibinfo{pages}{278--285}.
\newblock
\href{https://doi.org/10.1109/SBAC-PAD.2012.33}{doi:\nolinkurl{10.1109/SBAC-PAD.2012.33}}


\bibitem[Austin et~al\mbox{.}(2002)]%
        {austin2002simplescalar}
\bibfield{author}{\bibinfo{person}{Todd Austin}, \bibinfo{person}{Eric Larson},
  {and} \bibinfo{person}{Dan Ernst}.} \bibinfo{year}{2002}\natexlab{}.
\newblock \showarticletitle{{SimpleScalar}: An infrastructure for computer
  system modeling}.
\newblock \bibinfo{journal}{\emph{Computer}} \bibinfo{volume}{35},
  \bibinfo{number}{2} (\bibinfo{year}{2002}), \bibinfo{pages}{59--67}.
\newblock
\href{https://doi.org/10.1109/2.982917}{doi:\nolinkurl{10.1109/2.982917}}


\bibitem[Azevedo et~al\mbox{.}(2005)]%
        {azevedo2005archc}
\bibfield{author}{\bibinfo{person}{Rodolfo Azevedo}, \bibinfo{person}{Sandro
  Rigo}, \bibinfo{person}{Marcus Bartholomeu}, \bibinfo{person}{Guido Araujo},
  \bibinfo{person}{Cristiano Araujo}, {and} \bibinfo{person}{Edna Barros}.}
  \bibinfo{year}{2005}\natexlab{}.
\newblock \showarticletitle{The {ArchC} architecture description language and
  tools}.
\newblock \bibinfo{journal}{\emph{International Journal of Parallel
  Programming}} \bibinfo{volume}{33}, \bibinfo{number}{5}
  (\bibinfo{year}{2005}), \bibinfo{pages}{453--484}.
\newblock
\href{https://doi.org/10.1007/s10766-005-7301-0}{doi:\nolinkurl{10.1007/s10766-005-7301-0}}


\bibitem[Bachrach et~al\mbox{.}(2012)]%
        {bachrach2012chisel}
\bibfield{author}{\bibinfo{person}{Jonathan Bachrach}, \bibinfo{person}{Huy
  Vo}, \bibinfo{person}{Brian Richards}, \bibinfo{person}{Yunsup Lee},
  \bibinfo{person}{Andrew Waterman}, \bibinfo{person}{Rimas Avi{\v{z}}ienis},
  \bibinfo{person}{John Wawrzynek}, {and} \bibinfo{person}{Krste
  Asanovi{\'c}}.} \bibinfo{year}{2012}\natexlab{}.
\newblock \showarticletitle{Chisel: constructing hardware in a {Scala} embedded
  language}. In \bibinfo{booktitle}{\emph{Proceedings of the 49th Annual Design
  Automation Conference}}. ACM, \bibinfo{pages}{1216--1225}.
\newblock
\href{https://doi.org/10.1145/2228360.2228584}{doi:\nolinkurl{10.1145/2228360.2228584}}


\bibitem[Bartholomeu et~al\mbox{.}(2004)]%
        {bartholomeu2004optimizations}
\bibfield{author}{\bibinfo{person}{Marcus Bartholomeu},
  \bibinfo{person}{Rodolfo Azevedo}, \bibinfo{person}{Sandro Rigo}, {and}
  \bibinfo{person}{Guido Araujo}.} \bibinfo{year}{2004}\natexlab{}.
\newblock \showarticletitle{Optimizations for compiled simulation using
  instruction type information}. In \bibinfo{booktitle}{\emph{16th Symposium on
  Computer Architecture and High Performance Computing}}. IEEE,
  \bibinfo{pages}{74--81}.
\newblock
\href{https://doi.org/10.1109/SBAC-PAD.2004.28}{doi:\nolinkurl{10.1109/SBAC-PAD.2004.28}}


\bibitem[Bedichek(1990)]%
        {bedichek1990some}
\bibfield{author}{\bibinfo{person}{Robert Bedichek}.}
  \bibinfo{year}{1990}\natexlab{}.
\newblock \bibinfo{title}{Some Efficient Architecture Simulation Techniques,
  Winter 1990 {USENIX} Conference}.
\newblock


\bibitem[Bell(1973)]%
        {bell1973threaded}
\bibfield{author}{\bibinfo{person}{James~R Bell}.}
  \bibinfo{year}{1973}\natexlab{}.
\newblock \showarticletitle{Threaded code}.
\newblock \bibinfo{journal}{\emph{Commun. ACM}} \bibinfo{volume}{16},
  \bibinfo{number}{6} (\bibinfo{year}{1973}), \bibinfo{pages}{370--372}.
\newblock
\href{https://doi.org/10.1145/362248.362270}{doi:\nolinkurl{10.1145/362248.362270}}


\bibitem[Bellard(2005)]%
        {bellard2005qemu}
\bibfield{author}{\bibinfo{person}{Fabrice Bellard}.}
  \bibinfo{year}{2005}\natexlab{}.
\newblock \showarticletitle{{QEMU}, a {F}ast and {P}ortable {D}ynamic
  {T}ranslator}. In \bibinfo{booktitle}{\emph{USENIX Annual Technical
  Conference, FREENIX Track}} (Anaheim, CA). USENIX, \bibinfo{pages}{41--46}.
\newblock
\urldef\tempurl%
\url{https://www.usenix.org/legacy/event/usenix05/tech/freenix/full_papers/bellard/bellard.pdf}
\showURL{%
\tempurl}


\bibitem[Beringer(2022)]%
        {Beringer2022}
\bibfield{author}{\bibinfo{person}{Lennart Beringer}.}
  \bibinfo{year}{2022}\natexlab{}.
\newblock \bibinfo{booktitle}{\emph{Functional Representations of {SSA}}}.
\newblock \bibinfo{publisher}{Springer International Publishing},
  \bibinfo{address}{Cham}, \bibinfo{pages}{63--88}.
\newblock
\showISBNx{978-3-030-80515-9}
\href{https://doi.org/10.1007/978-3-030-80515-9_6}{doi:\nolinkurl{10.1007/978-3-030-80515-9_6}}


\bibitem[Bettini(2016)]%
        {Bettini2016XtextXtend}
\bibfield{author}{\bibinfo{person}{Lorenzo Bettini}.}
  \bibinfo{year}{2016}\natexlab{}.
\newblock \bibinfo{booktitle}{\emph{{I}mplementing {D}omain {S}pecific
  {L}anguages with {X}text and {X}tend - {S}econd {E}dition}
  (\bibinfo{edition}{2nd} ed.)}.
\newblock \bibinfo{publisher}{Packt Publishing}.
\newblock
\showISBNx{1786464969}
\urldef\tempurl%
\url{https://dl.acm.org/doi/10.5555/3074444}
\showURL{%
\tempurl}


\bibitem[Binkert et~al\mbox{.}(2011)]%
        {binkert2011gem5}
\bibfield{author}{\bibinfo{person}{Nathan Binkert}, \bibinfo{person}{Bradford
  Beckmann}, \bibinfo{person}{Gabriel Black}, \bibinfo{person}{Steven~K
  Reinhardt}, \bibinfo{person}{Ali Saidi}, \bibinfo{person}{Arkaprava Basu},
  \bibinfo{person}{Joel Hestness}, \bibinfo{person}{Derek~R Hower},
  \bibinfo{person}{Tushar Krishna}, \bibinfo{person}{Somayeh Sardashti},
  {et~al\mbox{.}}} \bibinfo{year}{2011}\natexlab{}.
\newblock \showarticletitle{The gem5 simulator}.
\newblock \bibinfo{journal}{\emph{ACM SIGARCH Computer Architecture News}}
  \bibinfo{volume}{39}, \bibinfo{number}{2} (\bibinfo{year}{2011}),
  \bibinfo{pages}{1--7}.
\newblock
\href{https://doi.org/10.1145/2024716.2024718}{doi:\nolinkurl{10.1145/2024716.2024718}}


\bibitem[Brandner(2009)]%
        {Brandner2009phdthesis}
\bibfield{author}{\bibinfo{person}{Florian Brandner}.}
  \bibinfo{year}{2009}\natexlab{}.
\newblock \emph{\bibinfo{title}{Compiler backend generation from structural
  processor models}}.
\newblock \bibinfo{thesistype}{Ph.\,D. Dissertation}.
  \bibinfo{school}{Technische Universit{\"a}t Wien}.
\newblock
\urldef\tempurl%
\url{https://repositum.tuwien.at/handle/20.500.12708/12553}
\showURL{%
\tempurl}


\bibitem[Brandner et~al\mbox{.}(2007)]%
        {brandner2007compiler}
\bibfield{author}{\bibinfo{person}{Florian Brandner}, \bibinfo{person}{Dietmar
  Ebner}, {and} \bibinfo{person}{Andreas Krall}.}
  \bibinfo{year}{2007}\natexlab{}.
\newblock \showarticletitle{Compiler generation from structural architecture
  descriptions}. In \bibinfo{booktitle}{\emph{Proceedings of the 2007
  international conference on Compilers, architecture, and synthesis for
  embedded systems}}. ACM, \bibinfo{pages}{13--22}.
\newblock
\href{https://doi.org/10.1145/1289881.1289886}{doi:\nolinkurl{10.1145/1289881.1289886}}


\bibitem[Brandner et~al\mbox{.}(2009)]%
        {brandner2009fast}
\bibfield{author}{\bibinfo{person}{Florian Brandner}, \bibinfo{person}{Andreas
  Fellnhofer}, \bibinfo{person}{Andreas Krall}, {and} \bibinfo{person}{David
  Riegler}.} \bibinfo{year}{2009}\natexlab{}.
\newblock \showarticletitle{{F}ast and {A}ccurate {S}imulation using the {LLVM}
  {C}ompiler {F}ramework}. In \bibinfo{booktitle}{\emph{Proceedings of the 1st
  Workshop on Rapid Simulation and Performance Evaluation: Methods and Tools,
  RAPIDO}}, Vol.~\bibinfo{volume}{9}. \bibinfo{pages}{1--6}.
\newblock
\urldef\tempurl%
\url{https://www.complang.tuwien.ac.at/andi/papers/rapido_09.pdf}
\showURL{%
\tempurl}


\bibitem[Brandner et~al\mbox{.}(2013)]%
        {brandner2013dsp}
\bibfield{author}{\bibinfo{person}{Florian Brandner}, \bibinfo{person}{Nigel
  Horspool}, {and} \bibinfo{person}{Andreas Krall}.}
  \bibinfo{year}{2013}\natexlab{}.
\newblock \showarticletitle{DSP instruction set simulation}.
\newblock \bibinfo{journal}{\emph{Handbook of Signal Processing Systems}}
  (\bibinfo{year}{2013}), \bibinfo{pages}{945--974}.
\newblock
\href{https://doi.org/10.1007/978-1-4614-6859-2_29}{doi:\nolinkurl{10.1007/978-1-4614-6859-2_29}}


\bibitem[Brandner et~al\mbox{.}(2012)]%
        {Brandner2012spe}
\bibfield{author}{\bibinfo{person}{Florian Brandner}, \bibinfo{person}{Viktor
  Pavlu}, {and} \bibinfo{person}{Andreas Krall}.}
  \bibinfo{year}{2012}\natexlab{}.
\newblock \showarticletitle{Automatic generation of compiler backends}.
\newblock \bibinfo{journal}{\emph{Software Practice \& Experience}}
  \bibinfo{volume}{43}, \bibinfo{number}{2} (\bibinfo{date}{Feb.}
  \bibinfo{year}{2012}), \bibinfo{pages}{207–240}.
\newblock
\showISSN{0038-0644}
\href{https://doi.org/10.1002/spe.2106}{doi:\nolinkurl{10.1002/spe.2106}}


\bibitem[Briggs et~al\mbox{.}(1997)]%
        {briggs1997value}
\bibfield{author}{\bibinfo{person}{Preston Briggs}, \bibinfo{person}{Keith~D
  Cooper}, {and} \bibinfo{person}{L~Taylor Simpson}.}
  \bibinfo{year}{1997}\natexlab{}.
\newblock \showarticletitle{{V}alue {N}umbering}.
\newblock \bibinfo{journal}{\emph{Software: Practice and Experience}}
  \bibinfo{volume}{27}, \bibinfo{number}{6} (\bibinfo{year}{1997}),
  \bibinfo{pages}{701--724}.
\newblock


\bibitem[Cain et~al\mbox{.}(2002)]%
        {cain2002precise}
\bibfield{author}{\bibinfo{person}{Harold~W Cain}, \bibinfo{person}{Kevin~M
  Lepak}, \bibinfo{person}{Brandon~A Schwartz}, {and} \bibinfo{person}{Mikko~H
  Lipasti}.} \bibinfo{year}{2002}\natexlab{}.
\newblock \showarticletitle{Precise and accurate processor simulation}. In
  \bibinfo{booktitle}{\emph{Workshop on Computer Architecture Evaluation using
  Commercial Workloads, HPCA}}, Vol.~\bibinfo{volume}{8}.
\newblock
\urldef\tempurl%
\url{https://pharm.ece.wisc.edu/papers/caecw_2002_final.pdf}
\showURL{%
\tempurl}


\bibitem[Chaitin et~al\mbox{.}(1981)]%
        {chaitin1981graphcoloring}
\bibfield{author}{\bibinfo{person}{Gregory~J. Chaitin},
  \bibinfo{person}{Marc~A. Auslander}, \bibinfo{person}{Ashok~K. Chandra},
  \bibinfo{person}{John Cocke}, \bibinfo{person}{Martin~E. Hopkins}, {and}
  \bibinfo{person}{Peter~W. Markstein}.} \bibinfo{year}{1981}\natexlab{}.
\newblock \showarticletitle{{R}egister {A}llocation {V}ia {C}oloring}.
\newblock \bibinfo{journal}{\emph{Computer Languages}} \bibinfo{volume}{6},
  \bibinfo{number}{1} (\bibinfo{year}{1981}), \bibinfo{pages}{47--57}.
\newblock
\showISSN{0096-0551}
\href{https://doi.org/10.1016/0096-0551(81)90048-5}{doi:\nolinkurl{10.1016/0096-0551(81)90048-5}}


\bibitem[Charv{\'a}t et~al\mbox{.}(2012)]%
        {Charvat2012CodAL}
\bibfield{author}{\bibinfo{person}{Luk{\'a} Charv{\'a}t}, \bibinfo{person}{Ale
  Smrcka}, {and} \bibinfo{person}{Tom{\'a} Vojnar}.}
  \bibinfo{year}{2012}\natexlab{}.
\newblock \showarticletitle{Automatic Formal Correspondence Checking of ISA and
  RTL Microprocessor Description}. In \bibinfo{booktitle}{\emph{2012 13th
  International Workshop on Microprocessor Test and Verification (MTV)}}.
  \bibinfo{pages}{6--12}.
\newblock
\href{https://doi.org/10.1109/MTV.2012.19}{doi:\nolinkurl{10.1109/MTV.2012.19}}


\bibitem[Chattopadhyay et~al\mbox{.}(2022)]%
        {Chattopadhyay2022ADL}
\bibfield{author}{\bibinfo{person}{Anupam Chattopadhyay},
  \bibinfo{person}{Zheng Wang}, {and} \bibinfo{person}{Grant" Martin}.}
  \bibinfo{year}{2022}\natexlab{}.
\newblock \bibinfo{booktitle}{\emph{Architecture Description Languages}}.
\newblock \bibinfo{publisher}{Springer Nature Singapore},
  \bibinfo{address}{Singapore}, \bibinfo{pages}{1--34}.
\newblock
\showISBNx{978-981-15-6401-7}
\href{https://doi.org/10.1007/978-981-15-6401-7_18-1}{doi:\nolinkurl{10.1007/978-981-15-6401-7_18-1}}


\bibitem[Cmelik and Keppel(1994)]%
        {cmelik1994dynbintrans}
\bibfield{author}{\bibinfo{person}{Bob Cmelik} {and} \bibinfo{person}{David
  Keppel}.} \bibinfo{year}{1994}\natexlab{}.
\newblock \showarticletitle{{S}hade: {A} {F}ast {I}nstruction-{S}et {S}imulator
  for {E}xecution {P}rofiling}.
\newblock \bibinfo{journal}{\emph{SIGMETRICS Perform. Eval. Rev.}}
  \bibinfo{volume}{22}, \bibinfo{number}{1} (\bibinfo{date}{May}
  \bibinfo{year}{1994}), \bibinfo{pages}{128–137}.
\newblock
\showISSN{0163-5999}
\href{https://doi.org/10.1145/183019.183032}{doi:\nolinkurl{10.1145/183019.183032}}


\bibitem[Cocke(1970)]%
        {cocke1970gcse}
\bibfield{author}{\bibinfo{person}{John Cocke}.}
  \bibinfo{year}{1970}\natexlab{}.
\newblock \showarticletitle{{G}lobal {C}ommon {S}ubexpression {E}limination}.
  In \bibinfo{booktitle}{\emph{Proceedings of a Symposium on Compiler
  Optimization}} (Urbana-Champaign, Illinois). \bibinfo{publisher}{Association
  for Computing Machinery}, \bibinfo{address}{New York, NY, USA},
  \bibinfo{pages}{20–24}.
\newblock
\showISBNx{9781450373869}
\href{https://doi.org/10.1145/800028.808480}{doi:\nolinkurl{10.1145/800028.808480}}


\bibitem[Cocke and Kennedy(1977)]%
        {cocke1977strengthreduction}
\bibfield{author}{\bibinfo{person}{John Cocke} {and} \bibinfo{person}{Ken
  Kennedy}.} \bibinfo{year}{1977}\natexlab{}.
\newblock \showarticletitle{{A}n {A}lgorithm for {R}eduction of {O}perator
  {S}trength}.
\newblock \bibinfo{journal}{\emph{Commun. ACM}} \bibinfo{volume}{20},
  \bibinfo{number}{11} (\bibinfo{date}{Nov.} \bibinfo{year}{1977}),
  \bibinfo{pages}{850–856}.
\newblock
\showISSN{0001-0782}
\href{https://doi.org/10.1145/359863.359888}{doi:\nolinkurl{10.1145/359863.359888}}


\bibitem[Contributors(2026)]%
        {OpenXLA2026}
\bibfield{author}{\bibinfo{person}{OpenXLA Contributors}.}
  \bibinfo{year}{2026}\natexlab{}.
\newblock \bibinfo{booktitle}{\emph{OpenXLA Project}}.
\newblock
\urldef\tempurl%
\url{https://openxla.org/}
\showURL{%
\tempurl}


\bibitem[Cooper and Torczon(2011)]%
        {cooper2011engineering}
\bibfield{author}{\bibinfo{person}{Keith~D Cooper} {and} \bibinfo{person}{Linda
  Torczon}.} \bibinfo{year}{2011}\natexlab{}.
\newblock \bibinfo{booktitle}{\emph{Engineering a compiler}}.
\newblock \bibinfo{publisher}{Elsevier}.
\newblock


\bibitem[Cytron et~al\mbox{.}(1991)]%
        {cytron1991ssa}
\bibfield{author}{\bibinfo{person}{Ron Cytron}, \bibinfo{person}{Jeanne
  Ferrante}, \bibinfo{person}{Barry~K. Rosen}, \bibinfo{person}{Mark~N.
  Wegman}, {and} \bibinfo{person}{F.~Kenneth Zadeck}.}
  \bibinfo{year}{1991}\natexlab{}.
\newblock \showarticletitle{{E}fficiently {C}omputing {S}tatic {S}ingle
  {A}ssignment {F}orm and the {C}ontrol {D}ependence {G}raph}.
\newblock \bibinfo{journal}{\emph{ACM Trans. Program. Lang. Syst.}}
  \bibinfo{volume}{13}, \bibinfo{number}{4} (\bibinfo{date}{Oct.}
  \bibinfo{year}{1991}), \bibinfo{pages}{451–490}.
\newblock
\showISSN{0164-0925}
\href{https://doi.org/10.1145/115372.115320}{doi:\nolinkurl{10.1145/115372.115320}}


\bibitem[Dewar(1975)]%
        {dewar1975indirect}
\bibfield{author}{\bibinfo{person}{Robert~BK Dewar}.}
  \bibinfo{year}{1975}\natexlab{}.
\newblock \showarticletitle{Indirect threaded code}.
\newblock \bibinfo{journal}{\emph{Commun. ACM}} \bibinfo{volume}{18},
  \bibinfo{number}{6} (\bibinfo{year}{1975}), \bibinfo{pages}{330--331}.
\newblock
\urldef\tempurl%
\url{https://dl.acm.org/doi/pdf/10.1145/360825.360849}
\showURL{%
\tempurl}


\bibitem[Dreesen(2011)]%
        {Dreesen2011Generating}
\bibfield{author}{\bibinfo{person}{Ralf Dreesen}.}
  \bibinfo{year}{2011}\natexlab{}.
\newblock \emph{\bibinfo{title}{Generating Processors from Specifications of
  Instruction Sets}}.
\newblock \bibinfo{thesistype}{Ph.\,D. Dissertation}.
  \bibinfo{school}{University of Paderborn}.
\newblock
\urldef\tempurl%
\url{https://digital.ub.uni-paderborn.de/hsx/content/titleinfo/315766}
\showURL{%
\tempurl}


\bibitem[Dreesen(2012)]%
        {Dreesen2012ViDL}
\bibfield{author}{\bibinfo{person}{Ralf Dreesen}.}
  \bibinfo{year}{2012}\natexlab{}.
\newblock \showarticletitle{{ViDL: A} Versatile {ISA} Description Language}. In
  \bibinfo{booktitle}{\emph{2012 IEEE 19th International Conference and
  Workshops on Engineering of Computer-Based Systems}}.
  \bibinfo{pages}{222--231}.
\newblock
\href{https://doi.org/10.1109/ECBS.2012.49}{doi:\nolinkurl{10.1109/ECBS.2012.49}}


\bibitem[Duboscq et~al\mbox{.}(2013)]%
        {duboscq2013graalir}
\bibfield{author}{\bibinfo{person}{Gilles Duboscq}, \bibinfo{person}{Lukas
  Stadler}, \bibinfo{person}{Thomas Würthinger}, \bibinfo{person}{Doug Simon},
  \bibinfo{person}{Christian Wimmer}, {and} \bibinfo{person}{Hanspeter
  Mössenböck}.} \bibinfo{year}{2013}\natexlab{}.
\newblock \showarticletitle{Graal {IR}: {A}n {E}xtensible {D}eclarative
  {I}ntermediate {R}epresentation}. In \bibinfo{booktitle}{\emph{Proceedings of
  the Asia-Pacific Programming Languages and Compilers Workshop}}.
  \bibinfo{pages}{1--9}.
\newblock
\urldef\tempurl%
\url{https://lafo.ssw.uni-linz.ac.at/pub/papers/2013_APPLC_GraalIR.pdf}
\showURL{%
\tempurl}


\bibitem[Duboscq et~al\mbox{.}(2013)]%
        {duboscq2013anintermediate}
\bibfield{author}{\bibinfo{person}{Gilles Duboscq}, \bibinfo{person}{Thomas
  Würthinger}, \bibinfo{person}{Lukas Stadler}, \bibinfo{person}{Christian
  Wimmer}, \bibinfo{person}{Doug Simon}, {and} \bibinfo{person}{Hanspeter
  Mössenböck}.} \bibinfo{year}{2013}\natexlab{}.
\newblock \showarticletitle{An intermediate representation for speculative
  optimizations in a dynamic compiler}. In
  \bibinfo{booktitle}{\emph{Proceedings of the 7th {ACM} workshop on Virtual
  machines and intermediate languages}} (New York, {NY}, {USA}, 2013-10-28)
  \emph{(\bibinfo{series}{{VMIL} '13})}. \bibinfo{publisher}{Association for
  Computing Machinery}, \bibinfo{pages}{1--10}.
\newblock
\showISBNx{978-1-4503-2601-8}
\href{https://doi.org/10.1145/2542142.2542143}{doi:\nolinkurl{10.1145/2542142.2542143}}


\bibitem[Ertl(2001)]%
        {ertl2001threaded}
\bibfield{author}{\bibinfo{person}{M~Anton Ertl}.}
  \bibinfo{year}{2001}\natexlab{}.
\newblock \showarticletitle{Threaded code variations and optimizations}. In
  \bibinfo{booktitle}{\emph{EuroForth 2001 Conference Proceedings}}. EuroForth,
  \bibinfo{pages}{49--55}.
\newblock
\urldef\tempurl%
\url{http://www.euroforth.org/ef01/ertl01.pdf}
\showURL{%
\tempurl}


\bibitem[Ertl and Gregg(2001)]%
        {ertl2001behavior}
\bibfield{author}{\bibinfo{person}{M~Anton Ertl} {and} \bibinfo{person}{David
  Gregg}.} \bibinfo{year}{2001}\natexlab{}.
\newblock \showarticletitle{The behavior of efficient virtual machine
  interpreters on modern architectures}. In \bibinfo{booktitle}{\emph{Euro-Par
  2001 Parallel Processing: 7th International Euro-Par Conference Manchester,
  UK, August 28--31, 2001 Proceedings 7}}. \bibinfo{publisher}{Springer},
  \bibinfo{pages}{403--413}.
\newblock
\urldef\tempurl%
\url{https://link.springer.com/chapter/10.1007/3-540-44681-8_59}
\showURL{%
\tempurl}


\bibitem[Eysholdt and Behrens(2010)]%
        {eysholdt2010xtext}
\bibfield{author}{\bibinfo{person}{Moritz Eysholdt} {and}
  \bibinfo{person}{Heiko Behrens}.} \bibinfo{year}{2010}\natexlab{}.
\newblock \showarticletitle{{X}text: {I}mplement {Y}our {L}anguage {F}aster
  than the {Q}uick and {D}irty {W}ay}. In \bibinfo{booktitle}{\emph{Proceedings
  of the ACM international conference companion on Object oriented programming
  systems languages and applications companion}}. ACM,
  \bibinfo{pages}{307--309}.
\newblock
\href{https://doi.org/10.1145/1869542.1869625}{doi:\nolinkurl{10.1145/1869542.1869625}}


\bibitem[Farfeleder et~al\mbox{.}(2005)]%
        {Farfeleder2005Ultra}
\bibfield{author}{\bibinfo{person}{Stefan Farfeleder}, \bibinfo{person}{Andreas
  Krall}, {and} \bibinfo{person}{Nigel Horspool}.}
  \bibinfo{year}{2005}\natexlab{}.
\newblock \showarticletitle{Ultra Fast Cycle-Accurate Compiled Emulation of
  Inorder Pipelined Architectures}. In \bibinfo{booktitle}{\emph{Embedded
  Computer Systems: Architectures, Modeling, and Simulation}},
  \bibfield{editor}{\bibinfo{person}{Timo~D. H{\"a}m{\"a}l{\"a}inen},
  \bibinfo{person}{Andy~D. Pimentel}, \bibinfo{person}{Jarmo Takala}, {and}
  \bibinfo{person}{Stamatis Vassiliadis}} (Eds.). \bibinfo{publisher}{Springer
  Berlin Heidelberg}, \bibinfo{address}{Berlin, Heidelberg},
  \bibinfo{pages}{222--231}.
\newblock
\showISBNx{978-3-540-31664-0}


\bibitem[Farfeleder et~al\mbox{.}(2007)]%
        {farfeleder2007ultra}
\bibfield{author}{\bibinfo{person}{Stefan Farfeleder}, \bibinfo{person}{Andreas
  Krall}, {and} \bibinfo{person}{Nigel Horspool}.}
  \bibinfo{year}{2007}\natexlab{}.
\newblock \showarticletitle{Ultra fast cycle-accurate compiled emulation of
  inorder pipelined architectures}.
\newblock \bibinfo{journal}{\emph{Journal of Systems Architecture}}
  \bibinfo{volume}{53}, \bibinfo{number}{8} (\bibinfo{year}{2007}),
  \bibinfo{pages}{501--510}.
\newblock
\href{https://doi.org/10.1016/j.sysarc.2006.11.003}{doi:\nolinkurl{10.1016/j.sysarc.2006.11.003}}


\bibitem[Farfeleder et~al\mbox{.}(2006)]%
        {Farfeleder2006Effective}
\bibfield{author}{\bibinfo{person}{Stefan Farfeleder}, \bibinfo{person}{Andreas
  Krall}, \bibinfo{person}{Edwin Steiner}, {and} \bibinfo{person}{Florian
  Brandner}.} \bibinfo{year}{2006}\natexlab{}.
\newblock \showarticletitle{Effective compiler generation by architecture
  description}. In \bibinfo{booktitle}{\emph{Proceedings of the 2006 ACM
  SIGPLAN/SIGBED Conference on Language, Compilers, and Tool Support for
  Embedded Systems}} (Ottawa, Ontario, Canada) \emph{(\bibinfo{series}{LCTES
  '06})}. \bibinfo{publisher}{ACM}, \bibinfo{address}{New York, NY, USA},
  \bibinfo{pages}{145–152}.
\newblock
\href{https://doi.org/10.1145/1134650.1134671}{doi:\nolinkurl{10.1145/1134650.1134671}}


\bibitem[{Fauth} et~al\mbox{.}(1995)]%
        {fauth1995nml}
\bibfield{author}{\bibinfo{person}{Andreas {Fauth}}, \bibinfo{person}{Johan
  {Van Praet}}, {and} \bibinfo{person}{Markus {Freericks}}.}
  \bibinfo{year}{1995}\natexlab{}.
\newblock \showarticletitle{Describing instruction set processors using {nML}}.
  In \bibinfo{booktitle}{\emph{Proceedings the European Design and Test
  Conference. ED\&TC 1995}}. IEEE, \bibinfo{pages}{503--507}.
\newblock
\href{https://doi.org/10.1109/EDTC.1995.470354}{doi:\nolinkurl{10.1109/EDTC.1995.470354}}


\bibitem[Fischer et~al\mbox{.}(2001)]%
        {FischerTeichWeper2001}
\bibfield{author}{\bibinfo{person}{Dirk Fischer}, \bibinfo{person}{J\"{u}rgen
  Teich}, \bibinfo{person}{Ralph Weper}, \bibinfo{person}{Uwe Kastens}, {and}
  \bibinfo{person}{Michael Thies}.} \bibinfo{year}{2001}\natexlab{}.
\newblock \showarticletitle{Design space characterization for
  architecture/compiler co-exploration}. In
  \bibinfo{booktitle}{\emph{Proceedings of the 2001 International Conference on
  Compilers, Architecture, and Synthesis for Embedded Systems}} (Atlanta,
  Georgia, USA) \emph{(\bibinfo{series}{CASES '01})}.
  \bibinfo{publisher}{Association for Computing Machinery},
  \bibinfo{address}{New York, NY, USA}, \bibinfo{pages}{108–115}.
\newblock
\showISBNx{1581133995}
\href{https://doi.org/10.1145/502217.502234}{doi:\nolinkurl{10.1145/502217.502234}}


\bibitem[Franke(2008)]%
        {franke2008fast}
\bibfield{author}{\bibinfo{person}{Bj{\"o}rn Franke}.}
  \bibinfo{year}{2008}\natexlab{}.
\newblock \showarticletitle{Fast cycle-approximate instruction set simulation}.
  In \bibinfo{booktitle}{\emph{Proceedings of the 11th international workshop
  on Software \& compilers for embedded systems}}. ACM,
  \bibinfo{pages}{69--78}.
\newblock
\href{https://doi.org/10.1145/1361096.1361109}{doi:\nolinkurl{10.1145/1361096.1361109}}


\bibitem[Fraser(1991)]%
        {fraser1991lcc}
\bibfield{author}{\bibinfo{person}{Christopher~W. Fraser}.}
  \bibinfo{year}{1991}\natexlab{}.
\newblock \showarticletitle{{A Retargetable Compiler for ANSI C}}.
\newblock \bibinfo{journal}{\emph{SIGPLAN Notices}} \bibinfo{volume}{26},
  \bibinfo{number}{10} (\bibinfo{date}{Oct.} \bibinfo{year}{1991}),
  \bibinfo{pages}{29–43}.
\newblock
\showISSN{0362-1340}
\href{https://doi.org/10.1145/122616.122621}{doi:\nolinkurl{10.1145/122616.122621}}


\bibitem[{Freericks}(1991)]%
        {freericks1991nml}
\bibfield{author}{\bibinfo{person}{Markus {Freericks}}.}
  \bibinfo{year}{1991}\natexlab{}.
\newblock \bibinfo{booktitle}{\emph{The {nML} machine description formalism}}.
\newblock \bibinfo{type}{{T}echnical {R}eport}. \bibinfo{institution}{TU
  Berlin, Berlin, Germany}.
\newblock
\urldef\tempurl%
\url{https://archive.air.in.tum.de/Main/Publications/Freericks1991a.pdf}
\showURL{%
\tempurl}


\bibitem[Freitag et~al\mbox{.}(2025)]%
        {freitag2025OpenVADL}
\bibfield{author}{\bibinfo{person}{Florian Freitag}, \bibinfo{person}{Linus
  Halder}, \bibinfo{person}{Benedikt Huber}, \bibinfo{person}{Benjamin Kasper},
  \bibinfo{person}{Michael Nestler}, \bibinfo{person}{Kevin Per},
  \bibinfo{person}{Matthias Raschhofer}, \bibinfo{person}{Alexander Ripar},
  \bibinfo{person}{Johannes Zottele}, {and} \bibinfo{person}{Andreas Krall}.}
  \bibinfo{year}{2025}\natexlab{}.
\newblock \showarticletitle{{OpenVADL}: An Open Source Implementation of the
  {Vienna Architecture Description Language}}. In
  \bibinfo{booktitle}{\emph{ARCS 2025 -- 38th GI/ITG International Conference
  on Architecture of Computing Systems}} (Kiel, Germany),
  \bibfield{editor}{\bibinfo{person}{Sven Tomforde}, \bibinfo{person}{Christian
  Krupitzer}, \bibinfo{person}{St{\'e}phane Vialle}, \bibinfo{person}{Estela
  Suarez}, {and} \bibinfo{person}{Thilo Pionteck}} (Eds.).
  \bibinfo{publisher}{Springer Nature Switzerland}, \bibinfo{pages}{156--171}.
\newblock
\href{https://doi.org/10.1007/978-3-032-03281-2_11}{doi:\nolinkurl{10.1007/978-3-032-03281-2_11}}


\bibitem[Fuchs(1992)]%
        {Fuchs1992}
\bibfield{author}{\bibinfo{person}{Norbert~E. Fuchs}.}
  \bibinfo{year}{1992}\natexlab{}.
\newblock \showarticletitle{Specifications are (preferably) executable}.
\newblock \bibinfo{journal}{\emph{Software Engineering Journal}}
  \bibinfo{volume}{7} (\bibinfo{year}{1992}), \bibinfo{pages}{323--334}.
\newblock
Issue 5.
\href{https://doi.org/10.1049/sej.1992.0033}{doi:\nolinkurl{10.1049/sej.1992.0033}}


\bibitem[Giorgi and Mariotti(2019)]%
        {giorgi2019webrisc}
\bibfield{author}{\bibinfo{person}{Roberto Giorgi} {and}
  \bibinfo{person}{Gianfranco Mariotti}.} \bibinfo{year}{2019}\natexlab{}.
\newblock \showarticletitle{{WebRISC-V}: a Web-Based Education-Oriented
  {RISC-V} Pipeline Simulation Environment}. In
  \bibinfo{booktitle}{\emph{Proceedings of the Workshop on Computer
  Architecture Education}}. \bibinfo{publisher}{ACM}, \bibinfo{pages}{1--6}.
\newblock
\href{https://doi.org/10.1145/3338698.3338894}{doi:\nolinkurl{10.1145/3338698.3338894}}


\bibitem[Goel et~al\mbox{.}(2017)]%
        {goel2017engineering}
\bibfield{author}{\bibinfo{person}{Shilpi Goel}, \bibinfo{person}{Warren~A
  Hunt}, {and} \bibinfo{person}{Matt Kaufmann}.}
  \bibinfo{year}{2017}\natexlab{}.
\newblock \bibinfo{booktitle}{\emph{Engineering a formal, executable x86 ISA
  simulator for software verification}}.
\newblock \bibinfo{publisher}{Springer International Publishing},
  \bibinfo{address}{Cham}, \bibinfo{pages}{173--209}.
\newblock
\showISBNx{978-3-319-48628-4}
\href{https://doi.org/10.1007/978-3-319-48628-4_8}{doi:\nolinkurl{10.1007/978-3-319-48628-4_8}}


\bibitem[Gonzalez(2000)]%
        {Gonzalez2000Xtensa}
\bibfield{author}{\bibinfo{person}{Ricardo~E. Gonzalez}.}
  \bibinfo{year}{2000}\natexlab{}.
\newblock \showarticletitle{Xtensa: A Configurable and Extensible Processor}.
\newblock \bibinfo{journal}{\emph{IEEE Micro}} \bibinfo{volume}{20},
  \bibinfo{number}{2} (\bibinfo{year}{2000}), \bibinfo{pages}{60--70}.
\newblock
\href{https://doi.org/10.1109/40.848473}{doi:\nolinkurl{10.1109/40.848473}}


\bibitem[Graf(2021)]%
        {graf2021compiler}
\bibfield{author}{\bibinfo{person}{Alexander Graf}.}
  \bibinfo{year}{2021}\natexlab{}.
\newblock \emph{\bibinfo{title}{Compiler Backend Generation using the VADL
  Processor Description Language}}.
\newblock \bibinfo{thesistype}{Master's\ thesis}. \bibinfo{school}{Technische
  Universit{\"a}t Wien}.
\newblock
\href{https://doi.org/10.34726/hss.2021.79221}{doi:\nolinkurl{10.34726/hss.2021.79221}}


\bibitem[Gray et~al\mbox{.}(2015)]%
        {gray2015Sail}
\bibfield{author}{\bibinfo{person}{Kathryn~E. Gray}, \bibinfo{person}{Gabriel
  Kerneis}, \bibinfo{person}{Dominic Mulligan}, \bibinfo{person}{Christopher
  Pulte}, \bibinfo{person}{Susmit Sarkar}, {and} \bibinfo{person}{Peter
  Sewell}.} \bibinfo{year}{2015}\natexlab{}.
\newblock \showarticletitle{An integrated concurrency and {core-ISA}
  architectural envelope definition, and test oracle, for {IBM POWER}
  multiprocessors}. In \bibinfo{booktitle}{\emph{Proceedings of the 48th
  International Symposium on Microarchitecture}} (Waikiki, Hawaii)
  \emph{(\bibinfo{series}{MICRO-48})}. \bibinfo{publisher}{Association for
  Computing Machinery}, \bibinfo{address}{New York, NY, USA},
  \bibinfo{pages}{635–646}.
\newblock
\showISBNx{9781450340342}
\href{https://doi.org/10.1145/2830772.2830775}{doi:\nolinkurl{10.1145/2830772.2830775}}


\bibitem[Grun et~al\mbox{.}(1998)]%
        {grun1998expression}
\bibfield{author}{\bibinfo{person}{Peter Grun}, \bibinfo{person}{Ashok
  Halambi}, \bibinfo{person}{Asheesh Khare}, \bibinfo{person}{Vijay Ganesh},
  \bibinfo{person}{Nikil Dutt}, {and} \bibinfo{person}{Alexandru Nicolau}.}
  \bibinfo{year}{1998}\natexlab{}.
\newblock \bibinfo{booktitle}{\emph{EXPRESSION: An {ADL} for system level
  design exploration}}.
\newblock \bibinfo{type}{{T}echnical {R}eport}.
  \bibinfo{institution}{University of California, Irvine}.
\newblock
\urldef\tempurl%
\url{http://www.ics.uci.edu/~express/pubs/expr_tr.ps}
\showURL{%
\tempurl}


\bibitem[Hachtel and Somenzi(2005)]%
        {hachtel2005logic}
\bibfield{author}{\bibinfo{person}{Gary~D Hachtel} {and} \bibinfo{person}{Fabio
  Somenzi}.} \bibinfo{year}{2005}\natexlab{}.
\newblock \bibinfo{booktitle}{\emph{Logic synthesis and verification
  algorithms}}.
\newblock \bibinfo{publisher}{Springer Science \& Business Media}.
\newblock


\bibitem[Hadjiyiannis and Devadas(2003)]%
        {hadjiyiannis2003isdl}
\bibfield{author}{\bibinfo{person}{George Hadjiyiannis} {and}
  \bibinfo{person}{Srinivas Devadas}.} \bibinfo{year}{2003}\natexlab{}.
\newblock \showarticletitle{Techniques for accurate performance evaluation in
  architecture exploration}.
\newblock \bibinfo{journal}{\emph{IEEE Transactions on Very Large Scale
  Integration (VLSI) Systems}} \bibinfo{volume}{11}, \bibinfo{number}{4}
  (\bibinfo{year}{2003}), \bibinfo{pages}{601--615}.
\newblock
\href{https://doi.org/10.1109/TVLSI.2003.812290}{doi:\nolinkurl{10.1109/TVLSI.2003.812290}}


\bibitem[Hadjiyiannis et~al\mbox{.}(1997)]%
        {hadjiyiannis1997isdl}
\bibfield{author}{\bibinfo{person}{George Hadjiyiannis},
  \bibinfo{person}{Silvina Hanono}, {and} \bibinfo{person}{Srinivas Devadas}.}
  \bibinfo{year}{1997}\natexlab{}.
\newblock \showarticletitle{{ISDL: An Instruction Set Description Language for
  Retargetability}}. In \bibinfo{booktitle}{\emph{Proceedings of the 34th
  Annual Design Automation Conference}} (Anaheim, California, USA)
  \emph{(\bibinfo{series}{DAC '97})}. \bibinfo{publisher}{ACM},
  \bibinfo{address}{New York, NY, USA}, \bibinfo{pages}{299–302}.
\newblock
\showISBNx{0897919203}
\href{https://doi.org/10.1145/266021.266108}{doi:\nolinkurl{10.1145/266021.266108}}


\bibitem[Hadjiyiannis et~al\mbox{.}(2000)]%
        {hadjiyiannis2000isdl}
\bibfield{author}{\bibinfo{person}{George Hadjiyiannis},
  \bibinfo{person}{Silvina Hanono}, {and} \bibinfo{person}{Srinivas Devadas}.}
  \bibinfo{year}{2000}\natexlab{}.
\newblock \showarticletitle{{ISDL: An Instruction Set Description Language for
  Retargetability and Architecture Exploration}}.
\newblock \bibinfo{journal}{\emph{Design Automation for Embedded Systems}}
  \bibinfo{volume}{6} (\bibinfo{year}{2000}), \bibinfo{pages}{39--69}.
\newblock
\href{https://doi.org/10.1023/A:1008937425064}{doi:\nolinkurl{10.1023/A:1008937425064}}


\bibitem[Halambi et~al\mbox{.}(2008)]%
        {halambi2008expression}
\bibfield{author}{\bibinfo{person}{Ashok Halambi}, \bibinfo{person}{Peter
  Grun}, \bibinfo{person}{Vijay Ganesh}, \bibinfo{person}{Asheesh Khare},
  \bibinfo{person}{Nikil Dutt}, {and} \bibinfo{person}{Alex Nicolau}.}
  \bibinfo{year}{2008}\natexlab{}.
\newblock \showarticletitle{{EXPRESSION}: A language for architecture
  exploration through compiler/simulator retargetability}. In
  \bibinfo{booktitle}{\emph{Design, Automation, and Test in Europe}}. EDAA,
  \bibinfo{pages}{31--45}.
\newblock
\urldef\tempurl%
\url{https://dl.acm.org/doi/pdf/10.1145/307418.307549}
\showURL{%
\tempurl}


\bibitem[Halder et~al\mbox{.}(2026)]%
        {halder2026ASIP}
\bibfield{author}{\bibinfo{person}{Linus Halder}, \bibinfo{person}{Matthias
  Raschhofer}, {and} \bibinfo{person}{Andreas Krall}.}
  \bibinfo{year}{2026}\natexlab{}.
\newblock \showarticletitle{{ Generation of Scalar ASIPs with OpenVADL }}. In
  \bibinfo{booktitle}{\emph{2026 IEEE 37th International Conference on
  Application-specific Systems, Architectures and Processors (ASAP)}}.
  \bibinfo{publisher}{IEEE Computer Society}, \bibinfo{address}{Los Alamitos,
  CA, USA}, \bibinfo{pages}{167--168}.
\newblock
\href{https://doi.org/10.1109/ASAP71856.2026.00039}{doi:\nolinkurl{10.1109/ASAP71856.2026.00039}}


\bibitem[Hayes and Jones(1989)]%
        {HayesJones1989}
\bibfield{author}{\bibinfo{person}{I.J. Hayes} {and} \bibinfo{person}{C.B.
  Jones}.} \bibinfo{year}{1989}\natexlab{}.
\newblock \showarticletitle{Specifications are not (necessarily) executable}.
\newblock \bibinfo{journal}{\emph{Software Engineering Journal}}
  \bibinfo{volume}{4} (\bibinfo{year}{1989}), \bibinfo{pages}{330--339}.
\newblock
Issue 6.
\href{https://doi.org/10.1049/sej.1989.0045}{doi:\nolinkurl{10.1049/sej.1989.0045}}


\bibitem[Hennessy and Patterson(2011)]%
        {hennessy2011computer}
\bibfield{author}{\bibinfo{person}{John~L. Hennessy} {and}
  \bibinfo{person}{David~A. Patterson}.} \bibinfo{year}{2011}\natexlab{}.
\newblock \bibinfo{booktitle}{\emph{Computer Architecture: A Quantitative
  Approach}}.
\newblock \bibinfo{publisher}{Elsevier}.
\newblock


\bibitem[Himmelbauer(2024)]%
        {Himmelbauer2024}
\bibfield{author}{\bibinfo{person}{Simon Himmelbauer}.}
  \bibinfo{year}{2024}\natexlab{}.
\newblock \emph{\bibinfo{title}{Atomic Instruction and Cache-Support for
  VADL}}.
\newblock Masters's Thesis. \bibinfo{school}{Technische Universit{\"a}t Wien}.
\newblock
\href{https://doi.org/10.34726/hss.2024.113945}{doi:\nolinkurl{10.34726/hss.2024.113945}}


\bibitem[Hochrainer and Krall(2023)]%
        {HochrainerKrall23}
\bibfield{author}{\bibinfo{person}{Christoph Hochrainer} {and}
  \bibinfo{person}{Andreas Krall}.} \bibinfo{year}{2023}\natexlab{}.
\newblock \showarticletitle{{A} {P}red-{LL}(*) {P}arsable {T}yped
  {H}igher-{O}rder {M}acro {S}ystem for {A}rchitecture {D}escription
  {L}anguages}. In \bibinfo{booktitle}{\emph{Proceedings of the 22nd ACM
  SIGPLAN International Conference on Generative Programming: Concepts and
  Experiences}} (Cascais, Portugal) \emph{(\bibinfo{series}{GPCE 2023})}.
  \bibinfo{publisher}{Association for Computing Machinery},
  \bibinfo{address}{New York, NY, USA}, \bibinfo{pages}{29–41}.
\newblock
\showISBNx{9798400704062}
\href{https://doi.org/10.1145/3624007.3624052}{doi:\nolinkurl{10.1145/3624007.3624052}}


\bibitem[Hohenauer and Leupers(2009)]%
        {hohenauer2009ccompilers}
\bibfield{author}{\bibinfo{person}{Manuel Hohenauer} {and}
  \bibinfo{person}{Rainer Leupers}.} \bibinfo{year}{2009}\natexlab{}.
\newblock \bibinfo{booktitle}{\emph{C Compilers for ASIPs}}.
\newblock \bibinfo{publisher}{Springer}, \bibinfo{address}{New York, NY}.
\newblock
\href{https://doi.org/10.1007/978-1-4419-1176-6}{doi:\nolinkurl{10.1007/978-1-4419-1176-6}}


\bibitem[Holzner(2004)]%
        {Holzner2004Eclipse}
\bibfield{author}{\bibinfo{person}{Steve Holzner}.}
  \bibinfo{year}{2004}\natexlab{}.
\newblock \bibinfo{booktitle}{\emph{Eclipse}}.
\newblock \bibinfo{publisher}{O'Reilly Media, Inc.}
\newblock


\bibitem[Hus{\'a}r et~al\mbox{.}(2011)]%
        {husar2011Automatic}
\bibfield{author}{\bibinfo{person}{Adam Hus{\'a}r}, \bibinfo{person}{Miloslav
  Trma{\v{c}}}, \bibinfo{person}{Jan Hran{\'a}{\v{c}}},
  \bibinfo{person}{Tom{\'a}{\v{s}} Hru{\v{s}}ka}, {and} \bibinfo{person}{Karel
  Masa{\v{r}}{\'\i}k}.} \bibinfo{year}{2011}\natexlab{}.
\newblock \showarticletitle{{Automatic C Compiler Generation from Architecture
  Description Language ISAC}}. In \bibinfo{booktitle}{\emph{Sixth Doctoral
  Workshop on Mathematical and Engineering Methods in Computer Science
  (MEMICS'10) -- Selected Papers}} \emph{(\bibinfo{series}{Open Access Series
  in Informatics (OASIcs)}, Vol.~\bibinfo{volume}{16})},
  \bibfield{editor}{\bibinfo{person}{Ludek Matyska}, \bibinfo{person}{Michal
  Kozubek}, \bibinfo{person}{Tomas Vojnar}, \bibinfo{person}{Pavel Zemcik},
  {and} \bibinfo{person}{David Antos}} (Eds.). \bibinfo{publisher}{Schloss
  Dagstuhl -- Leibniz-Zentrum f{\"u}r Informatik}, \bibinfo{address}{Dagstuhl,
  Germany}, \bibinfo{pages}{47--53}.
\newblock
\href{https://doi.org/10.4230/OASIcs.MEMICS.2010.47}{doi:\nolinkurl{10.4230/OASIcs.MEMICS.2010.47}}


\bibitem[Hwang et~al\mbox{.}(2008)]%
        {hwang2008cycle}
\bibfield{author}{\bibinfo{person}{Yonghyun Hwang}, \bibinfo{person}{Samar
  Abdi}, {and} \bibinfo{person}{Daniel Gajski}.}
  \bibinfo{year}{2008}\natexlab{}.
\newblock \showarticletitle{Cycle-approximate retargetable performance
  estimation at the transaction level}. In
  \bibinfo{booktitle}{\emph{Proceedings of the conference on Design, Automation
  and Test in Europe}}. EDAA, \bibinfo{pages}{3--8}.
\newblock
\href{https://doi.org/10.1145/1403375.1403380}{doi:\nolinkurl{10.1145/1403375.1403380}}


\bibitem[Izraelevitz et~al\mbox{.}(2017)]%
        {izraelevitz2017reusability}
\bibfield{author}{\bibinfo{person}{Adam Izraelevitz}, \bibinfo{person}{Jack
  Koenig}, \bibinfo{person}{Patrick Li}, \bibinfo{person}{Richard Lin},
  \bibinfo{person}{Angie Wang}, \bibinfo{person}{Albert Magyar},
  \bibinfo{person}{Donggyu Kim}, \bibinfo{person}{Colin Schmidt},
  \bibinfo{person}{Chick Markley}, \bibinfo{person}{Jim Lawson},
  {et~al\mbox{.}}} \bibinfo{year}{2017}\natexlab{}.
\newblock \showarticletitle{Reusability is {FIRRTL} ground: Hardware
  construction languages, compiler frameworks, and transformations}. In
  \bibinfo{booktitle}{\emph{2017 IEEE/ACM International Conference on
  Computer-Aided Design (ICCAD)}}. IEEE/ACM, \bibinfo{pages}{209--216}.
\newblock
\href{https://doi.org/10.1109/ICCAD.2017.8203780}{doi:\nolinkurl{10.1109/ICCAD.2017.8203780}}


\bibitem[Jain(2026)]%
        {TAIDL2026Source}
\bibfield{author}{\bibinfo{person}{Devansh Jain}.}
  \bibinfo{year}{2026}\natexlab{}.
\newblock \bibinfo{booktitle}{\emph{TAIDL: Tensor Accelerator ISA Definition
  Language}}.
\newblock
\urldef\tempurl%
\url{https://github.com/act-compiler/taidl}
\showURL{%
\tempurl}


\bibitem[Jain et~al\mbox{.}(2025)]%
        {taidl-test2025taidliss}
\bibfield{author}{\bibinfo{person}{Devansh Jain}, \bibinfo{person}{Marco
  Frigo}, \bibinfo{person}{Jai Arora}, \bibinfo{person}{Akash Pardeshi},
  \bibinfo{person}{Zhihao Wang}, \bibinfo{person}{Krut Patel}, {and}
  \bibinfo{person}{Charith Mendis}.} \bibinfo{year}{2025}\natexlab{}.
\newblock \showarticletitle{{TAIDL}: Tensor Accelerator {ISA} Definition
  Language with Auto-generation of Scalable Test Oracles}. In
  \bibinfo{booktitle}{\emph{Proceedings of the 2025 58th IEEE/ACM International
  Symposium on Microarchitecture}} \emph{(\bibinfo{series}{MICRO '25})}.
  \bibinfo{publisher}{Association for Computing Machinery},
  \bibinfo{address}{New York, NY, USA}, \bibinfo{pages}{1316--1333}.
\newblock
\showISBNx{9798400715730}
\href{https://doi.org/10.1145/3725843.3756075}{doi:\nolinkurl{10.1145/3725843.3756075}}


\bibitem[Jain et~al\mbox{.}(2026)]%
        {taidl-act2026taidlcomp}
\bibfield{author}{\bibinfo{person}{Devansh Jain}, \bibinfo{person}{Akash
  Pardeshi}, \bibinfo{person}{Marco Frigo}, \bibinfo{person}{Kaustubh Khulbe},
  \bibinfo{person}{Krut Patel}, \bibinfo{person}{Saatvik Lochan},
  \bibinfo{person}{Jai Arora}, {and} \bibinfo{person}{Charith Mendis}.}
  \bibinfo{year}{2026}\natexlab{}.
\newblock \showarticletitle{Automatically Generating {ML} Compiler Backends
  from Tensor Accelerator {ISA} Descriptions}.
\newblock \bibinfo{journal}{\emph{Proc. ACM Program. Lang.}}
  \bibinfo{volume}{10}, \bibinfo{number}{OOPSLA2}, Article
  \bibinfo{articleno}{325} (\bibinfo{date}{Oct.} \bibinfo{year}{2026}),
  \bibinfo{numpages}{32}~pages.
\newblock
\href{https://doi.org/10.1145/3839457}{doi:\nolinkurl{10.1145/3839457}}


\bibitem[Kabrick(2022)]%
        {Kabrick2022Rapid}
\bibfield{author}{\bibinfo{person}{Ryan Kabrick}.}
  \bibinfo{year}{2022}\natexlab{}.
\newblock \emph{\bibinfo{title}{Rapid Prototyping Framework for
  Hardware-Software Co-Design With Advanced Vector Architectures}}.
\newblock \bibinfo{thesistype}{Master's\ thesis}. \bibinfo{school}{University
  of Delaware}.
\newblock
\urldef\tempurl%
\url{https://udspace.udel.edu/handle/19716/31573}
\showURL{%
\tempurl}


\bibitem[Kahng and Spyrou(2021)]%
        {kahng2021openroad}
\bibfield{author}{\bibinfo{person}{Andrew~B Kahng} {and} \bibinfo{person}{Tom
  Spyrou}.} \bibinfo{year}{2021}\natexlab{}.
\newblock \showarticletitle{The {OpenROAD} project: Unleashing hardware
  innovation}. In \bibinfo{booktitle}{\emph{Proc. GOMAC}}.
  \bibinfo{pages}{1--6}.
\newblock


\bibitem[Kappes et~al\mbox{.}(2023)]%
        {Kappes2023Effective}
\bibfield{author}{\bibinfo{person}{Johannes Kappes}, \bibinfo{person}{Robert
  Kunzelmann}, \bibinfo{person}{Karsten Emrich}, \bibinfo{person}{Conrad Foik},
  \bibinfo{person}{Daniel M\"{u}ller-Gritschneder}, {and}
  \bibinfo{person}{Wolfgang Ecker}.} \bibinfo{year}{2023}\natexlab{}.
\newblock \showarticletitle{Effective Processor Model Generation from
  Instruction Set Simulator to Hardware Design}. In
  \bibinfo{booktitle}{\emph{2023 IEEE Nordic Circuits and Systems Conference
  (NorCAS)}}. \bibinfo{pages}{1--7}.
\newblock
\href{https://doi.org/10.1109/NorCAS58970.2023.10305465}{doi:\nolinkurl{10.1109/NorCAS58970.2023.10305465}}


\bibitem[Keppel(2009)]%
        {keppel2009detectselfmodifying}
\bibfield{author}{\bibinfo{person}{David Keppel}.}
  \bibinfo{year}{2009}\natexlab{}.
\newblock \showarticletitle{{How to Detect Self-Modifying Code During
  Instruction-Set Simulation}}.
\newblock
\urldef\tempurl%
\url{http://www.xsim.com/papers/sim-smc-amasbt2009.pdf}
\showURL{%
\tempurl}


\bibitem[Klint(1981)]%
        {klint1981interpretation}
\bibfield{author}{\bibinfo{person}{Paul Klint}.}
  \bibinfo{year}{1981}\natexlab{}.
\newblock \showarticletitle{Interpretation Techniques}.
\newblock \bibinfo{journal}{\emph{Software: Practice and Experience}}
  \bibinfo{volume}{11}, \bibinfo{number}{9} (\bibinfo{year}{1981}),
  \bibinfo{pages}{963--973}.
\newblock
\href{https://doi.org/10.1002/spe.4380110908}{doi:\nolinkurl{10.1002/spe.4380110908}}


\bibitem[Knoop et~al\mbox{.}(1992)]%
        {knoop1992lazycodemotion}
\bibfield{author}{\bibinfo{person}{Jens Knoop}, \bibinfo{person}{Oliver
  R\"{u}thing}, {and} \bibinfo{person}{Bernhard Steffen}.}
  \bibinfo{year}{1992}\natexlab{}.
\newblock \showarticletitle{{L}azy {C}ode {M}otion}.
\newblock \bibinfo{journal}{\emph{SIGPLAN Not.}} \bibinfo{volume}{27},
  \bibinfo{number}{7} (\bibinfo{date}{July} \bibinfo{year}{1992}),
  \bibinfo{pages}{224–234}.
\newblock
\showISSN{0362-1340}
\href{https://doi.org/10.1145/143103.143136}{doi:\nolinkurl{10.1145/143103.143136}}


\bibitem[Krall et~al\mbox{.}(2004)]%
        {Krall2004xDSPcore}
\bibfield{author}{\bibinfo{person}{Andreas Krall}, \bibinfo{person}{Ivan
  Pryanishnikov}, \bibinfo{person}{Ulrich Hirnschrott}, {and}
  \bibinfo{person}{Christian Panis}.} \bibinfo{year}{2004}\natexlab{}.
\newblock \showarticletitle{{xDSPcore}: a compiler-based configurable digital
  signal processor}.
\newblock \bibinfo{journal}{\emph{IEEE Micro}} \bibinfo{volume}{24},
  \bibinfo{number}{4} (\bibinfo{year}{2004}), \bibinfo{pages}{67--78}.
\newblock
\href{https://doi.org/10.1109/MM.2004.40}{doi:\nolinkurl{10.1109/MM.2004.40}}


\bibitem[Kroening and Paul(2001)]%
        {kroening2001}
\bibfield{author}{\bibinfo{person}{Daniel Kroening} {and}
  \bibinfo{person}{Wolfgang~J. Paul}.} \bibinfo{year}{2001}\natexlab{}.
\newblock \showarticletitle{Automated Pipeline Design}. In
  \bibinfo{booktitle}{\emph{Proceedings of the 38th Annual Design Automation
  Conference}} (Las Vegas, Nevada, USA) \emph{(\bibinfo{series}{DAC '01})}.
  \bibinfo{publisher}{Association for Computing Machinery},
  \bibinfo{address}{New York, NY, USA}, \bibinfo{pages}{810–815}.
\newblock
\showISBNx{1581132972}
\href{https://doi.org/10.1145/378239.379071}{doi:\nolinkurl{10.1145/378239.379071}}


\bibitem[Lattner and Adve(2004)]%
        {lattner2004llvm}
\bibfield{author}{\bibinfo{person}{Chris Lattner} {and} \bibinfo{person}{Vikram
  Adve}.} \bibinfo{year}{2004}\natexlab{}.
\newblock \showarticletitle{{LLVM: A Compilation Framework for Lifelong Program
  Analysis \& Transformation}}. In \bibinfo{booktitle}{\emph{Proceedings of the
  International Symposium on Code Generation and Optimization:
  Feedback-Directed and Runtime Optimization}} (Palo Alto, California)
  \emph{(\bibinfo{series}{CGO '04})}. \bibinfo{publisher}{IEEE Computer
  Society}, \bibinfo{address}{USA}, \bibinfo{pages}{75}.
\newblock
\showISBNx{0769521029}
\href{https://doi.org/10.1109/CGO.2004.1281665}{doi:\nolinkurl{10.1109/CGO.2004.1281665}}


\bibitem[Leidel et~al\mbox{.}(2021)]%
        {Leidel2021Toward}
\bibfield{author}{\bibinfo{person}{John Leidel}, \bibinfo{person}{Ryan
  Kabrick}, {and} \bibinfo{person}{David Donofrio}.}
  \bibinfo{year}{2021}\natexlab{}.
\newblock \showarticletitle{Toward an Automated Hardware Pipelining LLVM Pass
  Infrastructure}. In \bibinfo{booktitle}{\emph{2021 IEEE/ACM 7th Workshop on
  the LLVM Compiler Infrastructure in HPC (LLVM-HPC)}}.
  \bibinfo{pages}{39--49}.
\newblock
\href{https://doi.org/10.1109/LLVMHPC54804.2021.00010}{doi:\nolinkurl{10.1109/LLVMHPC54804.2021.00010}}


\bibitem[Leidel(2019)]%
        {leidel2019StoneCutter}
\bibfield{author}{\bibinfo{person}{John~D. Leidel}.}
  \bibinfo{year}{2019}\natexlab{}.
\newblock \bibinfo{booktitle}{\emph{StoneCutter Language Specification Version
  0.5}}.
\newblock \bibinfo{type}{{T}echnical {R}eport}. \bibinfo{institution}{Tactical
  Computing Laboratories, LLC}.
\newblock
\urldef\tempurl%
\url{http://www.systemarchitect.tech/releases/StoneCutterLang/stonecutterspec.0.5.pdf}
\showURL{%
\tempurl}


\bibitem[Leidel et~al\mbox{.}(2020)]%
        {Leidel2020StoneCutter}
\bibfield{author}{\bibinfo{person}{John~D. Leidel}, \bibinfo{person}{David
  Donofrio}, {and} \bibinfo{person}{Frank Conlon}.}
  \bibinfo{year}{2020}\natexlab{}.
\newblock \showarticletitle{{StoneCutter}: a very high level instruction set
  design language}. In \bibinfo{booktitle}{\emph{Proceedings of the 17th ACM
  International Conference on Computing Frontiers}} (Catania, Sicily, Italy)
  \emph{(\bibinfo{series}{CF '20})}. \bibinfo{publisher}{ACM},
  \bibinfo{address}{New York, NY, USA}, \bibinfo{pages}{233–236}.
\newblock
\href{https://doi.org/10.1145/3387902.3394029}{doi:\nolinkurl{10.1145/3387902.3394029}}


\bibitem[Leung and George(1999)]%
        {leung1999static}
\bibfield{author}{\bibinfo{person}{Allen Leung} {and} \bibinfo{person}{Lal
  George}.} \bibinfo{year}{1999}\natexlab{}.
\newblock \showarticletitle{Static single assignment form for machine code}.
\newblock \bibinfo{journal}{\emph{ACM SIGPLAN Notices}} \bibinfo{volume}{34},
  \bibinfo{number}{5} (\bibinfo{year}{1999}), \bibinfo{pages}{204--214}.
\newblock
\href{https://doi.org/10.1145/301631.301667}{doi:\nolinkurl{10.1145/301631.301667}}


\bibitem[Leupers and Marwedel(1998)]%
        {leupers1998retargetable}
\bibfield{author}{\bibinfo{person}{Rainer Leupers} {and} \bibinfo{person}{Peter
  Marwedel}.} \bibinfo{year}{1998}\natexlab{}.
\newblock \showarticletitle{Retargetable Code Generation Based on Structural
  Processor Description}.
\newblock \bibinfo{journal}{\emph{Design Automation for Embedded Systems}}
  \bibinfo{volume}{3}, \bibinfo{number}{1} (\bibinfo{date}{Jan.}
  \bibinfo{year}{1998}), \bibinfo{pages}{75–108}.
\newblock
\showISSN{0929-5585}
\href{https://doi.org/10.1023/A:1008807631619}{doi:\nolinkurl{10.1023/A:1008807631619}}


\bibitem[Leupers and Marwedel(2001)]%
        {leupers2001retargetable}
\bibfield{author}{\bibinfo{person}{Rainer Leupers} {and} \bibinfo{person}{Peter
  Marwedel}.} \bibinfo{year}{2001}\natexlab{}.
\newblock \bibinfo{booktitle}{\emph{Retargetable Compiler Technology for
  Embedded Systems: Tools and Applications}}.
\newblock \bibinfo{publisher}{Springer Science$+$Business Media}.
\newblock
\href{https://doi.org/10.1007/978-1-4757-6420-8}{doi:\nolinkurl{10.1007/978-1-4757-6420-8}}


\bibitem[Lezuo and Krall(2013)]%
        {Lezuo2013CASM}
\bibfield{author}{\bibinfo{person}{Roland Lezuo} {and} \bibinfo{person}{Andreas
  Krall}.} \bibinfo{year}{2013}\natexlab{}.
\newblock \showarticletitle{Using the CASM language for simulator synthesis and
  model verification}. In \bibinfo{booktitle}{\emph{Proceedings of the 2013
  Workshop on Rapid Simulation and Performance Evaluation: Methods and Tools}}
  (Berlin, Germany) \emph{(\bibinfo{series}{RAPIDO '13})}.
  \bibinfo{publisher}{Association for Computing Machinery},
  \bibinfo{address}{New York, NY, USA}, Article \bibinfo{articleno}{6},
  \bibinfo{numpages}{8}~pages.
\newblock
\showISBNx{9781450315395}
\href{https://doi.org/10.1145/2432516.2432522}{doi:\nolinkurl{10.1145/2432516.2432522}}


\bibitem[Lezuo et~al\mbox{.}(2014)]%
        {Lezuo2014CASM}
\bibfield{author}{\bibinfo{person}{Roland Lezuo}, \bibinfo{person}{Philipp
  Paulweber}, {and} \bibinfo{person}{Andreas Krall}.}
  \bibinfo{year}{2014}\natexlab{}.
\newblock \showarticletitle{{CASM}: optimized compilation of abstract state
  machines}. In \bibinfo{booktitle}{\emph{Proceedings of the 2014
  SIGPLAN/SIGBED Conference on Languages, Compilers and Tools for Embedded
  Systems}} (Edinburgh, United Kingdom) \emph{(\bibinfo{series}{LCTES '14})}.
  \bibinfo{publisher}{Association for Computing Machinery},
  \bibinfo{address}{New York, NY, USA}, \bibinfo{pages}{13–22}.
\newblock
\showISBNx{9781450328777}
\href{https://doi.org/10.1145/2597809.2597813}{doi:\nolinkurl{10.1145/2597809.2597813}}


\bibitem[Lockhart et~al\mbox{.}(2015)]%
        {lockhart2015pydgin}
\bibfield{author}{\bibinfo{person}{Derek Lockhart}, \bibinfo{person}{Berkin
  Ilbeyi}, {and} \bibinfo{person}{Christopher Batten}.}
  \bibinfo{year}{2015}\natexlab{}.
\newblock \showarticletitle{{P}ydgin: {G}enerating {F}ast {I}nstruction {S}et
  {S}imulators from {S}imple {A}rchitecture {D}escriptions with
  {M}eta-{T}racing {JIT} {C}ompilers}. In \bibinfo{booktitle}{\emph{2015 IEEE
  International Symposium on Performance Analysis of Systems and Software
  (ISPASS)}}. IEEE, \bibinfo{pages}{256--267}.
\newblock
\href{https://doi.org/10.1109/ISPASS.2015.7095811}{doi:\nolinkurl{10.1109/ISPASS.2015.7095811}}


\bibitem[L{\'o}pez-Parad{\'\i}s et~al\mbox{.}(2021)]%
        {lopez2021gem5}
\bibfield{author}{\bibinfo{person}{Guillem L{\'o}pez-Parad{\'\i}s},
  \bibinfo{person}{Adri{\`a} Armejach}, {and} \bibinfo{person}{Miquel
  Moret{\'o}}.} \bibinfo{year}{2021}\natexlab{}.
\newblock \showarticletitle{Gem5+ rtl: A framework to enable rtl models inside
  a full-system simulator}. In \bibinfo{booktitle}{\emph{Proceedings of the
  50th International Conference on Parallel Processing}}.
  \bibinfo{publisher}{ACM}, \bibinfo{pages}{1--11}.
\newblock
\href{https://doi.org/10.1145/3472456.3472461}{doi:\nolinkurl{10.1145/3472456.3472461}}


\bibitem[Lowe-Power et~al\mbox{.}(2020)]%
        {lowe2020gem5}
\bibfield{author}{\bibinfo{person}{Jason Lowe-Power},
  \bibinfo{person}{Abdul~Mutaal Ahmad}, \bibinfo{person}{Ayaz Akram},
  \bibinfo{person}{Mohammad Alian}, \bibinfo{person}{Rico Amslinger},
  \bibinfo{person}{Matteo Andreozzi}, \bibinfo{person}{Adri{\`a} Armejach},
  \bibinfo{person}{Nils Asmussen}, \bibinfo{person}{Brad Beckmann},
  \bibinfo{person}{Srikant Bharadwaj}, {et~al\mbox{.}}}
  \bibinfo{year}{2020}\natexlab{}.
\newblock \showarticletitle{The gem5 simulator: Version 20.0+}.
\newblock \bibinfo{journal}{\emph{arXiv preprint arXiv:2007.03152}}
  (\bibinfo{year}{2020}).
\newblock
\href{https://doi.org/10.48550/arXiv.2007.03152}{doi:\nolinkurl{10.48550/arXiv.2007.03152}}


\bibitem[Magnusson(1997)]%
        {magnusson1997efficient}
\bibfield{author}{\bibinfo{person}{Peter~S Magnusson}.}
  \bibinfo{year}{1997}\natexlab{}.
\newblock \showarticletitle{Efficient instruction cache simulation and
  execution profiling with a threaded-code interpreter}. In
  \bibinfo{booktitle}{\emph{Proceedings of the 29th conference on Winter
  simulation}}. \bibinfo{pages}{1093--1100}.
\newblock
\urldef\tempurl%
\url{https://dl.acm.org/doi/pdf/10.1145/268437.268745}
\showURL{%
\tempurl}


\bibitem[{Magnusson} et~al\mbox{.}(2002)]%
        {magnussion2002simics}
\bibfield{author}{\bibinfo{person}{Peter~S. {Magnusson}},
  \bibinfo{person}{Magnus {Christensson}}, \bibinfo{person}{Jesper {Eskilson}},
  \bibinfo{person}{Daniel {Forsgren}}, \bibinfo{person}{Gustav {H{\aa}llberg}},
  \bibinfo{person}{Johan {H{\"o}gberg}}, \bibinfo{person}{Fredrik {Larsson}},
  \bibinfo{person}{Andreas {Moestedt}}, {and} \bibinfo{person}{Bengt
  {Werner}}.} \bibinfo{year}{2002}\natexlab{}.
\newblock \showarticletitle{Simics: A full system simulation platform}.
\newblock \bibinfo{journal}{\emph{Computer}} \bibinfo{volume}{35},
  \bibinfo{number}{2} (\bibinfo{year}{2002}), \bibinfo{pages}{50--58}.
\newblock
\href{https://doi.org/10.1109/2.982916}{doi:\nolinkurl{10.1109/2.982916}}


\bibitem[Malicky(2026)]%
        {Malicky2026}
\bibfield{author}{\bibinfo{person}{Florian Malicky}.}
  \bibinfo{year}{2026}\natexlab{}.
\newblock \emph{\bibinfo{title}{Optimizing the {OpenVADL} {ISS} generator}}.
\newblock Bachelor's Thesis. \bibinfo{school}{Technische Universit{\"a}t Wien}.
\newblock
\urldef\tempurl%
\url{https://www.complang.tuwien.ac.at/vadl/papers/MalickyFinal.pdf}
\showURL{%
\tempurl}


\bibitem[Marwedel(1984)]%
        {marwedel1984mimola}
\bibfield{author}{\bibinfo{person}{Peter Marwedel}.}
  \bibinfo{year}{1984}\natexlab{}.
\newblock \showarticletitle{{The MIMOLA design system: Tools for the design of
  digital processors}}. In \bibinfo{booktitle}{\emph{21st Design Automation
  Conference Proceedings}}. IEEE, \bibinfo{pages}{587--593}.
\newblock
\href{https://doi.org/10.1109/DAC.1984.1585857}{doi:\nolinkurl{10.1109/DAC.1984.1585857}}


\bibitem[Masar{\'\i}k et~al\mbox{.}(2006)]%
        {Masarik2006Language}
\bibfield{author}{\bibinfo{person}{Karel Masar{\'\i}k},
  \bibinfo{person}{Tom{\'a}{\v{s}} Hru{\v{s}}ka}, {and}
  \bibinfo{person}{Du{\v{s}}an Kol{\'a}{\v{r}}}.}
  \bibinfo{year}{2006}\natexlab{}.
\newblock \showarticletitle{Language and development environment for
  microprocessor design of embedded systems}.
\newblock \bibinfo{journal}{\emph{IFAC Proceedings Volumes}}
  \bibinfo{volume}{39}, \bibinfo{number}{21} (\bibinfo{year}{2006}),
  \bibinfo{pages}{120--125}.
\newblock
\href{https://doi.org/10.1016/S1474-6670(17)30170-2}{doi:\nolinkurl{10.1016/S1474-6670(17)30170-2}}


\bibitem[Matsuda and Wang(2013)]%
        {matsuda2013flippr}
\bibfield{author}{\bibinfo{person}{Kazutaka Matsuda} {and}
  \bibinfo{person}{Meng Wang}.} \bibinfo{year}{2013}\natexlab{}.
\newblock \showarticletitle{FliPpr: A prettier invertible printing system}. In
  \bibinfo{booktitle}{\emph{Programming Languages and Systems: 22nd European
  Symposium on Programming, ESOP 2013, Held as Part of the European Joint
  Conferences on Theory and Practice of Software, ETAPS 2013, Rome, Italy,
  March 16-24, 2013. Proceedings 22}}. \bibinfo{publisher}{Springer},
  \bibinfo{pages}{101--120}.
\newblock
\href{https://doi.org/10.1007/978-3-642-37036-6_6}{doi:\nolinkurl{10.1007/978-3-642-37036-6_6}}


\bibitem[Mihaylov(2023)]%
        {Mihaylov2023Optimized}
\bibfield{author}{\bibinfo{person}{Hristo Mihaylov}.}
  \bibinfo{year}{2023}\natexlab{}.
\newblock \emph{\bibinfo{title}{Optimized processor simulation with VADL}}.
\newblock \bibinfo{thesistype}{Master's\ thesis}. \bibinfo{school}{Technische
  Universit{\"a}t Wien}.
\newblock
\href{https://doi.org/10.34726/hss.2023.102629}{doi:\nolinkurl{10.34726/hss.2023.102629}}


\bibitem[Mills et~al\mbox{.}(1991)]%
        {mills1991compiled}
\bibfield{author}{\bibinfo{person}{Christopher Mills},
  \bibinfo{person}{Stanley~C Ahalt}, {and} \bibinfo{person}{Jim Fowler}.}
  \bibinfo{year}{1991}\natexlab{}.
\newblock \showarticletitle{Compiled instruction set simulation}.
\newblock \bibinfo{journal}{\emph{Software: Practice and Experience}}
  \bibinfo{volume}{21}, \bibinfo{number}{8} (\bibinfo{year}{1991}),
  \bibinfo{pages}{877--889}.
\newblock
\href{https://doi.org/10.1002/spe.4380210807}{doi:\nolinkurl{10.1002/spe.4380210807}}


\bibitem[Mischkulnig(2025)]%
        {NiklasMischkulnig2025}
\bibfield{author}{\bibinfo{person}{Niklas Mischkulnig}.}
  \bibinfo{year}{2025}\natexlab{}.
\newblock \emph{\bibinfo{title}{Instruction Set Simulation of {VLIW}
  Architectures in {VADL}}}.
\newblock \bibinfo{thesistype}{Master's\ thesis}. \bibinfo{school}{Technische
  Universit{\"a}t Wien}.
\newblock
\href{https://doi.org/10.34726/hss.2025.110828}{doi:\nolinkurl{10.34726/hss.2025.110828}}


\bibitem[Mishra and Dutt(2005)]%
        {Mishra2005architecture}
\bibfield{author}{\bibinfo{person}{Prabhat Mishra} {and} \bibinfo{person}{Nikil
  Dutt}.} \bibinfo{year}{2005}\natexlab{}.
\newblock \showarticletitle{Architecture description languages for programmable
  embedded systems}.
\newblock \bibinfo{journal}{\emph{IEE Proceedings - Computers and Digital
  Techniques}}  \bibinfo{volume}{152} (\bibinfo{year}{2005}),
  \bibinfo{pages}{285--297}.
\newblock
Issue 3.
\urldef\tempurl%
\url{https://digital-library.theiet.org/doi/abs/10.1049/ip-cdt%3A20045071}
\showURL{%
\tempurl}


\bibitem[Mishra and Dutt(2008)]%
        {mishra2008processor}
\bibfield{author}{\bibinfo{person}{Prabhat Mishra} {and} \bibinfo{person}{Nikil
  Dutt}.} \bibinfo{year}{2008}\natexlab{}.
\newblock \bibinfo{booktitle}{\emph{Processor Description Languages}}.
\newblock \bibinfo{publisher}{Elsevier}.
\newblock
\href{https://doi.org/10.1016/B978-0-12-374287-2.X5001-0}{doi:\nolinkurl{10.1016/B978-0-12-374287-2.X5001-0}}


\bibitem[M{\"o}ssenb{\"o}ck et~al\mbox{.}(2018)]%
        {CocoR}
\bibfield{author}{\bibinfo{person}{Hanspeter M{\"o}ssenb{\"o}ck},
  \bibinfo{person}{Markus L{\"o}berbauer}, {and} \bibinfo{person}{Albrecht
  W{\"o}{\ss}}.} \bibinfo{year}{2018}\natexlab{}.
\newblock \bibinfo{booktitle}{\emph{{The Compiler Generator Coco/R}}}.
\newblock
\urldef\tempurl%
\url{https://ssw.jku.at/Research/Projects/Coco/}
\showURL{%
\tempurl}


\bibitem[Muchnick(1997)]%
        {muchnick1997advanced}
\bibfield{author}{\bibinfo{person}{Steven Muchnick}.}
  \bibinfo{year}{1997}\natexlab{}.
\newblock \bibinfo{booktitle}{\emph{{A}dvanced {C}ompiler {D}esign and
  {I}mplementation}}.
\newblock \bibinfo{publisher}{Morgan kaufmann}.
\newblock


\bibitem[Nestler(2024)]%
        {Nestler2024}
\bibfield{author}{\bibinfo{person}{Michael Nestler}.}
  \bibinfo{year}{2024}\natexlab{}.
\newblock \emph{\bibinfo{title}{{E}fficient {P}arsing of {OpenVADL}}}.
\newblock Bachelor's Thesis. \bibinfo{school}{Technische Universit{\"a}t Wien}.
\newblock
\urldef\tempurl%
\url{https://www.complang.tuwien.ac.at/vadl/papers/NestlerFinal.pdf}
\showURL{%
\tempurl}


\bibitem[Oppermann(2021)]%
        {CoreDSL2}
\bibfield{author}{\bibinfo{person}{Julian Oppermann}.}
  \bibinfo{year}{2021}\natexlab{}.
\newblock \bibinfo{booktitle}{\emph{CoreDSL 2 programmer's manual}}.
\newblock
\urldef\tempurl%
\url{https://github.com/Minres/CoreDSL/wiki/CoreDSL-2-programmer's-manual}
\showURL{%
\tempurl}


\bibitem[Oppermann et~al\mbox{.}(2024)]%
        {Oppermann2024Longnail}
\bibfield{author}{\bibinfo{person}{Julian Oppermann},
  \bibinfo{person}{Brindusa~Mihaela Damian-Kosterhon}, \bibinfo{person}{Florian
  Meisel}, \bibinfo{person}{Tammo M\"{u}rmann}, \bibinfo{person}{Eyck
  Jentzsch}, {and} \bibinfo{person}{Andreas Koch}.}
  \bibinfo{year}{2024}\natexlab{}.
\newblock \showarticletitle{Longnail: High-Level Synthesis of Portable Custom
  Instruction Set Extensions for {RISC-V} Processors from Descriptions in the
  Open-Source {CoreDSL} Language}. In \bibinfo{booktitle}{\emph{Proceedings of
  the 29th ACM International Conference on Architectural Support for
  Programming Languages and Operating Systems (ASPLOS)}} (La Jolla, CA, USA)
  \emph{(\bibinfo{series}{ASPLOS '24})}. \bibinfo{publisher}{ACM},
  \bibinfo{address}{New York, NY, USA}, \bibinfo{pages}{591–606}.
\newblock
\href{https://doi.org/10.1145/3620666.3651375}{doi:\nolinkurl{10.1145/3620666.3651375}}


\bibitem[Panda(2001)]%
        {panda2001systemc}
\bibfield{author}{\bibinfo{person}{Preeti~Ranjan Panda}.}
  \bibinfo{year}{2001}\natexlab{}.
\newblock \showarticletitle{{SystemC}: A Modeling Platform Supporting Multiple
  Design Abstractions}. In \bibinfo{booktitle}{\emph{Proceedings of the 14th
  International Symposium on Systems Synthesis}} (Montr\'{e}al, P.Q., Canada)
  \emph{(\bibinfo{series}{ISSS '01})}. \bibinfo{publisher}{Association for
  Computing Machinery}, \bibinfo{address}{New York, NY, USA},
  \bibinfo{pages}{75–80}.
\newblock
\showISBNx{1581134185}
\href{https://doi.org/10.1145/500001.500018}{doi:\nolinkurl{10.1145/500001.500018}}


\bibitem[Parr and Fisher(2011)]%
        {ParrFischer2011}
\bibfield{author}{\bibinfo{person}{Terence Parr} {and}
  \bibinfo{person}{Kathleen Fisher}.} \bibinfo{year}{2011}\natexlab{}.
\newblock \showarticletitle{{LL}(*) the foundation of the {ANTLR} parser
  generator}.
\newblock \bibinfo{journal}{\emph{ACM SIGPLAN Notices}} \bibinfo{volume}{46},
  \bibinfo{number}{6} (\bibinfo{year}{2011}), \bibinfo{pages}{425--436}.
\newblock
\href{https://doi.org/10.1145/1993316.1993548}{doi:\nolinkurl{10.1145/1993316.1993548}}


\bibitem[Parr and Quong(1994)]%
        {ParrQuong1994}
\bibfield{author}{\bibinfo{person}{Terence~J Parr} {and}
  \bibinfo{person}{Russell~W Quong}.} \bibinfo{year}{1994}\natexlab{}.
\newblock \showarticletitle{Adding semantic and syntactic predicates to
  {LL}(k): {pred-LL}(k)}. In \bibinfo{booktitle}{\emph{Compiler Construction:
  5th International Conference, CC'94 Edinburgh, UK, April 7--9, 1994
  Proceedings 5}}. Springer, \bibinfo{pages}{263--277}.
\newblock
\href{https://doi.org/10.1007/3-540-57877-3_18}{doi:\nolinkurl{10.1007/3-540-57877-3_18}}


\bibitem[Patterson and Hennessy(2017)]%
        {patterson2017comporganddesign}
\bibfield{author}{\bibinfo{person}{David~A. Patterson} {and}
  \bibinfo{person}{John~L. Hennessy}.} \bibinfo{year}{2017}\natexlab{}.
\newblock \bibinfo{booktitle}{\emph{Computer Organization and Design RISC-V
  Edition: The Hardware Software Interface} (\bibinfo{edition}{1st} ed.)}.
\newblock \bibinfo{publisher}{Morgan Kaufmann Publishers Inc.},
  \bibinfo{address}{San Francisco, CA, USA}.
\newblock
\showISBNx{0128122757}


\bibitem[Pees et~al\mbox{.}(1999)]%
        {pees1999lisa}
\bibfield{author}{\bibinfo{person}{Stefan Pees}, \bibinfo{person}{Andreas
  Hoffmann}, \bibinfo{person}{Vojin Zivojnovic}, {and}
  \bibinfo{person}{Heinrich Meyr}.} \bibinfo{year}{1999}\natexlab{}.
\newblock \showarticletitle{{LISA} -- machine description language for
  cycle-accurate models of programmable {DSP} architectures}. In
  \bibinfo{booktitle}{\emph{Proceedings of the 36th annual ACM/IEEE Design
  Automation Conference}}. \bibinfo{pages}{933--938}.
\newblock
\urldef\tempurl%
\url{https://dl.acm.org/doi/pdf/10.1145/309847.310101}
\showURL{%
\tempurl}


\bibitem[Per(2025)]%
        {PerKevin2025lcb}
\bibfield{author}{\bibinfo{person}{Kevin Per}.}
  \bibinfo{year}{2025}\natexlab{}.
\newblock \emph{\bibinfo{title}{{LLVM} compiler generator in {OpenVADL}}}.
\newblock \bibinfo{thesistype}{Master's\ thesis}. \bibinfo{school}{Technische
  Universit{\"a}t Wien}.
\newblock
\href{https://doi.org/10.34726/hss.2025.122088}{doi:\nolinkurl{10.34726/hss.2025.122088}}


\bibitem[Pryanishnikov et~al\mbox{.}(2007)]%
        {Pryanishnikov2007Compiler}
\bibfield{author}{\bibinfo{person}{Ivan Pryanishnikov},
  \bibinfo{person}{Andreas Krall}, {and} \bibinfo{person}{Nigel Horspool}.}
  \bibinfo{year}{2007}\natexlab{}.
\newblock \showarticletitle{Compiler optimizations for processors with {SIMD}
  instructions}.
\newblock \bibinfo{journal}{\emph{Software: Practice and Experience}}
  \bibinfo{volume}{37}, \bibinfo{number}{1} (\bibinfo{year}{2007}),
  \bibinfo{pages}{93--113}.
\newblock
\href{https://doi.org/10.1002/spe.751}{doi:\nolinkurl{10.1002/spe.751}}


\bibitem[P\v{r}ikryl(2014)]%
        {Prikryl2014CodAL}
\bibfield{author}{\bibinfo{person}{Zden\v{e}k P\v{r}ikryl}.}
  \bibinfo{year}{2014}\natexlab{}.
\newblock \showarticletitle{Fast Simulation of Pipeline in {ASIP} Simulators}.
  In \bibinfo{booktitle}{\emph{2014 15th International Microprocessor Test and
  Verification Workshop}}. \bibinfo{pages}{10--15}.
\newblock
\href{https://doi.org/10.1109/MTV.2014.18}{doi:\nolinkurl{10.1109/MTV.2014.18}}


\bibitem[P\v{r}ikryl et~al\mbox{.}(2011)]%
        {prikryl2011design}
\bibfield{author}{\bibinfo{person}{Zden{\v{e}}k P\v{r}ikryl},
  \bibinfo{person}{Jakub Kroustek}, \bibinfo{person}{Tom{\'a}{\v{s}}
  Hru{\v{s}}ka}, \bibinfo{person}{Du{\v{s}}an Kol{\'a}{\v{r}}},
  \bibinfo{person}{Karel Masa{\v{r}}{\'\i}k}, {and} \bibinfo{person}{Adam
  Hus{\'a}r}.} \bibinfo{year}{2011}\natexlab{}.
\newblock \showarticletitle{Design and Simulation of High Performance Parallel
  Architectures Using the {ISAC} Language.}
\newblock \bibinfo{journal}{\emph{GSTF Journal on Computing}}
  \bibinfo{volume}{1}, \bibinfo{number}{2} (\bibinfo{year}{2011}),
  \bibinfo{pages}{97--106}.
\newblock
\urldef\tempurl%
\url{http://dl6.globalstf.org/index.php/joc/article/download/894/2241}
\showURL{%
\tempurl}


\bibitem[P\v{r}ikryl et~al\mbox{.}(2009)]%
        {prikryl2009fast}
\bibfield{author}{\bibinfo{person}{Zden\v{e}k P\v{r}ikryl},
  \bibinfo{person}{Karel Masar{\'\i}k}, \bibinfo{person}{Tom{\'a}{\v{s}}
  Hru{\v{s}}ka}, {and} \bibinfo{person}{Adam Hus{\'a}r}.}
  \bibinfo{year}{2009}\natexlab{}.
\newblock \showarticletitle{Fast cycle-accurate interpreted simulation}. In
  \bibinfo{booktitle}{\emph{2009 10th International Workshop on Microprocessor
  Test and Verification}}. IEEE, \bibinfo{pages}{9--14}.
\newblock
\href{https://doi.org/10.1109/MTV.2009.11}{doi:\nolinkurl{10.1109/MTV.2009.11}}


\bibitem[Rastello(2016)]%
        {rastello2022ssa}
\bibfield{author}{\bibinfo{person}{Fabrice Rastello}.}
  \bibinfo{year}{2016}\natexlab{}.
\newblock \bibinfo{booktitle}{\emph{{SSA}-based {C}ompiler {D}esign}
  (\bibinfo{edition}{1st} ed.)}.
\newblock \bibinfo{publisher}{Springer Publishing Company, Incorporated}.
\newblock
\showISBNx{1441962018}


\bibitem[Ratsiambahotra et~al\mbox{.}(2009)]%
        {ratsiambahotra2009versatile}
\bibfield{author}{\bibinfo{person}{Tahiry Ratsiambahotra},
  \bibinfo{person}{Hugues Cass{\'e}}, {and} \bibinfo{person}{Pascal Sainrat}.}
  \bibinfo{year}{2009}\natexlab{}.
\newblock \showarticletitle{A versatile generator of instruction set simulators
  and disassemblers}. In \bibinfo{booktitle}{\emph{2009 International Symposium
  on Performance Evaluation of Computer \& Telecommunication Systems}},
  Vol.~\bibinfo{volume}{41}. IEEE, \bibinfo{pages}{65--72}.
\newblock
\urldef\tempurl%
\url{https://ieeexplore.ieee.org/abstract/document/5224142}
\showURL{%
\tempurl}


\bibitem[Reshadi et~al\mbox{.}(2003)]%
        {reshadi2003instruction}
\bibfield{author}{\bibinfo{person}{Mehrdad Reshadi}, \bibinfo{person}{Prabhat
  Mishra}, {and} \bibinfo{person}{Nikil Dutt}.}
  \bibinfo{year}{2003}\natexlab{}.
\newblock \showarticletitle{Instruction set compiled simulation: A technique
  for fast and flexible instruction set simulation}. In
  \bibinfo{booktitle}{\emph{Proceedings of the 40th Annual Design Automation
  Conference}}. ACM/IEEE, \bibinfo{pages}{758--763}.
\newblock
\href{https://doi.org/10.1145/775832.776026}{doi:\nolinkurl{10.1145/775832.776026}}


\bibitem[Rezunov et~al\mbox{.}(2025)]%
        {Rezunov2025CustumInstructionDesign}
\bibfield{author}{\bibinfo{person}{Evgenii Rezunov}, \bibinfo{person}{Niko
  Zurstraßen}, \bibinfo{person}{Lennart~M. Reimann}, {and}
  \bibinfo{person}{Rainer Leupers}.} \bibinfo{year}{2025}\natexlab{}.
\newblock \showarticletitle{Automatic Microarchitecture-Aware Custom
  Instruction Design for {RISC-V} Processors}. In
  \bibinfo{booktitle}{\emph{2025 IEEE/ACM International Conference On Computer
  Aided Design (ICCAD)}}. \bibinfo{pages}{1--9}.
\newblock
\href{https://doi.org/10.1109/ICCAD66269.2025.11240781}{doi:\nolinkurl{10.1109/ICCAD66269.2025.11240781}}


\bibitem[Rigo et~al\mbox{.}(2003)]%
        {Rigo2003ArchC}
\bibfield{author}{\bibinfo{person}{Sandro Rigo}, \bibinfo{person}{Rodolfo
  Azevedo}, \bibinfo{person}{Guido Araujo}, {and} \bibinfo{person}{Marcio
  Bartholomeu}.} \bibinfo{year}{2003}\natexlab{}.
\newblock \bibinfo{booktitle}{\emph{ArchC: A SystemC-Based Architecture
  Description Language}}.
\newblock \bibinfo{type}{Technical Report} IC-03-31.
  \bibinfo{institution}{Institute of Computing, University of Campinas
  (UNICAMP)}, \bibinfo{address}{Campinas, SP, Brazil}.
\newblock
\urldef\tempurl%
\url{https://www.ic.unicamp.br/~reltech/2003/03-15.pdf}
\showURL{%
\tempurl}


\bibitem[{RISC-V International}({[n.\,d.]})]%
        {riscvArchTests}
\bibfield{author}{\bibinfo{person}{{RISC-V International}}.}
  \bibinfo{year}{[n.\,d.]}\natexlab{}.
\newblock \bibinfo{title}{{RISC-V} Architectural Tests}.
\newblock \bibinfo{howpublished}{GitHub repository}.
\newblock
\urldef\tempurl%
\url{https://github.com/riscv/riscv-arch-test}
\showURL{%
\tempurl}


\bibitem[Schliebusch et~al\mbox{.}(2002)]%
        {schliebusch2002architecture}
\bibfield{author}{\bibinfo{person}{Oliver Schliebusch},
  \bibinfo{person}{Andreas Hoffmann}, \bibinfo{person}{Achim Nohl},
  \bibinfo{person}{Gunnar Braun}, {and} \bibinfo{person}{Heinrich Meyr}.}
  \bibinfo{year}{2002}\natexlab{}.
\newblock \showarticletitle{Architecture implementation using the machine
  description language {LISA}}. In \bibinfo{booktitle}{\emph{Proceedings of
  ASP-DAC/VLSI Design 2002. 7th Asia and South Pacific Design Automation
  Conference and 15h International Conference on VLSI Design}}.
  \bibinfo{pages}{239--244}.
\newblock
\href{https://doi.org/10.1109/ASPDAC.2002.994928}{doi:\nolinkurl{10.1109/ASPDAC.2002.994928}}


\bibitem[Schoeberl(2025)]%
        {schoeberl2025wildcat}
\bibfield{author}{\bibinfo{person}{Martin Schoeberl}.}
  \bibinfo{year}{2025}\natexlab{}.
\newblock \showarticletitle{Wildcat: Educational {RISC-V} microprocessors}. In
  \bibinfo{booktitle}{\emph{ARCS 2025 -- 38th GI/ITG International Conference
  on Architecture of Computing Systems}} (Kiel, Germany),
  \bibfield{editor}{\bibinfo{person}{Sven Tomforde}, \bibinfo{person}{Christian
  Krupitzer}, \bibinfo{person}{St{\'e}phane Vialle}, \bibinfo{person}{Estela
  Suarez}, {and} \bibinfo{person}{Thilo Pionteck}} (Eds.).
  \bibinfo{publisher}{Springer Nature Switzerland}, \bibinfo{pages}{189--202}.
\newblock
\href{https://doi.org/10.1007/978-3-032-03281-2_13}{doi:\nolinkurl{10.1007/978-3-032-03281-2_13}}


\bibitem[Schuiki et~al\mbox{.}(2020)]%
        {schuiki2020llhd}
\bibfield{author}{\bibinfo{person}{Fabian Schuiki}, \bibinfo{person}{Andreas
  Kurth}, \bibinfo{person}{Tobias Grosser}, {and} \bibinfo{person}{Luca
  Benini}.} \bibinfo{year}{2020}\natexlab{}.
\newblock \showarticletitle{LLHD: A multi-level intermediate representation for
  hardware description languages}. In \bibinfo{booktitle}{\emph{Proceedings of
  the 41st ACM SIGPLAN Conference on Programming Language Design and
  Implementation}}. ACM, \bibinfo{pages}{258--271}.
\newblock
\href{https://doi.org/10.1145/3385412.3386024}{doi:\nolinkurl{10.1145/3385412.3386024}}


\bibitem[Sch{\"u}tzenh{\"o}fer(2020)]%
        {schutzenhofer2020cycle}
\bibfield{author}{\bibinfo{person}{Hermann Sch{\"u}tzenh{\"o}fer}.}
  \bibinfo{year}{2020}\natexlab{}.
\newblock \emph{\bibinfo{title}{Cycle-Accurate simulator generator for the VADL
  processor description language}}.
\newblock \bibinfo{thesistype}{Master's\ thesis}. \bibinfo{school}{Technische
  Universit{\"a}t Wien}.
\newblock
\href{https://doi.org/10.34726/hss.2021.78460}{doi:\nolinkurl{10.34726/hss.2021.78460}}


\bibitem[Schwarzinger(2022)]%
        {schwarzinger2022flexible}
\bibfield{author}{\bibinfo{person}{Tobias Schwarzinger}.}
  \bibinfo{year}{2022}\natexlab{}.
\newblock \emph{\bibinfo{title}{Flexible generation of low-level developer
  tools with VADL}}.
\newblock \bibinfo{thesistype}{Master's\ thesis}. \bibinfo{school}{Technische
  Universit{\"a}t Wien}.
\newblock
\href{https://doi.org/10.34726/hss.2023.103246}{doi:\nolinkurl{10.34726/hss.2023.103246}}


\bibitem[Shen and Lipasti(2013)]%
        {shen2013modern}
\bibfield{author}{\bibinfo{person}{John~Paul Shen} {and}
  \bibinfo{person}{Mikko~H Lipasti}.} \bibinfo{year}{2013}\natexlab{}.
\newblock \bibinfo{booktitle}{\emph{Modern processor design: fundamentals of
  superscalar processors}}.
\newblock \bibinfo{publisher}{Waveland Press}.
\newblock


\bibitem[Singer(2003)]%
        {singer2003static}
\bibfield{author}{\bibinfo{person}{Jeremy Singer}.}
  \bibinfo{year}{2003}\natexlab{}.
\newblock \showarticletitle{Static single information from a functional
  perspective.}
\newblock \bibinfo{journal}{\emph{Trends in Functional Programming}}
  \bibinfo{volume}{4} (\bibinfo{year}{2003}), \bibinfo{pages}{63--78}.
\newblock


\bibitem[Sites(1993)]%
        {SitesAlpha93}
\bibfield{author}{\bibinfo{person}{Richard~L. Sites}.}
  \bibinfo{year}{1993}\natexlab{}.
\newblock \showarticletitle{{Alpha AXP Architecture}}.
\newblock \bibinfo{journal}{\emph{Commun. ACM}} \bibinfo{volume}{36},
  \bibinfo{number}{2} (\bibinfo{year}{1993}), \bibinfo{pages}{33–44}.
\newblock
\showISSN{0001-0782}
\href{https://doi.org/10.1145/151220.151226}{doi:\nolinkurl{10.1145/151220.151226}}


\bibitem[Stallman et~al\mbox{.}(2020)]%
        {stallman2020gcc}
\bibfield{author}{\bibinfo{person}{Richard Stallman} {et~al\mbox{.}}}
  \bibinfo{year}{2020}\natexlab{}.
\newblock \bibinfo{booktitle}{\emph{{Using the GNU Compiler Collection}}}.
\newblock \bibinfo{publisher}{Gnu Press Boston}.
\newblock


\bibitem[Steinberg et~al\mbox{.}(2008)]%
        {Steinberg2008EMF}
\bibfield{author}{\bibinfo{person}{Dave Steinberg}, \bibinfo{person}{Frank
  Budinsky}, \bibinfo{person}{Ed Merks}, {and} \bibinfo{person}{Marcelo
  Paternostro}.} \bibinfo{year}{2008}\natexlab{}.
\newblock \bibinfo{booktitle}{\emph{EMF: eclipse modeling framework}}.
\newblock \bibinfo{publisher}{Pearson Education}.
\newblock


\bibitem[Team(2016)]%
        {ArchC2016Source}
\bibfield{author}{\bibinfo{person}{ArchC Team}.}
  \bibinfo{year}{2016}\natexlab{}.
\newblock \bibinfo{booktitle}{\emph{ArchC}}.
\newblock
\urldef\tempurl%
\url{https://github.com/ArchC/ArchC}
\showURL{%
\tempurl}


\bibitem[Team(2026a)]%
        {CoreDSL2026Source}
\bibfield{author}{\bibinfo{person}{CoreDSL Team}.}
  \bibinfo{year}{2026}\natexlab{a}.
\newblock \bibinfo{booktitle}{\emph{CoreDSL}}.
\newblock
\urldef\tempurl%
\url{https://github.com/Minres/CoreDSL}
\showURL{%
\tempurl}


\bibitem[Team(2008)]%
        {Expression2008Source}
\bibfield{author}{\bibinfo{person}{Expression Team}.}
  \bibinfo{year}{2008}\natexlab{}.
\newblock \bibinfo{booktitle}{\emph{Expression ADL Toolkit}}.
\newblock
\urldef\tempurl%
\url{https://github.com/EPIC-CSU/expression}
\showURL{%
\tempurl}


\bibitem[Team(2026b)]%
        {OpenVADL2026Source}
\bibfield{author}{\bibinfo{person}{OpenVADL Team}.}
  \bibinfo{year}{2026}\natexlab{b}.
\newblock \bibinfo{booktitle}{\emph{OpenVADL}}.
\newblock
\urldef\tempurl%
\url{https://github.com/OpenVADL/openvadl}
\showURL{%
\tempurl}


\bibitem[Tomasulo(1967)]%
        {tomasulo1967}
\bibfield{author}{\bibinfo{person}{R.~M. Tomasulo}.}
  \bibinfo{year}{1967}\natexlab{}.
\newblock \showarticletitle{An Efficient Algorithm for Exploiting Multiple
  Arithmetic Units}.
\newblock \bibinfo{journal}{\emph{IBM Journal of Research and Development}}
  \bibinfo{volume}{11}, \bibinfo{number}{1} (\bibinfo{year}{1967}),
  \bibinfo{pages}{25--33}.
\newblock
\href{https://doi.org/10.1147/rd.111.0025}{doi:\nolinkurl{10.1147/rd.111.0025}}


\bibitem[Van~Kempen et~al\mbox{.}(2024)]%
        {VanKempen2024Seal5}
\bibfield{author}{\bibinfo{person}{Philipp Van~Kempen}, \bibinfo{person}{Mathis
  Salmen}, \bibinfo{person}{Daniel M\"{u}ller-Gritschneder}, {and}
  \bibinfo{person}{Ulf Schlichtmann}.} \bibinfo{year}{2024}\natexlab{}.
\newblock \showarticletitle{Seal5: Semi-Automated {LLVM} Support for {RISC-V}
  {ISA} Extensions Including Autovectorization}. In
  \bibinfo{booktitle}{\emph{2024 27th Euromicro Conference on Digital System
  Design (DSD)}}. \bibinfo{pages}{335--342}.
\newblock
\href{https://doi.org/10.1109/DSD64264.2024.00052}{doi:\nolinkurl{10.1109/DSD64264.2024.00052}}


\bibitem[Wagstaff(2015)]%
        {wagstaff2015high}
\bibfield{author}{\bibinfo{person}{Harry Wagstaff}.}
  \bibinfo{year}{2015}\natexlab{}.
\newblock \emph{\bibinfo{title}{From High Level Architecture Descriptions to
  Fast Instruction Set Simulators}}.
\newblock \bibinfo{thesistype}{Ph.\,D. Dissertation}. \bibinfo{school}{The
  University of Edinburgh}.
\newblock
\urldef\tempurl%
\url{http://hdl.handle.net/1842/14162}
\showURL{%
\tempurl}


\bibitem[Wang et~al\mbox{.}(2001)]%
        {Wang2001Hardware}
\bibfield{author}{\bibinfo{person}{Albert Wang}, \bibinfo{person}{Earl
  Killian}, \bibinfo{person}{Dror Maydan}, {and} \bibinfo{person}{Chris
  Rowen}.} \bibinfo{year}{2001}\natexlab{}.
\newblock \showarticletitle{Hardware/software instruction set configurability
  for system-on-chip processors}. In \bibinfo{booktitle}{\emph{Proceedings of
  the 38th Annual Design Automation Conference}} (Las Vegas, Nevada, USA)
  \emph{(\bibinfo{series}{DAC '01})}. \bibinfo{publisher}{ACM},
  \bibinfo{address}{New York, NY, USA}, \bibinfo{pages}{184–188}.
\newblock
\href{https://doi.org/10.1145/378239.378460}{doi:\nolinkurl{10.1145/378239.378460}}


\bibitem[Wegner(1972)]%
        {wegnerVDL72}
\bibfield{author}{\bibinfo{person}{Peter Wegner}.}
  \bibinfo{year}{1972}\natexlab{}.
\newblock \showarticletitle{{The Vienna Definition Language}}.
\newblock \bibinfo{journal}{\emph{Comput. Surveys}} \bibinfo{volume}{4},
  \bibinfo{number}{1} (\bibinfo{year}{1972}), \bibinfo{pages}{5–63}.
\newblock
\showISSN{0360-0300}
\href{https://doi.org/10.1145/356596.356598}{doi:\nolinkurl{10.1145/356596.356598}}


\bibitem[Wolf and Glaser(2013)]%
        {wolf2013yosys}
\bibfield{author}{\bibinfo{person}{Clifford Wolf} {and} \bibinfo{person}{Johann
  Glaser}.} \bibinfo{year}{2013}\natexlab{}.
\newblock \showarticletitle{Yosys-a free {Verilog} synthesis suite}. In
  \bibinfo{booktitle}{\emph{Proceedings of the 21st Austrian Workshop on
  Microelectronics (Austrochip)}}. \bibinfo{pages}{6}.
\newblock
\urldef\tempurl%
\url{https://yosyshq.net/yosys/files/yosys-austrochip2013.pdf}
\showURL{%
\tempurl}


\bibitem[W{\"o}{\ss} et~al\mbox{.}(2003)]%
        {Woess2003CocoR}
\bibfield{author}{\bibinfo{person}{Albrecht W{\"o}{\ss}},
  \bibinfo{person}{Markus L{\"o}berbauer}, {and} \bibinfo{person}{Hanspeter
  M{\"o}ssenb{\"o}ck}.} \bibinfo{year}{2003}\natexlab{}.
\newblock \showarticletitle{{LL}(1) {C}onflict {R}esolution in a {R}ecursive
  {D}escent {C}ompiler {G}enerator}. In \bibinfo{booktitle}{\emph{Modular
  Programming Languages}}, \bibfield{editor}{\bibinfo{person}{L{\'a}szl{\'o}
  B{\"o}sz{\"o}rm{\'e}nyi} {and} \bibinfo{person}{Peter Schojer}} (Eds.).
  \bibinfo{publisher}{Springer Berlin Heidelberg}, \bibinfo{address}{Berlin,
  Heidelberg}, \bibinfo{pages}{192--201}.
\newblock
\showISBNx{978-3-540-45213-3}
\urldef\tempurl%
\url{https://ssw.jku.at/Research/Papers/Woe03/WoeLoeMoe03.pdf}
\showURL{%
\tempurl}


\bibitem[Xiao and Liu(2023)]%
        {Xiao2023ISADL}
\bibfield{author}{\bibinfo{person}{Xin Xiao} {and} \bibinfo{person}{Zhong
  Liu}.} \bibinfo{year}{2023}\natexlab{}.
\newblock \showarticletitle{{ISADL}: An Instruction Set Architecture
  Description Language for {VLIW}}. In \bibinfo{booktitle}{\emph{2023 IEEE 29th
  International Conference on Parallel and Distributed Systems (ICPADS)}}.
  \bibinfo{pages}{92--99}.
\newblock
\href{https://doi.org/10.1109/ICPADS60453.2023.00022}{doi:\nolinkurl{10.1109/ICPADS60453.2023.00022}}


\bibitem[{Yi} et~al\mbox{.}(2005)]%
        {yi2005characterizing}
\bibfield{author}{\bibinfo{person}{Joshua~J. {Yi}},
  \bibinfo{person}{Sreekumar~V. {Kodakara}}, \bibinfo{person}{Resit {Sendag}},
  \bibinfo{person}{David~J. {Lilja}}, {and} \bibinfo{person}{Douglas~M.
  {Hawkins}}.} \bibinfo{year}{2005}\natexlab{}.
\newblock \showarticletitle{Characterizing and comparing prevailing simulation
  techniques}. In \bibinfo{booktitle}{\emph{11th International Symposium on
  High-Performance Computer Architecture}}. \bibinfo{pages}{266--277}.
\newblock
\href{https://doi.org/10.1109/HPCA.2005.8}{doi:\nolinkurl{10.1109/HPCA.2005.8}}


\bibitem[{Yi} and {Lilja}(2006)]%
        {yi2006simulation}
\bibfield{author}{\bibinfo{person}{Joshua~J. {Yi}} {and}
  \bibinfo{person}{David~J. {Lilja}}.} \bibinfo{year}{2006}\natexlab{}.
\newblock \showarticletitle{Simulation of computer architectures: simulators,
  benchmarks, methodologies, and recommendations}.
\newblock \bibinfo{journal}{\emph{IEEE Trans. Comput.}} \bibinfo{volume}{55},
  \bibinfo{number}{3} (\bibinfo{year}{2006}), \bibinfo{pages}{268--280}.
\newblock
\href{https://doi.org/10.1109/TC.2006.44}{doi:\nolinkurl{10.1109/TC.2006.44}}


\bibitem[Zimmermann(1976)]%
        {zimmermann1976mimola}
\bibfield{author}{\bibinfo{person}{Gerhard Zimmermann}.}
  \bibinfo{year}{1976}\natexlab{}.
\newblock \showarticletitle{{Eine Methode zum Entwurf von Digitalrechnern mit
  der Programmiersprache M I M O L A}}. In \bibinfo{booktitle}{\emph{GI --- 6.
  Jahrestagung}}, \bibfield{editor}{\bibinfo{person}{Erich~J. Neuhold}} (Ed.).
  \bibinfo{publisher}{Springer Berlin Heidelberg}, \bibinfo{address}{Berlin,
  Heidelberg}, \bibinfo{pages}{465--478}.
\newblock
\showISBNx{978-3-642-95289-0}
\href{https://doi.org/10.1007/978-3-642-95289-0_34}{doi:\nolinkurl{10.1007/978-3-642-95289-0_34}}


\bibitem[Zivojnovic et~al\mbox{.}(1996)]%
        {zivojnovic1996lisa}
\bibfield{author}{\bibinfo{person}{Vojin Zivojnovic}, \bibinfo{person}{Stefan
  Pees}, {and} \bibinfo{person}{Heinrich Meyr}.}
  \bibinfo{year}{1996}\natexlab{}.
\newblock \showarticletitle{LISA -- machine description language and generic
  machine model for HW/SW co-design}. In \bibinfo{booktitle}{\emph{VLSI Signal
  Processing, IX}}. IEEE, \bibinfo{pages}{127--136}.
\newblock
\href{https://doi.org/10.1109/VLSISP.1996.558311}{doi:\nolinkurl{10.1109/VLSISP.1996.558311}}


\bibitem[Zottele et~al\mbox{.}(2025)]%
        {zottele2025QEMUOpenVADL}
\bibfield{author}{\bibinfo{person}{Johannes Zottele}, \bibinfo{person}{Matthias
  Raschhofer}, \bibinfo{person}{Benedikt Huber}, {and} \bibinfo{person}{Andreas
  Krall}.} \bibinfo{year}{2025}\natexlab{}.
\newblock \showarticletitle{Generation of a {QEMU} based instruction set
  simulator from a processor description in {OpenVADL}}. In
  \bibinfo{booktitle}{\emph{Embedded Computer Systems: Architectures, Modeling,
  and Simulation}} (Samos, Greece), \bibfield{editor}{\bibinfo{person}{Jarmo
  Takala}, \bibinfo{person}{Ioannis Sourdis}, {and} \bibinfo{person}{Francesco
  Regazzoni}} (Eds.). \bibinfo{publisher}{Springer Nature Switzerland},
  \bibinfo{pages}{4}.
\newblock
\urldef\tempurl%
\url{https://link.springer.com/book/9783032285393}
\showURL{%
\tempurl}


\end{thebibliography}
